\documentclass[aps,reprint,nofootinbib,nobibnotes,notitlepage,superscriptaddress,onecolumn,prd,
 amsmath,amssymb
]{revtex4-2}

\usepackage[caption=false]{subfig}
\usepackage{braket}
\usepackage{float}
\usepackage{lipsum}
\usepackage{graphicx}
\usepackage{dcolumn}
\usepackage{bm}
\usepackage{natbib}
\usepackage{hyperref}
\hypersetup{
	colorlinks = true,
    linkcolor = Red,
    urlcolor  = Red,
    citecolor = Red
}

\usepackage{microtype}
\usepackage{gensymb}

\usepackage{verbatim}
\usepackage{siunitx}
\usepackage{tabularx}
\usepackage[dvipsnames]{xcolor}
\usepackage{wasysym}

\usepackage[skip=4pt plus1pt, indent=10pt]{parskip}

\usepackage{tikz}

\def\be{\begin{equation}}
\def\ee{\end{equation}}

\usepackage{color}
\definecolor{darkgreen}{RGB}{0,120,0}

\definecolor{darkgreen}{RGB}{0,120,0}

\newcommand{\hMpc}{h\,\mathrm{Mpc}^{-1}}

\newcommand{\av}[1]{\left\langle{#1}\right\rangle} 

\newcommand{\vk}{\vec k}
\newcommand{\hk}{\hat{\vec k}}
\newcommand{\vK}{\vec K}

\newcommand{\vx}{\vec x}

\newcommand{\Si}{\mathsf{S}^{-1}}
\newcommand{\F}{\mathcal{F}}

\newcommand{\R}{\mathcal{R}}

\newcommand{\hn}{\hat{\vec n}}

\newcommand{\hK}{\hat{\vec K}}

\newcommand{\tj}[6]{\begin{pmatrix} {#1} & {#2} & {#3}\\ {#4} & {#5} & {#6}\end{pmatrix}}

\newcommand{\gnldotdot}{g_{\rm NL}^{\dot{\sigma}^4}}
\newcommand{\gnldotdel}{g_{\rm NL}^{\dot{\sigma}^2(\partial\sigma)^2}}
\newcommand{\gnldeldel}{g_{\rm NL}^{(\partial\sigma)^4}}
\newcommand{\gestdotdot}{\widehat{g}_{\rm NL}^{\dot{\sigma}^4}}
\newcommand{\gestdotdel}{\widehat{g}_{\rm NL}^{\dot{\sigma}^2(\partial\sigma)^2}}
\newcommand{\gestdeldel}{\widehat{g}_{\rm NL}^{(\partial\sigma)^4}}
\newcommand{\fnl}{f_{\rm NL}^{\rm loc}}
\newcommand{\gnl}{g_{\rm NL}^{\rm loc}}
\newcommand{\taunl}{\tau_{\rm NL}^{\rm loc}}
\newcommand{\gcon}{g_{\rm NL}^{\rm con}}
\newcommand{\gest}{\widehat{g}_{\rm NL}}
\newcommand{\tauest}{\widehat{\tau}_{\rm NL}}
\definecolor{darkgreen}{RGB}{0,120,0}
\newcommand{\resub}[1]{{#1}}

\DeclareSymbolFont{toneletters}{T1}{\familydefault}{m}{it}
\SetSymbolFont{toneletters}{bold}{T1}{\familydefault}{bx}{it}

\DeclareMathSymbol\edth{\mathord}{toneletters}{"F0}

\usepackage{adjustbox}
\usepackage{dashbox}
\usepackage[normalem]{ulem}

\def\beq{\begin{eqnarray}}
\def\eeq{\end{eqnarray}}

\let\vec\mathbf

\usepackage{empheq}

\def\mnras{\rm{MNRAS}}

\def\aap{\rm{A\&A}}

\def\apj{\rm{ApJ}}                 
\def\jcap{\rm{JCAP}}

\defcitealias{Philcox4pt1}{Paper 1}
\defcitealias{Philcox4pt2}{Paper 2}

\usepackage{xspace}
\newcommand{\paperone}{\citetalias{Philcox4pt1}\xspace}
\newcommand{\papertwo}{\citetalias{Philcox4pt2}\xspace}
\newcommand{\polyspec}{\textsc{PolySpec}\xspace}

\begin{document}

\setlength{\parskip}{0pt}

\title{\texorpdfstring{\Large Searching for Inflationary Physics with the CMB Trispectrum:\\
\large
3. Constraints from \textit{Planck}}{Searching for Inflationary Physics with the CMB Trispectrum: 3. Constraints from \textit{Planck}}}

\author{Oliver~H.\,E.~Philcox}
\email{ohep2@cantab.ac.uk}
\affiliation{Simons Society of Fellows, Simons Foundation, New York, NY 10010, USA}
\affiliation{Center for Theoretical Physics, Columbia University, New York, NY 10027, USA}
\affiliation{Department of Physics,
Stanford University, Stanford, CA 94305, USA}

\begin{abstract} 
    \noindent
    Is there new physics hidden in the four-point function of the cosmic microwave background (CMB)? We conduct a detailed analysis of the \textit{Planck} PR4 temperature and polarization trispectrum for $\ell\in[2,2048]$. Using the theoretical and computational tools developed in \paperone and \papertwo, we search for 33 template amplitudes, encoding a variety of effects from inflationary self-interactions to particle exchange. We find no evidence for primordial non-Gaussianity and set stringent constraints on both phenomenological amplitudes and couplings in the inflationary Lagrangian. Due to the use of optimal estimators and polarization data, our constraints are highly competitive. For example, we find $\sigma(g_{\rm NL}^{\rm loc})=4.8\times 10^4$ and $\tau_{\rm NL}^{\rm loc} <1500$ (95\% CL), a factor of two improvement on Effective Field Theory amplitudes, and a $43\sigma$ detection of gravitational lensing. Many templates are analyzed for the first time, such as direction-dependent trispectra and the collapsed limit of the `cosmological collider', across a range of masses and spins. We perform a variety of validation tests; whilst our results are stable, the most relevant systematics are found to be lensing bias, residual foregrounds, and mismatch between simulations and data. The techniques discussed in this series can be extended to future datasets, allowing the primordial Universe to be probed at even higher sensitivity. 
\end{abstract}

\maketitle
\setlength{\parskip}{4pt}

\section{Introduction}
\noindent What has the Cosmic Microwave Background (CMB) taught us about inflation? Since the advent of anisotropy surveys such as COBE and WMAP, we have learned a lot about what the early Universe \textit{isn't}; an understanding of what it \textit{is}, however, remains elusive. Two key examples are tensor modes and primordial non-Gaussianity. Whilst a convincing detection of either would be a strong indicator of particular inflationary phenomena, such as gravitational waves or particle interactions, current data gives only upper bounds on the relevant amplitudes, which rule out many inflationary models. A major goal for the future is to either detect such signatures or push upper bounds towards the expected minimum values (for example $\fnl\sim 1$ for multi-field inflation). To learn more about inflation, we have two options: (a) repeat the standard analyses using more precise data; (b) search for new signatures in old data. In this series, we explore the second option.

Under the simplest inflationary assumptions (single-field slow-roll with canonical reheating), both the gauge-invariant curvature perturbations and the primary CMB anisotropies are Gaussian distributed and described only by their two-point function (or power spectrum) \citep[e.g.,][]{Guth:1980zm,Linde:1981mu,1982PhLB..116..335L,Ratra:1987rm,Starobinsky:1982ee,Mukhanov:1990me}. One can search for departures from the standard paradigm using higher-point functions whose amplitudes would usually be zero \citep[e.g.,][]{Maldacena:2002vr,Bartolo:2004if,Komatsu:2010hc,Komatsu:2003iq}. This program has been very successfully implemented for the three-point function (or bispectrum), which describes the skewness of the primordial curvature distribution. Since WMAP, CMB surveys have constrained various phenomenological bispectrum models (see \citep{2006JCAP...05..004C,Creminelli:2005hu,Senatore:2009gt,Philcox:2023xxk,Komatsu:2001wu,Philcox:2024wqx,2003MNRAS.341..623S,2014A&A...571A..24P,Planck:2015zfm,Planck:2019kim,Sohn:2024xzd,Smidt:2010ra} and many other works), such as the `local', `equilateral' and `orthogonal' $f_{\rm NL}$ amplitudes, which encode multi-field inflation and self-interactions. Moreover, several works have searched for signatures suggested by high-energy inflationary physics, such as the single-field Effective Field Theory of Inflation (EFTI), which map to the equilateral and orthogonal shapes \citep[e.g.,][]{Senatore:2009gt,2006JCAP...05..004C}. More recently, the first constraints have been placed on the `cosmological collider' signatures arising from additional massive (and possibly spinning) inflationary particles \citep{Sohn:2024xzd}. Whilst no \textit{bona fide} signal has yet arisen, bounds have been set on a wide variety of models, with particularly extensive searches performed in the official \textit{Planck} non-Gaussianity papers \citep{2014A&A...571A..24P,Planck:2015zfm,Planck:2019kim}, recent tensor non-Gaussianity studies \citep{Philcox:2024wqx}, and the modal analyses using \textsc{cmb-best} \citep{Sohn:2024xzd}.

Although many three-point models remain to be fully explored (such as dissipative models, tachyonic features, equilateral colliders and beyond \citep[e.g.,][]{Salcedo:2024smn,McCulloch:2024hiz,Jazayeri:2023xcj}), it is important to ask what's next. If the non-Gaussianity sourced by inflation is strong, a cascade of higher-point functions will be generated. Examples of this include extremely massive particle production and `heavy tails', such as from ultra-slow-roll inflation \citep[e.g.,][]{Flauger:2016idt,Kim:2021ida,Martin:2012pe,Tada:2021zzj}. In this regime, the optimal statistics are strongly non-Gaussian, and many analyses are best suited to position-space searches \citep[e.g.,][]{Munchmeyer:2019wlh,Kim:2021ida,Philcox:2024jpd,Coulton:2024vot}. Here, we instead focus on perturbative non-Gaussianity, in particular that sourcing a primordial four-point function. In many scenarios one can generate a large trispectrum without a large bispectrum: examples include theories protected by symmetries such as $\mathbb{Z}_2$ and effects arising from new particles uncoupled to the scalar inflaton \citep[e.g.,][]{Senatore:2010wk,Arkani-Hamed:2015bza,2015arXiv150200635S,Bartolo:2010di,Bartolo:2013eka,Arroja:2008ga,Arroja:2009pd,Chen:2022vzh,Chen:2009bc,Meerburg:2016zdz}. As such, there is much to gain from extending our analyses from three-point to four-point.

In this series of works, we perform a detailed analysis of the CMB trispectrum. This is certainly not the first time such statistics have been utilized: the local shapes $\gnl$ and $\taunl$ have been the subject of many previous studies \citep[e.g.,][]{Okamoto:2002ik,Chen:2009bc,Lee:2020ebj,2011MNRAS.412.1993M,2015arXiv150200635S,Bartolo:2010di,Kogo:2006kh,Hu:2001fa,Chen:2006dfn,Regan:2010cn,Smidt:2010ra,Kamionkowski:2010me,Fergusson:2010gn,2011MNRAS.412.1993M,Feng:2015pva,Munshi:2010bh,Vielva:2009jz,Hikage:2012bs,Sekiguchi:2013hza,WMAP:2006bqn,Smidt:2010ra,2015arXiv150200635S,Fergusson:2010gn,Marzouk:2022utf,Kunz:2001ym}, and the seminal paper \citep{2015arXiv150200635S} introduced estimators for the EFTI trispectra, which were used in the official \textit{Planck} PR2 and PR3 analyses \citep{Planck:2015zfm,Planck:2019kim}. Despite these notable examples, the four-point function has been studied far less than the three-point function, owing the inherent difficulties in its prediction, estimation, and implementation. Our analyses differ in several ways from the former: (1) we consider a much wider variety of primordial templates, both phenomenological and physics-motivated, including novel cosmological collider templates; (2) we build quasi-optimal estimators for all statistics that allow us to maximally utilize the available data and carefully account for the effects of lensing and masking; (c) we include polarization in all studies (which was previously featured only in \citep{Marzouk:2022utf}), noting that this will dominate future non-Gaussianity constraints; (d) we use the latest full-sky data from \textit{Planck} \citep{Planck:2020olo}, featuring improved treatment of noise and systematics compared to previous previous analyses \citep{2014A&A...571A..24P,Planck:2015zfm,Planck:2019kim}. 

This paper represents the third and final part of this series: application to \textit{Planck}. In \paperone we presented a review of primordial four-point physics and introduced our suite of thirteen trispectrum estimators (some of which depend on additional parameters such as mass), which are built upon the framework introduced in \citep{Sekiguchi:2013hza,2015arXiv150200635S}. Much of this work was devoted to recasting the estimators in an efficient form: a na\"ive implementation requires $\mathcal{O}(\ell_{\rm max}^8)$ operations, but can be reduced to just $\mathcal{O}(\ell_{\rm max}^2\log\ell_{\rm max})$ with careful use of spherical harmonic transforms and various computational tricks. \papertwo implemented each estimator in the efficient and flexible \textsc{python}/\textsc{c} code \textsc{PolySpec}. This further performed extensive validation of each estimators using both Gaussian and (where possible) non-Gaussian simulations, and forecast the distinguishability of a variety of models. In this work, we apply the estimators at scale to the latest \textit{Planck} PR4 / \textsc{npipe} dataset, obtaining constraints on a wide variety of templates. We additionally perform an extensive set of consistency tests and use our constraints to bound various physical models of inflation. 

\vskip 8pt
The remainder of this paper proceeds as follows. We begin in \S\ref{sec: models} with a brief overview of the primordial models used in this work. In \S\ref{sec: methods}, we discuss the CMB dataset, simulations, and analysis pipeline, before presenting our main results, including a battery of consistency tests, in \S\ref{sec: results}. Our bounds are placed into experimental and theoretical context in \S\ref{sec: discussion} before we conclude in \S\ref{sec: conclusion}. Appendix \ref{app: templates} lists the various trispectrum templates explicitly, and we discuss contributions to the estimator noise properties in Appendix \ref{app: variance-breakdown}. Finally, we perform a scale-dependent non-Gaussian analysis of local non-Gaussianity in Appendix \ref{app: tauNL-Lmax}. The code used to perform all the analyses in this work is publicly available at \href{https://github.com/oliverphilcox/PolySpec}{GitHub.com/OliverPhilcox/PolySpec}, and the \textit{Planck} PR4 dataset can be accessed on \textsc{nersc}. 

\section{Non-Gaussianity Models}\label{sec: models}

\noindent To begin, we present a brief overview of the various types of non-Gaussianity discussed in this work. Full details of these models and their theoretical motivations can be found in \paperone, and the associated curvature trispectra, $T_\zeta(\vk_1,\vk_2,\vk_3,\vk_4) \equiv \av{\zeta(\vk_1)\zeta(\vk_2)\zeta(\vk_3)\zeta(\vk_4)}_c$ for Fourier-space curvature perturbation $\zeta(\vk)$, are listed in Appendix \ref{app: templates}. 
\begin{itemize}
    \item \textbf{Local Non-Gaussianity}: $\gnl,\,\taunl$ (\ref{eq: template-gloc}\,\&\,\ref{eq: template-tauloc}). These are the most well-studied primordial shapes \citep[e.g.][]{Komatsu:2010hc,Okamoto:2002ik,Chen:2009zp}, and are generated from local transformations of the primordial curvature. The $\gnl$ template is dominated by squeezed configurations (with momenta $k_1\ll k_2,k_3,k_4$) with a curvature trispectrum $T_\zeta \sim P_\zeta(k_1)P_\zeta(k_2)P_\zeta(k_3)$. The $\taunl$ shape peaks in the collapsed regime ($K\equiv |\vk_1+\vk_2|\ll k_1,k_3$) with $T_\zeta \sim P_\zeta(k_1)P_\zeta(k_3)P_\zeta(K)$, and satisfies the Suyama-Yamaguchi inequality $\taunl\geq (\tfrac{6}{5}\fnl)^2$ \citep{Suyama:2007bg}.
    \item \textbf{Constant Non-Gaussianity}: $g_{\rm NL}^{\rm con}$ \eqref{eq: template-con}. This is a phenomenological template introduced in \citep{Regan:2010cn,Fergusson:2010gn} with $T_\zeta \sim [P_\zeta(k_1)P_\zeta(k_2)P_\zeta(k_3)P_\zeta(k_4)]^{3/4}$. This corresponds to a constant primordial shape function, with features arising only from CMB transfer functions.
    \item \textbf{Self-Interactions and Effective Field Theory Shapes}: $\gnldotdot,\,\gnldotdel,\,\gnldeldel$ \eqref{eq: template-efti}. These templates are generated by self-interactions in single- and multi-field inflation, as obtained from the Effective Field Theory of Inflation \citep[e.g.,][]{Chen:2006nt,Senatore:2009gt,Chen:2009zp,Senatore:2010jy,Cheung:2007st,2015arXiv150200635S,Senatore:2010wk}. They peak in the equilateral configuration with $k_1\sim k_2\sim k_3\sim k_4$, and can be used to constrain a number of primordial models including DBI inflation \citep{Arroja:2009pd,Mizuno:2009mv,Langlois:2008qf,Langlois:2008wt}.
    \item \textbf{Direction-Dependent Non-Gaussianity}: $\tau_{\rm NL}^{n_1n_3n},\,\tau_{\rm NL}^{n,{\rm even}},\,\tau_{\rm NL}^{n,{\rm odd}}$ (\ref{eq: template-direc},\,\ref{eq: template-even},\,\ref{eq: template-odd}). These generalize the $\taunl$ shape by incorporating dependence on the relative angles of $\vk_1$, $\vk_3$ and $\vK$. There are three flavors: parity-even $\tau_{\rm NL}^{n,{\rm even}}$ \citep{Shiraishi:2013oqa}, parity-odd $\tau_{\rm NL}^{n,{\rm odd}}$ \citep{Shiraishi:2016mok} and generalized $\tau_{\rm NL}^{n_1n_3n}$ \citep{Philcox4pt1}; these map to several models including solid inflation and inflationary gauge fields.
    \item \textbf{Collider Non-Gaussianity}: $\tau_{\rm NL}^{\rm light}(s,\nu_s),\,\tau_{\rm NL}^{\rm heavy}(s,\mu_s)$ (\ref{eq: template-light},\,\ref{eq: template-heavy}). These templates arise from the exchange of massive spin-$s$ particles in inflation, via the `cosmological collider' picture \citep[e.g.,][]{Arkani-Hamed:2015bza,Lee:2016vti,Chen:2009zp,Flauger:2016idt}. Since the full templates are both complex and model-dependent, we focus on the generic collapsed signatures and restrict our analysis to $K\lesssim k$. There are two regimes: (a) principal series / heavy, with mass parameter $\mu_0 = \sqrt{m^2/H^2-9/4}, \mu_{s>0} = \sqrt{m^2/H^2-(s-1/2)^2}$ and (b) complementary series / intermediate and light with mass parameter $\nu_s = i\mu_s$. The templates are analogous to $\taunl$ but feature an additional factors of $(K^2/k_1k_3)^{3/2-\nu_s}$ and $(K^2/k_1k_3)^{3/2}\cos\mu_s\log(K^2/k_1k_3)$, with additional angular dependence for $s>0$.
    \item \textbf{Gravitational Lensing}: $A_{\rm lens}$. Whilst this is a late-time source of non-Gaussianity, it is an important contaminant to primordial studies \citep[e.g.,][]{Hu:2001fa,Lewis:2006fu,2015arXiv150200635S,Babich:2004gb}. This also peaks in the quasi-collapsed regime and depends on the underlying lensing potential, which we model as $C_L^{\phi\phi} = A_{\rm lens}C_L^{\phi\phi,\rm fid}$.\footnote{This parameter is sometimes known as $A^{\phi}$ or $A^{\phi\phi}$. We use $A_{\rm lens}$ in this work, following the conventions of \citep{ACT:2023dou}. Similarly, $t_{\rm ps}$ is often known as $S_4$ \citep[e.g.,][]{Planck:2013mth}.}
    \item \textbf{Unresolved Point Sources}: $t_{\rm ps}$. This is another late-time contaminant, sourced by a Poisson distribution of bright sources in the late Universe \citep[e.g.,][]{Hobson:1998sc}. This has a flat trispectrum, and constraints are dominated by small-scales. Due to their non-detection in masked \textit{Planck} bispectrum analyses \citep{Planck:2019kim,Coulton:2022wln}, we ignore both polarized and clustered point sources.
\end{itemize}
This is not an exhaustive list of primordial four-point physics. Additional models include primordial oscillations and resonant features \citep{Wang:1999vf,Chen:2008wn,Flauger:2009ab,Flauger:2010ja}, partially massless states \citep{Baumann:2017jvh}, isocurvature modes \citep{Langlois:2008wt,Bartolo:2001cw}, parity-violating couplings \citep{Cabass:2022rhr,CyrilCS,Bartolo:2020gsh,Moretti:2024fzb}, supersymmetric fermionic contributions \citep{Alexander:2019vtb}, tachyonic features \citep{McCulloch:2024hiz}, scale-dependent non-Gaussianity \citep{Byrnes:2010ft,Wang:2022eop}, colliders with broken boost symmetries \citep[e.g.,][]{Wang:2022eop,Jazayeri:2022kjy}, non-standard inflationary vacua \citep{Meerburg:2009ys,Salcedo:2024smn,Mylova:2021eld,Chen:2006nt}, compensated isocurvature modes \citep{Grin:2013uya} and beyond; these could be profitable to study in the future. \resub{Furthermore, there are additional late-time sources of non-Gaussianity such as those sourced by Sunyaev-Zel'dovich sources and cross-correlations of CMB lensing with the integrated Sachs-Wolfe (ISW) effect \citep[e.g.,][]{Hill:2018ypf,Coulton:2022wln}. Whilst these can induce significant bias in three-point analyses \citep[e.g.,][]{Planck:2019kim}, their effect is far less relevant for the four-point function, particularly when multi-frequency cleaning methods are applied. This is demonstrated explicitly for the ISW-lensing trispectrum in \citep{Philcox:2025lxt}.}

\section{Data \& Methodology}\label{sec: methods}
\subsection{Dataset}\label{subsec: method-data}
\noindent In this work, we apply the trispectrum estimators developed in \paperone to observational data from \textit{Planck}. In particular, we utilize the latest (PR4) temperature and polarization measurements, processed using the \textsc{npipe} pipeline described in \citep{Planck:2020olo} (building on previous non-Gaussianity analyses \citep{PhilcoxCMB,Philcox:2023ypl,Philcox:2023xxk,Philcox:2024wqx,Marzouk:2022utf}). Whilst the raw data underlying this release is similar to that of its predecessor (PR3), \textsc{npipe} represents many enhancements in data processing, particularly with regards to large-scale noise and polarization, allowing a greater range of scales to be safely modeled. Furthermore, PR4 provides a maximally-independent split of the data into \textsc{npipe-a} and \textsc{npipe-b}, which allows for validation tests and estimation of the experimental noise properties. 

To separate primordial and late-time signals, we employ two component-separation algorithms: \textsc{sevem} and \textsc{smica} \citep[e.g.][]{Planck:2015mis,Planck:2018yye}. For \textsc{sevem}, we use the maps provided in the public PR4 release \citep{Planck:2020olo}; for \textsc{smica}, we perform the component-separation ourselves using the \textsc{npipe} frequency maps and the PR3 \textsc{smica} weights, following \citep{Carron:2022eyg}. As part of this pipeline, we use a diffusive inpainting scheme to excise the PR3 analysis mask and any $\text{SNR}>5$ point sources, and combine two temperature reconstructions, as in \citep{Planck:2018yye}.\footnote{The PR4 release also includes \textsc{commander} component-separated maps. We do not analyze these since the associated simulations do not contain joint temperature and polarization realizations.} In all cases, the processed datasets are \textsc{healpix} $T,Q,U$ maps at resolution $N_{\rm side}=2048$ \citep{Gorski:2004by}, which we downgrade to $N_{\rm side}=1024$ (shown to be sufficient in \papertwo). Whilst we will present results using both \textsc{smica} and \textsc{sevem}, most consistency tests are performed only using \textsc{sevem} to limit computational costs.

We require a suite of simulations to remove the Gaussian contributions to the trispectrum estimators and to validate our pipeline. For these purposes we use up to $200$ FFP10/\textsc{npipe} simulations (hereafter denoted FFP10) \citep{Planck:2015txa,Planck:2020olo}, divided into two sets of $100$.
These include most features relevant to CMB analyses, including spatially-varying noise (incorporating the noise alignment corrections described in \citep{Planck:2020olo}), residual foregrounds, the experimental beam, and the solar dipole, and are known to reproduce \textit{Planck} statistics fairly faithfully. As above, we employ two component separation algorithms, generating the \textsc{smica} simulations from the FFP10 frequency maps.

As noted in \citep{Carron:2022eyg,Marzouk:2022utf} and previous works, the FFP10 simulations exhibit a $\sim 3\%$ power deficit in temperature on small-scales. If uncorrected, this can severely bias the trispectrum estimators (cf.\,\S\ref{subsubsec: results-direc-sys}), since we require precise subtraction of the Gaussian contributions. To ameliorate this, we correct each simulation by adding a Gaussian field whose power spectrum is equal to the difference of \textit{Planck} and the mean of FFP10 (obtained using a Wiener filter), following the lensing and $\tau_{\rm NL}$ analyses of \citep{Carron:2022eyg} and \citep{Marzouk:2022utf} respectively.
\footnote{When analyzing the split \textsc{npipe-a}/\textsc{npipe-b} maps, we add the same contribution to both maps, since the deficit appears also the cross-spectrum.} This ensures that the power spectrum of the data and simulations match to high accuracy. 

\subsection{Analysis Components}\label{subsec: method-components}
\noindent Throughout this work, we assume the following model for the pixel-space observational dataset, $d$ (in vector form):
\beq\label{eq: data-model}
    d = \mathsf{P}a + n
\eeq
where $a$ is the underlying CMB realization (in harmonic-space), $n$ is a mean-zero noise component and $\mathsf{P}$ is the `pointing matrix'. As in previous works \citep[e.g.,][]{1999ApJ...510..551O,Smith:2007rg}, we assume $\mathsf{P}$ contains (a) an observational beam, (b) spherical harmonic synthesis and (c) a pixel-space mask. Explicitly, for harmonic indices $\ell,m$, spins $s$, and pixels $\hn$:
\beq
    {}_s[\mathsf{P}a](\hn) = {}_sW(\hn)\sum_{\ell mX}{}_sY_{\ell m}(\hn){}{}_s\R^XB_\ell^X a^X_{\ell m},
\eeq
where ${}_sW(\hn)$ is the (spin-dependent) mask, $B_\ell^X$ is the isotropic beam and the matrix ${}_s\R^X$ converts between spin and harmonic representations. Each of these ingredients is necessary to relate the observational data to the primordial templates given in Appendix \ref{app: templates}, and will be discussed below.

To remove contamination from point-sources and residual foregrounds, we use a mask formed by multiply the \textit{Planck} common component-separation mask with the common inpainting mask and the \textsc{gal70} mask (to suppress any residual foregrounds). We remove any holes containing $<1000$ pixels (at $N_{\rm side}=1024$) from the mask and smooth the residual using a $2\degree$ cosine apodization filter; this forms the fiducial ${}_sW(\hn)$. The removed pixels define an inpainting mask, which is utilized in the weighting operation.\footnote{Alternatively, we could retain the small holes in the mask, ensuring smoothness by enlarging the holes by two pixels and then applying a $0.3\degree$ cosine apodization. This mask is used for consistency tests in \S\ref{subsubsec: results-direc-sys}.} In total, we find a sky fraction of $f_{\rm sky} = 68.4\%$ and $68.2\%$ in temperature and polarization respectively. To test the dependence on residual foregrounds, we also create analysis masks using the \textsc{gal20}, \textsc{gal40}, \textsc{gal60}, and \textsc{gal80} regions (with the latter almost entirely subsumed by the component-separation mask). In the below, we will assume that all datasets have been multiplied by the mask (following \eqref{eq: data-model}). 

To model $B_\ell^X$, we use the effective isotropic beams provided for \textsc{sevem} and \textsc{smica}, additionally multiplying by the relevant \textsc{healpix} window function. These are approximately $5'$ Gaussian beams, but include additional features from the combination of multiple frequency channels with different resolutions. In polarization, we multiply the beam by a large-scale transfer function (at $\ell\leq 40$) to correct for the \textsc{npipe} calibration operations. This is measured by cross-correlating \textsc{npipe} simulations with the input CMB realizations, as described in \citep{Planck:2020olo}.

We additionally require a fiducial model for the two-point function, both to generate Gaussian simulations with known covariance (required to compute the estimator normalization), and as part of the weighting scheme (cf.\,\S\ref{subsec: method-weighting}). In harmonic-space, this is defined by
\beq\label{eq: cl-fid}
    C^{{\rm fid},XY}_\ell = B_\ell^X B_\ell^Y C_\ell^{XY} + \delta_{\rm K}^{XY}N_\ell^X,
\eeq
for $X\in\{T,E,B\}$,\footnote{This can be obtained from \eqref{eq: data-model} assuming unit mask and isotropic noise.} where $C_\ell^{XY}$ is a fiducial CMB power spectrum (\textit{i.e.}\ $\av{a_{\ell m}^Xa_{\ell m}^{Y*}}$) and $N_\ell^X$ is a translation-invariant noise model. The former is generated with \textsc{camb}, assuming the \textit{Planck} PR3 fiducial cosmology (which is similar to the FFP10 cosmology): $\{h = 0.6732, \omega_b = 0.02238, \omega_c = 0.1201, \tau_{\rm reio} = 0.05431, n_s = 0.9660, A_s =2.101\times 10^{-9}, \sum m_\nu = 0.06\,\mathrm{eV}\}$, assuming a single massive neutrino. This cosmology is also used to compute the primordial curvature power spectrum and the CMB transfer functions. The noise spectrum is estimated from the masked half-mission maps (employing the relevant component-separation method). 
Whilst this does not capture the spatial variation of the noise, it is a sufficient approximation, \resub{given that this spectrum} is not used to subtract \resub{disconnected contributions}.

\subsection{Weighting}\label{subsec: method-weighting}
\noindent An important ingredient in any CMB estimator is the weighting scheme applied to the data (here labeled by $\Si$). This sets the variance of the estimator and can be used to upweight low-noise regions of the sky as well as null those contaminated by foregrounds. Given the data-model of \eqref{eq: data-model}, and assuming negligible non-Gaussianity, the optimal weighting scheme is given by
\beq\label{eq: opt-weighting}
    \Si_{\rm opt}d \equiv \mathsf{P}^\dagger\mathsf{C}_{\rm tot}^{-1} d = \mathsf{P}^\dagger\left[\mathsf{P}\mathbb{C}\mathsf{P}^\dagger + \mathsf{N}\right]^{-1}
\eeq
\citep[e.g.,][]{Oh:1998sr,Philcox4pt1}, where $\mathsf{C}_{\rm tot}\equiv \av{dd^\dagger}$ is the full two-point covariance of $d$, comprising signal and noise contributions $\mathbb{C}\equiv\av{aa^\dagger}$ and $\mathsf{N}\equiv\av{nn^\dagger}$. This provides an optimal signal-plus-noise weighting and additionally deconvolves the experimental beam. In practice, the optimal weighting $\Si_{\rm opt}$ is difficult to implement, since the requisite matrices are extremely high-dimensional ($\mathcal{O}(10^{14})$ elements) and the covariance is dense and both harmonic- and pixel-space. 

In this paper, we will adopt two alternative weighting schemes to avoid full computation and inversion of $\mathsf{C}_{\rm tot}$. Firstly, we use \textit{idealized weights}, which are given explicitly by
\beq\label{eq: ideal-weighting}
    [\Si_{\rm ideal} d]_{\ell m}^X = B_\ell^X\sum_{Y}\left(C^{\rm fid}_\ell\right)^{-1,XY}\int d\hn\,{}_s\R^{\dagger,Y}{}_sY_{\ell m}^*(\hn)[\mathbb{I}d](\hn),
\eeq
where we (a) inpaint small holes via a linear diffusive inpainting operation (represented by $\mathbb{I}$), (b) perform spherical harmonic analysis, (c) divide by the fiducial power spectrum (including the noise and beam, cf.\,\eqref{eq: cl-fid}) and (d) apply the beam. This is closely related to the approach of \citep{2015arXiv150200635S}, also used in the trispectrum analyses of \citep{Planck:2015zfm,Planck:2019kim} (and is similar to that used in \citep{Philcox:2024wqx,PhilcoxCMB,Philcox:2023ypl,Philcox:2023xxk,Philcox:2024wqx}). This can be easily implemented using spherical harmonic transforms and is the default choice used in this work.

Secondly, we solve \eqref{eq: opt-weighting} approximately using conjugate gradient descent techniques \citep[e.g.,][]{Oh:1998sr} (see also \citep{Munchmeyer:2019kng,Costanza:2023cgt,Costanza:2024rut} for machine-learning based approaches). This is facilitated by rewriting \eqref{eq: opt-weighting}:
\beq\label{eq: cgd-weighting}
    \Si_{\rm opt}d \equiv \mathbb{C}^{-1}\left[\mathbb{C}^{-1}+\mathsf{P}^{\dagger}\mathsf{N}^{-1}\mathsf{P}\right]^{-1}\mathsf{P}^\dagger\mathsf{N}^{-1}d = \mathbb{C}^{-1}d_{\rm WF},
\eeq
where we have noted equivalence with the Wiener-filtered data, $d_{\rm WF}$.\footnote{This reduces to \eqref{eq: ideal-weighting} in the limit of translation invariant noise and a unit mask.} Given an explicit form for $\mathsf{N}^{-1}$, this can be solved by iteratively applying $\left[\mathbb{C}^{-1}+\mathsf{P}^{\dagger}\mathsf{N}^{-1}\mathsf{P}\right]$ to a given map, preconditioning via the idealized harmonic weighting scheme. Following previous works, we assume that $\mathsf{N}^{-1}$ is diagonal in pixel-space, and set by the inverse variance of the noise maps (obtained empirically from the processed simulations), or zero for masked pixels. We do not include mask apodization in $\mathsf{N}^{-1}$ to avoid upweighting pixels close to the boundary. The conjugate gradient descent is implemented using code derived from the \textsc{pixell} and \textsc{optweight} implementations,\footnote{Available at \href{https://github.com/simonsobs/pixell}{GitHub.com/SimonsObs/pixell} and \href{https://github.com/AdriJD/optweight}{GitHub.com/AdriJD/optweight}.} using $N_{\rm cgd}=50$ iterations, aiming for a convergence threshold of $10^{-5}$. 
Whilst this may yield somewhat improved constraints relative to the idealized weighting \eqref{eq: ideal-weighting}, this is still approximate since we have ignored any scale-dependence of the noise. 
Implementing more optimal weighting schemes is an interesting topic for future work.

\subsection{Estimator}\label{subsec: method-estimator}
\noindent To estimate the trispectrum amplitudes, we utilize the \href{https://github.com/oliverphilcox/PolySpec}{\textsc{PolySpec}} code described in \papertwo, which implements the quasi-optimal estimators described in \paperone.
This replaces the previous \textsc{PolyBin} code (see also \citep{Philcox:2024rqr} for 3D implementations), which was used to compute binned power spectra, bispectra, and trispectra in \citep{Philcox:2023uwe,Philcox:2023psd,PolyBin}. The code is written in a combination of \textsc{python}
and \textsc{c} (via \textsc{cython}) and makes extensive use of the \textsc{ducc} (based on \textsc{libsharp} \citep{Reinecke_2013}) to perform spherical harmonic transforms, as well as \textsc{healpy}/\textsc{healpix} package \citep{Gorski:2004by,Zonca:2019vzt} for data handling. In this work, we use \polyspec to compute the amplitudes of various primoridal trispectra, whose full forms are given in Appendix \ref{app: templates}. Whilst we refer the interested reader to \papertwo for a full description of the code, its usage, optimization and its validation, we highlight a few relevant parts below. The fiducial values of the hyperparameters described below are validated both in \papertwo and \S\ref{sec: results}.

The optimal estimators have two main components: the data-dependent numerator and the data-independent normalization (also known as the Fisher matrix, $\F$). Given a dataset $d$ and a set of simulations $\{\delta\}$ (whose two-point function matches that of the data) the numerator is given by
\beq\label{eq: numerator}
    \widehat{N}_\alpha &=& \widehat{\mathcal{N}}_\alpha[d,d,d,d] - \left(\av{\widehat{\mathcal{N}}_\alpha[d,d,\delta,\delta]}_\delta+\text{5 perms.}\right) + \left(\av{\widehat{\mathcal{N}}_\alpha[\delta,\delta,\delta',\delta']}_{\delta,\delta'}+\text{2 perms.}\right)\\\nonumber
    \widehat{\mathcal{N}}_\alpha[x,y,z,w] &=& \frac{1}{24}\sum_{\ell_im_iX_i}\frac{\partial T^{\ell_1\ell_2\ell_3\ell_4,X_1X_2X_3X_4}_{m_1m_2m_3m_4}}{\partial A_\alpha}\bigg([\Si x]^{X_1*}_{\ell_1m_1}[\Si y]^{X_2*}_{\ell_2m_2}[\Si z]^{X_3*}_{\ell_3m_3}[\Si w]^{X_4*}_{\ell_4m_4}\bigg)
\eeq
for some set of templates whose amplitudes $\{A_\alpha\}$ define the harmonic-space trispectrum $T$. The estimator combines four copies of the data, 
subtracting the disconnected contributions (\textit{i.e.}\ those arising from the $\av{d^2}\av{d^2}$ contributions to $\av{d^4}$). 
The trispectrum model depends on the beam and the CMB transfer functions; when estimating $A_{\rm lens}$, we additionally require the lensing potential and lensed CMB power spectra. Furthermore, we typically limit the analysis to some set of scales (\textit{i.e.}\ $\ell_{\rm min}\leq \ell\leq \ell_{\rm max}$) by setting the trispectrum template to zero outside these regimes.

For all the templates considered in this work, the numerator can be efficiently computed using spherical harmonic transforms and low-dimensional radial integrals (following a preprocessing step, which evaluates transfer function integrals using an array of $\approx 2\times 10^4$ sampling points in $k$). 
To evaluate the averages over simulations, we use $N_{\rm disc} = 100$ simulations (or $N_{\rm disc}=50$ when running the more expensive analyses on simulated data), noting that this induces an error scaling as $\sqrt{1+\kappa /N_{\rm disc}}$ where $\kappa$ is template-dependent but small. The radial integrals are computed as summations over a small number ($\sim 10-50$) of integration points, themselves obtained via an optimization algorithm (described in detail in \paperone, building on \citep{2011MNRAS.417....2S,2015arXiv150200635S}). This algorithm is run for each template and choice of scale cuts, and searches for a representation of the integral that recovers the Fisher matrix within $f_{\rm thresh} = 10^{-3}$. For the more expensive analyses (beyond $\gnl$, $\gcon$, and $\taunl$), we use a tighter tolerance ($f_{\rm thresh} = 10^{-4}$) but split the computation into four chunks for efficiency, as described in \citep{Philcox4pt1,2015arXiv150200635S}. These choices are validated in \S\ref{subsec: local-variants}.

The Fisher matrix, $\F$, is given explicitly by
\beq\label{eq: fisher}
    \F^*_{\alpha\beta} &=& \frac{1}{24}\sum_{\ell_im_iX_i\ell_i'm_i'X_i'}\frac{\partial T^{\ell_1\ell_2\ell_3\ell_4,X_1X_2X_3X_4*}_{m_1m_2m_3m_4}}{\partial A_\alpha}\left(\prod_{i=1}^4[\Si\mathsf{P}]_{\ell_im_i,\ell_i'm_i'}^{X_iX_i'}\right)\frac{\partial T^{\ell_1'\ell_2'\ell_3'\ell_4',X_1'X_2'X_3'X_4'}_{m_1'm_2'm_3'm_4'}}{\partial A_\beta},
\eeq
where $\mathsf{P}$ is the pointing matrix introduced in \eqref{eq: data-model}, encoding the 
beam and mask.\footnote{This is easily derived by inserting \eqref{eq: data-model} into \eqref{eq: numerator} and asserting that the $\widehat{A}_\alpha = [\F^{-1}\widehat{N}]_\alpha$ estimator is unbiased.} This (a) normalizes each estimator (fully accounting for the beam, mask, and weighting scheme), (b) accounts for correlations and leakage between templates, and (c) approximates the inverse covariance of the estimator (which is exact for optimally filtered Gaussian data, including with a mask). If $\Si\mathsf{P} = \mathsf{P}^\dag\mathsf{S}^{-\dag}$ (true for optimal weights), the matrix is symmetric. As shown in \paperone, the Fisher matrix can be efficiently computed using Monte Carlo techniques and spherical harmonic transforms, involving a sum over $N_{\rm fish}$ Gaussian random fields. Here, we usually adopt $N_{\rm fish}=20$ (or $N_{\rm fish}=10$ for the more complex templates), which induces negligible bias in practice \citep{Philcox4pt2}.

As argued in \paperone and demonstrated in \papertwo, the full estimator, $\widehat{A}_\alpha \equiv [\F^{-1}\widehat{N}]_\alpha$ is (a) free from noise-induced `false positives' provided that the covariance of the simulations matches that of the data (with a second-order error else), (b) unbiased for any choice of weighting scheme, (c) minimum-variance in the limit of optimal $\Si$. Moreover, it is closely related to the standard quadratic estimators used for measuring CMB lensing, and naturally includes recent methodological enhancements such as realization-dependent debiasing and optimal combination of temperature and polarization modes \citep[e.g.,][]{Namikawa:2012pe,Carron:2022edh,Maniyar:2021msb}. 

In practice, all analyses are performed using a high-performance computing cluster. Assuming the default scale-cuts $\ell_{\rm min}=2$ and $\ell_{\rm max}=2048$, computation of the local and lensing numerators required $90$ node-seconds per simulation (averaging across $100$ realizations, and storing the disconnected data-products to avoid recomputation). Similarly, the \polyspec optimization step required $20$ node-minutes (though only has to be run once), and computation of the Fisher matrix took $190$ node-seconds for each of $20$ Monte Carlo realizations. As shown in \papertwo, the run-times increase by up to $\sim 10\times$ when more complex templates, such as the higher-spin cosmological collider, are analyzed. 
Whilst great care has gone into ensuring that the \textsc{PolySpec} code is as efficient as possible, the combination of many validation checks, many data consistency tests, \sout{many mistakes,} and many templates has required a large computational budget -- the full project required a few $\times 10^5$ CPU-hours, equivalent to around one node-year.


\section{Results}\label{sec: results}
\noindent Next, we present the main results of our study: constraints on various forms of primordial non-Gaussianity from \textit{Planck} PR4 temperature and polarization dataset. 
Given that the local and lensing shapes are the most well-studied and the fastest to analyze, we pay closest attention to their analysis and perform a barrage of consistency tests to check for systematic contamination (most of which generalize to the other templates). Other shapes are more expensive to analyze, thus we consider fewer analysis variants, guided by the previous results. In the below, the `baseline' constraints include both temperature and polarization, $\ell\in[2,2048]$, the $68\%$ mask discussed in \S\ref{subsec: method-components}, and the idealized weighting scheme \eqref{eq: ideal-weighting}, which includes point source inpainting. By default, we use $N_{\rm disc}=100$ and $N_{\rm fish}=20$, 
and optimize the radial integrals with a tolerance of $f_{\rm thresh}=10^{-3}$.

We perform two types of analysis in this work. First, we analyze each template independently (fixing all others to zero); these measurements can be obtained from the vector of numerators $\widehat{N}$ and Fisher matrix $\F$ via $\widehat{A}_\alpha^{\rm single} = \widehat{N}_\alpha/\F_{\alpha\alpha}$. 
Second, we perform a joint analysis of multiple templates, setting $\widehat{A}_\alpha^{\rm joint} = \sum_\beta\F_{\alpha\beta}^{-1}\widehat{N}_\beta$. 
To account for biases induced by gravitational lensing, we can either perform a joint analysis of $A_\alpha$ with $A_{\rm lens}$ (which automatically accounts for their correlations), or subtract the expected bias from the estimator via $\Delta\widehat{A}_\alpha = \sum_\beta\F^{-1}_{\alpha\beta}\F_{\beta A_{\rm lens}}A^{\rm fid}_{\rm lens}$,\footnote{This is equivalent to performing a joint analysis with a tight prior around $A_{\rm lens} = A_{\rm lens}^{\rm fid}$.} where the $\F_{\alpha\beta}$ matrix does not include lensing. Given that the fiducial lensing signal is well-known and well-recovered, we will generally assume the second option.

Throughout this section, we report measurements in the form $\widehat{A}_\alpha\pm\sigma(\widehat{A}_\alpha)$, where $\sigma(\widehat{A}_\alpha)$ is the standard deviation measured from FFP10 simulations. Assuming a Gaussian sampling distribution for $\widehat{A}_\alpha$, these measurements can be interpreted as a $68\%$ confidence interval on $A_\alpha$, \textit{i.e.}\ $\widehat{A}_\alpha-\sigma(\widehat{A}_\alpha)\leq A_\alpha\leq \widehat{A}_\alpha+\sigma(\widehat{A}_\alpha)$.\footnote{For $\taunl$, we employ a different approach; this is discussed in \S\ref{subsec: results-local} and Appendix \ref{app: tauNL-Lmax}.} To aid comparison between the various results, we will scale all parameters by the same factor (e.g., reporting constraints on $10^{-2}\taunl$, regardless of the magnitude of $\taunl$).

\subsection{Local \& Lensing Non-Gaussianity}\label{subsec: results-local}

\noindent First, we analyze the local templates,  $\gnl$ and $\taunl$, in concert with two contaminants: lensing and unresolved point-sources. We use all $\ell$-modes with $\ell\in[2,2048]$ (given the increased robustness of the \textsc{npipe} release compared to previous versions), but restrict the $\taunl$ estimator to internal $L$-modes $L\in[1,30]$, noting that the signal-to-noise should be saturated by $L= 10$ \citep[e.g.,][]{Kogo:2006kh,Kalaja:2020mkq,Philcox4pt2}. As discussed in \citep{2014A&A...571A..24P}, the unprocessed $L=1$ modes are contaminated by the solar dipole: given that this effect is accounted for in the \textsc{npipe} pipeline and is also present in simulations \citep{Marzouk:2022utf}, we do not excise this mode from our analyses (though we test its removal in \S\ref{subsubsec: results-direc-sys}). For lensing, we include all modes with $L\in[2,2048]$, noting that the signal-to-noise peaks around $L\sim 100$ \citep{Carron:2022eyg}.

\newcommand{\res}[3]{${#1}$ & $\pm$ & ${#2}$ & (${#3}$)}
\begin{table}[!t]
\begin{tabular}{l||rclc|rclc|rclc}
    \textbf{Template} & \multicolumn{4}{c|}{$T$} & \multicolumn{4}{c|}{$E,B$} & \multicolumn{4}{c}{$T,E,B$}\\\hline
    $10^{-4}\gnl$ & \res{1.9}{6.6}{6.4} & \res{0}{33}{31} & \res{-1.1}{5.4}{4.5}\\
    $10^{-2}\taunl$ & \res{-0.3}{2.8}{2.7} & \res{2}{22}{22} & \res{-0.3}{2.4}{2.5}\\
    $A_{\rm lens}$ & \res{1.004}{0.028}{0.020} & \res{0.99}{0.14}{0.10} & \res{0.998}{0.024}{0.014}\\
    $10^{38}t_{\rm ps}$\,\,[$\mathrm{K}^4$] & \res{0.4}{6.3}{5.7} & \multicolumn{4}{c|}{--} & \res{0.3}{6.2}{5.6} 
\end{tabular}
\caption{\resub{Validation of the local estimators using FFP10 \textsc{sevem} simulations (which do not include primordial non-Gaussianity).} We show the mean and standard deviation of independently-analyzed trispectrum amplitudes across $100$ realizations, as well as the expected errors from an optimally-weighted Gaussian analysis (in parentheses). The columns show results obtained using different combinations of temperature and polarization (omitting the $E,B$ measurement of $t_{\rm ps}$, since we assume point-sources are unpolarized). All results are consistent with the fiducial values of $\gnl=\taunl=t_{\rm ps}=0$ and $A_{\rm lens}=1$, with largest deviations seen for $\gnl$ due to the finite number of simulations in the estimator numerator. The errorbars are comparable to the theoretical predictions (given the expected $7\%$ scatter), though the lensing analyses are suboptimal, due to the non-zero fiducial $A_{\rm lens}$. Analogous tests have been performed for every analysis considered in this work; for brevity we do not list them in this manuscript. 
}\label{tab: local-validation}
\end{table}

\vskip 8pt
\subsubsection{Tests on Simulations}
\noindent Before presenting the \textit{Planck} constraints, we validate the analysis using $100$ FFP10 \textsc{sevem} simulations. This tests (a) if the pipeline is biased by non-Gaussian noise and residual foregrounds present in the mocks, and (b) whether the estimator is close to optimal. In Tab.\,\ref{tab: local-validation} we show the mean and variance of $\widehat{g}_{\rm NL}^{\rm loc}$, $\widehat{\tau}_{\rm NL}^{\rm loc}$, $\widehat{A}_{\rm lens}$ and $\widehat{t}_{\rm ps}$ alongside the Fisher errors ($1/\sqrt{\F_{\alpha\alpha}}$) for three types of analysis: temperature-only, polarization-only and joint. In all cases, the mean and fiducial values agree to within $0.3\sigma$, indicating that our pipeline is unbiased. 
As demonstrated in \S\ref{subsubsec: results-direc-sys}, residual variations in the mean arise from the finite number of numerator simulations ($N_{\rm disc}=100$, common to all analyses), whose impact varies depending on the template.

We find fairly good agreement between the empirical and theoretical errorbars shown in Tab.\,\ref{tab: local-validation} for the primordial and point-source estimators (given the expected variation due to noise). For lensing, we find larger deviations, with a strong excess found in the temperature-plus-polarization analysis. This occurs due to the non-zero fiducial $A_{\rm lens}$ -- the Fisher matrix errors assume negligible non-Gaussianity, whilst lensing sources additional contributions to the estimator covariance. In general, \resub{the constraining power is set by a complex interplay of many phenomena: lensing, beams, masks, and noise}. In Appendix \ref{app: variance-breakdown}, we unpack the contribution of each effect to the local and lensing errorbars. For $\gnl$ and $\taunl$, the dominant features are the beam (leading to a $15\%$ and $30\%$ increase in the error) and the mask ($30\%$ and $38\%$), whilst for $A_{\rm lens}$, we find an additional $30\%$ loss of signal-to-noise due to lensing.

\newcommand{\ress}[2]{${#1}$ & $\pm$ & ${#2}$}
\newcommand{\ressb}[2]{$\mathbf{#1}$ & $\mathbf{\pm}$ & $\mathbf{#2}$}
\begin{table}
    \begin{tabular}{l||rcl|rcl|rcl||rcl|rcl|rcl}
        & \multicolumn{9}{c||}{\textit{Planck}-\textsc{sevem}} & \multicolumn{9}{c}{\textit{Planck}-\textsc{smica}}\\
        \textbf{Template} & \multicolumn{3}{c|}{$T$} & \multicolumn{3}{c|}{$E,B$} & \multicolumn{3}{c||}{$T,E,B$} & \multicolumn{3}{c|}{$T$} & \multicolumn{3}{c|}{$E,B$} & \multicolumn{3}{c}{$T,E,B$}\\\hline
$10^{-4}g_{\rm NL}^{\rm loc}$ & \ress{-2.3}{6.6} & \ress{16.5}{33.2} & \ressb{1.0}{5.4} & \ress{-5.6}{6.5} & \ress{36.7}{33.2} & \ressb{0.3}{4.8}\\
$10^{-2}\tau_{\rm NL}^{\rm loc}$ & \ress{2.6}{2.8} & \ress{-13.8}{21.9} & \ressb{2.9}{2.4} & \ress{-0.3}{2.8} & \ress{-58.2}{21.9} & \ressb{-0.2}{2.4}\\
$A_{\rm lens}$ & \ress{0.965}{0.028} & \ress{0.92}{0.14} & \ressb{1.001}{0.024} & \ress{0.961}{0.028} & \ress{0.89}{0.14} & \ressb{0.979}{0.023}\\
$10^{38}t_{\rm ps}$ & \ress{32.0}{6.3} && -- && \ressb{32.1}{6.2} & \ress{18.9}{5.9} && -- && \ressb{19.1}{5.9}\\
\end{tabular}
    \caption{\resub{\textit{Planck} PR4 constraints on local and late-time non-Gaussianity.} In each case, we give the \textit{Planck} measurement and the standard deviation measured from $100$ FFP10 simulations; assuming a Gaussian likelihood, this is equivalent to a $68\%$ \resub{confidence interval}. The left (right) panels show results obtained using the \textsc{sevem} (\textsc{smica}) component-separation pipeline, and we ignore correlations between templates.
    The bold columns indicate the fiducial results of this work. We report no detection of local primordial non-Gaussianity, with the baseline $\gnl$ ($\taunl$) constraints consistent with zero at $0.2\sigma$ ($1.2\sigma$). In contrast, we detect the lensing signal at $43\sigma$ (consistent with the fiducial amplitude) and unpolarized point-sources at up to $5.2\sigma$. In Fig.\,\ref{fig: local-corner} we perform a joint $\gnl-\taunl$ analysis, and show the results of various systematics tests in Fig.\,\ref{fig: local-systematics}.}\label{tab: local-results}
\end{table}

\vskip 8pt
\subsubsection{Local Non-Gaussianity Results}

\noindent \textit{Planck} constraints on local non-Gaussianity results are presented in Tab.\,\ref{tab: local-results}. From the baseline (temperature-plus-polarization) analyses, we find:
\beq
    \{10^{-4}\gest^{\rm loc}, 10^{-2}\tauest^{\rm loc}\} &=& \{1.0\pm5.4,2.9\pm2.4\}\quad(\textsc{sevem})\quad = \,\,\,\{\resub{0.3\pm4.8},-0.2\pm2.4\}\quad(\textsc{smica}),\nonumber
\eeq
subtracting lensing bias and ignoring correlations between templates. Although the polarization-only constraints are around \resub{$5\times$} weaker than the temperature-only bounds, the combination (which was not used in previous $\gnl$ analyses) sharpens the $\gnl$ constraints by up to $25\%$, though has a more modest impact for $\taunl$ \citep[cf.][]{Kalaja:2020mkq,Marzouk:2022utf}. All of the baseline and temperature-only results are consistent with zero within $1.2\sigma$, indicating no evidence for primordial non-Gaussianity. In the \textsc{smica} polarization-only analysis, $\widehat{\tau}_{\rm NL}^{\rm loc}$ is \resub{$2.7\sigma$} below zero: this is attributed to noise given (a) the weak constraining power of polarization alone, (b) the large shifts between \textsc{smica} and \textsc{sevem}, (c) the non-Gaussianity of the $\taunl$ posterior (as discussed below), and (d) the sign, noting that any physical $\taunl$ must be positive. In general, we find fairly large differences between \textsc{sevem} and \textsc{smica} (up to $1.3\sigma$ in the temperature-only and baseline analyses); some variation is expected given the different treatment of foregrounds and effective noise properties. This is consistent with the differences seen in simulated data (up to $1.5\sigma$), and those observed in the official \textit{Planck} $f_{\rm NL}$ analyses \citep{Planck:2019kim} (up to $1\sigma$).

\begin{figure}[!t]
    \includegraphics[width=0.8\linewidth]{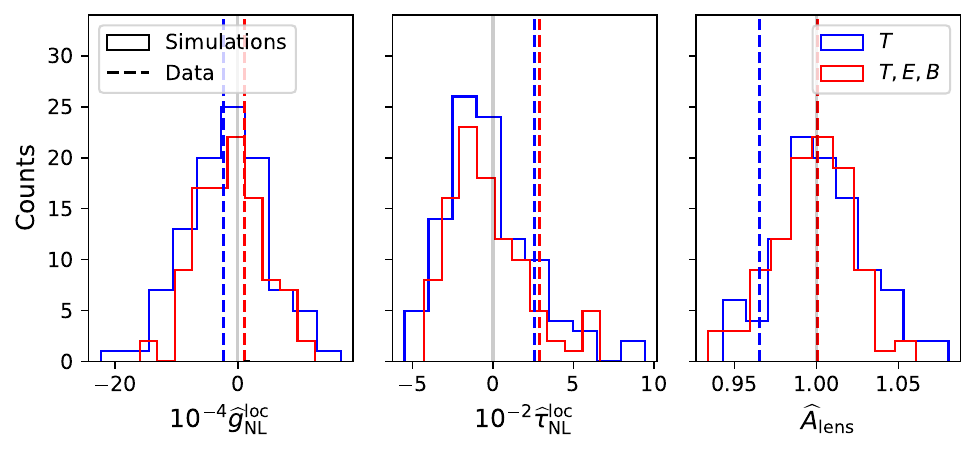}
    \caption{Comparison of the \textit{Planck} local and lensing non-Gaussianity amplitudes (dotted lines) to those from 100 FFP10 simulations (histograms) generated with $\gnl=\taunl=0$ and $A_{\rm lens}=1$. We show results from temperature-only (blue) and temperature-plus-polarization (red), adopting the \textsc{sevem} component-separation pipeline in all cases. We find good agreement of data and simulations, though note that the $\tauest^{\rm loc}$ histograms are markedly asymmetric. This is explored in Appendix \ref{app: tauNL-Lmax}, and yields a variety of $95\%$ limits on $\taunl$ with the strongest being $\taunl<1500$ from \textsc{smica} temperature and polarization.}\label{fig: local-pdf}
\end{figure}

In Fig.\,\ref{fig: local-pdf} we plot the empirical distributions of $\gest^{\rm loc}$ and $\tauest^{\rm loc}$ alongside the \textit{Planck} results. As above, we find good agreement between data and simulations: for the temperature-and-polarization dataset, the \textit{Planck} measurement of $\gnl/\taunl/A_{\rm lens}$ lie at the $65\mathrm{th}/89\mathrm{th}/ 50\mathrm{th}$ percentile of the FFP10 distribution respectively (or $67\mathrm{th}/65\mathrm{th}/38\mathrm{th}$ percentile for \textsc{smica}). The $\gest^{\rm loc}$ and $\widehat{A}_{\rm lens}$ histograms have a symmetric, Gaussian shape centered on the fiducial value -- this implies that the $68\%$ confidence interval on such parameters is equivalent to the $\widehat{A}\pm\sigma_A$ constraints quoted above. In contrast, the distribution of $\tauest^{\rm loc}$ is skewed with a clear tail to positive values. This arises since the $\taunl$ estimator involves a positive-definite statistic,\footnote{As discussed in \paperone, our estimator essentially computes the power spectrum of a large-scale modulation field from which a noise contribution is subtracted.} with contributions from individual $L$ modes following an approximate $\chi^2$ distribution \citep[e.g.,][]{Kamionkowski:2010me,Smith:2012ty,Marzouk:2022utf,2014A&A...571A..24P}. As a result, it is non-trivial to transform the the \textit{Planck} $\widehat{\tau}_{\rm NL}^{\rm loc}$ value to a $95\%$ confidence interval on $\taunl$. 

As discussed in \citep{2014A&A...571A..24P,Marzouk:2022utf}, we can obtain a bound on $\taunl$ by performing an $L$-by-$L$ reconstruction using the approximate likelihoods of \citep{Hamimeche:2008ai} (given that the sampling distribution of the compressed estimator is not well understood). This analysis is performed in Appendix \ref{app: tauNL-Lmax} and yields measurements consistent with zero at each mode with the possible exception of $L=3$ (matching \citep{2014A&A...571A..24P,Marzouk:2022utf}). We find the following constraints: 
\beq
    \taunl < 2360 \quad (\textsc{sevem}), \qquad \taunl < 1500 \quad (\textsc{smica})\nonumber
\eeq
from the baseline analysis or
\beq
    \taunl < 2460 \quad (\textsc{sevem}), \qquad \taunl < 1740 \quad (\textsc{smica})\nonumber
\eeq
from temperature alone. The \textsc{smica} constraints are significantly tighter than those of \textsc{sevem}: this occurs due to a downwards noise fluctuation in the $L=3$ mode (cf.\,Fig.\,\ref{fig: tauNL-L-dep}) and highlights the volatility of one-sided parameter constraints. 

\begin{figure}
    \includegraphics[width=0.45\linewidth]{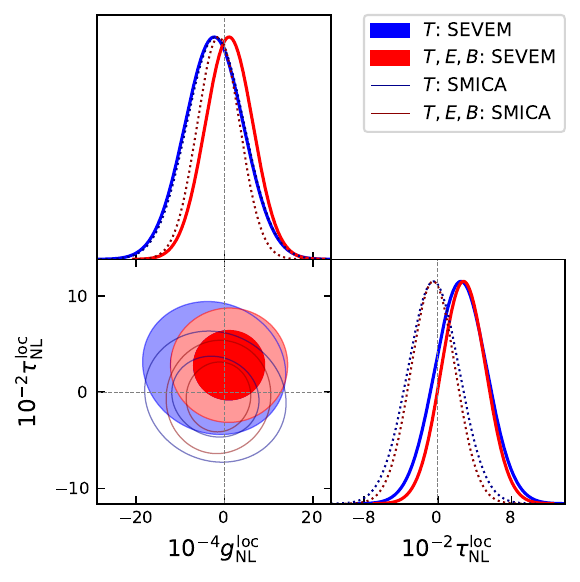}
    \caption{Joint constraints on the local non-Gaussianity parameters using \textit{Planck} PR4 data, processed using \textsc{sevem} (filled/solid) and \textsc{smica} (empty/dashed). We show results from both temperature-only (blue) and temperature-plus-polarization (red) datasets, assuming a Gaussian likelihood with covariances measured from $100$ FFP10 simulations. We include all modes with $\ell\in[2,2048]$ and $L\in[1,30]$ for $\taunl$, and \resub{subtract biases from CMB lensing}. We observe limited correlations between $\gnl$ and $\taunl$; as such, the marginal constraints are similar to the single-template results given in Tab.\,\ref{tab: local-results}.}\label{fig: local-corner}
\end{figure}

All of the above results have been obtained from independent analyses of $\gnl$ and $\taunl$ (\textit{i.e.}\ assuming only one is non-zero). In Fig.\,\ref{fig: local-corner}, we show the joint constraint from the baseline and temperature-only analyses, assuming a Gaussian likelihood with covariance measured from the FFP10 simulations (without enforcing $\taunl>0$ given above non-detection). This yields the joint constraints:
\beq
    \{10^{-4}\gest^{\rm loc}, 10^{-2}\tauest^{\rm loc}\} &=& \{1.0\pm5.4,2.8\pm2.4\}\quad(\textsc{sevem})\quad = \,\,\,\{0.3\pm4.8,-0.2\pm2.4\}\quad(\textsc{smica}),\nonumber
\eeq
from the baseline analysis; these are almost identical to the independent constraints due to the small correlations ($<10\%$) between the two templates.

\vskip 8pt
\subsubsection{Late-Time Non-Gaussianity Results}
\noindent As shown in Tab.\,\ref{tab: local-results} and Fig.\,\ref{fig: local-pdf}, we obtain a strong detection of lensing in \textit{Planck} PR4 with the 68\% limits
\beq
    \widehat{A}_{\rm lens}=1.001\pm0.024 \quad (\textsc{sevem}) \quad = \quad 0.979\pm0.023 \quad (\textsc{smica})\nonumber;
\eeq
this is fully consistent with the fiducial model. Notably, we find strong detections in each channel: the signal-to-noise ratio is $35\sigma$ ($35\sigma$) in temperature, \resub{$6.7\sigma$ ($7.6\sigma$)} in polarization, and $42\sigma$ ($43\sigma$) in the combined \textsc{sevem} (\textsc{smica}) dataset. Primordial non-Gaussianity notwithstanding, this represents an important result of this work and is competitive with previous analyses, as we discuss in \S\ref{sec: discussion}. 

Using the joint Fisher matrix, we can assess the lensing-induced bias on $\gnl$ and $\taunl$ (shown graphically in Fig.\,\ref{fig: local-systematics}): $\Delta \gnl = 2.7\times 10^4$ (corresponding to $0.5\sigma$) and $\Delta\taunl = 70$ ($0.3\sigma$) from the baseline \textsc{sevem} analysis. Due to our scale cuts, the lensing contributions to $\taunl$ are small \citep[cf.][]{2014A&A...571A..24P}, though they are larger for $\gnl$ and may dominate at larger $\ell_{\rm max}$. As discussed above, we could alternatively ameliorate lensing bias by performing a joint analysis of $A_{\rm lens}$ and the template of interest (without enforcing $A_{\rm lens}=1$); this yields the same shifts with a negligible change to the errorbars, given the tight constraints on $A_{\rm lens}$.

We additionally find a significant detection of point sources, with the baseline result
\beq
    10^{38}\widehat{t}_{\rm ps} = (32\pm6.2)\,\mathrm{K}^4 \quad (\textsc{sevem})\quad  =\quad (19.1\pm5.9)\,\mathrm{K}^4 \quad (\textsc{smica}). \nonumber
\eeq
As discussed in \S\ref{sec: discussion}, this is consistent with past results. 
Since we did not account for point-source polarization, we cannot measure $t_{\rm ps}$ from $E$- and $B$-modes, thus the baseline constraints are almost identical to those from $T$-only.\footnote{Slight variations arise due to the $TE$ correlations included in the weighting scheme.} Due to the different approaches to masking and spectral separation (which in part, aims to remove such contributions), we find a $2\sigma$ shift between \text{smica} and \textsc{sevem}. Unlike for lensing, point-sources lead to negligible bias in the primordial estimators (estimated via joint analyses of $t_{\rm ps}$ with the primordial shapes), with $\Delta\gnl = 140, \Delta\taunl = 13$, both of which are below $0.05\sigma$. For this reason, we do not include point-sources in the EFTI and exchange template analyses.

\subsubsection{Analysis Variants}\label{subsec: local-variants}
\begin{figure}[!t]
    \centering
    \includegraphics[width=0.8\linewidth]{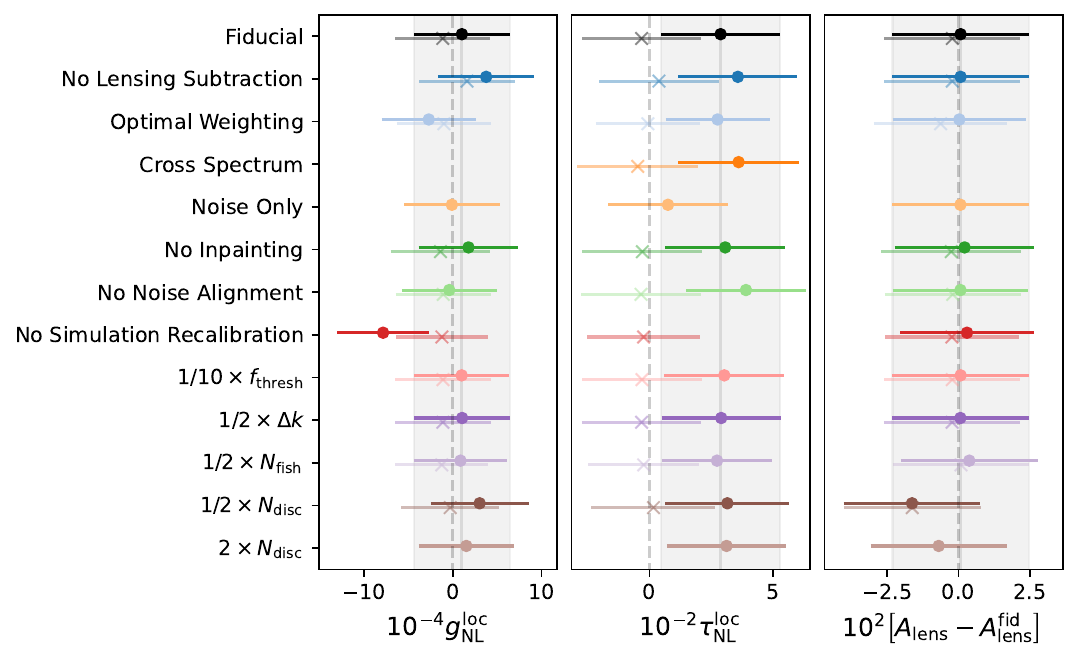}
    \caption{Summary of the main validation tests performed for the local and lensing \textit{Planck} non-Gaussianity analyses. Each point represents the measurement and $1\sigma$ error for $\gnl,\taunl,A_{\rm lens}$ (analyzed independently) obtained from an analysis varying a particular part of the pipeline. The dark circles indicate the data, whilst the light crosses show the mean of $100$ FFP10 simulations. In all cases, we analyze the baseline \textsc{sevem} dataset at $\ell_{\rm max}=2048$. Some points are omitted since the corresponding analyses were not carried out (e.g., $\gnl$ cross-spectra). We show results from the following tests: not removing lensing-induced biases (blue); optimal $\Si$ weighting (light blue); applying the estimators to the split $A/B$ or noise-only datasets (dark and light orange); including point-sources in the mask instead of inpainting (dark green); altering the FFP10 simulation noise and calibration (light green and red, with the latter $\taunl$ constraint outside the plotting region); changing the estimator hyperparameters such as optimization convergence (pink), $k$-integral resolution (purple), Fisher matrix iterations (light purple) and the number of simulations used to remove the disconnected trispectrum (brown). We find excellent agreement in almost all cases indicating that our pipeline is robust; the largest deviations are found when changing the weighting scheme (due to differing treatment of noise), analyzing the noise-only simulations (which null cosmic variance fluctuations), and dropping the simulation recalibration (which corrects for a $3\%$ power deficit). Additional tests elucidating the impact of scale- and sky-cuts are shown in Figs.\,\ref{fig: local-fsky}\,\&\,\ref{fig: local-lmax}.}
    \label{fig: local-systematics}
\end{figure}

\noindent Given the complexity of our pipeline, it is important to verify that our results are robust to modeling assumptions. To this end, we have performed a variety of additional $\gnl$ and $\taunl$ analyses modifying many aspects of the pipeline (focusing throughout on the \textsc{sevem} component-separation pipeline). Below, we describe these tests and their conclusions, with a graphical summary shown provided in Fig.\,\ref{fig: local-systematics}.

\vskip 8pt
\paragraph{Weighting} 
As discussed in \S\ref{subsec: method-weighting}, the variance of our trispectrum estimators depends on the choice of weighting scheme, $\Si$. Whilst our fiducial analyses use the idealized form of \eqref{eq: ideal-weighting}, we may also apply the conjugate-gradient scheme given in \eqref{eq: cgd-weighting}, which accounts for spatial variation of the noise and mask (though ignores the scale-dependence of the noise).
This relies on repeated harmonic transforms and is thus more expensive to compute; however, it may yield lower-noise trispectrum estimators. Applied to the baseline \textsc{sevem} dataset, the (quasi-)optimal weighting gives
\beq
    \{10^{-4}\gest^{\rm loc},10^{-2}\tauest^{\rm loc},\widehat{A}_{\rm lens},10^{38}\widehat{t}_{\rm ps}\}\,\,= \,\,\{-2.7\pm5.3,2.8\pm2.1,1.000\pm0.023,\resub{27.3\pm5.9}\} \quad (\textsc{sevem}), \nonumber
\eeq
which can be compared to the idealized-$\Si$ constraints in Tab.\,\ref{tab: local-results}. We find shifts of up to $0.8\sigma$ in the parameters, attributed to the differing noise properties, and a $1-15\%$ improvement in both empirical and theoretical errorbars, which leads to a slightly stronger ($43\sigma$) detection of lensing. We find largest improvements for $\sigma(\taunl)$; this suggests that the gains induced by optimal weighting schemes derive primarily from large-scales (which are highly contaminated by the mask). Given the significant increase in computation time required to implement such forms and the fairly modest improvements, we do not consider them further in this work.

\vskip 8pt
\paragraph{Data}
To assess the impact of (possibly non-Gaussian) noise on our parameter constraints we use the \textsc{npipe-a/b} data-splits. These are constructed to have maximally independent noise properties, such that the cross-spectrum only contains cosmological and foreground contributions. As discussed in \citep{Marzouk:2022utf,Philcox4pt1} (see also \citep{2014A&A...571A..24P,Feng:2015pva}), these can be used to build a `cross-field' estimator for exchange non-Gaussianity.\footnote{To understand this, we note that the exchange estimators can be written in terms of the power spectrum of some quadratic field $\Phi_{LM}[d,d]$ (see Appendix \ref{app: tauNL-Lmax}). Replacing $\Phi_{LM}[d,d]$ with a quadratic field built from independent data-splits, $\Phi_{LM}[d^A,d^B]$, results in a estimator that is less susceptible to noise, in part due to the removal of noise contributions from the mean-field term $\av{\Phi_{LM}[d^A,d^B]}$. Since the CMB signal is preserved, this does not require a change in the normalization.} This is implemented in a modified version of the \textsc{PolySpec} code, utilizing both the \textit{Planck} and FFP10 $\textsc{a/b}$ splits.\footnote{This approach cannot be applied to $\gnl$ since the corresponding estimator cannot be factorized as a product of quadratic fields. To remove disconnected noise contributions in this case, we would require four data-splits.} For the baseline (temperature-only) dataset, we find $10^{-2}\tauest^{\rm loc} = 3.6\pm2.5$ ($2.8\pm2.7$), corresponding to a shift of $0.3\sigma$ ($0.06\sigma$) with respect to the fiducial values, and a negligible change in the errorbars. This variation is consistent with the $0.3\sigma$ root-mean-square difference between full and data-split $\taunl$ analyses of FFP10 simulations.

Alternatively, we can apply the trispectrum estimators to the difference of the two data-splits. This isolates the effects of noise, nulling any physical signal. 
From the baseline analysis, we find $\{10^{-4}\gest^{\rm loc}, 10^{-2}\tauest^{\rm loc}, \widehat{A}_{\rm lens}, 10^{38}\widehat{t}_{\rm ps}\} = \{-0.1\pm5.4,0.8\pm2.4,0.00\pm0.024,-0.6\pm6.2\}$; these are all consistent with zero within $0.3\sigma$. This further indicates that the noise contributions to the $\gnl$ and $A_{\rm lens}$ analyses are subdominant.

\vskip 8pt
\paragraph{Simulations}
Our estimators require a suite of simulations that accurately capture the two-point function of the data. Although our estimators null any mismatch at leading order (due to the two-field $d^2\av{d^2}$ term in the estimators), the high-precision of \textit{Planck} implies that second-order effects can still lead to significant bias. To assess this, we first perform analyses without the FFP10 `noise alignment' maps, which were created to address a percent-level mismatch between observed and simulated high-frequency noise spectra at $\ell>100$ \citep{Planck:2020olo}.
This yields shifts of $\Delta\gnl = -14000, \Delta\taunl = 102$, corresponding to $-0.3\sigma$ and $0.4\sigma$ respectively, though negligible bias on $\widehat{A}_{\rm lens}$ or $\widehat{t}_{\rm ps}$. The variation in $\gest^{\rm loc}$ is much larger than the noise-only measurement reported in the previous section, implying that such a correction is important to ensure that the simulations faithfully represent the data.

As discussed in \S\ref{subsec: method-data}, the FFP10 simulations exhibit a $\approx 3\%$ deficit in $C_\ell^{TT}$ at $\ell>1000$ compared to the \textit{Planck} data (and associated theoretical models). Since this is also present in the cross-spectrum of the data-splits, it is physical in origin. By default, we correct for this mismatch by adding Gaussian noise to the simulations, such that the two power spectra agree (following \citep{Marzouk:2022utf}). Removing this correction leads to large parameter shifts: the baseline analysis gives $\Delta\gnl = -89000$, $\Delta\taunl = 1500$, and $\Delta t_{\rm ps} = 8.3\times 10^{-38}$, corresponding to $-1.7\sigma$, $6.4\sigma$, and $1.3\sigma$ respectively (with negligible change to $A_{\rm lens}$, matching \citep{Carron:2022eyg}). This corresponds to a (spurious) $7.6\sigma$ detection of $\taunl$, and highlights the importance of accurate simulations.

\vskip 8pt
\paragraph{Masking}

\begin{figure}[!t]
    \includegraphics[width=\linewidth]{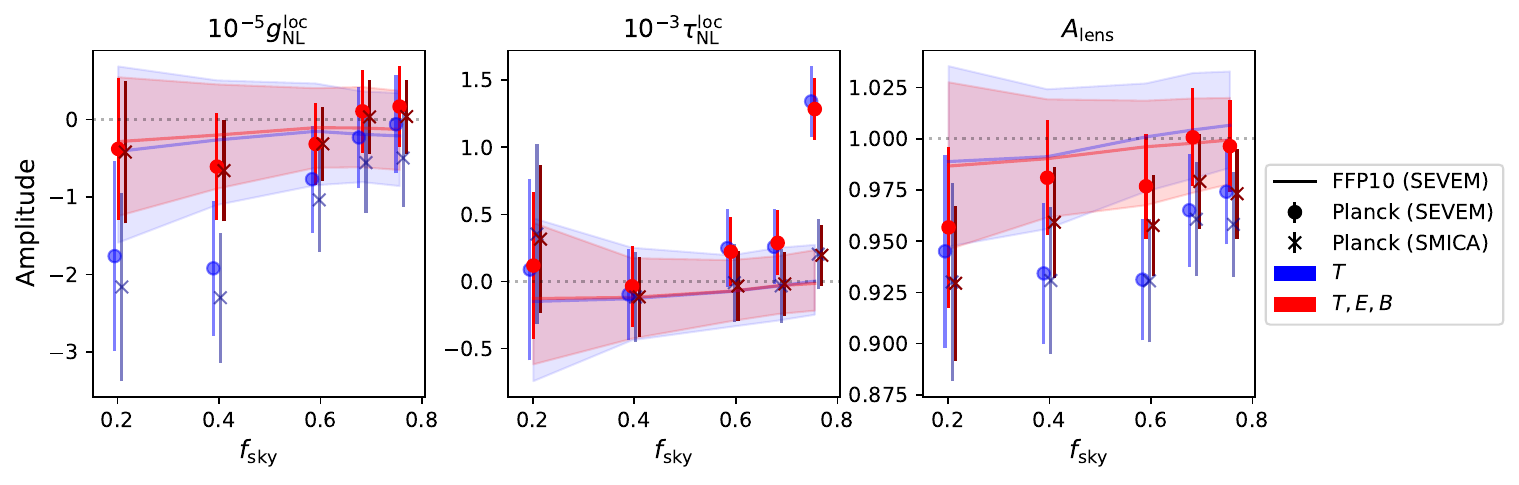}
    \caption{Dependence of the \textit{Planck} local and lensing non-Gaussianity constraints on the sky fraction, $f_{\rm sky}$. We present results for five choices of galactic mask (combining \textsc{gal020} to \textsc{gal080} with the common component-separation mask), using both \textsc{sevem} (light circles) and \text{smica} (dark crosses). The baseline and temperature-only results are displayed in red and blue respectively, and the shaded regions summarize the measurements from $100$ FFP10 simulations. We find consistent results for the $\gnl$ and $A_{\rm lens}$ analyses, given the noise fluctuations induced by the changing sky fraction. For $\taunl$, we find strong evidence for foreground contamination in the \textsc{sevem} using the \textsc{gal080} ($f_{\rm sky}\approx 0.75$) mask. To ameliorate this, we adopt the \textsc{gal070} mask throughout this paper.}\label{fig: local-fsky}
\end{figure}

Galactic dust and extragalactic point sources are strongly non-Gaussian, and, if unaccounted for, could severely bias our parameter constraints.
To test their impact, we perform a suite of analyses using different galactic masks (see \S\ref{subsec: method-components}). Since the impact of dust also depends on the efficacy of the component separation pipeline, we consider both \textsc{sevem} and \textsc{smica} in this section, analyzing the baseline and temperature-only datasets. 

Fig.\,\ref{fig: local-fsky} shows the dependence of our local and lensing constraints on the analysis mask.
For the $\gnl$ and $A_{\rm lens}$ analyses, we find consistent results for all masks and datasets, given the expected $1-2\sigma$ variation due to the large change in sky fraction, $f_{\rm sky}$. 
Furthermore, the results approximately converge by $f_{\rm sky}\approx 0.6$ (\textsc{gal060}), indicating that any foreground biases are small. In all $\taunl$ analyses, we find consistent results up to $f_{\rm sky}\approx 0.7$, with the \textsc{gal060} and \textsc{gal070} measurements consistent to within $0.25\sigma$. 

Increasing to $f_{\rm sky}\approx 0.75$ (via \textsc{gal080}), we find a $4\sigma$ shift in both the baseline and temperature-only \textsc{sevem} $\widehat{\tau}_{\rm NL}^{\rm loc}$ measurements, and a more modest $0.9\sigma$ shift in \textsc{smica}. Notably, the \textsc{sevem} measurements are $4.7\sigma$ away from zero, with $10^{-2}\tauest^{\rm loc} = 12.8\pm2.4$. The stark difference between the \textsc{gal070} and \textsc{gal080} measurements indicates contamination from Galactic foregrounds given that (a) we have added just $10\%$ more data (with the mocks exhibiting an average deviation of $0.3\sigma$), and (b) the shift depends strongly on the component-separation pipeline.
Since the $\taunl$ estimators are dominated by large-scale $L$-modes, this shift is not surprising, particularly given the known non-Gaussianity of Galactic dust \citep[e.g.,][]{Coulton:2019bnz}. For $f_{\rm sky}\lesssim 0.7$, all of the estimators are stable, and we find no evidence for residual foregrounds; this motivates the fiducial mask (\textsc{gal070}) used in this work and in previous analyses \citep{Carron:2022eyg,Marzouk:2022utf}, and implies that our main results are robust (particularly given the consistency of \textsc{sevem} and \textsc{smica}). 

Finally, we test the dependence of our constraints on the inpainting scheme. In the fiducial analyses, the mask includes only large-scale features, with small holes (from point sources or other residual foregrounds) infilled using a diffusive scheme. As discussed in \S\ref{subsec: method-components}, we can alternatively define a combined mask which removes both large- and small-scale features, obviating the need for inpainting. 
This leads to at most $0.1\sigma$ shifts in the baseline \textsc{sevem} constraints, 
implying that our inpainting procedure does not induce bias. 

\begin{figure}[!t]
    \includegraphics[width=\linewidth]{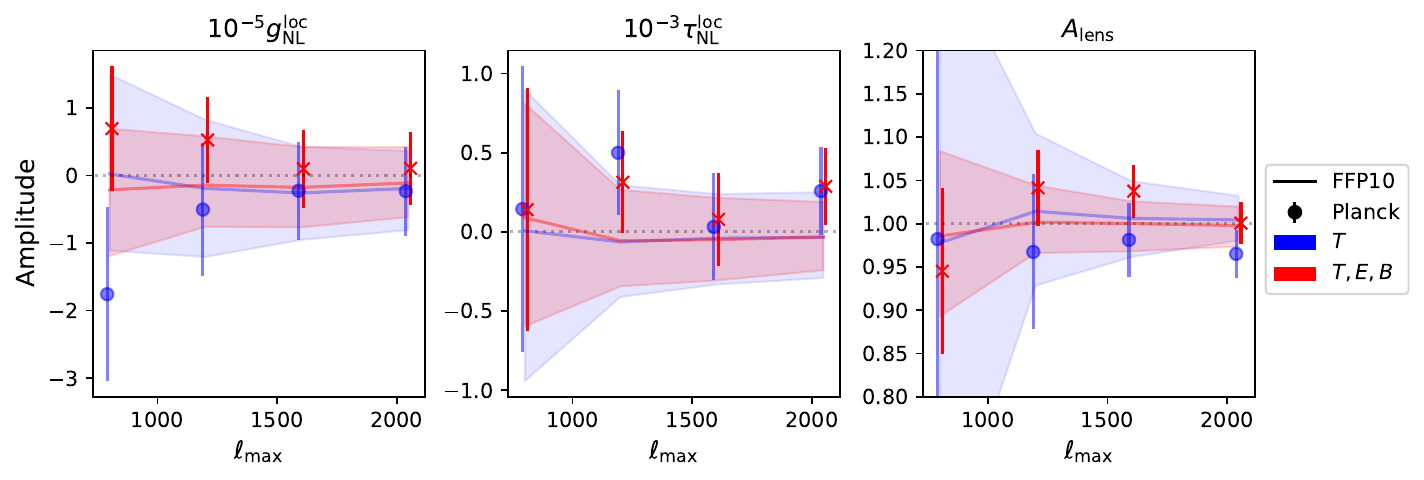}
    \caption{As Fig.\,\ref{fig: local-fsky}, but showing the dependence of local and lensing non-Gaussianity constraints on the maximum scale included in the analysis, $\ell_{\rm max}$. All results are obtained using the \textsc{sevem} component-separation pipeline. At every $\ell_{\rm max}$, the \textit{Planck} data is consistent with the expected distribution, yielding no evidence for scale-dependent bias nor primordial non-Gaussianity.}\label{fig: local-lmax}
\end{figure}

\vskip 8pt
\paragraph{Scale-Cuts}
In Fig.\,\ref{fig: local-lmax}, we assess the dependence on $\ell_{\rm max}$, the maximum harmonic mode included in the analysis. Although we find some variation in $\tauest^{\rm loc}$ and $\widehat{A}_{\rm lens}$ as $\ell_{\rm max}$ increases (due to the changing noise properties), the results are broadly consistent across scales, and closely match the fiducial values ($\gnl=\taunl=0$, $A_{\rm lens}=1$).
This implies that our analysis is robust to additional physics entering at high-$\ell$, such as small-scale noise and foregrounds, as well as CMB secondary effects.

As detailed in \citep{Kogo:2006kh,Kalaja:2020mkq}, the error on $\gest^{\rm loc}$ should scale as $\ell_{\rm max}^{-1}$, whilst that on $\tauest^{\rm loc}$ enjoys a stronger $\ell_{\rm max}^{-2}$ scaling. From the temperature-only analyses in Fig.\,\ref{fig: local-lmax}, we find the expected scaling for $\gnl$ at $\ell_{\rm max}\leq 1600$, but a saturation thereafter; when adding polarization, $\sigma(\gnl)$ is constant beyond $\ell_{\rm max}=1200$. This occurs since \textit{Planck} becomes noise-dominated at high-$\ell$ (particularly for polarization), thus the theoretical scalings do not apply.
For $\taunl$, we find similar results: strong scalings for $\ell_{\rm max}\leq 1200$, 
but saturation on smaller scales. Constraints on $A_{\rm lens}$ show particularly strong dependence on $\ell_{\rm max}$:
as noted in many previous analyses, lensing reconstruction is driven by small-scales and considerably aided by polarization information. The results indicate that the scale cuts used in this work are sufficient; moreover, we 
would find minimal loss of information if we reduced $\ell_{\rm max}$ to $1600$ (as in the \textit{Planck} analyses \citep{Planck:2015zfm,Planck:2019kim}).

$\taunl$ analyses require additional scale-cuts for the 
internal trispectrum leg, $L$.
As discussed in \citep{Kogo:2006kh,Kamionkowski:2010me,Kalaja:2020mkq,Philcox4pt2}, we expect constraints to be dominated by the lowest $L$ modes, principally $L=1$ and $L=2$. 
Repeat the baseline analysis with $L_{\rm max}=2$ ($L_{\rm max}=10$) results in a $-0.9\sigma$ ($0.2\sigma$) variation in $\widehat{\tau}_{\rm NL}^{\rm loc}$; this is attributed to noise fluctuations, as well as the (likely spurious) octopole signal discussed in Appendix \ref{app: tauNL-Lmax}.
Moreover, this leads to just $10\%$ ($<1\%$) inflation in the theoretical errorbars, confirming the previous forecasts and implying that our baseline choice ($L_{\rm max}=30$) is conservative.
We can also vary the minimum scale-cut, $L_{\rm min}$, to check for contamination from the $L=1$ kinematic dipole discussed in   
previous \textit{Planck} analyses \citep{2014A&A...571A..24P}.
Restricting to $L_{\rm min}=2$ significantly degrades the $\taunl$ constraints: we find $10^{-2}\tauest^{\rm loc}= 13.7\pm6.8$ from the baseline \textsc{sevem} dataset, almost $3\times$ weaker than the fiducial results shown in Tab.\,\ref{tab: local-results}. Regardless of whether the $L=1$ mode is included, we find no significant preference for non-zero $\taunl$, 
indicating that the \textsc{npipe} processing effectively removes the kinematic dipole.
Further discussion of the $L$-dependence of our $\taunl$ constraints (including an $L$-by-$L$ plot) can be found in Appendix \ref{app: tauNL-Lmax}.

\vskip 8pt
\paragraph{Hyperparameters}
As discussed in \S\ref{sec: methods}, our trispectrum estimators involve a number of hyperparameters and precision settings. Whilst these are extensively validated in \papertwo at $\ell_{\rm max}=512$, we perform some additional tests herein to ensure that our high-resolution results do not suffer from systematic contamination. These are visualized in Fig.\,\ref{fig: local-systematics}.
\begin{itemize}
    \item \textbf{Optimization}: The $f_{\rm thresh}$ parameter controls the tolerance in the \polyspec optimization algorithm \citep{Philcox4pt1,2011MNRAS.417....2S,2015arXiv150200635S}. More specifically, we require that the radial integrals contained in the $\gnl$ and $\taunl$ estimators are sufficiently well sampled such that the ideal Fisher matrices have an error below $f_{\rm thresh}$. 
    Reducing from $f_{\rm thresh}=10^{-3}$ to $f_{\rm thresh}=10^{-4}$ changes the baseline $\gest^{\rm loc}$ and $\tauest^{\rm loc}$ by $-0.005\sigma$ and $0.06\sigma$ respectively and the Fisher matrices by $<0.4\%$, indicating that the integrals are sufficiently converged.
    \item \textbf{Fourier-Space Resolution}: To convert from curvature perturbations to CMB anisotropies, we must perform $k$-space integrals over the transfer function and Bessel function. Since these integrals can be precomputed, 
    we typically use a fine $k$-grid with $N_k\approx 2\times 10^4$ points (as computed by \textsc{camb}). 
    Reducing to $N_k \approx 7000$ yields negligible changes to the constraints: the largest variation is in the $\taunl$ analysis, with a $0.25\%$ change in the Fisher matrix and a $0.01\sigma$ variation in the baseline constraint, indicating excellent convergence.
    \item \textbf{Fisher Matrix Convergence}: To avoid prohibitively high-dimensional sums, the Fisher matrices are computed using Monte Carlo summation.
    By default, we use $N_{\rm fish}=20$ realizations;
    halving this leads to variation in the Fisher matrix of up to $7\%$ (for $\taunl$), and shifts in the \textit{Planck} measurements by up to $-0.06\sigma$ ($\taunl$) and $0.13\sigma$ ($A_{\rm lens}$), which are consistent with the $0.1\sigma$ root-mean-square deviation 
    observed in simulations. Since these variations are a small fraction of a sigma and overestimate the $N_{\rm fish}=20$ error by $\sim \sqrt{2}$, we consider our Fisher matrices sufficiently converged.
    \item \textbf{Numerator Convergence}: Since the Gaussian contribution to the CMB four-point function typically dominates, we require a large value of $N_{\rm disc}$ to isolate the non-Gaussian contribution of interest. Repeating the baseline analysis with $N_{\rm disc}=50$ disconnected simulations leads to $\{0.4\sigma,0.1\sigma,-0.7\sigma,-0.1\sigma\}$ shifts in $\{\gest^{\rm loc},\tauest^{\rm loc},\widehat{A}_{\rm lens},\widehat{t}_{\rm ps}\}$, which are fairly significant. Whilst the scaling of this effect is known ($\sim 1/\sqrt{N_{\rm disc}}$), the amplitude differs between templates and is not easy to estimate \textit{a priori}, due to the differing overlap of the connected and disconnected contributions (though see \citep{Carron:2022eyg} for $\taunl$). Increasing to $N_{\rm disc}=200$ yields improved agreement, with shifts of $\{0.1\sigma,0.1\sigma,-0.3\sigma,0.1\sigma\}$. This corresponds to the baseline \textsc{sevem} constraints $\{10^{-4}\gest^{\rm loc}, 10^{-2}\tauest^{\rm loc}, \widehat{A}_{\rm lens}, \widehat{t}_{\rm ps}\} = \{1.5\pm5.4,3.1\pm2.4,0.993\pm0.024,32.9\pm6.2\}$. Since all shifts are within $0.3\sigma$ (and smaller for the primordial non-Gaussianity parameters, which are our main focus), we conclude that our analysis is robust to variations in $N_{\rm disc}$. In general, convergence can be assessed using the validation simulations: if $N_{\rm disc}$ is too small, the mean will be noticeably biased away from zero.
\end{itemize}

\vskip 8pt 
\paragraph{Conclusion}
In this section, we have performed detailed validation of the \textit{Planck} local and lensing non-Gaussianity constraints by performing a wide variety of supplementary analyses. Our main conclusion is that the fiducial constraints quoted in Tab.\,\ref{tab: local-results} are robust to modest variations in masking, noise properties, weighting, scale-cuts, resolution parameters, and beyond. The largest sources of error are: (a) mismatch between simulations and data; (b) residual foregrounds; 
(c) insufficient number of disconnected simulations. If unaccounted for, these effects could cause spurious detections of primordial non-Gaussianity; here, our fiducial analysis settings ensure that any such contamination is small.

\subsection{Constant Non-Gaussianity}\label{subsec: data-con}
\noindent Next, we consider the constant non-Gaussianity template introduced in \citep{Regan:2010cn,Fergusson:2010gn}. As discussed in \S\ref{sec: models}, this features a roughly constant amplitude in all trispectrum configurations, with features imprinted only by the transfer functions. Given the simple curvature four-point function (see Appendix \ref{app: templates}), the shape is straightforward to analyze and the estimator is very similar to that of $\gnl$. To obtain the observational constraints, we use the same hyperparameters as in the previous section and constrain the following amplitudes: $\{\gnl,\gcon,A_{\rm lens},t_{\rm ps}\}$, which can be analyzed independently or jointly.

\begin{table}[!t]
    \begin{tabular}{l||rcl|rcl|rcl||rcl|rcl|rcl}
        & \multicolumn{9}{c||}{\textit{Planck}-\textsc{sevem}} & \multicolumn{9}{c}{\textit{Planck}-\textsc{smica}}\\
        \textbf{Template} & \multicolumn{3}{c|}{$T$} & \multicolumn{3}{c|}{$E,B$} & \multicolumn{3}{c||}{$T,E,B$} & \multicolumn{3}{c|}{$T$} & \multicolumn{3}{c|}{$E,B$} & \multicolumn{3}{c}{$T,E,B$}\\\hline
        %
        %
        $10^{-4}g_{\rm NL}^{\rm loc}$ & \ress{-2.5}{6.6} & \ress{16.2}{33.3} & \ressb{1.0}{5.4} & \ress{-5.6}{6.5} & \ress{36.6}{32.5} & \ressb{0.3}{4.8}\\
$10^{-5}g_{\rm NL}^{\rm con}$ & \ress{-18.2}{11.0} & \ress{-3.9}{24.6} & \ressb{-4.3}{5.2} & \ress{-14.6}{11.0} & \ress{-13.4}{22.0} & \ressb{-4.8}{5.0}\\\hline
        %
    %
    $10^{-5}g_{\rm NL}^{\dot{\sigma}^4}$ & \ress{-20.2}{16.0} & \ress{-29.3}{45.7} & \ressb{-10.3}{8.0} & \ress{-15.3}{16.3} & \ress{-49.9}{38.0} & \ressb{-11.7}{7.7}\\
    $10^{-5}g_{\rm NL}^{\dot{\sigma}^2(\partial{\sigma})^2}$ & \ress{-23.7}{18.4} & \ress{-66.5}{52.6} & \ressb{-10.3}{9.5} & \ress{-21.3}{18.6} & \ress{-75.0}{47.0} & \ressb{-11.2}{8.9}\\
    $10^{-5}g_{\rm NL}^{(\partial{\sigma})^4}$ & \ress{-7.1}{4.4} & \ress{-22.5}{12.0} & \ressb{-1.3}{2.6} & \ress{-7.2}{4.4} & \ress{-20.6}{11.3} & \ressb{-1.2}{2.5}\\
    \end{tabular}
    \caption{\resub{\textit{Planck} PR4 constraints on contact non-Gaussianity.} We give results for the local shape $\gnl$ (matching Tab.\,\ref{tab: local-results}), the featureless constant template $g_{\rm NL}^{\rm con}$, and the three Effective Field Theory of Inflation templates $\gnldotdot,\gnldotdel,\gnldeldel$, all of which are analyzed independently. This table is analogous to Tab.\,\ref{tab: local-results}, and subtracts lensing bias from all results, with errorbars obtained from FFP10 simulations. Across all analyses, we find no detection of primordial non-Gaussianity, and observe that constraints tighten significantly ($\approx 2\times$) when polarization is included. Joint analyses featuring $\gcon$ and the EFTI templates are shown in Figs.\,\ref{fig: contact-corner}\,\&\,\ref{fig: efti-corner} respectively, with marginalized constraints given in the text.}\label{tab: contact-results}
\end{table}

In Tab.\,\ref{tab: contact-results}, we list the \textit{Planck} PR4 constraints on $\gcon$, with the baseline result:
\beq
    10^{-5}\gest^{\rm con} &=& \resub{-4.3\pm5.2\quad(\textsc{sevem})}\quad = \resub{-4.8\pm5.0\quad(\textsc{smica})},\nonumber
\eeq
subtracting lensing bias as before. Across all fields and component-separation methods, we find no evidence for primordial non-Gaussianity, with a maximum deviation of (-)$1.7\sigma$ for the temperature-only \textsc{sevem} analysis. Results for \textsc{sevem} and \textsc{smica} are consistent within $0.1\sigma$ (baseline), or $0.3\sigma$ (temperature-only), indicating no evidence for foreground contamination. Adding polarization reduces the errorbars by more than a factor of two; the improvement primarily stems from the inclusion of cross-spectra such as $TTTE$, and is larger than for $\gnl$ due to the differing shape of the template (with less focus on large-scales, where $E$-modes are noise-dominated).

\begin{figure}
    \includegraphics[width=0.4\linewidth]{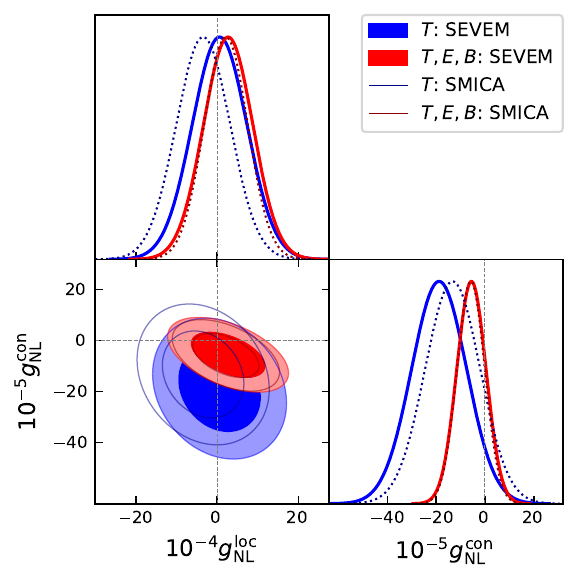}
    \caption{As Fig.\,\ref{fig: local-corner}, but plotting joint constraints on local cubic non-Gaussianity, $\gnl$, and the constant $\gcon$ template (assuming a Gaussian likleihood, as before). When polarization is included, we find significant correlations between parameters, yielding the joint \textsc{sevem} (\textsc{smica}) constraints $\{10^{-4}\widehat{g}_{\rm NL}^{\rm loc},10^{-5}\widehat{g}_{\rm NL}^{\rm con}\} = \{2.7\pm6.0,-5.6\pm6.0\}$ ($\{2.1\pm5.4,-5.8\pm5.7\}$), which are $10-15\%$ weaker than the single-template constraints listed in Tab.\,\ref{tab: local-results}.}\label{fig: contact-corner}
\end{figure}

A joint analysis of the $\gnl$ and $\gcon$ contact templates is shown in Fig.\,\ref{fig: contact-corner}. The corresponding constraints are consistent with zero within $1.7\sigma$, with the baseline values
\beq
    \{10^{-4}\gest^{\rm loc}, 10^{-5}\widehat{g}_{\rm NL}^{\rm con}\} &=& \{2.7\pm6.0,-5.6\pm6.0\}\quad(\textsc{sevem})\quad = \,\,\,\{2.1\pm5.4,-5.8\pm5.7\}\quad(\textsc{smica}),\nonumber
\eeq
with the $\gcon$ bound weakening by a factor of two when polarization channels are excluded.
As seen in the Figure, the baseline contours are fairly correlated (with $\mathrm{corr}(\gnl,\gcon) = 0.45$ measured from $100$ FFP10 simulations); this leads to a $15\%$ degradation in the joint errorbars compared to the independent results.

Lastly, we can perform some simple consistency checks. Analyzing the mean of $100$ FFP10 simulations returns results consistent with zero at $0.2\sigma$,
indicating no significant biases from the finite number of disconnected simulations, nor from the residual foregrounds and noise non-Gaussianity present in the simulation suite. Moreover, the errorbar on $\gcon$ is consistent with the expected value of $5.4\times 10^5$ obtained from the inverse Fisher matrix (given the $\mathcal{O}(10\%)$ variation expected from the finite number of simulations), implying that our pipeline is close to optimal. Performing a joint analysis of $\gcon$ and $t_{\rm ps}$ negligibly impacts our parameter constraints (with $\Delta\gcon = 1300$ in the baseline analysis); however, we find a large lensing-induced bias with $\Delta\gcon = 1.7\times 10^{6}$, corresponding to $3.3\sigma$ (both from the Fisher matrix prediction and through joint analyses).
If this bias is not subtracted, we would report a spurious detection of constant non-Gaussianity; since the mean-of-FFP10 analysis returns null results after we apply the correction, we conclude that any residual bias is small.

\subsection{EFTI Non-Gaussianity}\label{subsec: results-efti}
\noindent Our final set of contact templates are the EFTI shapes, encoding self-interactions in single- and multi-field inflation. As discussed in \paperone and \citep{2015arXiv150200635S}, these involve double integrals (over radius and conformal time), and are thus somewhat more expensive to compute.
Here, we analyze the three EFTI amplitudes -- $\gnldotdot,\gnldotdel,\gnldeldel$ -- in concert with the cubic local template, $\gnl$, and the lensing amplitude, $A_{\rm lens}$, to account for any correlations. To reduce computational costs, we use slightly different hyperparameters to \S\ref{subsec: results-local}:
(a) we set $N_{\rm disc}=50$ when analyzing simulations (retaining $N_{\rm disc}=100$ for the observational data); (b) we set $N_{\rm fish}=10$; (c) we use a stricter optimization tolerance, $f_{\rm thresh}=10^{-4}$, but perform the optimization in two stages, as described in \S\ref{subsec: method-estimator}; (d) we use a coarser $k$-integration grid (to avoid excess memory requirements); (e) we analyze only $50$ simulations. These choices 
are validated at the end of this section.

\subsubsection{Independent Constraints}
\noindent The lower half of Tab.\,\ref{tab: contact-results} presents our main results: indpendent constraints on the three EFTI parameters.
From the baseline analysis, we find
\beq
    \{10^{-5}\gestdotdot,10^{-5}\gestdotdel,10^{-5}\gestdeldel\} &=& \{-10\pm 8,-10\pm10,-1\pm3\}\quad (\textsc{sevem})\nonumber\\
    &=& \{-12\pm8, -11\pm9,-1\pm3\}\quad (\textsc{smica})\nonumber,
\eeq
all of which are consistent with zero within $1.3\sigma$ ($1.5\sigma$). Across all data combinations, we find no detection of EFTI non-Gaussianity, with a maximum deviation of \resub{$1.9\sigma$} (from the \textsc{smica} polarization-only analysis). The addition of polarization (which is new to this work) sharpens the constraints by a factor of $1.6-2.1$; this improvement is larger than for $\gnl$ or $\taunl$ due to the equilateral shape of the EFTI templates (noting that large- and small-scale polarization modes are noise dominated). As for the local non-Gaussianity analyses, constraints are primarily driven by temperature anisotropies and cross-spectra: the polarization dataset yields $\approx 3\times$ weaker constraints than the temperature dataset. In the baseline analyses, we find excellent agreement between the two component-separation methods (within $0.2\sigma$), though the polarization-only results show larger shifts (up to $0.5\sigma$). As for $\gnl$, the empirical distribution of $\widehat{g}_{\rm NL}$ estimates from 50 FFP10 simulations appear with a Gaussian distribution, and, moreover, we find good agreement between empirical and theoretical error-bars ($5-15\%$ in the baseline analysis, noting the expected $10\%$ scatter).
This implies that our pipeline is close to optimal.

\begin{figure}[!t]
    \includegraphics[width=0.7\linewidth]{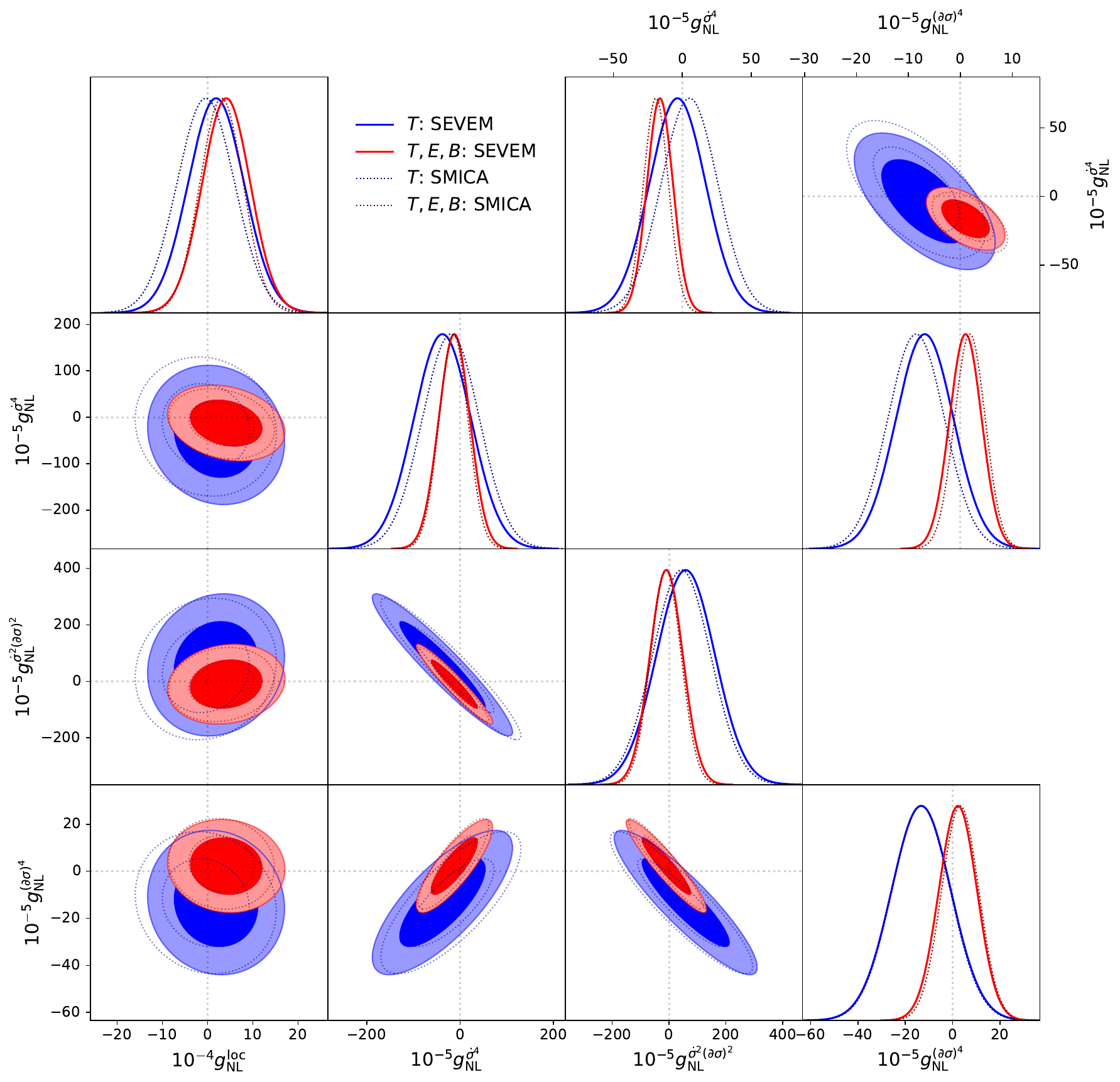}
    \caption{As Fig.\,\ref{fig: local-corner}, but plotting joint constraints on the local and EFTI parameters: $\{\gnl,\gnldotdot,\gnldotdel,\gnldeldel\}$. The bottom left plots show the results of a joint analysis of all four templates, whilst the top right plots restrict to $\{\gnldotdot,\gnldeldel\}$ allowing comparison with literature constraints. The addition of polarization (which is new to this work) leads to significantly tighter constraints on EFTI parameters. For the two-dimensional analysis, we find the marginalized temperature-plus-polarization constraints $\{10^{-5}\gnldotdot,10^{-5}\gnldotdel\} = \{-16.2\pm9.2,1.0\pm3.1\}$ from \textsc{sevem}, or $\{-19.1\pm9.0,1.8\pm2.9\}$ from \textsc{smica}, which are only mildly weaker than the constraints given in Tab.\,\ref{tab: contact-results}.
    }\label{fig: efti-corner}
\end{figure}

\subsubsection{Joint Constraints}
\noindent As noted in \citep{2015arXiv150200635S}, the three EFTI templates describe similar physics, thus it is profitable to perform joint analyses in addition to the above. In the bottom left panels of Fig.\,\ref{fig: efti-corner}, we show the joint constraints on $\gnldotdot$, $\gnldotdel$, $\gnldeldel$, and $\gnl$ 
(noting that many models source both EFTI and local non-Gaussianity).
Whilst $\gnl$ is roughly uncorrelated with the other parameters ($\lesssim 10\%$, or $\lesssim 5\%$ without polarization), there are clear degeneracies in the triplet of EFTI parameters. The inverse Fisher matrix predicts the correlation coefficients 
\beq
    \mathrm{corr}(\gnldotdot,\gnldotdel) = -0.94,\quad \mathrm{corr}(\gnldotdot,\gnldeldel) = 0.78,\quad \mathrm{corr}(\gnldotdel,\gnldeldel) = -0.91 \nonumber
\eeq
(from the baseline analysis, though similar from temperature-alone); these agree with the correlations extracted from the FFP10 simulations within $5\%$. As in the single-template analyses, we find no evidence for non-Gaussianity, with the joint constraints:
\beq
    \{10^{-4}\gest^{\rm loc},10^{-5}\gestdotdot,10^{-5}\gestdotdel,10^{-5}\gestdeldel\} &=& \{4.1\pm 5.3, -12\pm33,-9\pm57,2.2\pm8.0\} \quad (\textsc{sevem})\nonumber\\
    &=&\{3.1\pm4.7,-14\pm31,-11\pm53,3.2\pm7.8\}\quad(\textsc{smica}),\nonumber
\eeq
from the baseline analysis. 
As before, these are highly consistent between \textsc{sevem} and \textsc{smica}, and weaken by a factor of two if one excludes polarization.

Due to the strong correlations between templates, the joint constraints on the EFTI parameters are $3-6\times$ weaker than the independent constraints reported in Tab.\,\ref{tab: contact-results}. One way to ameliorate this is to switch basis, defining a decorrelated set of EFTI parameters using a Gram-Schmidt orthogonalization of the theoretical covariance matrix. For the baseline analysis, this yields
\beq
    \{10^{-5}\widehat{g}_{\rm NL}^{\rm EFTI,1},10^{-5}\widehat{g}_{\rm NL}^{\rm EFTI,2},10^{-5}\widehat{g}_{\rm NL}^{\rm EFTI,3}\} &=& \{-12\pm33,-29\pm17,-1.5\pm2.6\} \quad (\textsc{sevem})\nonumber\\
    &=&\{-14\pm31,-33\pm16,-1.0\pm2.5\}\quad(\textsc{smica}),\nonumber
\eeq
defining $g_{\rm NL}^{\rm EFTI,1} = \gnldotdot$, $g_{\rm NL}^{\rm EFTI,2} = 1.57\gnldotdot+\gnldotdel,$ $g_{\rm NL}^{\rm EFTI,3} = 0.14\gnldotdot+0.19\gnldotdel+\gnldeldel$. Alternatively, we can restrict our attention to a subset of the parameters, setting $\gnldotdel=\gnl=0$ as in \citep{2015arXiv150200635S,Planck:2015zfm,Planck:2019kim}. This leads to the constraints shown in the top right panels of Fig.\,\ref{fig: efti-corner}, with the baseline result:
with temperature-only, or
\beq
    \{10^{-5}\gestdotdot,10^{-5}\gestdeldel\} &=& \{-16.2\pm9.3,1.0\pm3.1\} \quad (\textsc{sevem})\quad = \,\,\,\{-19.2\pm9.0,1.8\pm2.9\}\quad(\textsc{smica}).\nonumber
\eeq
These are consistent with zero within $2.1\sigma$, and only $15\%$ weaker than the single-template constraints. 
Here, the utility of polarization is abundantly clear: the posterior volume of the \textsc{sevem} (\textsc{smica}) baseline constraints is $3.6\times$ ($4.0\times$) smaller than for the temperature-only analyses.

\subsubsection{Consistency Tests}
\noindent Given the complex form of the EFTI trispectra (cf.\,Appendix \ref{app: templates}) and the less stringent hyperparameter settings used above, it is important to validate our pipeline, particularly the optimization step.
Reducing the optimization tolerance from $f_{\rm thresh}=10^{-4}$ to $10^{-3}$ has only a minor impact: the baseline \textsc{sevem} measurements shift by up to $0.2\sigma$, and the errorbars broaden by up to $10\%$. We find similar shifts when doubling the resolution of the initial $r,\tau$ integration grid by a factor of two, but negligible ($<10^{-3}\sigma$) variation when halving the number of $k$ sampling points. Whilst not insignificant, these shifts are not a cause for concern given that they overestimate the error in the $f_{\rm thresh}=10^{-4}$ analysis (relative to some $f_{\rm thresh}\to 0$ limit). 
Moreover, the use of an approximate template cannot lead to bias: in essence, we are searching for a slightly simplified shape 
that is highly correlated to the target (within $\sim f^{1/2}_{\rm thresh}$).

We find fairly good convergence with the number of Monte Carlo simulations. Halving $N_{\rm fish}$ leads to consistent results within $0.05\sigma$, whilst reducing to $N_{\rm disc}$ to $50$ leads to shifts of $0.1-0.3\sigma$ (smaller than for the local templates), implying that the fiducial analysis (at $N_{\rm disc}=100$) is converged to within $\lesssim 0.3\sigma$. An end-to-end test of our pipeline is obtained by analyzing $50$ FFP10 simulations; the mean values are consistent with zero at $0.1-0.3\sigma$, with largest deviations found for $\gnldeldel$ (which has greatest sensitivity to $N_{\rm disc}$). 

In all the quoted results, we have subtracted lensing bias: for the baseline \textsc{sevem} analysis, this is given by $\{\Delta\gnldotdot,\Delta\gnldotdel,\Delta\gnldeldel\} = \{3.5\times 10^6, 4.3\times 10^6, 1.2\times 10^6\}$, corresponding to $\{4.4\sigma,4.5\sigma,4.5\sigma\}$. These deviations are very large (and exceed $6\sigma$ in the temperature-only dataset), and are sourced by the combination of a weak ($\sim 5\%$) correlation between the EFTI and lensing templates and the strong ($>40\sigma$) detection of lensing. These biases vary considerably with the $\ell$-cuts and noise profiles: \citep{2015arXiv150200635S} found only $0.3\sigma$ shifts in a WMAP temperature-only analysis. These results further indicate that future constraints on non-Gaussianity could be considerably aided by delensing \citep[cf.,][]{Coulton:2019odk,Shiraishi:2019yux,Green:2016cjr,Trendafilova:2023xtq}.

\subsection{Direction-Dependent Non-Gaussianity}\label{subsec: results-direc}
\noindent Having analyzed the main contact trispectrum templates available in the literature, we now turn to exchange trispectra,.
In this section, we focus on direction-dependent generalizations of $\taunl$ before considering the cosmological collider shapes in \S\ref{subsec: results-coll}. We utilize three-types of templates: the parity-even and parity-odd forms discussed in \citep{Shiraishi:2013oqa,Shiraishi:2016mok} with amplitudes $\tau_{\rm NL}^{n,\rm even},\tau_{\rm NL}^{n,\rm odd}$, and the generalized shapes introduced in \paperone, specified by $\tau_{\rm NL}^{n_1n_3n}$ (for non-negative integers $n_1,n_3,n$). When analyzing the parity-even/parity-odd models, we restrict to the following amplitudes:
$\{\tau_{\rm NL}^{0,\rm even},\tau_{\rm NL}^{2,\rm even},\tau_{\rm NL}^{1,\rm odd}\}$; noting that higher-order terms are more difficult to constrain, and that
even (odd) templates with odd (even) 
$n$ cannot be meaningfully constrained from CMB anisotropy experiments, as shown in \papertwo. Following the same restrictions, we will constrain the generalized templates $\{\tau_{\rm NL}^{000},\tau_{\rm NL}^{220},\tau_{\rm NL}^{222},\tau_{\rm NL}^{022},\tau_{\rm NL}^{221}\}$. These sets are highly correlated, with non-zero values of the parity-even/parity-odd amplitudes generating the following:
\beq\label{eq: tau-direc-relations}
    \tau_{\rm NL}^{000} &&= (4\pi)^{3/2}\tau_{\rm NL}^{0, \rm even} = (4\pi)^{3/2}\taunl, \qquad \tau_{\rm NL}^{220} \supset \frac{1}{3}\frac{(4\pi)^{3/2}}{\sqrt{5}}\tau_{\rm NL}^{2, \rm even}, \qquad \tau_{\rm NL}^{022} \supset \frac{1}{3}\frac{(4\pi)^{3/2}}{\sqrt{5}}\tau_{\rm NL}^{2, \rm even}\\\nonumber
    \tau_{\rm NL}^{221}&\supset&\frac{\sqrt{2}}{9\sqrt{5}}(4\pi)^{3/2}\left(\tau_{\rm NL}^{1,\rm odd}-\frac{3}{7}\tau_{\rm NL}^{3,\rm odd}\right),
\eeq
(though we ignore $\tau_{\rm NL}^{3,\rm odd}$ in this work). We adopt the same analysis settings as in \S\ref{subsec: results-efti} the EFTI analyses, restricting to the $L$-range $[1,30]$ following \citep{Philcox4pt2,Shiraishi:2013oqa,Shiraishi:2016mok}. 

\begin{table}[!t]
    \begin{tabular}{l||rcl|rcl|rcl||rcl|rcl|rcl}
        & \multicolumn{9}{c||}{\textit{Planck}-\textsc{sevem}} & \multicolumn{9}{c}{\textit{Planck}-\textsc{smica}}\\
        \textbf{Template} & \multicolumn{3}{c|}{$T$} & \multicolumn{3}{c|}{$E,B$} & \multicolumn{3}{c||}{$T,E,B$} & \multicolumn{3}{c|}{$T$} & \multicolumn{3}{c|}{$E,B$} & \multicolumn{3}{c}{$T,E,B$}\\\hline
        %
%
$10^{-4}\widehat{\tau}_{\rm NL}^{000}$ & \ress{1.0}{1.1} & \ress{-6.3}{10.6} & \ressb{1.3}{1.1} & \ress{-0.1}{1.1} & \ress{-29.4}{9.8} & \ressb{-0.1}{1.0}\\
$10^{-4}\widehat{\tau}_{\rm NL}^{220}$ & \ress{5.4}{2.2} & \ress{-14.0}{18.0} & \ressb{4.2}{2.1} & \ress{2.0}{2.2} & \ress{-29.1}{15.4} & \ressb{0.8}{1.9}\\
$10^{-4}\widehat{\tau}_{\rm NL}^{222}$ & \ress{2.9}{4.5} & \ress{9.1}{33.4} & \ressb{4.3}{3.6} & \ress{2.0}{4.5} & \ress{38.2}{28.3} & \ressb{1.9}{3.6}\\
$10^{-4}\widehat{\tau}_{\rm NL}^{022}$ & \ress{-15.1}{6.2} & \ress{39.0}{41.6} & \ressb{-10.4}{5.0} & \ress{-12.6}{5.9} & \ress{52.6}{38.5} & \ressb{-8.8}{4.7}\\
$10^{-4}\widehat{\tau}_{\rm NL}^{221}$ & \ress{-26.5}{6.8} & \ress{-8.3}{50.2} & \ressb{-3.7}{4.8} & \ress{-27.2}{6.5} & \ress{-15.1}{44.2} & \ressb{-4.6}{5.0}\\\hline
%
%
$10^{-2}\tau_{\rm NL}^{0,\rm even}$ & \ress{2.3}{2.3} & \ress{-14.6}{23.8} & \ressb{2.9}{2.3} & \ress{-0.3}{2.5} & \ress{-65.4}{22.0} & \ressb{-0.1}{2.2}\\
$10^{-3}\tau_{\rm NL}^{2,\rm even}$ & \ress{1.9}{2.6} & \ress{-5.2}{21.4} & \ressb{0.9}{2.5} & \ress{-1.0}{2.8} & \ress{-18.2}{19.8} & \ressb{-1.9}{2.4}\\
$10^{-4}\tau_{\rm NL}^{1,\rm odd}$ & \ress{-8.4}{2.1} & \ress{-2.3}{16.0} & \ressb{-1.2}{1.4} & \ress{-8.6}{2.1} & \ress{-4.5}{14.1} & \ressb{-1.5}{1.6}\\
   \end{tabular}
    \caption{\resub{\textit{Planck} constraints on direction-dependent non-Gaussianity,} analyzing each template independently. The top panel shows measurements of the general $\tau_{\rm NL}^{n_1n_3n}$ parameters defined in \paperone, whilst the bottom panel gives results for the parity-even and parity-odd compressions used in \citep{Shiraishi:2013oqa,Shiraishi:2016mok}. We restrict to simple angular dependencies (involving spherical harmonics up to order two), and drop any templates that are difficult to constrain with the CMB (e.g., $\tau_{\rm NL}^{111}, \tau_{\rm NL}^{1,\rm even}$ \citep{Philcox4pt2}). Whilst $\tau_{\rm NL}^{000}$ and $\tau_{\rm NL}^{0,\rm even}$ are fully degenerate with $\taunl$, we include them here for comparison. In Fig.\,\ref{fig: direc-pdf}, we show the empirical distribution of $\tauest^{n_1n_3n}$, and give the joint constraints on all parameters in Fig.\,\ref{fig: direc-corner}. The baseline constraints are consistent with zero within $2.1\sigma$; when excluding polarization, we find deviations up to $4.2\sigma$, though note that this does not account for posterior non-Gaussianity (which Fig.\,\ref{fig: direc-pdf} indicates to be significant), it is not robust to variations in $f_{\rm sky}$, and is not reproduced in the combined dataset.}\label{tab: tau-direc-results}
\end{table}

\subsubsection{\texorpdfstring{$\tau_{\rm NL}^{n_1n_3n}$ Constraints}{Generalized Constraints}}
\noindent In the top panel of Tab.\,\ref{tab: tau-direc-results}, we report the measured values and errorbars for the generalized direction-dependent trispectrum amplitudes, $\tau_{\rm NL}^{n_1n_3n}$. These include the baseline results:\footnote{Although $\tau_{\rm NL}^{000}$ and $\taunl$ are formally equivalent, the corresponding constraints vary slightly due to (a) a different optimization algorithm and choice of $N_{\rm fish}$ (which yields $0.03\sigma$ shifts), and (b) the different number of simulations analyzed (leading to a $7\%$ change in the errorbars).}
\beq
    10^{-4}\{\tauest^{000},\tauest^{220},\tauest^{222},\tau_{\rm NL}^{022},\tauest^{221}\} &=& \{1.3\pm1.1,4.2\pm2.1,4.3\pm3.6,-10.4\pm5.0,-3.7\pm4.8\} \quad (\textsc{sevem})\\\nonumber
    &=&\{-0.1\pm1.0,0.8\pm1.9,1.9\pm3.6,-8.8\pm4.7,-4.6\pm5.0\}\quad(\textsc{smica}),\nonumber
\eeq
where the $\tau_{\rm NL}^{000}$ constraint is just a rescaling of the $\taunl$ result of \S\ref{subsec: results-local}. Each constraint is broadly consistent with zero; the largest deviations occur for \textsc{sevem} $\tau_{\rm NL}^{220}$ ($2.0\sigma$) and $\tau_{\rm NL}^{022}$ ($-2.1\sigma$); these reduce to $0.4\sigma$ and $-1.9\sigma$ with \textsc{smica}. As for the local and EFTI shapes, we report no significant detection of primordial non-Gaussianity. In general, constraints on $\tau_{\rm NL}^{n_1n_3n}$ weaken as $n_1,n_3,n$ increases, which results from the inherent difficulty in constraining more complex direction-dependence (with, for example, the angle-averaged trispectrum vanishing if $n>0$). Constraints from \textsc{sevem} and \textsc{smica} differ by $0.2-1.8\sigma$; as discussed in \S\ref{subsec: results-local}, this is broadly consistent with noise variations (and seen also in the \textit{Planck} $f_{\rm NL}$ analyses), but may indicate small residual foreground contributions. For all values of $n_1,n_3,n$, we find excellent agreement of the empirical and theoretical errorbars (within the expected $10\%$ scatter), indicating that our estimators are close to optimal.

\begin{figure}
    \centering
    \includegraphics[width=0.9\linewidth]{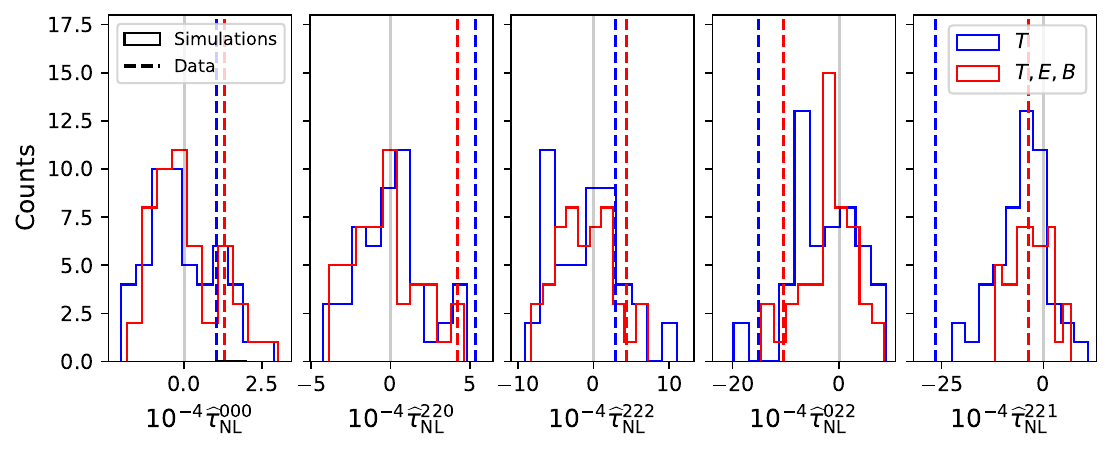}
    \caption{As Fig.\,\ref{fig: local-pdf}, but showing the distribution of direction-dependent non-Gaussianity amplitudes, $\widehat{\tau}_{\rm NL}^{n_1n_3n}$, from \textit{Planck} and 50 FFP10 simulations. 
    We use the \textsc{sevem} component-separation pipeline in all cases and analyze each template separately, noting that the $\widehat{\tau}_{\rm NL}^{000}$ distribution is analogous to $\tauest^{\rm loc}$ in Fig.\,\ref{fig: local-pdf}. For the baseline analyses (red) we find generally good agreement between data (vertical lines) and simulations (histograms); in temperature-only analyses, we find some deviations for $\tau_{\rm NL}^{220}$ ($2.4\sigma$) and $\tau_{\rm NL}^{221}$ ($-3.9\sigma$). The empirical histograms display clear non-Gaussianity (with a tail to negative $\widehat{\tau}_{\rm NL}^{221}$, for example), which is a likely culprit for these deviations, given that they are not reproduced in the combined analysis.}
    \label{fig: direc-pdf}
\end{figure}

In \S\ref{subsec: results-local}, we noted that the null distribution of $\widehat{\tau}_{\rm NL}^{\rm loc}$ is non-Gaussian and takes a non-trivial form in the presence of primordial non-Gaussianity (\textit{i.e.}\ $\taunl\neq 0$). The same is true for our direction-dependent forms, which are also dominated by a small number of low-$L$ modes (though the importance of this effect reduces as $n$ increases). Unlike for $\taunl$, however, the direction-dependent estimators cannot be written in terms of a large-scale power spectrum of quadratic fields (cf.\,Appendix \ref{app: tauNL-Lmax}),\footnote{In the language of Appendix \ref{app: tauNL-Lmax}, the generalized $\tau_{\rm NL}^{n_1n_3n}$ estimator computes two (complex) quadratic fields, $\Phi_{LM}^{(n_1)}, \Phi_{L'M'}^{(n_3)}$, which are combined by summing over $L'\in[L-n,L+n]$, weighting by Gaunt factors. This is not positive definite (except for $n=0$); moreover, it is less clear how to partition the estimator into individual $L$-modes, which is necessary to build the $\taunl$ posterior.} making it unclear how to derive an approximate form for the distribution $P(\tau_{\rm NL}^{n_1n_3n}|\widehat{\tau}_{\rm NL}^{n_1n_3n})$ and thus compute the non-Gaussian confidence interval on $\tau_{\rm NL}^{n_1n_3n}$. 
In Fig.\,\ref{fig: direc-pdf}, we plot the empirical distributions of $\widehat{\tau}_{\rm NL}^{n_1n_3n}$ from 50 FFP10 simulations, alongside the measurements from \textit{Planck}. We find some evidence for non-Gaussianity in the baseline PDFs, with, for example, $\widehat{\tau}_{\rm NL}^{000}$ ($\widehat{\tau}_{\rm NL}^{221}$) displaying a prominent tail to positive (negative) values. In the absence of a theoretical model for the non-Gaussian likelihood (and in the absence of a strong motivation, given the above lack of anomalies), we will proceed by simply quoting the measured value and $1\sigma$ error, though we caution that the tails of the distribution are not well understood.

As for the contact and EFTI analyses, \textit{Planck} polarization data alone is unable to place strong constraints on $\tau_{\rm NL}^{n_1n_3n}$, as evidenced by the $(7-10)\times$ larger values of $\sigma(\tau_{\rm NL}^{n_1n_3n})$ for polarization-only compared to the temperature-plus-polarization dataset. If we instead restrict to temperature-alone, we find fairly strong constraints, with $\sigma(\tau_{\rm NL}^{n_1n_3n})$ increased by $8-36\%$ relative to the baseline, with greatest loss of information found for the parity-odd template $\tau_{\rm NL}^{221}$ (as in \papertwo). This is clearly seen in Fig.\,\ref{fig: direc-pdf}, where the $\tau_{\rm NL}^{000}$ PDF is almost unchanged by the excision of polarization, whilst that of $\tau_{\rm NL}^{221}$ inflates considerably, notably with a notably enhanced tail to negative values.

Whilst the baseline constraints are broadly consistent with the null hypothesis, we find some discrepancies when analyzing only temperature or polarization. In particular, the \textsc{sevem} (\textsc{smica}) temperature-only measurement of $\tau_{\rm NL}^{221}$ lies $3.9\sigma$ ($4.2\sigma$) below zero, and that of $\tau_{\rm NL}^{000}$ is \resub{$3.0\sigma$} below zero in the \textsc{smica} polarization channel (but consistent with zero in \textsc{sevem}, as discussed in \S\ref{subsec: results-local}). Taken at face value, these results may appear alarming: however, they do not indicate novel inflationary physics since (a) they are not found in the baseline temperature-plus-polarization dataset and (b) they do not account for the non-Gaussianity of the $\widehat{\tau}_{\rm NL}^{n_1n_3n}$ distribution. For $\tau_{\rm NL}^{221}$, the addition of polarization sharpens constraints by $\approx 40\%$; if the temperature-only value were physical, we would expect to detect it at $\approx 5.5\sigma$ in the baseline analysis (but instead find consistency with zero at $0.9\sigma$). Similarly, the polarization-only value of $\tau_{\rm NL}^{000}$ would be detected at almost \resub{$30\sigma$}. These considerations suggest that
the anomalous signals are a statistical fluctuation or some late-time residual (such as a chiral dust signature not included in the FFP10 simulations) -- to further explore this, we perform a variety of systematic checks in \S\ref{subsubsec: results-direc-sys}.

\begin{figure}[!t]
    \centering
    \includegraphics[width=0.8\linewidth]{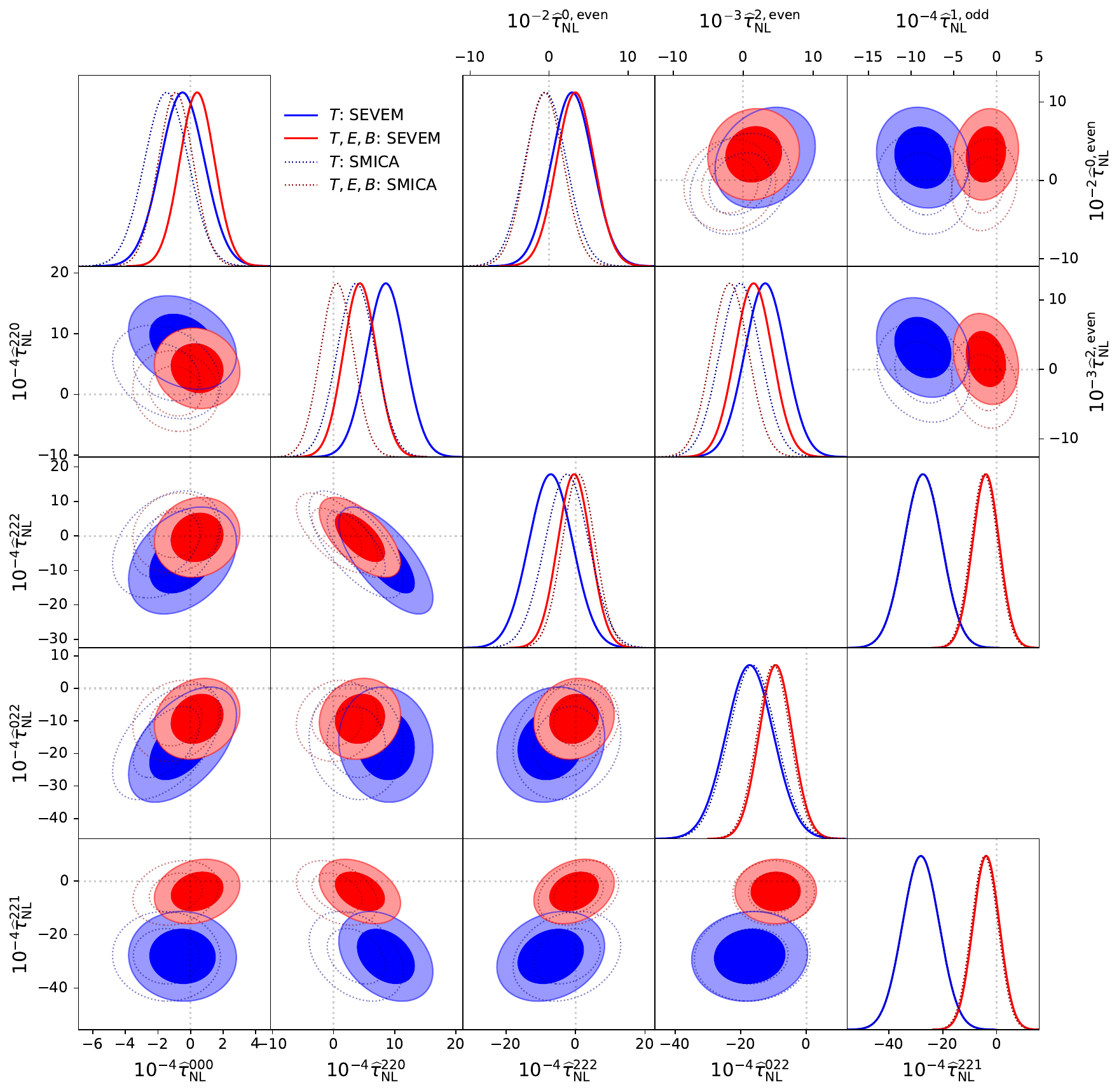}
    \caption{As Fig.\,\ref{fig: local-corner}, but displaying the joint constraints on direction-dependent exchange non-Gaussianity. The bottom left panels show the $\tau_{\rm NL}^{n_1n_3n}$ parameters, whilst the 
    top right panels show the $\tau_{\rm NL}^{n,\rm even}$ and $\tau_{\rm NL}^{n,\rm odd}$ parameters, 
    restricting to amplitudes that can be meaningfully and efficiently constrained using CMB data.
    Whilst most constraints are broadly consistent with zero, the temperature-only analyses find negative $\tau_{\rm NL}^{221}$ and $\tau_{\rm NL}^{1,\rm odd}$ (which are more than $99.9\%$ correlated).
    We do not take this as a sign of novel inflationary physics since the deviation vanishes in the joint analysis as well as analyses excluding $L=1$, and these contours do not account for the (likely significant) non-Gaussianity of the posterior. Single-template constraints are given in Tab.\,\ref{tab: tau-direc-results}.}
    \label{fig: direc-corner}
\end{figure}

To explore the correlations between the various direction-dependent templates, we perform a joint analysis of all five shapes, leading to the corner plot shown in Fig.\,\ref{fig: direc-corner} (again assuming a Gaussian likelihood). Some parameters are fairly correlated: the inverse Fisher matrix predicts
\beq
    \mathrm{corr}(\tau_{\rm NL}^{000},\tau_{\rm NL}^{220}) = -0.07, \quad \mathrm{corr}(\tau_{\rm NL}^{000},\tau_{\rm NL}^{222}) = 0.08,\quad \mathrm{corr}(\tau_{\rm NL}^{000},\tau_{\rm NL}^{022}) = 0.36,\quad 
    \mathrm{corr}(\tau_{\rm NL}^{220},\tau_{\rm NL}^{222}) = -0.57,\nonumber
\eeq
for the baseline analyses, with all other correlations below $5\%$. Removing polarization leads to slightly stronger correlations, reaching $-62\%$ for the $\tauest^{220},\tauest^{222}$ pair, though we find $\tauest^{221}$ to be $<5\%$ correlated with the other parameters in all cases. These are broadly consistent with the values obtained from the FFP10 simulations, given the scatter, and imply that (a) we can meaningfully distinguish between templates with different angular dependence and (b) the parity-even (even $n_1+n_3+n$) and parity-odd (odd $n_1+n_3+n$) sectors are largely decoupled, as expected. The joint constraints from the baseline analyses are given by
\beq
    10^{-4}\{\tauest^{000},\tauest^{220},\tauest^{222},\tauest^{022},\tau_{\rm NL}^{221}\} &=& \{0.4\pm1.1,4.3\pm2.7,-0.4\pm4.7,-9.4\pm5.0,-3.9\pm4.9\} \quad (\textsc{sevem})\\\nonumber
    &=&\{-0.9\pm1.0,0.6\pm2.4,0.9\pm4.6,-10.0\pm4.7,-4.4\pm5.0\}\quad(\textsc{smica}),\nonumber
\eeq
which are up to $25\%$ weaker than the single template analyses, though they remain consistent with zero within $2.1\sigma$. From the temperature-only analyses, we find the same anomalously low $\tauest^{221}$ as before; this is expected given the lack of correlation between the parity-even and parity-odd templates. The inclusion of polarization leads to $10-40\%$ improvements in constraining power, with larger improvements seen for the more complex angular dependencies (matching \papertwo), and only a minor gain for $\tau_{\rm NL}^{0,\rm even}$ (as in \S\ref{subsec: results-local}), as discussed in \citep{Kalaja:2020mkq}. 

\subsubsection{Parity-Even \& Parity-Odd Constraints} 
\noindent Constraints on the individual parity-even/parity-odd direction-dependent amplitudes, $\tau_{\rm NL}^{n,\rm even/odd}$, are given in the bottom panel of Tab.\,\ref{tab: tau-direc-results}, with the baseline results
\beq
    \{10^{-2}\tauest^{0,\rm even},10^{-3}\tauest^{2,\rm even},10^{-4}\tauest^{1,\rm odd}\} &=& \{2.9\pm2.3,0.9\pm2.5,-1.2\pm1.4\} \quad (\textsc{sevem})\\\nonumber
    &=&\{-0.1\pm2.2,-1.9\pm2.4,-1.5\pm1.6\}\quad(\textsc{smica}),\nonumber
\eeq
with a maximal $1.3\sigma$ deviation from zero. The constraining power varies significantly with order: $\sigma(\tau_{\rm NL}^{1,\rm odd})\approx 6\sigma(\tau_{\rm NL}^{2,\rm even})\approx 60\sigma(\tau_{\rm NL}^{0,\rm even})$ (with $\tau_{\rm NL}^{0,\rm even}\equiv \taunl$); this is stronger than for $\tau_{\rm NL}^{n_1n_3n}$ due to the different choice of basis.
The \textsc{sevem} and \textsc{smica} constraints differ by up to $1.3\sigma$ (with largest deviation for $\tau_{\rm NL}^{0,\rm even}\equiv \taunl$), though this is not unexpected, as discussed above.

As in \eqref{eq: tau-direc-relations}, the $\tau_{\rm NL}^{n,\rm even/odd}$ templates are closely related to the generalized $\tau_{\rm NL}^{n_1n_3n}$ shapes.\footnote{More specifically, $\tauest^{n,\rm even}$ is a weighted sum of $\widehat{\tau}_{\rm NL}^{nn0}$ and $\widehat{\tau}_{\rm NL}^{0nn}$, whilst $\widehat{\tau}_{\rm NL}^{n,\rm odd}$ involves $\widehat{\tau}_{\rm NL}^{1mm'}, \widehat{\tau}_{\rm NL}^{mm'1}$ with $m,m'=n\pm1$.} Consequently, we observe that the two sets of analyses have similar dependence on polarization, with $10-30\%$ weaker constraints obtained by restricting to temperature alone. The temperature-only constraints on $\tau_{\rm NL}^{1,\rm odd}$ using \textsc{sevem} (\textsc{smica}) lie $4.0\sigma$ ($4.2\sigma$) below zero, with a \resub{$3.0\sigma$} decrement for $\tau_{\rm NL}^{0,\rm even}$ in polarization-only \textsc{sevem} analyses. These are simply a restatement of the above anomalies (given the $99.97\%$ correlation of $\widehat{\tau}_{\rm NL}^{221}$ and $\widehat{\tau}_{\rm NL}^{1,\rm odd}$) and are likely insignificant, given the significant posterior non-Gaussianity and the lack of detection in the combined analysis.

The upper right panels of Fig.\,\ref{fig: direc-corner} show the joint constraints on $\{\tau_{\rm NL}^{0,\rm even},\tau_{\rm NL}^{2,\rm even},\tau_{\rm NL}^{1,\rm odd}\}$ from the \textsc{sevem} and \textsc{smica} analyses. We observe fairly small covariances between the templates, with the theoretical correlations
\beq
    \mathrm{corr}(\tau_{\rm NL}^{0,\rm even},\tau_{\rm NL}^{2,\rm even}) = 0.19,\quad \mathrm{corr}(\tau_{\rm NL}^{0,\rm even},\tau_{\rm NL}^{1,\rm odd}) = -0.002,\quad \mathrm{corr}(\tau_{\rm NL}^{2,\rm even},\tau_{\rm NL}^{1,\rm odd}) = -0.03 \nonumber
\eeq
in the baseline analysis, consistent with the empirical FFP10 values, and demonstrating the approximate decoupling of the parity-even and parity-odd sectors. Due to the weak correlations, the marginalized constraints are similar to those of the single-template analyses (and consistent with zero within $1.4\sigma$):
\beq
    \{10^{-2}\tauest^{0,\rm even},10^{-3}\tauest^{2,\rm even},10^{-4}\tauest^{1,\rm odd}\} &=& \{3.3\pm2.4,1.5\pm2.6,-1.3\pm1.5\} \quad (\textsc{sevem})\\\nonumber
    &=&\{-0.6\pm2.2,-1.9\pm2.4,-1.4\pm1.6\}\quad(\textsc{smica});\nonumber
\eeq
moreover, we again find a low value of $\tau_{\rm NL}^{1,\rm odd}$ from temperature-only analyses. The inclusion of polarization leads to sharper constraints, with $\sigma(\tau_{\rm NL})$ reducing by $\{9\%,19\%,36\%\}$ for $\{\tau_{\rm NL}^{0,\rm even},\tau_{\rm NL}^{2,\rm even},\tau_{\rm NL}^{1,\rm odd}\}$.

\subsubsection{Consistency Tests}\label{subsubsec: results-direc-sys}
\noindent We now present various consistency tests, both to assess the dependence of our constraints on hyperparameter choices and to further explore the mildly anomalous temperature-only constraints. Throughout this section, we will utilize the \textsc{sevem} dataset, and work with the compressed $\tau_{\rm NL}^{n,\rm even/odd}$ parameters. 

First, we test the optimization algorithm.
As in the EFTI analyses, we adopt a two-step procedure, partitioning the starting grid into four pieces for facilitate faster computation.
Switching to a global optimization scheme has a minimal impact on our constraints, with a maximal shift of $0.13\sigma$ in the \textit{Planck} temperature measurements. 
Similarly, changing the optimization tolerance to $f_{\rm thresh}=10^{-3}$ or increasing the number of points in the initial integration grid by $50\%$ yields consistent results within $0.02\sigma$. We find similar insensitivity to the $k$-space integration grid (which \papertwo found to be a key test for the direction-dependent templates). 
Due to the low value of $L_{\rm max}$, our results are only minorly affected by gravitational lensing: we find shifts of $\{\Delta\tau_{\rm NL}^{0,\rm even},\Delta\tau_{\rm NL}^{2,\rm even},\Delta\tau_{\rm NL}^{1,\rm odd}\} = \{60,1900,100\}$, corresponding to $\{0.25\sigma,0.71\sigma,0.01\sigma\}$. We further report that point-sources induce bias below $0.05\sigma$.

Reducing to $N_{\rm fish}=5$ leads to variation of up to $0.14\sigma$ in $\widehat{\tau}_{\rm NL}^{n,\rm even/odd}$, implying that the fiducial ($N_{\rm fish}=10$) analyses are sufficiently converged. Setting $N_{\rm disc}=50$ yields larger shifts, reaching $0.3\sigma$ for the baseline $\widehat{\tau}_{\rm NL}^{1, \rm odd}$ measurement, or $0.8\sigma$ from temperature-alone.\footnote{In the idealized limit, $\widehat{\tau}_{\rm NL}^{n,\rm odd}$ is independent of $N_{\rm disc}$, since the disconnected trispectrum is parity-even and thus uncorrelated with the parity-odd template. Observational effects such as the mask lead to leakage between parity-even and parity-odd modes 
which are here found to be important.
} The temperature-only shift is fairly large, and suggests that residual disconnected components may lead to small biases in the $N_{\rm disc}=100$ posterior. To further understand this, we analyze $50$ FFP10 simulations using $N_{\rm disc}=100$ (noting that we previously used $N_{\rm disc}=50$ for simulations, but $N_{\rm disc}=100$ for data). The mean $\tauest^{1,\rm odd}$ (which primarily traces residual disconnected contributions) shifts from $-0.8\sigma$ to $-0.07\sigma$, indicting that the residual disconnected contributions are subdominant.

Following \S\ref{subsec: local-variants}, we can additionally perform consistency tests by isolating particular features of the data. Repeating the temperature-only analyses using the \textsc{npipe-a/b} splits instead of the full data-vector yields consistent results to within $0.3\sigma$ (or $0.1\sigma$ for the even templates), indicating that bias from the intrinsic noise four-point function is small. This is confirmed by analyzing the difference of the data-splits, which is consistent with zero within 
$<0.1\sigma$. 
The parity-odd results are also stable under removing the simulation calibration factors discussed in \S\ref{subsec: method-data}.

As for $\taunl$, constraints on $\tau_{\rm NL}^{n,\rm even/odd}$ are dominated by the largest scales included in the analysis, \textit{i.e.}\ low $L$. Restricting to $L_{\rm min}=2$ inflates the errorbars by $2-3\times$, and yields a temperature-only measurement of $10^{-4}\tauest^{1,\rm odd} = -2.9\pm6.2$. This is consistent with zero at $0.5\sigma$, indicating that the aforementioned anomaly is contained within the $L=1$ mode. Finally, we can vary the galactic mask as in Fig.\,\ref{fig: local-fsky}. For $\tau_{\rm NL}^{n,\rm even}$, we find consistent temperature-only results for $f_{\rm sky}\lesssim 0.7$, as in \S\ref{subsec: local-variants}, whilst, for $\tau_{\rm NL}^{1,\rm odd}$, we find null results for $f_{\rm sky}\lesssim 0.4$, then a $3\sigma$ jump to negative values for $f_{\rm sky}\gtrsim0.6$.
Our conclusion is that the low value of $\tau_{\rm NL}^{1, \rm odd}$ reported in the temperature channel is likely a statistical fluctuation (given the heavy tail of the posterior, which implies that a $-4\sigma$ result is not highly unlikely) and/or a residual foreground that is not present in the simulations. Since we do not find any deviation in the combined temperature-plus-polarization analysis, we do not consider this a sign of new physics.

\subsection{Collider Non-Gaussianity}\label{subsec: results-coll}

\noindent Finally, we present constraints on the `cosmological collider', searching for the characteristic signatures of massive spinning particle exchange in inflation. As discussed in \paperone, the full collider trispectra are both model-dependent and difficult to derive, thus we here focus on the collapsed limits, which are 
set by symmetry considerations \citep{Arkani-Hamed:2015bza}. 
To isolate these contributions, we restrict the primordial templates (defined in 
Appendix \ref{app: templates}) to trispectrum contributions with $k>k_{\rm coll}$ and $K<K_{\rm coll}$, \textit{i.e.}\ hard external momenta and soft internal momenta (for suitably chosen $k_{\rm coll},K_{\rm coll}$).
The models are described by two characteristic amplitudes, $\tau_{\rm NL}^{\rm light}(s,\nu_s)$ and $\tau_{\rm NL}^{\rm heavy}(s,\mu_s)$, which describe intermediate/light and heavy particles (or more technically, those in the complementary and principal series respectively), depending on the spin and mass parameters $s$ and $\nu_s\equiv i\mu_s$.

In this section, we will constrain the light spin-zero amplitudes with $\nu_0\in\{3/2,1,1/2,0\}$, as well as the spin-1 and spin-2 amplitudes with $\nu_s\in\{1/2,0\}$. We also consider the heavy spin-$s$ templates with $\mu_s\in\{1,2,3\}$ and $s\in\{0,1,2\}$. These ranges are set by the following considerations: (a) the Higuchi bound demands $\nu_s\leq 1/2$ for spin $s>0$, or $\nu_s\leq 3/2$ else \citep{Higuchi:1986py}; (b) light templates with $\nu_s = 0$ are equivalent to heavy templates with $\mu_s = 0$; (c) higher spins have more complex angular dependence that is both more difficult and more expensive to constrain (as in \S\ref{subsec: results-direc}); (d) templates with larger $\mu_s$ have fast oscillations that are difficult to measure in the CMB; (e) less separated values of $\nu_s$ and $\mu_s$ are more correlated and thus harder to distinguish. These parameters include the massless ($\nu_s = 3/2$, relevant only for $s=0$), and conformally coupled ($\nu_s = 0$) limits. Whilst it would be interesting to marginalize over the mass parameters, this is difficult in a template analysis, thus we adopt a frequentist approach, computing both independent and joint constraints on each amplitude of interest. 

Following preliminary Fisher forecasts, we set $k_{\rm coll}=K_{\rm coll}=0.03\,\mathrm{Mpc}^{-1}$, which restricts to quasi-collapsed trispectra with minimal loss of signal-to-noise (across all spins and masses). To further reduce contamination from non-collapsed configurations, we marginalize over the three (equilateral) EFTI templates, and set $L_{\rm max}=\ell_{\rm max}/2=1024$, though note that the constraints are unchanged with $L_{\rm max}=512$.\footnote{As discussed in \papertwo, the collider templates are less peaked at low $L$ compared to the local shapes (except in the $\nu_s\to3/2$ limit), thus we require larger $L_{\rm max}$.}
We adopt the same analysis settings as for the EFTI and direction-dependent analyses.

\subsubsection{Spin-Zero Constraints}\label{subsec: results-coll-spin0}

\begin{table}[!t]
    \begin{tabular}{l||rcl|rcl|rcl||rcl|rcl|rcl}
        & \multicolumn{9}{c||}{\textit{Planck}-\textsc{sevem}} & \multicolumn{9}{c}{\textit{Planck}-\textsc{smica}}\\
        \textbf{Template} & \multicolumn{3}{c|}{$T$} & \multicolumn{3}{c|}{$E,B$} & \multicolumn{3}{c||}{$T,E,B$} & \multicolumn{3}{c|}{$T$} & \multicolumn{3}{c|}{$E,B$} & \multicolumn{3}{c}{$T,E,B$}\\\hline
        \multicolumn{19}{c}{\textit{Spin-0}}\\\hline
$10^{-2}\tau_{\rm NL}^{\rm light}(0,3/2)$ & \ress{2.2}{2.6} & \ress{-23.2}{29.5} & \ressb{2.9}{2.6} & \ress{-0.4}{2.5} & \ress{-85.8}{24.6} & \ressb{-0.0}{2.3}\\
$10^{-4}\tau_{\rm NL}^{\rm light}(0,1)$ & \ress{1.7}{8.9} & \ress{-48.2}{66.1} & \ressb{-0.0}{8.2} & \ress{-4.0}{8.6} & \ress{-182.2}{54.0} & \ressb{-7.5}{7.3}\\
$10^{-6}\tau_{\rm NL}^{\rm light}(0,1/2)$ & \ress{-0.8}{1.9} & \ress{1.5}{9.3} & \ressb{-1.0}{1.4} & \ress{-0.9}{1.9} & \ress{-3.3}{8.1} & \ressb{-1.7}{1.3}\\
$10^{-6}\tau_{\rm NL}^{\rm light}(0,0)$ & \ress{-0.8}{8.9} & \ress{12.4}{26.8} & \ressb{-1.8}{4.8} & \ress{-0.6}{8.8} & \ress{9.8}{24.3} & \ressb{-4.4}{4.7}\\\hline
$10^{-6}\tau_{\rm NL}^{\rm heavy}(0,1)$ & \ress{-13.2}{17.1} & \ress{6.1}{71.8} & \ressb{-5.8}{7.9} & \ress{-11.1}{16.6} & \ress{-4.9}{57.7} & \ressb{-0.3}{7.5}\\
$10^{-6}\tau_{\rm NL}^{\rm heavy}(0,2)$ & \ress{12.1}{21.7} & \ress{-49.6}{48.5} & \ressb{9.5}{9.7} & \ress{7.4}{20.8} & \ress{-30.3}{43.5} & \ressb{4.1}{9.1}\\
$10^{-7}\tau_{\rm NL}^{\rm heavy}(0,3)$ & \ress{-1.2}{2.7} & \ress{3.2}{5.3} & \ressb{-0.8}{1.1} & \ress{-1.2}{2.6} & \ress{2.2}{4.8} & \ressb{-0.4}{1.1}\\\hline
\multicolumn{19}{c}{\textit{Spin-1}}\\\hline
$10^{-7}\tau_{\rm NL}^{\rm light}(1,1/2)$ & \ress{-1.7}{4.5} & \ress{-4.7}{8.8} & \ressb{-0.7}{2.1} & \ress{-2.8}{4.5} & \ress{-4.3}{8.5} & \ressb{-0.8}{2.0}\\
$10^{-7}\tau_{\rm NL}^{\rm light}(1,0)$ & \ress{2.7}{4.8} & \ress{1.7}{10.0} & \ressb{3.3}{2.6} & \ress{3.3}{4.8} & \ress{6.3}{8.8} & \ressb{3.7}{2.4}\\\hline
$10^{-7}\tau_{\rm NL}^{\rm heavy}(1,1)$ & \ress{3.2}{8.2} & \ress{-12.6}{14.9} & \ressb{1.7}{3.3} & \ress{0.9}{8.4} & \ress{-6.3}{14.4} & \ressb{1.6}{3.4}\\
$10^{-7}\tau_{\rm NL}^{\rm heavy}(1,2)$ & \ress{-3.3}{7.9} & \ress{25.2}{13.4} & \ressb{-4.3}{3.3} & \ress{-0.7}{8.0} & \ress{11.2}{12.2} & \ressb{-5.3}{3.3}\\
$10^{-7}\tau_{\rm NL}^{\rm heavy}(1,3)$ & \ress{-0.4}{8.1} & \ress{-9.0}{14.8} & \ressb{4.3}{3.4} & \ress{-2.5}{8.2} & \ress{-0.2}{14.1} & \ressb{5.6}{3.4}\\\hline
\multicolumn{19}{c}{\textit{Spin-2}}\\\hline
$10^{-6}\tau_{\rm NL}^{\rm light}(2,1/2)$ & \ress{3.0}{2.9} & \ress{13.9}{13.4} & \ressb{2.0}{1.8} & \ress{3.2}{2.9} & \ress{12.1}{12.1} & \ressb{1.7}{1.8}\\
$10^{-7}\tau_{\rm NL}^{\rm light}(2,0)$ & \ress{-2.3}{1.7} & \ress{1.2}{5.1} & \ressb{-0.4}{1.1} & \ress{-2.4}{1.6} & \ress{1.7}{4.5} & \ressb{-0.25}{1.00}\\\hline
$10^{-7}\tau_{\rm NL}^{\rm heavy}(2,1)$ & \ress{-3.1}{5.7} & \ress{5.7}{13.7} & \ressb{0.8}{2.9} & \ress{-1.3}{5.7} & \ress{7.3}{12.0} & \ressb{1.6}{2.7}\\
$10^{-7}\tau_{\rm NL}^{\rm heavy}(2,2)$ & \ress{-1.1}{6.4} & \ress{0.5}{16.6} & \ressb{-2.7}{3.0} & \ress{-4.1}{6.2} & \ress{1.1}{13.8} & \ressb{-2.4}{2.8}\\
$10^{-7}\tau_{\rm NL}^{\rm heavy}(2,3)$ & \ress{8.8}{7.1} & \ress{-0.6}{18.4} & \ressb{4.0}{3.6} & \ress{10.0}{6.6} & \ress{2.1}{15.7} & \ressb{3.4}{3.4}
   \end{tabular}
    \caption{\resub{\textit{Planck} PR4 constraints on spin-zero, spin-one, and spin-two cosmological collider non-Gaussianity.} Each row gives the results from a single template analysis of $\tau_{\rm NL}^{\rm light}(s,\nu_s)$ or $\tau_{\rm NL}^{\rm heavy}(s,\mu_s)$, for spin $s$ and mass parameter $\nu_s\equiv i\mu_s$ (in the same format as Tab.\,\ref{tab: local-results}). We include results for a range of mass parameters satisfying physical and computational constraints (including the massless and conformal limits, which correspond to $\nu_0 = 3/2$ and $\nu_s = 0$ respectively). In all cases, we focus on the collapsed limit of the corresponding trispectrum, removing equilateral information by (a) excising scales in Fourier-space, (b) restricting to $L\lesssim \ell/2$, and (c) marginalizing over the equilateral EFTI amplitudes. Across all seventeen analyses, we find no detection of primordial non-Gaussianity, with a maximum detection significance of $1.6\sigma$ in the baseline analyses. These results are visualized in Fig.\,\ref{fig: coll-spin-results}, with joint constraints on the spin-zero (all spin) templates displayed in Fig.\,\ref{fig: coll0-corner} (Fig.\,\ref{fig: coll-spin-corner}).}\label{tab: coll-results}
\end{table}

\noindent First, we present constraints on inflationary spin-zero particles. Analyzing each template independently (though marginalizing over the EFTI shapes), we obtain the bounds given in the first panel of Tab.\,\ref{tab: coll-results}, including the following baseline results for light particles (with mass $m<3H/2$):
\beq
    \{10^{-2}\tauest^{\rm light}(0,3/2),10^{-4}\tauest^{\rm light}(0,1),10^{-6}\tauest^{\rm light}(0,1/2)\} &=& \{2.9\pm2.6,0.0\pm8.2,-1.0\pm1.4\} \quad (\textsc{sevem})\nonumber\\
    &=&\{0.0\pm2.3,-7.5\pm7.3,-1.7\pm1.3\}\quad(\textsc{smica}),\nonumber
\eeq
conformally coupled particles ($m=3H/2$):
\beq
    10^{-6}\tauest^{\rm light}(0,0) &=& -1.8\pm4.8 \quad (\textsc{sevem})\quad = \quad -4.4\pm4.7\quad(\textsc{smica}),\nonumber
\eeq
and heavy particles ($m>3H/2$):
\beq
    \{10^{-6}\tauest^{\rm heavy}(0,1),10^{-6}\tauest^{\rm heavy}(0,2),10^{-7}\tauest^{\rm heavy}(0,3)\} &=& \{-5.8\pm7.9,9.5\pm9.7,-0.8\pm1.1\} \quad (\textsc{sevem})\nonumber\\
    &=&\{-0.3\pm7.5,4.1\pm9.1,-0.4\pm1.1\}\quad(\textsc{smica}).\nonumber
\eeq
We find no evidence for primordial non-Gaussianity, with a maximum deviation of $-1.3\sigma$ for \textsc{smica} $\nu_0 = 1/2$. Furthermore, the \textsc{sevem} and \textsc{smica} results are broadly consistent, with sub-$\sigma$ agreement found for all templates except for massless particles ($\nu_0 = 3/2$, approximately equivalent to $\taunl$),\footnote{The constraints on $\taunl$ and $\tau_{\rm NL}^{\rm heavy}(0,3/2)$ differ slightly due to the scale-cuts imposed on collider templates, marginalization of the EFTI shapes, the larger $L_{\rm max}$, and the smaller number of simulations used to estimate the errorbar.} which shows a $1.1\sigma$ difference, as in \S\ref{subsec: results-local}. The constraining power varies considerably with $\nu_0,\mu_0$; for example, the errorbar on conformally coupled particles is $\approx 2\times10^4$ times weaker than that on massless particles. This occurs since templates with $\nu_0\to 3/2$ exhibit a strong divergence in the collapsed limit (see Appendix \ref{app: templates}), and our shapes are normalized in equilateral regimes.

As discussed in \S\ref{subsec: results-local}\,\&\,\ref{subsec: results-direc}, the distribution of $\widehat{\tau}_{\rm NL}^{\rm loc}$ can be significantly non-Gaussian if the estimator is dominated by low-$L$ modes. For collider signatures with large $\nu_0$, we expect a similar effect (noting that the two estimators coincide at $\nu_0 = 3/2$).
For heavier fields, however, information is sourced by a wider range of scales (as demonstrated in \S\ref{subsec: data-coll-consistency}), thus the degree of non-Gaussianity is significantly lessened. This can be verified using the empirical distributions obtained from the FFP10 simulations. Since the non-Gaussianity is important only for very large $\nu_0$ (where the constraints reprise those of $\taunl$), we do not attempt to derive a non-Gaussian posterior on $\tau_{\rm NL}^{\rm light/heavy}$ in this section, instead quoting only the measured value and FFP10 variance, as above. For both \textsc{smica} and \textsc{sevem}, the empirical errors closely match the theoretical predictions (within $18\%$, with an expected $10\%$ scatter). This indicates that our spin-zero cosmological collider estimators are essentially optimal.

As shown in Tab.\,\ref{tab: coll-results}, the estimators extract significant information from both temperature and polarization. In particular, the temperature-only analyses (all of which are consistent with zero within $0.9\sigma$) find weaker constraints by up to $130\%$ compared to the baseline analyses, with largest deviations found for heavy particles.
This is expected, since templates with $\nu_0<3/2$ are less dependent on large scales, which are noise-dominated in polarization. Removing temperature from the analysis degrades the constraints by a factor of \resub{$4-12$} (with largest effects for $\nu_0=3/2$), and we find some mild anomalies, with 
the \textsc{smica} $\nu_0 = 3/2$ and $\nu_0 = 1$ measurements around $3\sigma$ below zero (with the former measurement reproducing the $\taunl$ result). These effects are not reproduced in \textsc{sevem}, and are likely noise fluctuations given the minimal constraining power of the polarization dataset, the correlations between the two templates, the negative sign, and the likelihood non-Gaussianity.

\begin{figure}[!t]
    \centering
    \includegraphics[width=\linewidth]{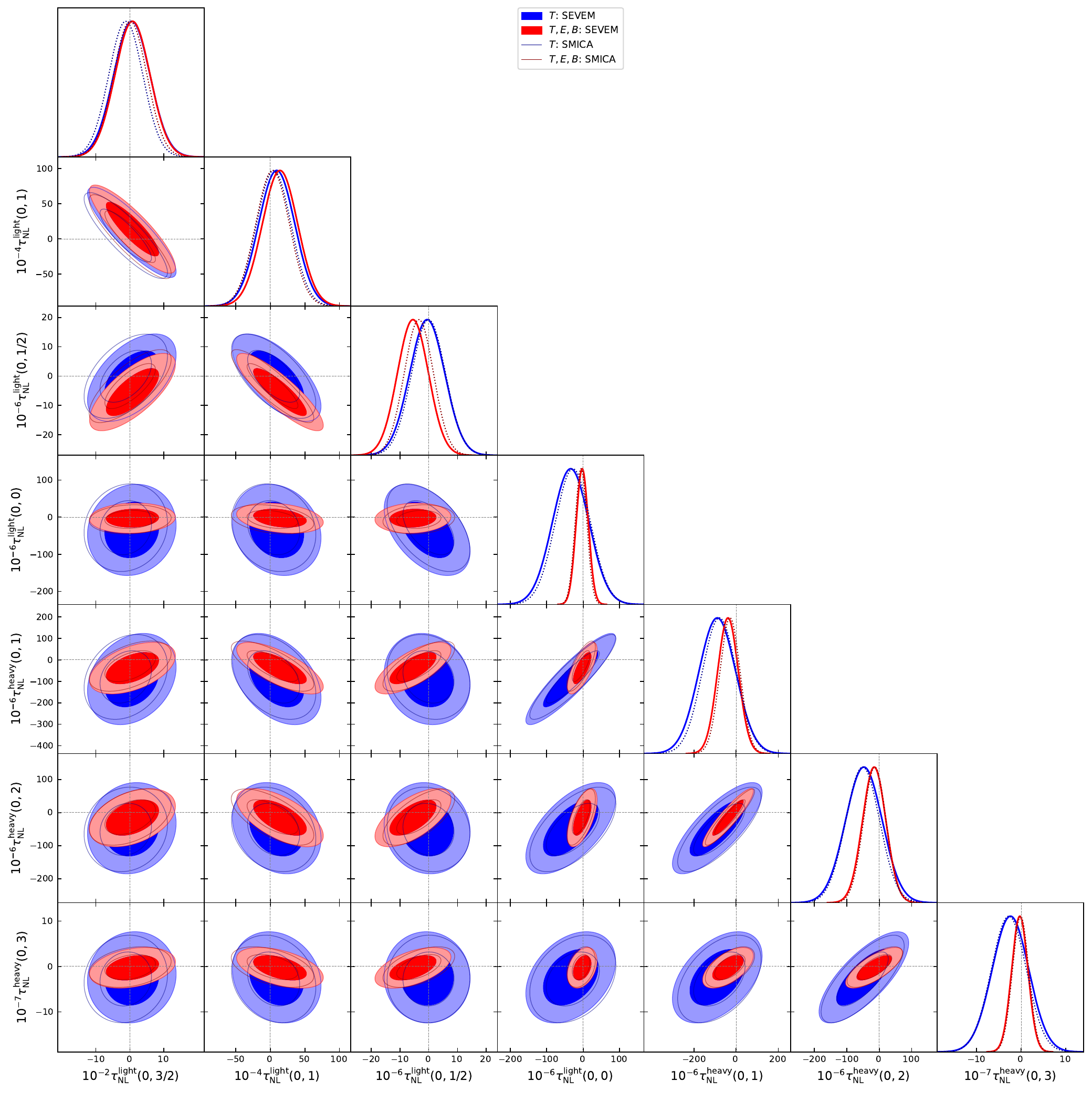}
    \caption{As Fig.\,\ref{fig: local-corner}, but displaying the joint constraints on spin-zero cosmological collider non-Gaussianity (assuming a Gaussian likelihood). \resub{We present results} for seven choices of mass, ranging from massless (first column; $\nu_0 = 3/2$) to conformally coupled (fourth column; $m = 3H/2, \nu_0=0$) to very massive (last column; $m = \sqrt{45}H/2, \mu_0 = 3$), and marginalize over the three EFTI amplitudes in all cases. We find no evidence for spin-zero collider non-Gaussianity, though find fairly strong degeneracies between templates with similar mass parameters. Constraints from \textsc{sevem} and \textsc{smica} are highly consistent, and, for $\nu_0\lesssim1$, are significantly enhanced by including polarization data.}
    \label{fig: coll0-corner}
\end{figure}

Next, we perform a joint analysis of the seven spin-zero amplitudes, displaying the results in Fig.\,\ref{fig: coll0-corner}. Some parameters exhibit fairly strongly correlations, particularly $\nu_0 = 3/2/$ with $\nu_0 = 1$ ($-83\%$), $\nu_0 = 1$ with $\nu_0 = 1/2$ ($-82\%$), and $\mu_0 = 1$ with $\mu_0 = 2$ ($+77\%$). This matches the conclusion of \papertwo; it is difficult to distinguish between exchange templates with similar mass. Due to the degeneracies, the joint parameter constraints are significantly weaker than those in Tab.\,\ref{tab: coll-results}, with the baseline measurements
\beq
    \{10^{-2}\tauest^{\rm light}(0,3/2),10^{-4}\tauest^{\rm light}(0,1),10^{-6}\tauest^{\rm light}(0,1/2)\} &=& \{0.8\pm5.1,14\pm26,-5.5\pm5.4\} \quad (\textsc{sevem})\nonumber\\
    &=&\{0.1\pm4.9,3\pm24,-3.4\pm5.0\}\quad(\textsc{smica}),\nonumber
\eeq
for light particles, 
\beq
    10^{-6}\tauest^{\rm light}(0,0) &=& -3\pm16 \quad (\textsc{sevem})\quad = \quad -6\pm16\quad(\textsc{smica}),\nonumber
\eeq
for conformally coupled particles, and 
\beq
    \{10^{-6}\tauest^{\rm heavy}(0,1),10^{-6}\tauest^{\rm heavy}(0,2),10^{-7}\tauest^{\rm heavy}(0,3)\} &=& \{-39\pm49,-16\pm36,-0.3\pm1.8\} \quad (\textsc{sevem})\nonumber\\
    &=&\{-29\pm47,-14\pm34,-0.1\pm1.8\}\quad(\textsc{smica}).\nonumber
\eeq
for heavy particles. These are inflated by a factor $(1.6-6.3)$, with largest effects seen around $\nu_0 = \mu_0 = 0$. Except for massless particles, the two-dimensional posteriors shrink significantly when polarization is included, with the two-dimensional figures-of-merit improving by up to $4.6\times$ -- this implies that future surveys can significantly enhance cosmological collider constraints. Finally, we note that the heavy-template \textsc{sevem} and \textsc{smica} results are highly consistent (within $0.2\sigma$), indicating that foreground residuals do not project well onto the oscillatory collider templates.

\subsubsection{Higher-Spin Constraints}

\noindent Higher-spin particles can be analyzed similarly, albeit with a fair increase in computational costs. The single-template results are shown in the second and third sections of Tab.\,\ref{tab: coll-results} for spin-one (spin-two) particles respectively. As before, we highlight the baseline constraints on light and conformally-coupled particles ($m \leq (s-1/2)H$):
\beq
    \{10^{-7}\tauest^{\rm light}(1,1/2),10^{-7}\tauest^{\rm light}(1,0)\}  &=& \{-0.7\pm2.1,3.3\pm2.6\} \quad (\textsc{sevem})\nonumber\\
    &=&\{-0.8\pm2.0,3.7\pm2.4\}\quad(\textsc{smica})\nonumber\\
    \{10^{-6}\tauest^{\rm light}(2,1/2),10^{-7}\tauest^{\rm light}(2,0)\}  &=& \{2.0\pm1.8,-0.4\pm1.1\} \quad (\textsc{sevem})\nonumber\\
    &=&\{1.7\pm1.8,-0.3\pm1.0\}\quad(\textsc{smica}),\nonumber
\eeq
and on heavy particles:
\beq
    \{10^{-7}\tauest^{\rm heavy}(1,1),10^{-7}\tauest^{\rm heavy}(1,2),10^{-7}\tauest^{\rm heavy}(1,3)\} &=& \{1.7\pm3.3,-4.3\pm3.3,4.3\pm3.4\} \quad (\textsc{sevem})\nonumber\\
    &=&\{1.6\pm3.4,-5.3\pm3.3,5.6\pm3.4\}\quad(\textsc{smica})\nonumber\\
    \{10^{-7}\tauest^{\rm heavy}(2,1),10^{-7}\tauest^{\rm heavy}(2,2),10^{-7}\tauest^{\rm heavy}(2,3)\} &=& \{0.8\pm2.9,-2.7\pm3.0,4.0\pm3.6\} \quad (\textsc{sevem})\nonumber\\
    &=&\{1.6\pm2.7,-2.4\pm2.8,3.4\pm3.4\}\quad(\textsc{smica}).\nonumber
\eeq
In all cases, we find no evidence for primordial non-Gaussianity, with the ten \textsc{smica} (\textsc{sevem}) amplitudes consistent with zero within $1.3\sigma$  ($1.6\sigma$), with the largest deviations seen for the spin-one $\mu_1 = 3$ template. For all templates, we obtain consistent results for \textsc{sevem} and \textsc{smica} within $0.4\sigma$, indicating that residual foregrounds are subdominant. We further find no evidence for non-Gaussianity in the temperature-only and polarization-only analyses, with a maximum deviation of \resub{$1.9\sigma$} ($1.5\sigma$) for \textsc{sevem} (\textsc{smica}), and $1.3-2.6\times$ larger errors in the former case.

\begin{figure}[!t]
    \centering
    \includegraphics[width=0.9\linewidth]{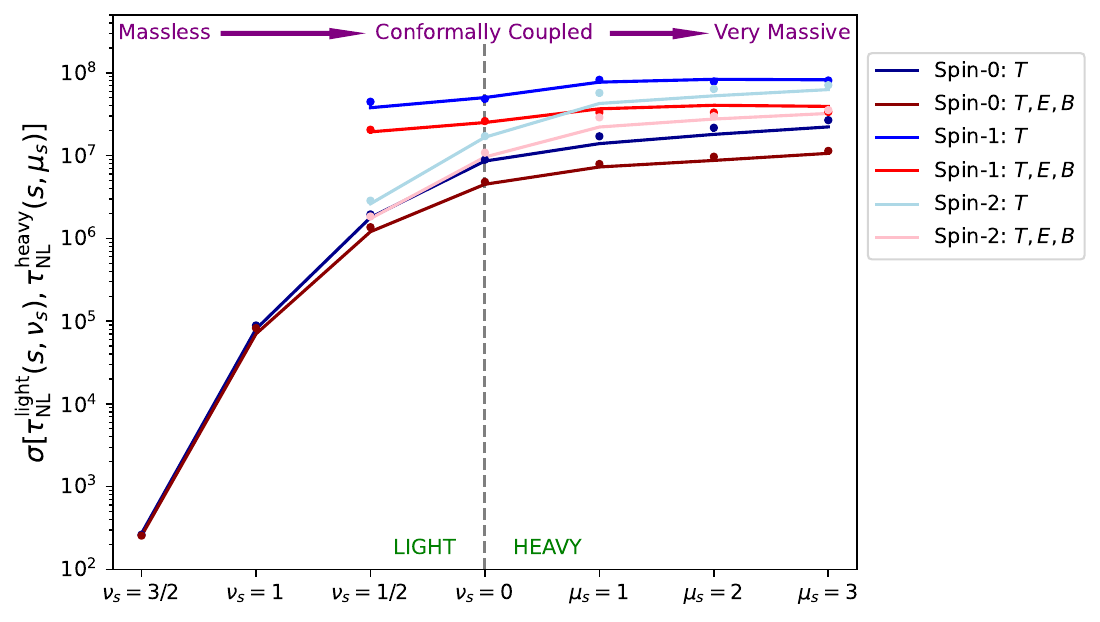}
    \caption{$1\sigma$ errors on the collider non-Gaussianity amplitudes as a function of the mass parameter $\nu_s \equiv i\mu_s$ and spin $s$. Solid lines indicate the expected errorbars on $\tau_{\rm NL}^{\rm light}(s,\nu_s)$ (left) and $\tau_{\rm NL}^{\rm heavy}(s,\mu_s)$ (right) from the inverse Fisher matrices, whilst the points indicate the empirical estimates obtained using 50 FFP10 simulations. We analyze each template individually, but marginalize over the three EFTI parameters in all cases. Dark, medium, and light lines show results for spin-zero, one, and two respectively, with temperature-plus-polarization (temperature-only) \textsc{sevem} results shown in red (blue). We do not include values with $\nu_s<1/2$ for $s>0$, since these violate the Higuchi bound. The individual constraints are tabulated in Tab.\,\ref{tab: coll-results}. Notably, the constraints are a strong function of $\nu_s$ but fairly insensitive to $\mu_s$, and the high-mass constraints are similar for all spins, though the bounds on light spin-one particles are weak. With the exception of massless scalars, polarization significantly sharpens all constraints.}
    \label{fig: coll-spin-results}
\end{figure}

To facilitate comparison between the various collider constraints, we plot $\sigma(\tau_{\rm NL})$ as a function of spin and mass parameter in Fig.\,\ref{fig: coll-spin-results}. We find good agreement between the empirical and theoretical errorbars (within $20\%$, except for a $31\%$ offset for $s=2, \mu_2 = 1$, with an expected $10\%$ scatter), indicating that our estimators are close to optimal, regardless of spin. As noted above, the spin-zero constraints exhibit strong dependence on $\nu_s$; for higher-spins, this is more subtle, since unitarity restricts us to $\nu_s\leq 1/2$ (though see \citep{Bordin:2019tyb} for a counterexample). For spin-one particles, we find only minor dependence on mass (with at most a factor of $1.7$ variation), whilst for spin-two particles, the variation is larger, with a $5.9\times$ weaker ($3.3\times$ stronger) bound on $\mu_2=3$ ($\nu_2=1/2$) compared to the conformal limit of $\nu_2 = 0$. 
Switching from $s=0$ to $s=2$, constraints weaken by a factor between $1.5$ (low-mass) and $3$ (high-mass); in contrast, switching from $s=0$ to $s=1$ weakens constraints by a factor of $16$ (low-mass) to $4$ (high-mass). This is expected since models with large $\nu_s$ are dominated by highly squeezed configurations, and the collapsed limit is suppressed by a factor of $K^2/k_1k_3$ if $s$ is odd \citep{Arkani-Hamed:2015bza} (as forecast previously \citep{Philcox4pt2,Bordin:2019tyb,Kalaja:2020mkq,MoradinezhadDizgah:2018ssw}).

\begin{figure}[!t]
    \centering
    \includegraphics[width=\linewidth]{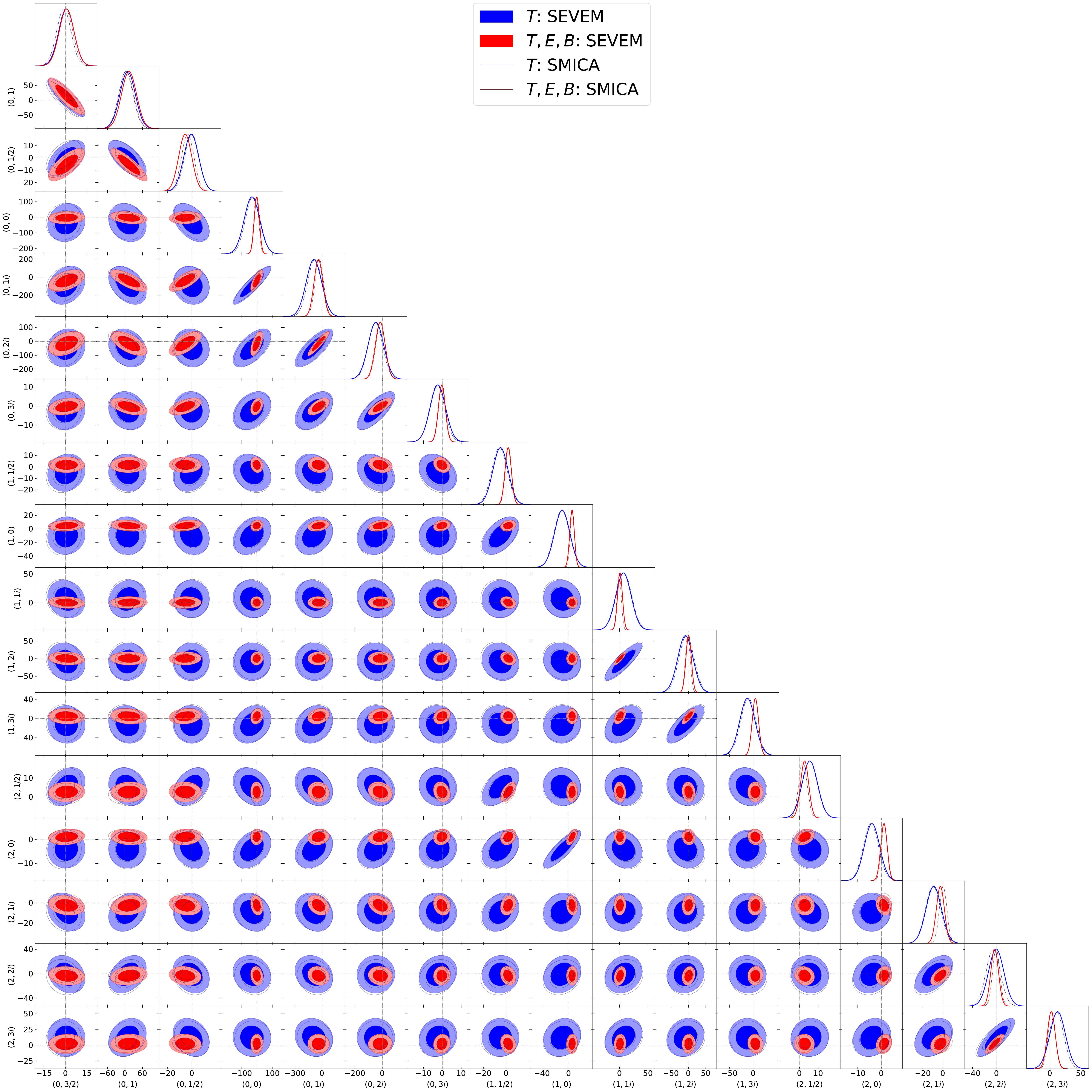}
    \caption{As Fig.\,\ref{fig: coll0-corner}, but displaying the joint constraints on cosmological collider non-Gaussianity for spin $s\in\{0,1,2\}$. We show results for each of the seventeen $\tau_{\rm NL}^{\rm light/heavy}$ amplitudes analyzed in this work (marginalized over $\gnldotdot,\gnldotdel,\gnldeldel$), with the labels indicating the value of $(s,\nu_s)$, with complex $\nu_s$ indicating heavy particles. As for the spin-zero templates, we find generally good agreement between \textsc{sevem} and \textsc{smica}. Most analyses strongly benefit from the inclusion of polarization information, particularly large with large $s$ and/or large masses. Despite the large number of templates, most of the correlations are fairly weak (see Fig.\,\ref{fig: coll_spin_correlation}, such that the various regimes can be independently constrained from data.}
    \label{fig: coll-spin-corner}
\end{figure}

\resub{In Fig.\,\ref{fig: coll-spin-corner},} we present joint constraints on collider non-Gaussianity from spins zero, one, and two. Due to the large number of templates considered in this work, this is a fairly formidable plot: the major messages are (a) we find no evidence for primordial non-Gaussianity in the joint analysis, with a maximum detection significance of $1.7\sigma$ (for $s=1, \nu_1 = 0$), (b) results from \textsc{sevem} and \textsc{smica} are highly consistent, as discussed above, (c) including polarization leads to much sharper two-dimensional constraints (increasing the two-dimensional figure-of-merit by up to $6.6\times$), and (d) \resub{some templates} exhibit fairly strong correlations. The final point implies that the joint $\tau_{\rm NL}$ constraints will be significantly weaker than the single-template analyses: indeed, we find degradation ranging from $1.4\times$ to $3.7\times$, with largest effects found for heavy spin-one templates. 

\begin{figure}[!t]
    \centering
    \includegraphics[width=0.8\linewidth]{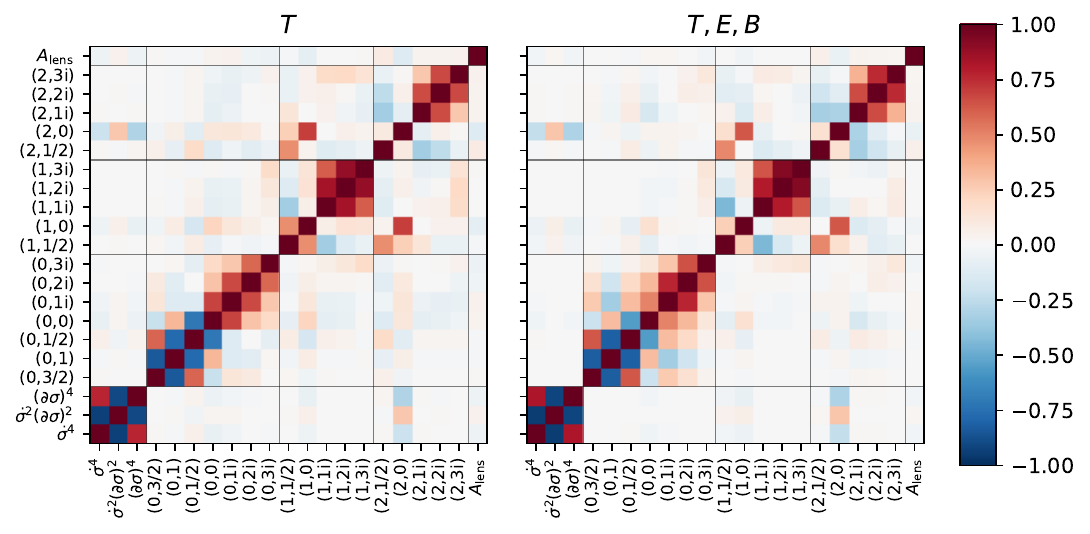}
    \caption{Theoretical correlation matrix for the cosmological collider analysis. We show the normalized covariances between the three EFTI parameters (leftmost columns), the seventeen collider non-Gaussianity amplitudes (middle columns), and the CMB lensing amplitude (right column), all obtained from the inverse Fisher matrix. The labels indicate the template with collider shapes specified by $(s,\nu_s)$ as in Fig.\,\ref{fig: coll-spin-corner}. The left and right plots show results for the temperature-only and baseline analyses respectively. We find significant correlations between templates with similar mass parameters (close to the diagonal), but only weak correlations between templates with different spin. Although correlations with lensing are weak, they can severely bias the analysis if not accounted for.}
    \label{fig: coll_spin_correlation}
\end{figure}

In Fig.\,\ref{fig: coll_spin_correlation}, we plot the theoretical correlation matrices between all collider templates, the EFTI shapes, and CMB lensing. We find an analogous correlation structure for both the temperature-only and baseline analyses, dominated by correlations between templates with similar mass parameters (close to the diagonal).
At low masses, the various spin-one and spin-two templates are weakly correlated, but we find significant degeneracies at large $\mu_s$, with correlations up to $92\%$ between $\mu_1 = 2$ and $\mu_1=3$. This leads to the significant broadening of $\sigma(\tau_{\rm NL})$ found above, and implies that particles with even higher masses will be challenging to constrain. The EFTI parameters are only weakly correlated with the collider templates (with largest overlap found for the spin-two conformally-coupled correlator); this suggests that our estimators extract little information from close-to-equilateral regimes (as desired). 
In general, we find only weak correlations between the different spins; this justifies our approach, since it demonstrates that the higher spin analyses are adding new information, rather than just reprocessing that present in spin-zero correlators. 

\subsubsection{Consistency Tests}\label{subsec: data-coll-consistency}
\noindent As in previous sections, we conclude by assessing the dependence of the measurements on our analysis choices. Given the increased complexity of these analyses, we perform only a limited set of tests: further validation can be found in \papertwo. First, we note good convergence with $N_{\rm fish}$: reducing to $N_{\rm fish}=5$ changes results by at most $0.15\sigma$. For $N_{\rm disc}$, we observe that the mean of the 50 FFP10 measurements is consistent with zero within $0.5\sigma$ for all templates; these are generated using $N_{\rm disc}=50$ and provide an upper bound for the Monte Carlo error in the real data analyses (which use $N_{\rm disc}=100$). Despite the weak correlations seen in Fig.\,\ref{fig: coll_spin_correlation}, we find that gravitational lensing induces large distortions to our collider measurements; after marginalizing over EFTI parameters, the spin-zero and spin-two shifts can exceed $3\sigma$ (reaching (-)$5.1\sigma$ for $s=2$ and $\nu_s = 1/2$). This is an important systematic that must be accounted for in all current and future analyses.

Finally, we assess the dependence of our constraints on the maximum internal mode, $L_{\rm max}$. Reducing from the fiducial $L_{\rm max}=1024$ to $L_{\rm max}=512$ leads to negligible variation in the constraints: this is due to our physical $K$-space restrictions, which null any contributions to the estimator from $L\gtrsim 400$. Setting $L_{\rm max}=256$ leads to fairly significant loss of constraining power; whilst the massless limit is unchanged (as in the $\taunl$ analysis), constraints on the most massive particles degrade by up to $\approx 3\times$ (or $\approx 2\times$ when excluding spin-one templates). This matches the conclusion of \papertwo, and implies that the $\tau_{\rm NL}^{\rm heavy}$ estimators extract significant information from only partially-collapsed triangles. This is expected given the suppressed $K\to 0$ limit of spin-one and high-mass templates, and does not indicate a breakdown of our model, since the $K$-space cuts practically restrict to $L\lesssim \ell/4$.

\section{Discussion}\label{sec: discussion}
\noindent In \S\ref{sec: results}, we presented constraints on various models of primordial and late-time non-Gaussianity using \textit{Planck} PR4 data. Below, we compare these to previous bounds, both from the CMB and other observables, which demonstrate the strengths of our approach. 
As discussed in \paperone, the template amplitudes map to a variety of inflationary models: using our results, we will place constraints on many such scenarios. 

\subsection{Local and Contact Non-Gaussianity}

\begin{figure}[!t]
    \centering
    \includegraphics[width=\linewidth]{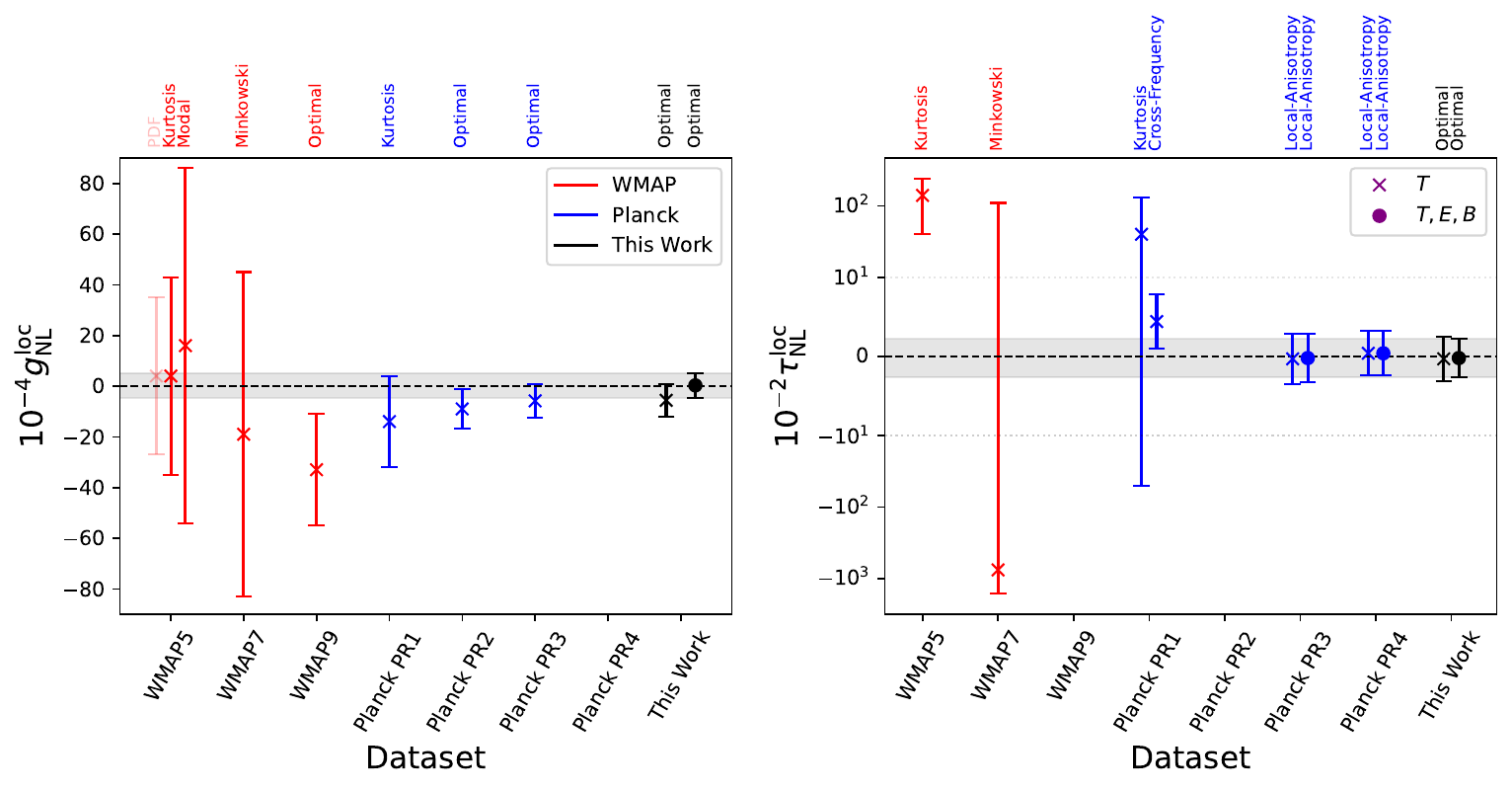}
    \caption{Literature constraints on local non-Gaussianity parameters obtained using WMAP (red) and \textit{Planck} (blue) data, alongside the main results presented in this work (black, adopting the \textsc{smica} component-separation pipeline). Crosses and circles represent results from temperature-only and temperature-plus-polarization analyses respectively, and we plot the measured amplitudes and simulated errorbars in all cases. The top labels indicate the analysis method: PDFs of the temperature maps, kurtosis power spectra, modal decompositions, cross-frequency estimators, local-anisotropy (`quadratic') estimators, and (quasi-)optimal quartic estimators. 
    In the $\taunl$ case, we switch to logarithmic axes for $10^{-2}|\taunl|>100$. Further discussion and references can be found in the text.}
    \label{fig: local-literature}
\end{figure}

\subsubsection{\texorpdfstring{CMB Constraints on $\gnl$}{CMB Constraints on Cubic Local Non-Gaussianity}}
\noindent Since the early WMAP analyses \citep{WMAP:2006bqn}, experiments have placed increasingly precise bounds on the cubic non-Gaussianity parameter $\gnl$, as shown in Fig.\,\ref{fig: local-literature}.
Using WMAP5, \citep{Vielva:2009jz} reported $10^{-4}\gest^{\rm loc} = 4\pm31$ from the $N_{\rm side}=32$ temperature PDF (which is significantly tighter than the theoretical Fisher bound), whilst \citep{Smidt:2010ra} found $10^{-4}\gest^{\rm loc} = 4\pm39$ using a kurtosis power spectrum estimator \citep{Munshi:2009wy} at $\ell_{\rm max}=600$,\footnote{As noted in \citep{Fergusson:2010gn}, this analysis omits a number of systematic effects such as anisotropic noise.}  and \citep{Fergusson:2010gn} applied modal trispectrum estimators to obtain $10^{-4}\gest^{\rm loc} = 16 \pm 70$ at $\ell_{\rm max}=500$, within $20\%$ of the Fisher bound.
From WMAP7, \citep{Hikage:2012bs} constrained $\gnl$ using Minkowski functionals, obtaining $10^{-4}\gest^{\rm loc} = -19\pm64$; due to the choice of statistic, this is a fairly weak constraint. The first application of optimal template estimators was in \citep{Sekiguchi:2013hza}, which obtained $10^{-4}\gest^{\rm loc} = -33\pm22$ from WMAP9 at $\ell_{\rm max}=1024$, including a full treatment of the mask, inhomogeneous noise and weighting; an independent analysis by \citep{2015arXiv150200635S} found a similar constraint: $10^{-4}\gest^{\rm loc} = -38\pm22$.

Due to its higher-resolution, \textit{Planck} data can place significantly stronger bounds on $\gnl$ than WMAP data (with \citep{Sekiguchi:2013hza} forecasting $10^{-4}\sigma(\gnl) = 6.7$ in an optimal analysis). Furthermore, tighter constraints are possible if one includes polarization: despite the discussion in \citep{Munshi:2010bh}, this combination has not been analyzed in any previous works.
Using \textit{Planck} PR1 data at $\ell_{\rm max}=1000$, \citep{Feng:2015pva} found $10^{-4}\gest^{\rm loc} = -14\pm18$ using the kurtosis estimator of \citep{Munshi:2009wy}, which represents a significant improvement with respect to the analogous WMAP5 analysis. Based on the optimal pipeline discussed in \citep{2015arXiv150200635S}, the official \textit{Planck} PR2 and PR3 analyses found $10^{-4}\gest^{\rm loc} = -9.0\pm7.7$ and $-5.8\pm6.5$ respectively, working at $\ell_{\rm max}=1600$ and adopting the \textsc{smica} component-separation pipeline \citep{Planck:2015zfm,Planck:2019kim}. Analyzing PR4 data at $\ell_{\rm max}=2048$, we have obtained a temperature-only \textsc{smica} measurement of $10^{-4}\gest^{\rm loc} = -5.6\pm6.5$, which sharpen to $0.3\pm4.8$ with polarization. Our temperature constraints are in excellent agreement with the \textit{Planck} PR3 results, validating our analysis pipeline;
our polarization constraints represent the tightest bounds on $\gnl$ obtained to date, by $\approx 35\%$ (cf.\,Fig.\,\ref{fig: local-literature}).

\subsubsection{\texorpdfstring{CMB Constraints on $\taunl$}{CMB Constraints on Quadratic Local Non-Gaussianity}}
\noindent Analyses of $\taunl$ a have a longer history than those of $\gnl$, stretching back to \citep{Kunz:2001ym}, which probed exchange trispectra using COBE-DMR data at $\ell_{\rm max}=20$, yielding a constraining power of $10^{-2}\sigma(\taunl)\sim 10^6$ \cite[cf.][]{Smidt:2010ra}.\footnote{Throughout this section, we use the same scaling factors as in \S\ref{sec: results} for ease of comparison, leading to the somewhat inconvenient units.}
More recently (see Fig.\,\ref{fig: local-literature}), \citep{Smidt:2010ra} analyzed the WMAP5 dataset using kurtosis estimators at $\ell_{\rm max}=600$, finding $10^{-2}\tauest^{\rm loc} = 140\pm100$; 
this was refined to $40\pm90$ using frequency cross-spectra from \textit{Planck} PR1 in \citep{Feng:2015pva}, working at $\ell_{\rm max}=1000$ and performing a joint analysis with $\gnl$. Moreover, \citep{Hikage:2012bs} obtained $10^{-2}\tauest^{\rm loc} = -760\pm 870$ by analyzing WMAP7 data with Minkowski functionals; due to the suboptimal analysis method, this bound is comparatively weak.

The \textit{Planck} PR1 non-Gaussianity paper analyzed the trispectrum using local anisotropy estimators \citep[e.g.,][]{Hanson:2009gu}:\footnote{As shown in \paperone, these are approximately equivalent to the optimal estimator used in this work.} this found $10^{-2}\tauest^{\rm loc} = 4.4\pm3.4$ from temperature alone using $L_{\rm max}=10$ \citep{2014A&A...571A..24P}. Although not featured in the official \textit{Planck} PR2 or PR3 papers, $\taunl$ was recently reanalyzed using PR3 and PR4 data in \citep{Marzouk:2022utf}, using a similar approach to \citep{2014A&A...571A..24P}. This led to the temperature-only \textsc{smica} results $10^{-2}\tauest^{\rm loc} = -0.3\pm3.2$ (PR3) and $0.4\pm2.8$ (PR4), with the inclusion of polarization yielding $-0.2\pm3.1$ (PR3) and $0.4\pm2.8$ (PR4). In this work, we obtained the \textsc{smica} temperature-only result $10^{-2}\tauest^{\rm loc} = -0.3\pm2.8$, and a temperature-plus-polarization constraint $-0.2\pm2.4$ (cf.\,Tab.\,\ref{tab: local-results}). Our temperature-only results are similar to those of \citep{Marzouk:2022utf}, though we find somewhat stronger constraints (by $\approx 15\%$) when including polarization (as in Fig.\,\ref{fig: local-literature}). This could be caused by differences in the estimator (including, for example, auto-spectra, finite-recombination effects and the full mask-dependent normalization factor).

As discussed in \S\ref{subsec: results-local}, most analyses quote a $95\%$ upper bound on $\taunl$ in addition to the measured value and $1\sigma$ error. 
This requires an $L$-by-$L$ analysis, and leads to the \textsc{smica} temperature-only $95\%$ bounds $\taunl<2800$, $\taunl<2800$ and $\taunl<1900$ from \textit{Planck} PR1, PR3, and PR4 respectively \citep{2014A&A...571A..24P,Marzouk:2022utf}. Adding polarization tightens the latter two to $\taunl<2300$ and $\taunl<1700$ respectively, assuming $L_{\rm max}=50$ in all cases. In Appendix \ref{app: tauNL-Lmax}, we found $\taunl<1700$ from the \textsc{smica} PR4 temperature dataset at $L_{\rm max}=30$, or $\taunl<1500$ with polarization. Whilst these are somewhat stronger than the bounds of \citep{Marzouk:2022utf}, we caution that they are strongly affected by noise fluctuations (with the \textsc{sevem} bounds weaker by almost $50\%$, due to an upwards fluctuation in $\widehat{\tau}_{\rm NL}^{\rm loc}$). Finally, we note that the individual $\tauest^{\rm loc}(L)$ results are very similar between ours and previous analyses, featuring the same weak excess of power at $L=3$.

\subsubsection{\texorpdfstring{CMB Constraints on $g_{\rm NL}^{\rm con}$}{CMB Constraints on Constant Non-Gaussianity}}
\noindent 
The constant template parametrized by $g_{\rm NL}^{\rm con}$ has been the subject of limited previous study. In the modal trispectrum analyses of \citep{Fergusson:2010gn}, the authors constrained the $t_{\rm NL}^{\rm const}$ parameter, equivalent to our $g_{\rm NL}^{\rm con}$ up to slow-roll effects, finding $10^{-5}t_{\rm NL}^{\rm const} = -13\pm36$ from WMAP5 data at $\ell_{\rm max}=500$, consistent with the (simplified) Fisher expectation within $30-40\%$. In \S\ref{subsec: data-con}, we obtained the temperature-only measurement $10^{-5}\gest^{\rm con}=-18\pm11$ (from \textsc{sevem}), which tightens to $-4.3\pm5.2$ when including polarization. These constraints are a factor of $3.3$ and $7.0$ tighter than previous bounds, which reflects the significant increase in resolution (with \textit{Planck} being signal-dominated up to $\ell\approx 1600$), and the utility of polarization.


\subsubsection{Constraints from Other Probes}
\noindent Primordial non-Gaussianity leaves signatures in a wide variety of observables, not just the CMB. Notable examples include the halo mass function and galaxy bias, which undergo predictable distortions in the presence of $\gnl$- and $\taunl$-type non-Gaussianity. This has been discussed in a number of works \citep[e.g.,][]{Desjacques:2009jb,Smith:2011ub,LoVerde:2011iz,Jeong:2009vd} (see also \citep{Chongchitnan:2010xz} for a discussion of void statistics). In \citep{Desjacques:2009jb}, the scale-dependent bias effect was used to obtain the $95\%$ constraint $-35<10^{-4}\gnl<82$ from SDSS observations (mainly quasars), building on the $\fnl$ constraints reported in \citep{Slosar:2008hx}. A later analysis of SDSS quasars found $-27<10^{-4}\gnl<19$ \citep{Leistedt:2014zqa}, and \citep{Giannantonio:2013uqa} obtained $-56<10^{-4}\gnl<51$ at $95\%$ by considering the correlations of BOSS galaxies and the integrated Sachs-Wolfe effect. These results are roughly consistent with the WMAP constraints discussed above, and are expected to improve significantly in the future: \citep{Ferraro:2014jba} (see also \citep{Desjacques:2009jb}) predict that a $25h^{-3}\mathrm{Gpc}^3$ survey could reach $10^{-4}\sigma(\gnl) \approx 10^{-2}\sigma(\taunl) \approx 10$.

Beyond scale-dependent bias, galaxy clustering statistics can be used as a detailed probe of primordial non-Gaussianity. Whilst this has been demonstrated theoretically \citep[e.g.,][]{Jeong:2009vd,Biagetti:2012xy,Assassi:2015fma,Assassi:2015jqa}, constraints on $\gnl$ and $\taunl$ from the full shape of galaxy correlators have yet to be obtained. That said, recent work has placed competitive constraints on $\fnl$ \citep{Cabass:2022ymb,DAmico:2022gki}, and studies have forecast the use of non-perturbative methods to measure $\taunl$ from dark matter and weak lensing \citep{Goldstein:2024bky}. We can additionally cross-correlate with other probes: for example, \citep{AnilKumar:2022flx} demonstrated that $10^{-2}\sigma(\taunl)<1$ could be obtained by combining DESI galaxies with kinematic Sunyaev-Zeldovich measurements from the Simons Observatory. 

More futuristically, primordial $\mu$-distortions could probe $\gnl$ and $\taunl$ on small-scales, with \citep{Bartolo:2015fqz} suggesting constraints of $10^{-4}\sigma(\gnl) = 4\times 10^{-5}, 10^{-2}\sigma(\taunl) = 0.4$ from a cosmic-variance-limited $TT\mu$ bispectrum analysis, though constraints from current and near-future surveys will be many orders of magnitude weaker (\citep[cf.][]{Khatri:2015tla}, which found $10^{-2}\taunl<1.4\times 10^9$ from \textit{Planck} $\mu\mu$ spectra). Finally, `dark ages' neutral hydrogen surveys may prove the ultimate probe of local non-Gaussianity due to their large dynamic range and linear evolution: \citep{Floss:2022grj} predicted $10^{-4}\sigma(\gnl) = 2\times 10^{-6}$, $10^{-2}\sigma(\taunl) = 2\times 10^{-9}$ from a cosmic-variance limited experiment up to $k_{\rm max}=300\hMpc$ at $z=30-100$. 

From these results, it is clear that the bounds on $\gnl$ and $\taunl$ presented in this work are much tighter than those currently possible with other probes. That said, galaxy surveys could soon overtake the CMB as the dominant probe of primordial physics \citep[cf.][]{Sailer:2021yzm,Cabass:2022epm}, and, in the distant future, the crown may be given to other observables, such as spectral distortions and 21cm surveys.

-

\subsubsection{Implications}
\noindent The $\fnl$, $\gnl$ and $\taunl$ parameters encode local distortions of a Gaussian field and could be created by a wide variety of primordial effects, such as modulated reheating, curvatons, bouncing cosmologies, and beyond (see \paperone). Below, we review a selection of these scenarios in light of our measurements, coupled with the \textit{Planck} bounds on $\fnl$. To facilitate comparison with previous works, we will assume the \textsc{smica} component-separation method where necessary.

A popular generator of large non-Gaussianity is the `curvaton' scenario \citep[e.g.,][]{Bartolo:2003jx,Sasaki:2006kq}, in which an additional light field, subdominant during inflation, sources late-time density perturbations instead of the inflaton. Defining $r_D \equiv 3\rho_\chi/[\rho_\chi+4\rho_{\rm rad}]$ as the fractional energy density of the curvaton at the time of its decay into radiation, the simplest adiabatic models predict
\beq\label{eq: curvaton}
    \fnl &=& \frac{5}{6}\left(\frac{3}{2r_D}-2-r_D\right), \qquad 
    \gnl = \frac{25}{54}\left(-\frac{9}{r_D}+\frac{1}{2}+10r_D+3r_D^2\right), \qquad \taunl=\left(\frac{3}{2r_D}-2-r_D\right)^2,
\eeq
with $0<r_D<1$ \citep{Bartolo:2003jx,Sasaki:2006kq}. This has been constrained in many previous analyses: \citep{Planck:2019kim} found $r_D\geq 0.22$ ($r_D\geq 0.19$) using \textit{Planck} PR3 temperature-plus-polarization (temperature-only) bispectra at $95\%$ CL. Our $\gnl$ and $\taunl$ results indicate $r_D\geq 0.10$ ($r_D\geq 0.09$); this is driven entirely by $\taunl$ (with $r_D\geq 0.05$ from $\gnl$ alone, reproducing the uniform prior \citep[cf.][]{Planck:2019kim}) and is not competitive with the $\fnl$ constraint -- this is expected given the asymptotic relation $\gnl \approx -(10/3)\fnl$. Modifications to this scenario can lead to larger $\gnl$ as discussed in \citep[e.g.,][]{Byrnes:2010em,Huang:2013yla}.

Local non-Gaussianity can also be generated in non-inflationary settings, such as the ekpyrotic/cyclic scenario. Whilst the simplest ekpyrotic conversion models were already ruled out by \textit{Planck} PR1 \citep{2014A&A...571A..24P}, the kinetic conversion model remains feasible, as emphasized in \citep{Lehners:2013cka}. This predicts 
\beq\label{eq: ekpyrosis-model}
    \fnl = \pm 5+ \frac{3}{2}\kappa_3\sqrt{\epsilon},\qquad  \gnl = \left(-40+\frac{5}{3}\kappa_4+\frac{5}{4}\kappa_3^2\right)\epsilon, \qquad \taunl = \left(\frac{6}{5}\fnl\right)^2,
\eeq
where $\kappa_{3,4},\epsilon$ encode the potential of the two scalar fields during the ekpyrotic phase, with $\epsilon\gtrsim 50$ required to attain ekpyrosis. Assuming $\epsilon=100$, we can obtain constraints on $\kappa_3$ and $\kappa_4$ by combining the PR3 bispectrum results with our trispectrum constraints: from temperature-alone, we find $-1.1<\kappa_3<0.38$, $-1073<\kappa_4<453$, assuming a $+$ sign for $\fnl$, or $-0.44<\kappa_3<1.05$, $-1080<\kappa_4<454$ with a $-$ sign. Adding polarization, these tighten to $-1.07<\kappa_3<0.28$, $-522<\kappa_4<602$ and $-0.40<\kappa_3<0.94$, $-520<\kappa_4<610$ respectively. The $\kappa_3$ constraints match the $\fnl$-only analyses of \citep{Planck:2019kim}, with the $\kappa_4$ far larger than the theoretical estimate of $\kappa_4\approx 4$ \citep{Lehners:2013cka}. Given the $\fnl$ bounds, the above model predicts $\gnl \approx -35\epsilon<-1700$, which will require far higher-precision data to probe. The same is true for non-minimally coupled ekpyrosis models, which predict $-1000\lesssim \gnl\lesssim -100$ \citep{Fertig:2015ola}.

As discussed in \citep{Suyama:2013nva}, light scalar fields in inflation present a plethora of mechanisms to produce $\fnl$, $\gnl$ and $\taunl$.\footnote{One can also generate non-Gaussianity using vector fields \citep[e.g.,][]{Dimopoulos:2006ms}. As discussed in \citep{Valenzuela-Toledo:2009bzd}, this sources $\taunl\lesssim 800(g_\zeta/0.01)^2$, where $g_\zeta$ is the power-spectrum anisotropy parameter. This is constrained by \textit{Planck}'s non-detection of $g_\zeta$ however \citep{Planck:2019evm}.} In many scenarios, such as modulated reheating, modulated trapping, hybrid and thermal inflation with an inhomogeneous ending, velocity modulation, and beyond, one produces similar values of $\fnl$ and $\gnl$. In these cases, a joint analysis of the bispectrum and trispectrum will negligibly improve errorbars compared to a bispectrum-only analysis due to the four-point function's significant cosmic variance penalty. Other models can produce $|\gnl|\gg |\fnl|$: these include variations of the light-field models discussed in \citep{Suyama:2013nva} and the general supersymmetric model analyzed in \S\ref{subsec: discussion-efti}. Whilst we do not present detailed analysis of each scenario herein (since each depends on a number of free parameters, and several require considerable fine-tuning), we note that the tight $\gnl$ constraints of this work significantly reduce the viable parameter-space of such models.

Under a general set of assumptions, the Suyama-Yamaguchi inequality implies that $\taunl \geq (\tfrac{6}{5}\fnl)^2$ \citep{Suyama:2007bg,Smith:2011if}. This can be simply understood by considering the curvature fluctuation, $\zeta_{\rm G}$, to be modulated by a light scalar field, $\sigma$:
\beq\label{eq: scalar-modulation}
    \zeta_{\rm G}(\vx)\to\zeta_{\rm G}(\vx)\left[1+A_{\sigma}\sigma(\vx)\right]\zeta_{\rm G}(\vx).
\eeq
Assuming that $\sigma$ and $\zeta_{\rm G}$ have the same power spectrum, we find:
\beq
    \fnl = \frac{5}{3}A_\sigma r_\sigma, \qquad \taunl = 2A_\sigma^2[1+r_\sigma^2] = \left(\frac{6}{5}\fnl\right)^2\frac{1+r_\sigma^2}{2r_\sigma^2}, \geq \left(\frac{6}{5}\fnl\right)^2,
\eeq
where $r_\sigma\in[0,1]$ is the correlation coefficient. As such, $\fnl$ can be sourced by light fields correlated to $\zeta$, whilst $\taunl$ is additionally sensitive to uncorrelated fields. Using this inequality, we can form a number of additional constraints:
\begin{itemize}
    \item Any measurement of $\widehat{\tau}_{\rm NL}^{\rm loc}$ provides an upper bound on $|\fnl|$. Using the non-Gaussian temperature-plus-polarization (temperature-only) likelihood, we find $|\fnl|<28$ ($|\fnl|<30$) at $95\%$ CL. These bounds are $\approx 2.5\times$ weaker than, though independent from, the \textit{Planck} bispectrum constraints.
    \item Assuming the Suyama-Yamaguchi inequality to be saturated, we can constrain $\taunl$ from observations of $\widehat{f}^{\rm loc}_{\rm NL}$. This yields $\left.\taunl\right|_{\rm SY}<149$ ($<175$), using \textit{Planck} PR3 measurements from \citep{Planck:2019kim}. These are $\approx 10\times$ stronger than the trispectrum bounds, but neglect uncorrelated fields.
    \item Finally, we can isolate the trispectrum-specific information by defining $\Delta\taunl \equiv  \taunl - (\tfrac{6}{5}\fnl)^2$. Combining our $\widehat{\tau}_{\rm NL}$ likelihood with the $\fnl$ measurements of \citep{Planck:2019kim} gives $\Delta\taunl<1450$ ($<1670$) including (excluding) polarization. This additionally constrains the $r_\sigma$ parameter of \eqref{eq: scalar-modulation}: $|r_\sigma|\geq 0.129$ ($\geq 0.127$) at 95\% CL, marginalizing over $\fnl$, which tightens to $|r_\sigma|\geq 0.20$ ($|r_\sigma|\geq 0.18$) if we fix $\fnl$ to the maximum value allowed by the \textit{Planck} $1\sigma$ bound.
\end{itemize}

\subsection{EFTI Non-Gaussianity}\label{subsec: discussion-efti}

\begin{figure}[!t]
    \centering
    \includegraphics[width=\linewidth]{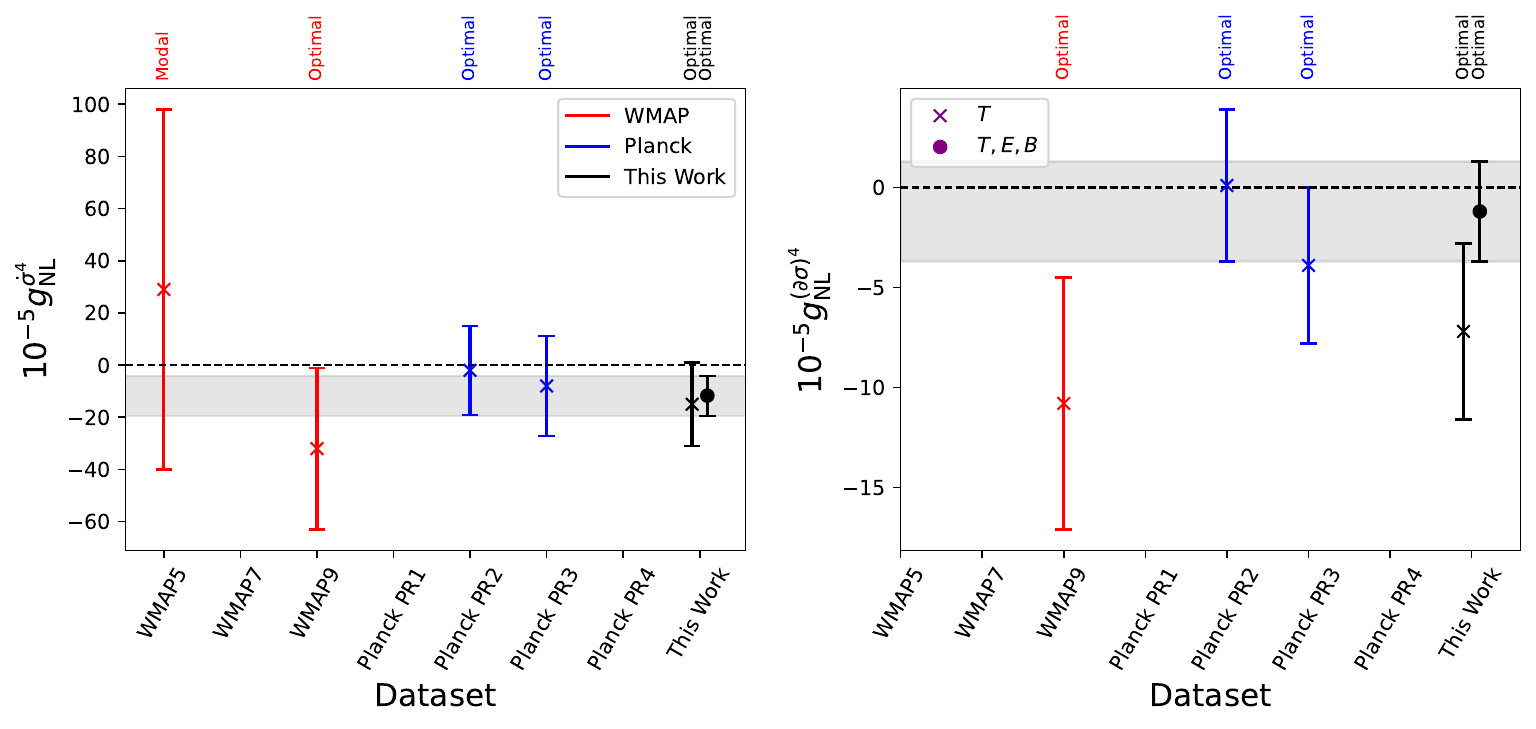}
    \caption{As Fig.\,\ref{fig: local-literature}, but showing the constraints on the EFTI parameters $\gnldotdot$ and $\gnldeldel$ (analyzed independently). Our constraints are significantly stronger than those of previous work, primarily due to the addition of polarization.}
    \label{fig: efti-literature}
\end{figure}

\subsubsection{Previous Constraints}
\noindent As shown in Fig.\,\ref{fig: efti-literature}, the EFTI amplitudes, $\gnldotdot,\gnldotdel,\gnldeldel$, have been constrained in a number of previous CMB analyses. Firstly, \citep{Fergusson:2010gn} searched for the equilateral trispectrum amplitude $t_{\rm NL}^{\rm equil} \equiv (27/25)\gnldotdot$ from WMAP5 data using the modal estimator, yielding the constraint $10^{-5}\gestdotdot = 29\pm69$ at $\ell_{\rm max}=500$ (which was translated into inflationary parameters in \citep{Izumi:2011di}).\footnote{Strictly, the equivalence between $t_{\rm NL}^{\rm equil}$ and $\gnldotdot$ holds only up to slow-roll factors of $1-n_s$.} The first optimal template analysis of the EFTI shapes was presented in \citep{2015arXiv150200635S} (using analogous estimators to this work), with the results $10^{-5}\gestdotdot =-32\pm31,10^{-5}\gestdeldel = -10.8\pm6.3$ using WMAP9 data at $\ell_{\rm max}=1024$, with $\gestdotdel$ omitted due to its correlations with the other parameters. The same pipeline was used in the official \textit{Planck} PR2 non-Gaussianity analyses to yield the temperature-only \textsc{smica} measurements $10^{-5}\gestdotdot = -2\pm17,10^{-5}\gestdeldel = 0.1\pm3.8$ using \textsc{smica} at $\ell_{\rm max}=1600$ \citep{Planck:2015zfm}, with only minor differences found in the PR3: $10^{-5}\gestdotdot = -8\pm19,10^{-5}\gestdeldel = -3.9\pm3.9$ \citep{Planck:2019kim}.

From Tab.\,\ref{tab: contact-results}, our temperature-only \textsc{smica} constraints are in good agreement with those from \textit{Planck} PR3: we find $10^{-5}\gestdotdot = -15\pm16, 10^{-5}\gestdeldel = -7.2\pm4.4$, with the variances matching to within $10\%$. When adding polarization, the constraining power doubles, with $10^{-5}\gestdotdot = -11.7\pm7.7, 10^{-5}\gestdeldel = -1.2\pm 2.5$; as such, our results represent the strongest bounds on EFTI templates obtained to date (as is clear from Fig.\,\ref{fig: efti-literature}).

As for local non-Gaussianity, constraints can also be wrought from other datasets. As discussed in \citep{Mizuno:2015qma}, galaxy bispectra are sensitive to the EFTI parameters, and their analysis could yield competitive constraints to the CMB; more generally, the above models could be constrained using galaxy trispectra (just as the cubic EFTI couplings have been probed using bispectra  \citep{Cabass:2022wjy,Cabass:2022epm}). Finally, analysis of 21cm fluctuations in a (highly) futuristic dark ages survey could place extremely strong constraints on the EFTI parameters: \citep{Floss:2022grj} predicts $10^{-5}\sigma(\gnldotdot) = 2\times 10^{-6}, 10^{-5}\sigma(\gnldeldel) = 6\times 10^{-7}$ from a cosmic-variance-limited experiment. If inflation has secrets, such analyses will surely reveal them.

\subsubsection{Implications}
\paragraph{Single-Field}
As discussed in \citep{2015arXiv150200635S,Philcox4pt1}, the EFTI parameters $\gnldotdot,\gnldotdel,\gnldeldel$ relate to the structure of the low-energy inflationary Lagrangian. Under a fairly weak set of assumptions, the most general action for single-field inflation is given by 
\beq\label{eq: efti-single}
    S_\pi &=& \int d^4x\sqrt{-g}\,\bigg\{-M_{\rm Pl}^2\dot{H}(\partial_\mu\pi)^2+2M_2^4\left[\dot{\pi}^2+\dot{\pi}^3-\dot{\pi}\frac{(\partial_i\pi)^2}{a^2}+(\partial_\mu\pi)^2(\partial_\nu\pi)^2+\cdots\right]\\\nonumber
    &&\qquad\qquad\qquad\,-\,\frac{M_3^4}{3!}\left[8\dot{\pi}^3+12\dot{\pi}^2(\partial_\mu\pi)^2+\cdots\right]+\frac{M_4^4}{4!}\left[16\dot{\pi}^4+\cdots\right]+\cdots\bigg\},
\eeq
where $\pi$ is the Goldstone mode of broken time-translations (\textit{i.e.}\ the inflationary clock), $H$ is the Hubble scale, $M_{\rm Pl}$ is the Planck mass and $\{M_n\}$ are coupling coefficients \citep{Cheung:2007st}. In addition to the kinetic term, this features both cubic and quartic interactions, with higher-order terms constrained by the non-linear realization of boosts. The cubic operators $\dot{\pi}(\partial\pi)^2$ and $\dot{\pi}^3$ generate bispectra with amplitudes:
\beq
    f_{\rm NL}^{\dot{\pi}(\partial\pi)^2}= \frac{85}{324}\left(1-\frac{1}{c_s^{2}}\right),\qquad f_{\rm NL}^{\dot{\pi}^3}=\frac{10}{243}\left(1-\frac{1}{c_s^{2}}\right)\left(\tilde{c}_3+\frac{3}{2}c_s^2\right),
\eeq
depending on the sound-speed parameter $c_s^{-2} \equiv 1-2M_2^4/(M_{\rm Pl}^2\dot{H})$ and $\tilde{c}_3 =-c_s^2M_3^4/M_2^4$ \citep[e.g.,][]{Cheung:2007st,Senatore:2009gt}. Similarly, the quartic operators in \eqref{eq: efti-single} source trispectra: as discussed in \citep{Senatore:2010jy}, only $\dot\pi^4$ is unconstrained given the strong observational bounds on inflationary bispectra. The resulting four-point function has the $\gnldotdot$ shape with amplitude
\beq
    g_{\rm NL}^{\dot{\pi}^4}&=&\frac{25}{288}\frac{M_4^4}{H^4}A_\zeta c_s^3.
\eeq
where $A_\zeta = k_\ast^3P_\zeta (k_\ast) = 2\pi^2 A_s\approx 4.2\times 10^{-8}$.

Using the \textsc{smica} constraints given in Tab.\,\ref{tab: contact-results}, we constrain
\beq
     -13.1 < 10^{-14} c_s^3(M_4^4/H^4) < 4.8\quad (T\text{-only}), \qquad     -7.5 < 10^{-14} c_s^3(M_4^4/H^4) < 1.0\quad (T,E,B)\nonumber
\eeq
at 95\% CL. Our baseline constraint is $\approx 2\times$ stronger than the \textit{Planck} PR3 bound ($-12.8 < 10^{-14}c_s^3(M_4^4/H^4) < 8.2$) due to the addition of polarization \citep{Planck:2019kim};\footnote{There is a typo in Eq.\,69 of \citep{Planck:2019kim}: $\gnldotdot$ constrains $(M_4^4/H^4)/c_s^3$ rather than $M_4^4/(c_s^3H^4)$. There are further errors in the $f_{\rm NL}^{\rm equil}$ and $f_{\rm NL}^{\rm ortho}$ definitions of \citep[Eq.\,58]{Planck:2015zfm} and (in the arXiv version only) \citep[Eq.\,87]{Planck:2019kim}: the correct relations are $f_{\rm NL}^{\rm equil} = (1-c_s^2)/c_s^2 [-0.275-0.0780c_s^2-(2/3)\times 0.0780\tilde{c}_3], f_{\rm NL}^{\rm ortho} = (1-c_s^2)/c_s^2 [0.0159+0.0167c_s^2+(2/3)\times 0.0167\tilde{c}_3]$.} furthermore, it is around $4\times$ tighter than the WMAP constraint.
Performing a joint analysis of $\gnldotdot$ with $f_{\rm NL}^{\rm equil}$ and $f_{\rm NL}^{\rm ortho}$ (taken from \citep{Planck:2019kim}), we can break the degeneracy with $c_s$. Assuming uniform priors on $c_s^{-2}$, $(c_s^{-2}-1)(\tfrac{2}{3}\tilde{c}_3+c_s^2)$ and $M_4/H^4 c_s^3$, we find 
\beq
    -4.7<10^{-20}\frac{M_4^4}{H^4}<0.3, \qquad c_s\geq 0.022, \qquad -8.1<\tilde{c}_3<-1.4, \nonumber
\eeq
from the baseline \textsc{smica} analysis.
To place these results in context, we note that unitarity requires $H/\Lambda_U\lesssim 1$, where the cut-off satisfies $\Lambda_U^4 \sim 16\pi^2(\dot{H} M_{\rm Pl}^2)^2M_4^{-4}$, for $A_\zeta = H^4/(4c_sM_{\rm Pl}^2|\dot{H}|)$ \citep{Senatore:2010jy,Cheung:2007st,2015arXiv150200635S}. Applied our set-up, we find (at $95\%$ CL)
\begin{alignat*}{3}
    &\textbf{Unitarity}: \qquad&& c_s^3(M_4^4/H^4) \lesssim (5\times 10^{15})c_s, \qquad&& A_\zeta |g_{\rm NL}^{\dot\pi^4}| \lesssim 0.9c_s\\\nonumber
    &\textbf{Data}: \qquad&& c_s^3(M_4^4/H^4) \lesssim 8\times 10^{14}, \qquad&& A_\zeta |g_{\rm NL}^{\dot\pi^4}| \lesssim 0.067.
\end{alignat*}
Assuming that $c_s$ is of order unity (which is suggested by the non-detection of bispectrum non-Gaussianity), our constraints are well-within the unitarity bounds, and are thus competitive.

\vskip 8pt
\paragraph{Multi-Field}
Additional fields present during inflation can source different shapes of four-point non-Gaussianity. As shown in \citep{Senatore:2010wk}, these can be modeled using the effective action:
\beq\label{eq: efti-multi}
    S_\sigma = \int d^4x\sqrt{-g}\left\{\frac{1}{2}(\partial_\mu\sigma)^2+\frac{1}{\Lambda_1^4}\dot{\sigma}^4+\frac{1}{\Lambda_2^4}\dot{\sigma}^2(\partial_i\sigma)^2+\frac{1}{\Lambda_3^4}(\partial_i\sigma)(\partial_j\sigma)^2+\frac{\mu^4}{\Lambda^4}\sigma^4+\cdots\right\},
\eeq
where $\mu$ and $\{\Lambda_n\}$ are coupling constants and $\sigma$ is a spectator field whose fluctuations source curvature via $\zeta = (2A_\zeta)^{1/2}\sigma/H+\cdots$. We have ignored any cubic terms: these could be suppressed by some symmetry such as $\sigma\to -\sigma$. Since $\sigma$ is not the inflaton (\textit{i.e.}\ the Goldstone boson), we are not restricted by diffeomorphism symmetries: as a result, we form four leading-order trispectra with amplitudes
\beq\label{eq: gnl-LI}
    \gnldotdot A_\zeta &=& \frac{25}{768}\frac{H^4}{\Lambda_1^4}, \qquad \gnldotdel A_\zeta = -\frac{325}{6912}\frac{H^4}{\Lambda_2^4}, \qquad \gnldeldel A_\zeta = \frac{2575}{20736}\frac{H^4}{\Lambda_3^4}, \qquad \gnl \approx -\frac{50}{27}\frac{\mu^4}{\Lambda^4}N_e,
\eeq
assuming inflation lasts for $N_e$ $e$-folds. \citep{2015arXiv150200635S} used the WMAP $\gnldotdot$ bounds to obtain the $\Lambda_1$ constraint $-12.6 < H^4/\Lambda_1^4< 4.0$; analyzing each term independently,
we find the 95\% bounds
\beq
    -3.6 < 10^{5}\frac{\mu^4}{\Lambda^4} < 3.5, \qquad -3.5 < \frac{H^4}{\Lambda_1^4} < 0.6, \qquad
-0.65 < \frac{H^4}{\Lambda_2^4} < 2.59, \qquad
-0.21 < \frac{H^4}{\Lambda_3^4} < 0.13 \nonumber
\eeq
from the baseline \textsc{smica} analysis, greatly enhancing the previous constraint. In practice, one should constrain the parameters jointly; this leads to slightly weaker constraints:
\beq
    -4.8 < 10^{5}\frac{\mu^4}{\Lambda^4} < 2.6, \qquad -10.0 < \frac{H^4}{\Lambda_1^4} < 6.3, \qquad -9.0 < \frac{H^4}{\Lambda_2^4} < 10.7, \qquad -0.45 <\frac{H^4}{\Lambda_3^4} < 0.65. \nonumber
\eeq
Physically, the quantity of relevance is the ratio of the de Sitter temperature, $H/(2\pi)$, to $\Lambda_i$ \citep{Senatore:2010wk}: our EFTI results bound these ratios to $\{H/(2\pi\Lambda_1),H/(2\pi\Lambda_2),H/(2\pi\Lambda_3)\} < \{0.27, 0.30, 0.15\}$ at 95\% CL (and are thus practically relevant). Asserting that the underlying theory is Lorentz invariant enforces $\Lambda_1^4=-2\Lambda_2^4=\Lambda_3^4 \equiv \Lambda_{\rm LI}^4$, corresponding to an interaction term $\Lambda_{\rm LI}^{-4}(\partial_\mu\sigma)^2(\partial_\nu\sigma)^2$ in \eqref{eq: efti-multi}. This leads to the 95\% bound $-0.184 < H^4/\Lambda_{\rm LI}^4 < 0.068$, equivalent to $H/(2\pi \Lambda_{\rm LI}) < 0.088$, which is twice as tight as the \textit{Planck} PR3 constraint.

\vskip 8pt
\paragraph{Other Models} Single-field $P(X,\phi)$ theories (see also $k$-inflation) provide another potential source of trispectrum non-Gaussianity \citep[e.g.,][]{Izumi:2011di,Planck:2019kim,Arroja:2009pd,Chen:2009bc,Chen:2013aj,Philcox4pt1,Babich:2004gb,Alishahiha:2004eh,Silverstein:2003hf,Langlois:2008qf,Langlois:2008wt}. These modify the kinetic term in the inflationary action: $X-V(\phi)\to P(X,\phi)$ where $X\equiv-(1/2) g^{\mu\nu}\partial_\mu \phi\partial_\nu\phi$ and $\phi$ is the inflaton. As discussed in \citep{Izumi:2011di,Philcox4pt1}, these generate contact trispectra with amplitudes\footnote{Exchange trispectra can also be formed \citep{Arroja:2009pd}: we neglect these for the purposes of this discussion.}
\beq
    \gnldotdot A_\zeta = -\frac{25}{768}\frac{H^4c_s}{P^2_{,X}}\beta_1, \qquad \gnldotdel A_\zeta = \frac{325}{6912}\frac{H^4}{c_sP_{,X}^2}\beta_2, \qquad \gnldeldel A_\zeta = -\frac{2575}{20736}\frac{H^4}{c_s^3P_{,X}^2}\beta_3,
\eeq
where the $\beta_i$ parameters can be expressed in terms of $\dot\phi$, the sound speed $c_s$, and derivatives of $P(X,\phi)$ \citep[e.g.,][]{Izumi:2011di}.\footnote{Although we have assumed single-field inflation, we can source all three EFTI trispectra. This occurs since we work in the $c_s\ll1$ limit, where higher order terms in \eqref{eq: efti-single} cannot be neglected.} Notably, these amplitudes are equivalent to \eqref{eq: gnl-LI} with the replacement $H^4/\Lambda^4_{i} \to -H^4/P^2_{,X}\times \beta_i c_s^{3-2i}$. We immediately find the joint constraints
\beq
    -6.3 < c_s\frac{H^4}{P_{,X}^2}\beta_1 < 10.0, \qquad -10.7 < \frac{1}{c_s}\frac{H^4}{P_{,X}^2}\beta_2 < 9.0, \qquad -0.65 <\frac{1}{c_s^3}\frac{H^4}{P_{,X}^2}\beta_3 < 0.45, \nonumber
\eeq
from the baseline \textsc{smica} analysis. A particularly well-studied $P(X,\phi)$ model is DBI inflation: this corresponds to $\beta_1 = (2c_s^7\dot\phi^2)^{-1}$, $\beta_2 = (4c_s^3\dot\phi^2)^{-1}$, $\beta_3 = -(8c_s\dot\phi^2)^{-1}$ \citep{Planck:2019kim,Arroja:2009pd,Chen:2009bc,Chen:2013aj}. For $c_s\ll 1$, this is dominated by the first EFTI shape with $\gnldotdot = -(25/768)c_s^{-4}$. Assuming a uniform prior for $c_s\in[0,0.2]$, we find $c_s\geq 0.018$ at 95\% CL from our \textsc{smica} $\gnldotdot$ measurement (or $(c_s^{-4} \leq 8.9\times 10^7$ if sampled directly); this is somewhat stronger than the \textit{Planck} PR3 bound of $c_s\geq 0.015$ \citep{Planck:2019kim}. The trispectrum bound is weaker than, but independent from, the bispectrum limit $c_s\geq 0.086$ \citep{Planck:2019kim}.

As shown in \citep{Izumi:2011di}, the $\gnldotdot$ parameter traces many inflationary models not covered by the above discussion, such as multi-field DBI inflation, exchange interactions in $k$-inflation, ghost inflation, and Lifshitz scalar theory. In many cases, the full trispectra take a different form to the EFTI templates, but can be probed indirectly due to the correlations between shapes; this implies that stronger constraints can be wrought if one analyzes the full trispectrum. A notable example is ghost inflation \citep{Arkani-Hamed:2003juy,Senatore:2009gt,Cheung:2007st}, for which the altered kinetic term leads to a proliferation of higher-order terms that can generate large non-Gaussianity (and other effects, such as parity-violation \citep{Cabass:2022rhr,Cabass:2022oap}). Directly constraining such models is beyond the scope of this work.





\subsection{Direction-Dependent Non-Gaussianity}
\noindent Whilst the direction-dependent $\tau_{\rm NL}^{n_1n_3n}$ parameters are new to this series, several previous works have considered the parity-even and parity-odd amplitudes $\tau_{\rm NL}^{n,\rm even}$ and $\tau_{\rm NL}^{n,\rm odd}$. Using idealized Fisher forecasts, \citep{Shiraishi:2013oqa} predicted $10^{-2}\sigma(\tau_{\rm NL}^{0,\rm even}) \equiv 10^{-2}\sigma(\taunl) = 1.6$, $10^{-3}\sigma(\tau_{\rm NL}^{2,\rm even}) = 0.63$ from a cosmic-variance limited experiment at $\ell_{\rm max}=2000$.\footnote{Previous works adopt the $d_n^{\rm even/odd}$ parameters, which are approximately equivalent to $\tau_{\rm NL}^{n,\rm odd/even}/6$, as discussed in \paperone.}
Similarly \citep{Shiraishi:2016mok} forecast $10^{-4}\sigma(\tau_{\rm NL}^{1,\rm odd}) = 0.38$, noting that the other parameters (e.g., $\tau_{\rm NL}^{0,\rm odd},\tau_{\rm NL}^{1,\rm even}$) are difficult to constrain with CMB data.
This setup is a rough proxy for our \textit{Planck} temperature-only analyses, which found $\{10^{-2}\tau_{\rm NL}^{0,\rm even}, 10^{-3}\tau_{\rm NL}^{2, \rm even}, 10^{-4}\tau_{\rm NL}^{1, \rm odd}\} = \{2.3\pm2.3, 1.9\pm2.6,-8.4\pm2.1\}$, using \textsc{sevem}. 
Though we find similar scalings (see \papertwo), our constraints on high-order parameters are around $5\times$ weaker constraints than the forecasts; this is unsurprising, given the various simplifications in \citep{Shiraishi:2013oqa,Shiraishi:2016mok} and the optimistic effective $\ell_{\rm max}$ adopted therein. We further obtain a $50\%$ improvement in $\tau_{\rm NL}^{1,\rm odd}$ when polarization is included.

The $\tau_{\rm NL}^{1,\rm odd}$ parameter has been previously constrained using the parity-odd four-point function. In \citep{Philcox:2023ypl} (see also \citep{PhilcoxCMB}) a binned CMB trispectrum was used to constrain Chern-Simons gauge fields, which translates to the 68\% bounds $10^{-4}\tau_{\rm NL}^{\rm 1,\rm odd} = 27\pm20$ from temperature-alone or $10^{-4}\tau_{\rm NL}^{1,\rm odd}=-1.9\pm5.9$ with polarization (assuming $\ell_{\rm max}=2000$).\footnote{\citep{Philcox:2023ypl} also places constraints on $\tau_{\rm NL}^{0,\rm odd}$, but these are weaker by around a factor of $10^5$, as discussed in \citep{Philcox4pt2}.} These bounds are an order of magnitude weaker than the constraints obtained in this work due to the significant loss of information incurred by the wide radial bins (as found in the bispectrum analyses of \citep{Philcox:2024wqx}).
The direction-dependent amplitudes can also be constrainted using galaxy correlators: \citep{Philcox:2022hkh} found $10^{-4}\tau_{\rm NL}^{1,\rm odd} = -8\pm 21$ and $10^{-3}\tau_{\rm NL}^{0,\rm odd} = -3.7\pm4.8$ using the parity-odd four-point function of BOSS galaxies. Whilst the constraining power of such datasets is still significantly less than the CMB (for $\tau_{\rm NL}^{1,\rm odd}$), 
they provide a promising avenue for future explorations of chiral primordial physics.

As shown in \citep{Shiraishi:2016hjd}, spectral distortions could provide an extremely strong constraint on the $\tau_{\rm NL}^{n,\rm even}$ parameters. Whilst legacy and near-term experiments would reach $10^{-4}\sigma(\tau_{\rm NL}^{2,\rm even})\approx 160$ (\textit{Planck}) and $\approx 16$ (PIXIE, assuming $\ell_{\rm max}=1000$), a cosmic-variance limited experiment could exceed CMB trispectrum bounds by a factor of $10^4$. A similar conclusion likely holds for dark-ages 21cm analyses.


\subsubsection{Implications}
\paragraph{Gauge Fields} 
A well-studied source of direction-dependent trispectrum non-Gaussianity is primordial gauge fields \citep[e.g.,][]{Bartolo:2015dga,Bartolo:2014hwa,Shiraishi:2013vja,Shiraishi:2013oqa,Naruko:2014bxa,Bartolo:2012sd,Dimastrogiovanni:2010sm,Shiraishi:2016mok,Bartolo:2017szm,Bartolo:2018elp}.\footnote{An additional source is gravitational wave exchange, through chiral scalar-tensor theories including Chern-Simons gravity \citep{CyrilCS,Bartolo:2020gsh,Moretti:2024fzb,Salvarese}.} These are described by the action
\beq\label{eq: action-gauge}
    S_{\phi,A_\mu} &=& \int d^4x\sqrt{-g}\bigg\{-\frac{1}{2}(\partial_\mu\phi)^2-V(\phi)+I^2(\phi)\left(-\frac{1}{4}F^{\mu\nu}F_{\mu\nu}+\frac{\gamma}{4}\tilde{F}^{\mu\nu}F_{\mu\nu}\right)\bigg\}.
\eeq
where $\phi$ is a pseudo-scalar inflaton, $A_\mu$ is a gauge field, $F_{\mu\nu}$ is the electromagnetic tensor and $\tilde{F}_{\mu\nu}$ is its Hodge dual.  Assuming $I^2(\phi)\sim a^{-2}$, this generates scale-invariant correlators, with chirality imparted by the Chern-Simons term $\gamma \tilde{F}F$ for constant $\gamma$. As discussed in \citep{Shiraishi:2013oqa,Shiraishi:2013vja,Bartolo:2015dga,Shiraishi:2016mok} (and summarized in \paperone), this sources various anisotropies, including a direction-dependent power spectrum, bispectrum, and trispectrum, with the first described by the parameter
\beq
    g_\ast \approx \frac{48N_e^2}{\epsilon}\frac{\rho_E^{\rm vev}}{\rho_\phi}f(\gamma), \qquad f(\gamma)=\begin{cases}1 & (\gamma=0)\\
    \frac{e^{4\pi|\gamma|}}{32\pi|\gamma|^3}& (\gamma\gg 1)\end{cases},
\eeq
where $\epsilon$ is the slow-roll parameter and $\rho_E^{\rm vev}/\rho_\phi$ is the fractional energy density in the gauge field's vacuum expectation value during inflation. Assuming $\gamma=0$ (\textit{i.e.}\ dropping the Chern-Simons term), we have:
\beq
    &&\tau_{\rm NL}^{000} = \frac{128}{3}\mathcal{A}(\gamma), \qquad \tau_{\rm NL}^{220}=\tau_{\rm NL}^{202}=\tau_{\rm NL}^{022}=\frac{64}{3\sqrt{5}}\mathcal{A}(\gamma), \qquad \qquad \tau_{\rm NL}^{222}=\frac{16\sqrt{14}}{45}\mathcal{A}(\gamma)
\eeq
for $\mathcal{A}(\gamma) \equiv (4\pi)^{3/2}N_e^2|g_\ast|f^2(\gamma)$, whilst $\gamma\gg1$ sources the additional terms:
\beq
    &&\tau_{\rm NL}^{110} = -\tau_{\rm NL}^{101}=\tau_{\rm NL}^{011}=\frac{64}{\sqrt{3}}\mathcal{A}(\gamma), \qquad \tau_{\rm NL}^{112}=-\tau_{\rm NL}^{121}=\tau_{\rm NL}^{211}=-16\sqrt{\frac{2}{3}}\mathcal{A}(\gamma)\\\nonumber
    &&\tau_{\rm NL}^{111} = 16\sqrt{2}\mathcal{A}(\gamma), \qquad \tau_{\rm NL}^{221} = -\tau_{\rm NL}^{212}=\tau_{\rm NL}^{122}=16\sqrt{\frac{2}{5}}\mathcal{A}(\gamma).
\eeq
as derived in \citep{Philcox4pt1}. Since the CMB can only constrain $\tau^{n_1n_3n}_{\rm NL}$ parameters for even $n_1,n_3$, several of the above amplitudes are not observable.

Using the non-Gaussian $\taunl$ likelihood discussed in \S\ref{subsec: results-local} and Appendix \ref{app: tauNL-Lmax}, we can constrain $\mathcal{A}(\gamma)$ from $\tau_{\rm NL}^{000}$: this yields $\mathcal{A}(\gamma)<1550$ ($<1810$) at 95\% CL using the \textsc{smica} baseline (temperature-only) dataset, which corresponds to $|g_\ast|<0.0096$ ($<0.011$) for $\gamma=0$ or $10^{28}|g_\ast|<1.3$ ($<1.5$) for $\gamma=3$. Whilst the higher-order direction-dependent amplitudes are not able to appreciably tighten the above constraints (due to their larger variances), they can be used to isolate characteristic features of the gauge-field model. Performing an `anisotropy-only' analysis using all $\tau_{\rm NL}^{n_1n_3n}$ except $\tau_{\rm NL}^{000}$ yields $\mathcal{A}(\gamma)<3790$ for $\gamma=0$ and $\mathcal{A}(\gamma)<2870$ for $\gamma=3$; these correspond to $|g_\ast|<0.0236$ and $10^{28}|g_\ast|<2.4$ respectively (using the baseline \textsc{smica} dataset in all cases).\footnote{As in \S\ref{subsec: results-direc}, we adopt a joint Gaussian likelihood for $\{\tau_{\rm NL}^{n_1n_3n}\}$. 
Including non-Gaussianities will lead to a broader posterior on $\mathcal{A}(\gamma)$.} Furthermore, restricting to $\tau_{\rm NL}^{221}$ directly probes the chiral part of the trispectrum: this gives 
$\mathcal{A}(\gamma)<7360$ and $10^{28}|g_\ast|<6.1$ at $\gamma=3$.

Our constraints on the anisotropy parameter $g_\ast$ can be compared to a number of previous forecasts. Assuming a \textit{Planck}-like experiment with $\ell_{\rm max}=2000$, \citep{Shiraishi:2013oqa} and \citep{Shiraishi:2016mok} predicted $\sigma(g_\ast)\approx 1\times 10^{-3}$ at $\gamma=0$ and $10^{28}\sigma(g_\ast) \approx 0.09$ at $\gamma=3$ from the parity-even and parity-odd amplitudes $\tau_{\rm NL}^{n,\rm even}$ and $\tau_{\rm NL}^{n,\rm odd}$ respectively (equivalent to $\sigma(\mathcal{A}(\gamma)) \approx 160$ and $\mathcal{A}(\gamma\gg 1)\approx 1000$).
These forecasts are comparable to our results given the optimistic choice of $\ell_{\rm max}$ and the non-Gaussianity of the $\taunl$ posterior.
Moreover, our bounds are much stronger than those obtained in the binned parity-odd trispectrum analysis of \citep{Philcox:2023ypl}: $\mathcal{A}(\gamma) = -4500\pm16720$ for $\gamma\gg 1$, equivalent to $10^{28}\sigma(g_\ast)\lesssim 14$ for $\gamma=3$. From the bispectrum, \citep{Shiraishi:2013vja,Bartolo:2015dga} predicted $\sigma(g_\ast)\approx 0.014$ at $\gamma=0$, or $10^{18}\sigma(g_\ast)\approx 1600$ at $\gamma=3$. Finally, anisotropy of the \textit{Planck} power spectrum was constrained in \citep{Planck:2018jri}, finding $g_\ast \sim 10^{-2}$. 

For $\gamma=0$, our bounds on $|g_\ast|$ are stronger than those predicted for bispectrum analyses (by almost $3\times$), and comparable with the power spectrum bounds. For $\gamma\gg1$, we find extremely strong constraints: this occurs since higher-order correlators are enhanced by factors of $N_ef(\gamma)\gg1$ relative to the power spectrum \citep{Shiraishi:2016mok}. Finally, we can relate these bounds to the fractional energy density in gauge fields: for $\gamma=0$ our $\taunl$ constraint gives the 95\% limit $\rho_E^{\rm vev}/\rho_\phi<5.6\times 10^{-10}$ at $\gamma=0$ or $\rho_E^{\rm vev}/\rho_\phi<8.5\times 10^{-49}$ at $\gamma=3$, with somewhat weaker bounds found from the anisotropy-only analyses. 
If gauge fields are present in the early Universe, they must have a very small vacuum expectation value.

\vskip 8pt
\paragraph{Solid Inflation} Another potential source of direction-dependent non-Gaussianity is solid inflation \citep{Endlich:2012pz,Endlich:2013jia,Gruzinov:2004ty,Bartolo:2014xfa}. This posits that inflation is driven by a triplet of scalar fields $\{\phi^I\}$ with the action:
\beq
    S_{\rm solid} = \int d^4x\sqrt{-g}\,F[\hat{X},\hat{Y},\hat{Z}]+\cdots, \qquad  \hat{X} = \mathrm{tr}\,\hat{B}, \quad \hat{X}^2\hat{Y}=\mathrm{tr}\,\hat{B}^2, \quad \hat{X}^3\hat{Z} = \mathrm{tr}\,\hat{B}^3
\eeq
where $\hat{B}^{IJ} \equiv  g^{\mu\nu}\hat{\phi}^I_\mu\hat{\phi}^J_\nu$. By suitable choice of $F$, this can be made shift-symmetric and rotationally invariant, and leads to various phenomenology including a (squared) anisotropic sound speed, $c_L^2\approx 1/3+(8/9)(F_{,Y}+F_{,Z})/XF_{,X}$. Although the full trispectrum has not yet been computed, \paperone demonstrated that the $\tau_{\rm NL}^{n_1n_3n}$ parameters have lower bounds
\beq\label{eq: tau-solid}
    \tau_{\rm NL}^{220} &=& -7\sqrt{\frac{7}{10}}\tau_{\rm NL}^{222} =\frac{1}{3}\sqrt{\frac{7}{2}}\tau_{\rm NL}^{224}\, \gtrsim\, (4\pi)^{3/2}\frac{16\sqrt{5}}{81}\left(\frac{F_{,Y}}{F}\frac{1}{\epsilon c_L^2}\right)^2,
\eeq
using the generalized Suyama-Yamaguchi inequality and focusing on the longitudinal mode.  Notably, the model is strongly direction-dependent and does not source $\tau_{\rm NL}^{000}$ at leading-order.
Performing a joint analysis of $\tau_{\rm NL}^{220}$ and $\tau_{\rm NL}^{222}$ 
we constrain $\left(F_{,Y}/(F\epsilon c_L^2)\right)^2<2630$ ($<3920$) from \textsc{smica} including (excluding) polarization, which corresponds to $F_{,Y}/(F\epsilon c_L^2)<51$ ($<63$) at 95\% CL. Thought our bounds are somewhat weaker than those from direction-dependent bispectra, which predict $\sigma(F_{,Y}/(F\epsilon c_L^2))\approx 5.9$ \citep{Bartolo:2014xfa}, we caution that \eqref{eq: tau-solid} represents only part of the template.

Primordial magnetic fields present another interesting source of direction-dependent non-Gaussianity, through the quadratic sourcing of a scalar curvature perturbation \citep{Shaw:2009nf}. This scenario has been discussed in a variety of previous works \citep[e.g.,][]{Shiraishi:2012sn,Shiraishi:2013vha,Shiraishi:2013vja,Planck:2015zrl,Shiraishi:2012rm}, and creates a variety of modifications to late-time correlators. Whilst a full computation of the magnetic field trispectrum (building on \citep{Trivedi:2011vt,Trivedi:2013wqa}) is beyond the scope of this work, it is interesting to note that the resulting trispectra can be used to probe both the non-helical and helical parts of the magnetic field \citep{Shiraishi:2012sn}.


\subsection{Collider Non-Gaussianity}
\noindent Despite significant theoretical interest \citep[e.g.,][]{Chen:2009zp,Arkani-Hamed:2015bza,Lee:2016vti,Flauger:2016idt,Wang:2022eop,Pimentel:2022fsc,Kumar:2019ebj,Reece:2022soh,Arkani-Hamed:2015bza,Alexander:2019vtb,Jazayeri:2023xcj,McCulloch:2024hiz,Liu:2019fag,Meerburg:2016zdz,Wang:2019gbi,Tong:2022cdz,Sohn:2024xzd,Pimentel:2022fsc,Cabass:2024wob,Chen:2016uwp,Chen:2018xck,Lu:2019tjj,Wang:2020ioa,Bodas:2020yho,Jazayeri:2022kjy,Kim:2019wjo,Lu:2021wxu,Cui:2021iie,Qin:2022lva,Werth:2023pfl,Chen:2022vzh,Xianyu:2023ytd,Pinol:2023oux,Chakraborty:2023qbp,Craig:2024qgy,Yin:2023jlv,Baumann:2011nk,Assassi:2012zq,Baumann:2017jvh,Arkani-Hamed:2018kmz,Cabass:2022rhr,Bordin:2018pca,Bordin:2019tyb,Jazayeri:2023kji,Chen:2018sce,Green:2023ids,Noumi:2012vr}, there have been few observational constraints on the cosmological collider paradigm. For the trispectrum, \citep{Bordin:2019tyb} forecast the detectability of spinning particles in the CMB; whilst these results are not directly comparable to ours, since they assume Higuchi-bound violating particles, our phenomenological conclusions are similar: constraints on higher-mass particles are not dominated by extremely low $L$, and errorbars rapidly degrade as the mass becomes larger, but more slowly with increasing spin. A specific spin-one coupling was constrained from the binned \textit{Planck} parity-odd CMB trispectrum in \citep{PhilcoxCMB} (and in \citep{Cabass:2022oap} using BOSS galaxies); this obtained limits of order the perturbativity bound, though these are difficult to translate to our formalism since they assume a chiral coupling not found in the basic collider picture of \citep{Arkani-Hamed:2015bza}.

The bispectrum signatures of massive particle non-Gaussianity have been studied in much greater depth. \citep{Cabass:2024wob} searched for heavy spin-zero collider non-Gaussianity by repurposing \textit{Planck} $f_{\rm NL}$ constraints, whilst \citep{Sohn:2024xzd} performed a detailed search for scalar and higher-spin non-Gaussianity, using a 
modal bispectrum formalism.
Utilizing approximate bootstrap methods, this was able to extend beyond the squeezed limit (corresponding to our collapsed limit), and included a number of regimes not considered in this work, such as equilateral and low-speed colliders (obtained by varying the sound-speed, $c_s$) 
and the `integrated-out' signatures of extremely massive particles. In all cases, there was no detection of primordial non-Gaussianity.

There has additionally been a slew of work devoted to large-scale structure probes of collider non-Gaussianity. As noted in \citep[e.g.,][]{MoradinezhadDizgah:2017szk,Cabass:2018roz}, heavy fields source novel contributions to galaxy power spectra and bispectra, which can be used to 
constrain the associated $f_{\rm NL}$ parameters \citep{Cabass:2024wob,Cabass:2022oap}. Light and intermediate fields predominantly source scale-dependent bias; 
recent work by \citep{Green:2023uyz,Goldstein:2024bky} demonstrates that this is a powerful avenue for constraining such phenomena. Information can also be extracted from galaxy shapes \citep{Schmidt:2015xka,Kogai:2020vzz} as well as weak lensing, though their analysis may require non-perturbative methods \citep{Goldstein:2023brb,Dimastrogiovanni:2015pla}. Most previous forecasts consider only three-point non-Gaussianity; the models discussed in this series, however, could be similarly probing using such observables. Finally, 21cm observations of the dark ages could yield extremely strong constraints on both bispectrum and trispectrum amplitudes, potentially reaching $\sigma(\tau_{\rm NL}^{\rm heavy})\sim 1$ as well as $\sigma(\tau_{\rm NL}^{\rm light})\sim 10^{-5} - 10^0$ (depending on $\nu_s$) \citep{Floss:2022grj}. 

\subsubsection{Implications}
\noindent Previous bispectrum studies raise an interesting question: is it worth searching for collider non-Gaussianity in the four-point function given that we have not detected it in the three-point function? To answer this, it is instructive to consider a simple set-up: the exchange of a spin-$s$ particle, $\sigma$, between gauge bosons, $\pi$. As discussed in \paperone (and \citep{Arkani-Hamed:2015bza}), the curvature bispectrum and trispectrum generated by such an interaction take the schematic form
\beq\label{eq: schematic-collider}
    B_\zeta(k_1,k_2,k_3) &\sim& B_{\pi\pi\sigma^{(0)}}(k_1,k_2,k_3)\times\frac{1}{P_{\sigma^{(0)}\sigma^{(0)}} (k_3)}P_{\pi\sigma^{(0)}}(k_3) \qquad (k_3\ll k_1)\\\nonumber
    T_\zeta(k_1,k_2,k_3,k_4,K) &\sim& \sum_{\lambda}B_{\pi\pi\sigma^{(\lambda)}}(k_1,k_2,K)\frac{1}{P_{\sigma^{(\lambda)}\sigma^{(\lambda)}}(K)}B_{\pi\pi\sigma^{(\lambda)}}(k_3,k_4,K) \qquad (K\ll k_1,k_2),
\eeq
where $B_{\pi\pi\sigma^{(\lambda)}}$ represents the three-point interaction involving two $\pi$ bosons and a $\sigma$ field in helicity state $\lambda$ \textit{et cetera}. Whilst the trispectrum depends on two cubic interaction vertices, the bispectrum requires both a cubic and a quadratic vertex, with the latter describing the coupling of $\sigma^{(\lambda)}$ to the inflaton; by symmetry, this requires $\lambda=0$. \eqref{eq: schematic-collider} leads to two important conclusions: (1) collider bispectra are sourced only by the longitudinal helicity mode; (2) they require a quadratic mixing with the inflaton. If either of these are not present (for example, if transverse states dominate \citep[e.g.,][]{Bordin:2018pca} or the quadratic coupling is protected by symmetry), we will not form an inflationary bispectrum. As such, the curvature trispectrum represents a much more natural testing ground for collider physics 
and can uncover a wide range of previously hidden physics.

Although none of the above models are forbidden experimentally, some are theoretically disfavored. In particular, the production rate of massive particles is suppressed by $e^{-\pi m/H}$, which restricts us to comparatively small $\mu_s$ (though see \citep[e.g.,][]{Bodas:2020yho,Kim:2021ida,Craig:2024qgy,Lee:2016vti} for exceptions to this). Practically, searching for particles with large $\mu_s$ becomes difficult, since the characteristic oscillations are washed out by the CMB projection integrals. Furthermore, in the limit of $m\gg H$, massive particles mimic inflationary self-interactions (albeit with a spin-dependent modification \citep[cf.][]{Sohn:2024xzd}). This sources equilateral trispectrum (generalizing the EFTI shapes), and requires a very different analysis methodology, since we can no longer restrict to the collapsed regime and must resort to different methods to compute the templates, such as bootstrap techniques \citep{Pimentel:2022fsc}. In general, implementing an everywhere-defined trispectrum template could yield significantly tighter constraints on conformally coupled and massive particles, 
and would be an interesting avenue for future work.

\subsection{Late-Time Non-Gaussianity}

\noindent Since its first measurement in \citep{Smith:2007rg}, there have been a large number of detections of gravitational lensing with increasing significance. Building on the PR1 and PR2 analyses \citep{Planck:2013mth,Planck:2015mym}, 
the \textit{Planck} collaboration used PR3 temperature and polarization data to obtain $A_{\rm lens} = 0.995 \pm 0.026$ across $8\leq L\leq 2048$ (their $A^{\phi\phi}$ parameter); this weakens to $1.004 \pm 0.033$ if one excludes polarization, or $1.011\pm0.028$ using the fiducial value of $L_{\rm max}=400$ \citep{Planck:2018lbu}. Recently, \citep{Carron:2022eyg} analyzed \textit{Planck} PR4 data, finding $A_{\rm lens}=1.004\pm0.024$ from temperature-plus-polarization at $L_{\rm max}=400$. In \S\ref{subsec: results-local}, we found $A_{\rm lens}=1.001\pm0.024$ from the combined \textsc{sevem} dataset (including polarization), with a slightly tighter error obtained from \textsc{smica} ($0.023$) or \textsc{sevem} with optimal weighting ($0.023$); our results are highly consistent with the previous measurements, and with the $\Lambda$CDM model. Ours are among the tightest constraints on lensing obtained from current data (matching the precision of ACT DR6 \citep{ACT:2023dou}). Much stronger constraints will soon be obtained with high-resolution data from ACT, SPT, the Simons Observatory and CMB-S4.

Our sensitivity to the point-source trispectrum is also comparable to that of previous works. \citep{Planck:2013mth} found $10^{38}t_{\rm ps} = (60\pm30)\,\mathrm{K}^4$ from \textit{Planck} PR1 data (their $S_4$ parameter), and \citep{Carron:2022eyg} analyzed \textit{Planck} PR4 \textsc{smica} maps, obtaining $10^{38}t_{\rm ps} = (16.7\pm4.4)\,\mathrm{K}^4$. In this work we obtain $10^{38}t_{\rm ps} = (32.1\pm6.2)\,\mathrm{K}^4$ ($(19.1\pm5.9)\,\mathrm{K}^4$) from \textsc{sevem} (\textsc{smica}), as shown in Tab.\,\ref{tab: local-results}. Whilst the individual bounds differ somewhat due to variations in masking and preprocessing (including the removal of bright point-sources), the empirical consistency is clear. 


\section{Conclusion}\label{sec: conclusion}
\noindent In this series, we have performed a detailed analysis of the inflationary trispectrum. In \paperone, we explored various models of non-Gaussian primordial physics and constructed a suite of separable templates that allow their physical signatures to be constrained. Using these forms, we derived quasi-optimal estimators for the trispectrum amplitudes (fully accounting for various observational effects such as the mask), which allow them to be efficiently measured from high-resolution CMB data. \papertwo introduced \href{https://github.com/oliverphilcox/PolySpec}{\polyspec}, an efficient \textsc{python}/\textsc{c} implementation of these estimators, and presented a wide variety of validation and performance tests, confirming that we can robustly extract trispectrum amplitudes from data. In this work we have applied our formalism to the latest \textit{Planck} temperature and polarization data, and presented constraints on $33$ trispectrum amplitudes (four of which are degenerate), spanning a wide range of physical regimes including single-field inflation, gravitational lensing, and the cosmological collider. Furthermore, we have demonstrated that our results are robust to various systematic effects (including Monte Carlo convergence, galactic foregrounds, and noise mismodeling), and translated our constraints into bounds on various inflationary scenarios.

Across all of the $>150$ single-template analyses performed in this work (dropping degenerate templates, but accounting for temperature and polarization differences), we have found no robust detection of non-Gaussianity. Due to the look-elsewhere effect \citep{look-elsewhere,Bayer:2020pva,Bayer:2021lhk}, the significance of any detection is reduced compared to the na\"ive expectation -- we do not attempt to model this effect in this work, since our templates and datasets are non-trivially correlated. We find largest deviations for the direction-dependent $\tau_{\rm NL}^{221}$ template, with temperature-only measurements from \textsc{smica} and \textsc{sevem} $\approx 4\sigma$ below zero (Tab.\,\ref{tab: tau-direc-results}). Following thorough investigation (see \S\ref{subsec: results-direc}), we conclude that this is not physical since we have not accounted for the (demonstrably significant) non-Gaussianity in the $\tau_{\rm NL}^{221}$ posterior, the result is not robust to masking variations, and we find no signal in the combined temperature-plus-polarization dataset. 
The remaining measurements are all consistent with zero at $<3\sigma$ except for the polarization-only $\taunl$ \textsc{smica} measurement, which carries negligible weight in the combined analysis.

Due to the use of optimal estimators and polarization data, many of our constraints are significantly tighter than those found in previous works: for example, our $\gnl$ result is $35\%$ stronger than the official \textit{Planck} PR3 result \citep{Planck:2019kim}, and our EFTI bounds improve on the literature values by $50-150\%$. Furthermore, our constraints on $\taunl$ and CMB lensing are consistent with (or slightly tighter than) previous works. In many cases, we perform the first analysis of the relevant templates, such as direction-dependent non-Gaussianity. 
An important outcome of this series is a set of templates describing the collapsed limit of cosmological collider non-Gaussianity \citep{Arkani-Hamed:2015bza}: by measuring the corresponding amplitudes, we have placed constraints on inflationary particles with various spins and masses, probing a range of phenomenology including oscillations and squeezed limit divergences. 

It is important to reflect on the limitations of our analysis. Whilst the main results presented in this work pass a barrage of consistency tests, we have had to carefully account for a number of systematic effects that could have sourced spurious detections. Below, we discuss a number of effects that will be important to account for in future analyses:
\begin{itemize}
    \item \textbf{Simulations}: Our estimators use a suite of mocks to subtract the disconnected (\textit{i.e.}\ Gaussian) contributions to the four-point function. Any mismatch between simulations and data can lead to a significant bias (even for the parity-odd templates due to mask-induced leakage): as demonstrated in Fig.\,\ref{fig: local-systematics}, a $3\%$ mismatch in the high-$\ell$ power spectrum leads to a $2\sigma$ bias in $\gnl$ and a $6\sigma$ bias in $\taunl$. Relatedly, we require a large number of simulations: doubling the number leads to $0.3\sigma$ shifts in the $A_{\rm lens}$ measurement, though we find smaller variation for the main primordial templates of interest. 
    \item \textbf{Dust}: As with any CMB analysis, the impact of residual foregrounds must be carefully explored. Whilst we find consistent results using $f_{\rm sky}\leq 70\%$, the \textsc{sevem} $\taunl$ measurements show large shifts when using a less conservative mask (Fig.\,\ref{fig: local-fsky}), which could easily be misinterpreted as evidence for primordial non-Gaussianity. 
    \item \textbf{Lensing}: Despite the weak correlations between the $A_{\rm lens}$ trispectrum and the primordial shapes, gravitational lensing can lead to significant biases in our constraints. For \textit{Planck}, we found shifts of up to $5\sigma$, with largest effects seen in the EFTI and collider analyses. This will become increasingly important as lensing in the future, but can be mollified using the techniques discussed in this series. 
    \item \textbf{Posteriors}: Interpreting the measured non-Gaussianity amplitudes can be difficult. For example, the sampling distribution of the $\taunl$ estimator is strongly non-Gaussian, which leads to significantly weaker bounds than expected given the empirical variances. Whilst an accurate likelihood approximation exists for $\widehat{\tau}_{\rm NL}^{\rm loc}$, this can be expensive to compute (requiring an $L$-by-$L$ analysis), and no similar approximation exists for the more nuanced templates, such as direction-dependent non-Gaussianity. 
    \item \textbf{Parameter Studies}: Many templates depend non-linearly on continuous hyperparameters, which are not known \textit{a priori}. An important example is the cosmological collider, which is a function of the mass parameter $\nu_s \equiv i\mu_s$. Ideally, one would perform a joint analysis of the amplitude and $\nu_s$, allowing computation of the marginalized posteriors. Given that our estimators themselves depend on $\nu_s$, this cannot be easily achieved in our framework. This limits us to frequentist studies (as in Tab.\,\ref{tab: coll-results}, see also \citep{Sohn:2024xzd}) or marginalization over not-too-correlated templates (as in Fig.\,\ref{fig: coll_spin_correlation}).     
\end{itemize}

What's next for primordial trispectrum analyses? As usual, the answer is more models and more data. Although we have discussed a wide variety of four-point functions in this work, our census is by no means complete: a selection of models worthy of future study include collider models with broken conformal invariance \citep{Lee:2016vti}, higher spin particles \citep{Bordin:2019tyb}, supersymmetric fermions \citep{Alexander:2019vtb}, tachyons \citep{McCulloch:2024hiz} partially massless states \citep{Baumann:2017jvh}, low-speed colliders \citep{Wang:2022eop,Jazayeri:2023xcj,Jazayeri:2023xcj}, parity-violating models \citep[e.g.,][]{Cabass:2022rhr}, thermal baths \citep[e.g.,][]{Salcedo:2024smn}, resonances \citep[e.g.,][]{Chen:2008wn,Flauger:2009ab,Flauger:2010ja}, isocurvature models \citep{Langlois:2008wt,Bartolo:2001cw,Grin:2013uya}, ghost inflation \citep{Izumi:2010wm,Huang:2010ab}, scale dependence non-Gaussianity \citep{Byrnes:2010ft,Wang:2022eop}, solid inflation \citep{Bartolo:2014xfa,Endlich:2013jia}, and primordial magnetic fields \citep{Trivedi:2011vt}. Some of these require simple modifications to the estimators of this series, whilst others will require significant methodological developments. 

The next generation of CMB observatories will yield significantly tighter bounds on non-Gaussianity (or, more optimistically, its first detection). Experiments such as LiteBIRD, the Simons Observatory, and CMB-S4 \citep{Hazumi:2019lys,SimonsObservatory:2018koc,CMB-S4:2016ple} will give complementary views of the CMB temperature and polarization anisotropies down to much smaller scales. Given the strong scaling of our constraints with $\ell_{\rm max}$, we can expect improvements across a range of non-Gaussianity analyses; much of the new information will arise from polarization, which we have discovered to yield encode significant primoprdial information. Future constraints may benefit from delensing \citep[e.g.,][]{Trendafilova:2023xtq}; whilst this would not significantly enhance the \textit{Planck} analyses (Fig.\,\ref{fig: variance-breakdown}), it will become increasingly important as $\ell_{\rm max}$ increases. Finally, we note that the tools presented in this series can be realistically scaled to future datasets. At the full \textit{Planck} resolution, computation of $100$ $\gnl$, $\taunl$ and $A_{\rm lens}$ numerators using \polyspec required under four minutes per simulation on a 64-core machine (though this increases for more complex templates). Whilst future datasets will require higher precision computations, combining the fast estimators presented in this series with future methodological and computational enhancements, we will be able to place tight constraints on many models of primordial non-Gaussianity in the coming years.



\acknowledgements
\vskip 8pt
{\small
\noindent We thank Giovanni Cabass, William Coulton, Adriaan Duivenvoorden, Sam Goldstein, Colin Hill and Maresuke Shiraishi for insightful discussions, as well as Colin Hill and Maresuke Shiraishi for comments on the manuscript. \resub{We additionally thank the anonymous referee for their careful reading of the manuscript and useful comments.}\begingroup\hypersetup{hidelinks} OHEP is a Junior Fellow of the Simons Society of Fellows, and thanks
\href{https://www.flickr.com/photos/198816819@N07/54319059246/}{Lleonardo da Vinci} for creative inspiration.
\endgroup OHEP would also like to thank the Center for Computational Astrophysics for their hospitality across the multiple years this set of papers took to write. 
The computations in this work were run at facilities supported by the Scientific Computing Core at the Flatiron Institute, a division of the Simons Foundation.

\appendix

\section{Primordial Templates}\label{app: templates}
\noindent In this appendix, we list the primordial templates used in this work. A detailed discussion of each can be found in \paperone; here, we give the explicit forms for reference. In all cases, we use the Fourier-space gauge-invariant curvature perturbation, $\zeta(\vk)$, whose two-point function is given by $\av{\zeta(\vk_1)\zeta(\vk_2)} = P_\zeta(k_1)\delta_{\rm D}(\vk_1+\vk_2)$ where $\delta_{\rm D}$ is the Dirac delta function.

\begin{itemize}
    \item \textbf{Local Non-Gaussianity}. This is defined by two four-point functions with amplitudes $\gnl$ and $\taunl$, which parametrize cubic and quadratic local transformations. Explicitly:
\beq\label{eq: template-gloc}    \av{\zeta(\vk_1)\zeta(\vk_2)\zeta(\vk_3)\zeta(\vk_4)}'_c &\supset& \frac{54}{25}g^{\rm loc}_{\rm NL}P_\zeta(k_1)P_\zeta(k_2)P_\zeta(k_3)+\text{3 perms.},
\eeq
and
\beq\label{eq: template-tauloc}
    \av{\zeta(\vk_1)\zeta(\vk_2)\zeta(\vk_3)\zeta(\vk_4)}'_c &\supset&\taunl P_\zeta(k_1)P_\zeta(k_3)P_\zeta(K)+\text{11 perms.}
\eeq
    where $K\equiv |\vk_1+\vk_2|$ is the exchange momentum.
    \item \textbf{Contact Non-Gaussianity}. This is an approximately featureless shape, defined by the amplitude $g_{\rm NL}^{\rm con}$ and the following four-point function:
\beq\label{eq: template-con}
    \av{\zeta(\vk_1)\zeta(\vk_2)\zeta(\vk_3)\zeta(\vk_4)}'_c \supset \frac{216}{25}g_{\rm NL}^{\rm con}\left[P_\zeta(k_1)P_\zeta(k_2)P_\zeta(k_3)P_\zeta(k_4)\right]^{3/4}.
\eeq
    \item \textbf{EFTI Non-Gaussianity}. The Effective Field Theory of Inflation predicts three trispectra with amplitude parameters $\gnldotdot$, $\gnldotdel$, and $\gnldeldel$. These can be expressed as integrals over conformal time, $\tau$:
\beq\label{eq: template-efti}
    \av{\zeta(\vk_1)\zeta(\vk_2)\zeta(\vk_3)\zeta(\vk_4)}'_c &\supset& \frac{9216}{25}\gnldotdot A_\zeta^3\int_{-\infty}^0d\tau\,\tau^4\left(\prod_{i}\frac{e^{k_i\tau}}{k_i}\right)\\\nonumber
    &&\,-\,\frac{13824}{325}\gnldotdel A_\zeta^3\int_{-\infty}^0d\tau\,\tau^2\frac{(1-k_3\tau)(1-k_4\tau)}{k_1k_2(k_3k_4)^3}(\vk_3\cdot\vk_4)e^{\sum_i k_i\tau}+\text{5 perms.}\\\nonumber
    &&\,+\,\frac{82944}{2575}\gnldeldel A_\zeta^3\int_{-\infty}^0d\tau\,\left(\prod_{i}\frac{(1-k_i\tau)e^{k_i\tau}}{k_i^3}\right)(\vk_1\cdot\vk_2)(\vk_3\cdot\vk_4)+\text{2 perms.},
\eeq
    where $A_\zeta$ is the power spectrum amplitude. In practice, this is replaced by $k^3P_\zeta(k)$ to ensure the correct $k^{-3(4-n_s)}$ scaling. 
\item \textbf{Direction-Dependent Non-Gaussianity}. The $\taunl$ template of \eqref{eq: template-tauloc} can be generalized by allowing for dependence on the angles between $\vk_1$, $\vk_3$ and $\vK$. A general expansion is given by
\beq\label{eq: template-direc}
    \av{\zeta(\vk_1)\zeta(\vk_2)\zeta(\vk_3)\zeta(\vk_4)}'_c &\supset& \frac{1}{2}\sum_{n_1n_3n}\tau_{\rm NL}^{n_1n_3n}\left[\sum_{m_1m_3m}\tj{n_1}{n_3}{n}{m_1}{m_3}{m}Y_{n_1m_1}(\hk_1)Y_{n_3m_3}(\hk_3)Y_{nm}(\hK)\right]\\\nonumber
    &&\qquad\,\times\,P_\zeta(k_1)P_\zeta(k_3)P_\zeta(K)+\text{23 perms.}
\eeq
    with amplitude parameters $\tau_{\rm NL}^{n_1n_3n}$ (for integer $n_i$), where $Y_{nm}$ is a spherical harmonic and the quantity in parentheses is a Wigner $3j$ symbol. A simplified decomposition is given by the parity-even and parity-odd expressions:
\beq\label{eq: template-even}
    \av{\zeta(\vk_1)\zeta(\vk_2)\zeta(\vk_3)\zeta(\vk_4)}'_c &\supset& \frac{1}{6}\sum_{n\geq 0}\tau_{\rm NL}^{n, \rm even}\left[\mathcal{L}_n(\hk_1\cdot\hk_3)+(-1)^n\mathcal{L}_n(\hk_1\cdot\hK)+\mathcal{L}_n(\hk_3\cdot\hK)\right]\\\nonumber
    &&\qquad\,\times\,P_\zeta(k_1)P_\zeta(k_3)P_\zeta(K)+\text{23 perms.}
\eeq
    and 
\beq\label{eq: template-odd}
    \av{\zeta(\vk_1)\zeta(\vk_2)\zeta(\vk_3)\zeta(\vk_4)}'_c &\supset&-\frac{i}{6}\sum_{n\geq 0}\tau_{\rm NL}^{n, \rm odd}\left[\mathcal{L}_n(\hk_1\cdot\hk_3)+(-1)^n\mathcal{L}_n(\hk_1\cdot\hK)+\mathcal{L}_n(\hk_3\cdot\hK)\right](\hk_1\times\hk_3\cdot\hK)\\\nonumber
    &&\qquad\,\times\,P_\zeta(k_1)P_\zeta(k_3)P_\zeta(K)+\text{23 perms.},
\eeq
    with characteristic amplitudes $\tau_{\rm NL}^{n,\rm even}$ and $\tau_{\rm NL}^{n,\rm odd}$, where $\mathcal{L}_n$ is a Legendre polynomial.
    \item \textbf{Collider Non-Gaussianity}. Massive particles present during inflation leave signatures in the collapsed limits of the inflationary four-point function (with $K\ll k_1,k_3$). These depend on the particle spin, $s$, and mass, $m$, which is usually parametrized by $\nu_s = \sqrt{(s-1/2)^2-m^2/H^2}$ for inflationary Hubble rate $H$ (with $\nu_0 = \sqrt{9/4-m^2/H^2}$ for spin-zero). We define the following template for light/intermediate mass particles (formally the complementary series): 
\beq\label{eq: template-light}
    \av{\zeta(\vk_1)\zeta(\vk_2)\zeta(\vk_3)\zeta(\vk_4)}'_c &\supset& \tau_{\rm NL}^{\rm light}(s,\nu_s)\sum_{S=0}^{2s}\mathcal{C}_s(S,i\nu_s)\sum_{\lambda_1\lambda_3\Lambda}\tj{s}{s}{S}{\lambda_1}{\lambda_3}{\Lambda}Y_{s\lambda_1}(\hk_1)Y_{s\lambda_3}(\hk_3)Y_{S\Lambda}(\hK)\\\nonumber
    &&\,\times\,\left(\frac{K^2}{k_1k_3}\right)^{3/2-\nu_s}P_\zeta(k_1)P_\zeta(k_3)P_\zeta(K)\\\nonumber
    &&\,\times\,\Theta_{\rm H}(k_1-k_{\rm coll})\Theta_{\rm H}(k_3-k_{\rm coll})\Theta_{\rm H}(K_{\rm coll}-K)+\text{11 perms.}
\eeq
    with amplitude parameter $\tau_{\rm NL}^{\rm light}(s,\nu_s)$. This depends on a coupling parameter $\mathcal{C}_s$ given in \paperone. In the last line, we restrict to $k_i\geq k_{\rm coll}$ and $K\leq K_{\rm coll}$ via the Heaviside functions $\Theta_{\rm H}$: for suitably chosen $k_{\rm coll},K_{\rm coll}$, these ensure that we remain in mildly collapsed regimes where the template is valid. This differs from the $\taunl$ template principally due to the angular dependence and the suppression by $(K^2/k_1k_3)^{3/2-\nu_s}$.
    
    For heavy masses $m>(s-1/2)H$, or $m>3H/2$ for $s=0$ (formally the principal series), we define the following template depending on mass parameter $\mu_s \equiv -i\nu_s$:
\beq\label{eq: template-heavy}
    \av{\zeta(\vk_1)\zeta(\vk_2)\zeta(\vk_3)\zeta(\vk_4)}'_c &\supset& \tau_{\rm NL}^{\rm heavy}(s,\mu_s)\sum_{S=0}^{2s}\left|\mathcal{C}_s(S,\mu_s)\right|\sum_{\lambda_1\lambda_3\Lambda}\tj{s}{s}{S}{\lambda_1}{\lambda_3}{\Lambda}Y_{s\lambda_1}(\hk_1)Y_{s\lambda_3}(\hk_3)Y_{S\Lambda}(\hK)\\\nonumber
    &&\,\times\,\frac{1}{2}\left[\left(\frac{K^2}{k_1k_3}\right)^{3/2+i\mu_s}e^{i\omega_s(S,\mu_s)}+\left(\frac{K^2}{k_1k_3}\right)^{3/2-i\mu_s}e^{-i\omega_s(S,\mu_s)}\right]\\\nonumber
    &&\,\times\,\Theta_{\rm H}(k_1-k_{\rm coll})\Theta_{\rm H}(k_3-k_{\rm coll})\Theta_{\rm H}(K_{\rm coll}-K)P_\zeta(k_1)P_\zeta(k_3)P_\zeta(K)+\text{11 perms.},
\eeq
    where $\tau_{\rm NL}^{\rm heavy}(s,\mu_s)$ is the characteristic amplitude parameter and $\omega_s$ is a phase defined in \paperone. This features oscillations with frequency $\mu_s$ in the collapsed limit.
\end{itemize}
In this work, we additionally late-time non-Gaussianity from gravitational lensing and unresolved point sources. The CMB trispectra arising from these effects are discussed in \paperone.

\section{Variance Breakdown}\label{app: variance-breakdown}
\noindent Fisher forecasts are a powerful tool to predict the constraining power of a given dataset. In many scenarios, one adds simplifying assumptions to make the forecast more tractable: these include translation-invariant noise, a unit mask, and a unit beam. In this appendix, we perform an array of forecasts to explore the dependence of the \textit{Planck} local and lensing non-Gaussianity constraints on these assumptions.

\begin{figure}
    \centering
    \includegraphics[width=0.8\linewidth]{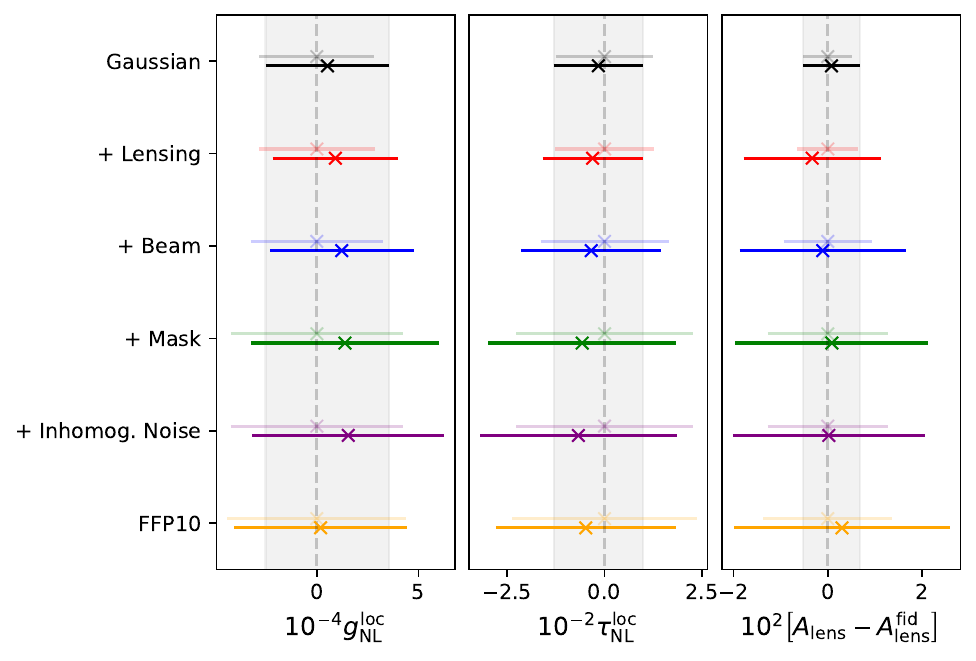}
    \caption{Forecasted constraints on local and lensing non-Gaussianity including various degrees of realism. We start from an analysis of Gaussian data with homogeneous Poisson noise (black) then successively add lensing distortions (red), the \textit{Planck} beam (blue), the observational mask (green), 
    inhomogeneous Poisson noise matching the \textit{Planck} scan strategy (purple), and the full FFP10 component-separation noise (yellow). All results are obtained with temperature and polarization data with $\ell\in[2,2048]$, $L\in[1,30]$. Solid lines indicate the mean result from analyzing $100$ simulations (shared between each test), whilst the faint lines give the predictions from the (numerically-computed) inverse Fisher matrix. In \textit{Planck}, the largest contributions to the primordial variance are the beam and mask, though their effects vary significantly across templates. In most cases, we find fairly good agreement between the empirical and theoretical errors, except for $A_{\rm lens}$, wherein lensing non-Gaussianities lead to significant broadening of the constraints.} 
    \label{fig: variance-breakdown}
\end{figure}

Fig.\,\ref{fig: variance-breakdown} shows the forecasted constraints on $\gnl$, $\taunl$, and $A_{\rm lens}$ from the baseline \textit{Planck} PR4 analysis computed using the Fisher matrix (obtained numerically, as described in \S\ref{subsec: method-estimator}), as well as $100$ mock simulations. In the simplest scenario, we analyze unlensed Gaussian CMB realizations without a beam or mask, adding translation-invariant Poisson noise. In this limit, we forecast tight constraints on all parameters with good agreement between the theoretical and empirical variances (given the expected $7\%$ scatter from finite-mock effects). Switching to lensed CMB realizations leads to a $25\%$ increase in the theoretical $\sigma(A_{\rm lens})$: in contrast, the empirical error is increased by a factor of $140\%$. This is a signature of lensing-induced non-Gaussianity, 
and signifies suboptimalities in the lensing estimator. The primordial constraints are only weakly affected by lensing (aside from a $1.0\sigma$ shift in the mean of $\gnl$ which we predict and subtract).

The next entry in Fig.\,\ref{fig: variance-breakdown} shows the effect of adding a \textit{Planck}-like beam to the mock data. The associated damping of high-$\ell$ modes leads to a minor inflation in the $\gnl$ constraint ($15\%$), but a larger increase for $\taunl$ and $A_{\rm lens}$ ($30-40\%$) -- this is expected due to the sharper dependence of these errorbars on $\ell_{\rm max}$. Repeating the analysis with an $f_{\rm sky}\approx=0.75$ mask degrades the constraints between $15\%$ ($A_{\rm lens}$) and $35\%$ ($\taunl$); these variations are not well captured by the $1/\sqrt{f_{\rm sky}}$ factor included in most idealized forecasts. 
Allowing for spatial variations in noise properties (via the FFP10 noise simulations) leads to negligible variation in the constraints, though we caution that this effect could be larger for less uniform surveys.

Finally, we compare the theoretical variances to those from 100 FFP10 simulations. These incorporate all the above features, in addition to realistic point-sources, artifacts and foreground residuals. For $\gnl$ and $\taunl$, we find excellent agreement between empirical and theoretical errors (within $5\%$), whilst the empirical $\sigma(A_{\rm lens})$ is larger than the numerical Fisher matrix prediction by $67\%$; notably, these errorbars can be significantly larger than those suggested by rescaling idealized forecasts by factors of $1/\sqrt{f_{\rm sky}}$. The conclusion from this exercise is the following: the precise errors on non-Gaussianity parameters reflect a complex interplay between various features, the most important of which are gravitational lensing, the instrumental beam, and observational mask. 

\section{\texorpdfstring{Scale-Dependent $\taunl$ Analysis}{Scale-Dependent Quadratic Local Non-Gaussianity Analysis}}\label{app: tauNL-Lmax}

\noindent As noted in Fig.\,\ref{fig: local-pdf}, the null distribution of $\widehat{\tau}_{\rm NL}^{\rm loc}$ is non-Gaussian with a tail towards positive values. 
In this appendix, we perform an alternative $\taunl$ analysis based on \citep{Marzouk:2022utf} (itself generalizing \citep{2014A&A...571A..24P}), which allows for construction of a physical bound on $\taunl$, accounting for the skewed likelihood.

To understand this approach, it is useful to rewrite the $\taunl$ estimator in a more familiar form.\footnote{These equations are purposefully schematic -- we refer the reader to \paperone for a more detailed discussion of the relation between our optimal estimators and standard `local anisotropy' forms.} First, we introduce a quadratic field, $\Phi_{LM}$ (analogous to the usual CMB lensing estimator) that measures some large-scale distortion field proportional to $\sqrt{\taunl}$ \citep[e.g.,][]{Hanson:2009gu}. For suitably defined $\Phi_{LM}$, we can rewrite the estimator of \eqref{eq: numerator}:
\beq
    \widehat{N}_{\taunl} &= & \sum_{LM}\Bigg\{\Phi_{LM}[d,d]\Phi^*_{LM}[d,d] - \bigg(\av{\Phi_{LM}[d,d]\Phi^*_{LM}[\delta,\delta]}_{\delta}+\text{5 perms.}\bigg)\\\nonumber
    &&\qquad\quad\,+\,\bigg(\av{\Phi_{LM}[\delta,\delta]\Phi^*_{LM}[\delta',\delta']}_{\delta,\delta'}+\text{2 perms.}\bigg)\Bigg\},
\eeq
again involving both the data and averages over simulations $\{\delta,\delta'\}$. Defining a `mean-field' $\bar\Phi_{LM}\equiv\av{\Phi_{LM}[\delta,\delta]}_{\delta}$, this can be recognized as an estimator for the power spectrum of $\Phi$:
\beq\label{eq: simple-lensing-estimator}
   \widehat{N}_{\taunl} &= & \sum_{LM}\Big\{\big|\Phi_{LM}[d,d]-\bar{\Phi}_{LM}\big|^2\Big\} - {N}^{(0)}[d,d]\\\nonumber
   {N}^{(0)}[d,d] &\equiv& \sum_{LM}\Big\{4\av{\Phi_{LM}[d,\delta]\Phi^*_{LM}[d,\delta]}_\delta-2\av{\Phi_{LM}[\delta,\delta']\Phi^*_{LM}[\delta,\delta']}_{\delta,\delta'}\Big\},
\eeq
including a `realization-dependent-noise' term, ${N}^{(0)}$, which subtracts off the Gaussian noise bias. The combination $\widehat{N}_{\taunl}+N^{(0)}$ is positive definite, leading to the physical constraint $\taunl>0$. It is straightforward to define estimators for each $L$-mode in turn by summing only over $M$: as such, we can build a suite of $\taunl$ estimators for $L\in[L_{\rm min},L_{\rm max}]$:
\beq
    \widehat{\tau}_{\rm NL}^{\rm loc}(L) = \sum_{L'=L_{\rm min}}^{L_{\rm max}}\F^{-1}(L,L')\widehat{N}_{\taunl}(L'),
\eeq
where $\F(L,L')$ is a normalization matrix, defined as in \eqref{eq: fisher} but restricted to a single pair of $L$-modes.\footnote{Due to the efficient computation scheme developed in \paperone, this can be computed in $\mathcal{O}(L_{\rm max})$ time. The full estimator is exactly recovered via the definition $\widehat{\tau}_{\rm NL}^{\rm loc} = \sum_L\widehat{N}_{\taunl}(L)/\sum_{LL'}\F(L,L') \equiv \sum_{LL'}\F(L,L')\widehat{\tau}_{\rm NL}^{\rm loc}(L')/\sum_{LL'}\F(L,L')$.}

As discussed in \citep{2014A&A...571A..24P,Marzouk:2022utf}, we can model the distribution of $\widehat{\tau}_{\rm NL}^{\rm loc}(L)$ using tools developed for cut-sky power spectrum analyses \citep{Hamimeche:2008ai}, since each estimator can be written in terms of a large-scale power spectrum. 
This gives the following (approximate) likelihood for the $\{\widehat{\tau}_{\rm NL}^{\rm loc}(L)\}$ datavector:
\beq\label{eq: tauNL-like}
    -2\log\mathcal{L}\left(\{\widehat{\tau}^{\rm loc}_{\rm NL}(L)\}|\taunl\right) &=& \sum_{LL'}N^{\rm fid}(L)g[x(L)]\times\mathcal{C}^{-1}(L,L')\times N^{\rm fid}(L')g[x(L')]\\\nonumber
    g(x) &=& \text{sign}(x-1)\sqrt{2(x-\log(x)-1)}\\\nonumber
    x(L) &=& \frac{\widehat{N}_{\taunl}(L)+{N}^{(0)}(L)}{\sum_{L'}\F(L,L')\taunl+N^{(\rm lens)}(L)+{N}^{(0)}(L)}
\eeq
where $N^{\rm fid}$ is a fiducial power spectrum (including noise) whose covariance is given by $\mathcal{C}$. Whilst this may look imposing, it simply involves the ratio of empirical ($\widehat{N}_{\taunl}(L)+{N}^{(0)}(L)$) and theoretical ($\sum_{L'}\F(L,L')\taunl+N^{(\rm lens)}+{N}^{(0)}(L)$) power spectra, noting the definition of $\F$ and including a lensing contribution. This generalizes the standard $\chi^2$ distribution obtained for a full-sky experiment, and asymptotes to a Gaussian likelihood for the combined estimator $\widehat{\tau}_{\rm NL}^{\rm loc}$ in the limit of large $N^{(0)}(L)$.

\begin{figure}
    \centering
    \includegraphics[width=0.5\linewidth]{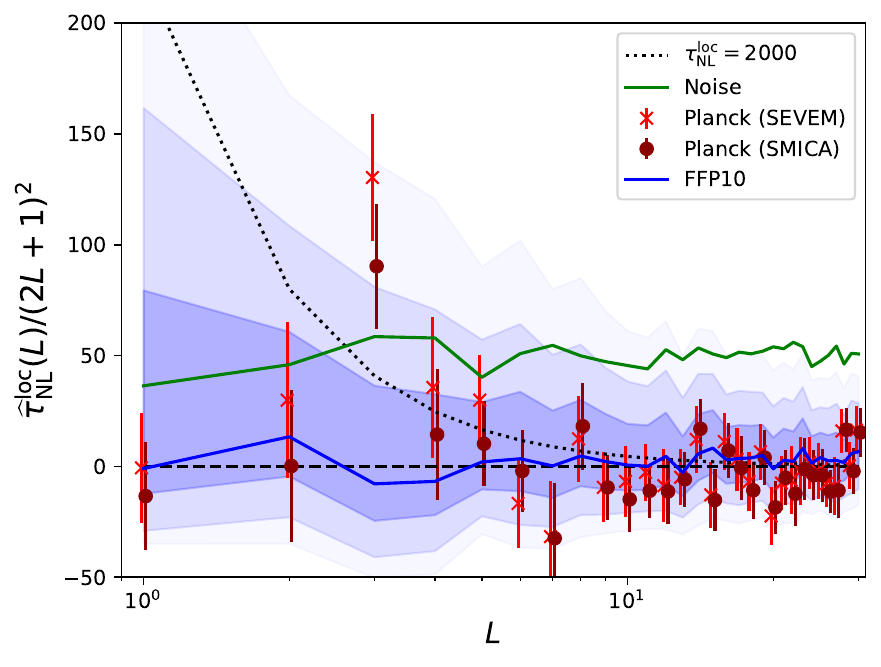}
    \caption{Measurements of $\tauest^{\rm loc}$ from the \textit{Planck} temperature-plus-polarization dataset, treating each $L$-mode independently (as in \citep{2014A&A...571A..24P,Marzouk:2022utf}). We show results for the \textsc{sevem} (\textsc{smica}) dataset in light red crosses (dark red circles), with noise contributions, $[\F^{-1}N^{(0)}](L)$, shown in green. The blue lines show results from the mean of 100 FFP10 simulations, and the bands show the $\{1,2,3\}\sigma$-equivalent confidence regions obtained from \eqref{eq: tauNL-like} under null assumptions. The noise, FFP10, and likelihood curves are almost identical for both component separation methods, thus we show only those for \textsc{sevem}. For reference, we plot the expected signal from a $\taunl=2000$ model as a dotted black line. With the possible exception of $L=3$, all measurements are consistent with zero, and we find the $95\%$ confidence intervals $\taunl<2360$ ($1500$) for \textsc{sevem} (\textsc{smica}) with the difference principally driven by the $L=3$ mode.}
    \label{fig: tauNL-L-dep}
\end{figure}

In practice, we implement the above estimators using a modified version of the \textsc{PolySpec} code, defining a set of new templates including contributions from only a single $L$-mode. In Fig.\,\ref{fig: tauNL-L-dep} we show the $\tauest^{\rm loc}(L)$ measurements for $L\in[1,30]$ using the \textit{Planck} temperature-plus-polarization dataset (cf.\,Fig.\,7 of \citep{Marzouk:2022utf}). The decomposition into $L$-modes allows for increased interpretability compared to the main estimators used in the text -- here, we observe that the measurements are broadly consistent with zero on all scales for both \textsc{sevem} and \textsc{smica}, with the constraints dominated by the largest $L$-modes. As observed in previous analyses \citep{2014A&A...571A..24P,Marzouk:2022utf}, the $L=3$ octopole is non-zero at $2-3\sigma$; whilst this may appear to be evidence for $\taunl>0$, we observe that a similar trend is not found at lower $L$, and the detection significance varies by $\approx 1\sigma$ between \textsc{smica} and \textsc{sevem}.

Given these results, we can obtain a posterior on the $\taunl$ parameter. In practice, this is computed by integrating \eqref{eq: tauNL-like} over $\taunl>0$, equating $\mathcal{C}^{-1}$ with the Fisher matrix prediction, 
with $N^{\rm fid}$ and $N^{(\rm lens)}$ set by the mean of $\widehat{N}_{\taunl}(L)+N^{(0)}(L)$ and $\widehat{N}_{\taunl}(L)$ (respectively) over 100 FFP10 simulations. From the temperature-only analysis, we find $\taunl<1740$ ($\taunl<2360$) from \textsc{smica} (\textsc{sevem}) at $95\%$ confidence; adding polarization information tightens this to $\taunl<1500$ ($\taunl<2460$). The \textsc{sevem} constraints are markedly weaker than those of \textsc{smica}: this is due to the fluctuations in $\widehat{\tau}_{\rm NL}^{\rm loc}$ observed in Fig.\,\ref{fig: tauNL-L-dep} (either from noise or residual foregrounds) coupled with the physical $\taunl>0$ restriction. 
Sampling noise leads to significant variations in the one-sided constraints: repeating the \textit{Planck} analysis for each of the $100$ \textsc{smica} (\textsc{sevem}) simulations yields $95\%$ upper bounds ranging from $\taunl=580$ ($540$) to $\taunl=4700$ ($4320$) with a mean of $1605$ ($1600$). Following the preference discussed in \citep{Carron:2022eyg,Marzouk:2022utf} for \textsc{smica} over \textsc{sevem}, 
we will regard the \textsc{smica} results as the fiducial constraints from our analysis.

\bibliographystyle{apsrev4-1}
\bibliography{refs}

\begin{thebibliography}{244}%
\makeatletter
\providecommand \@ifxundefined [1]{%
 \@ifx{#1\undefined}
}%
\providecommand \@ifnum [1]{%
 \ifnum #1\expandafter \@firstoftwo
 \else \expandafter \@secondoftwo
 \fi
}%
\providecommand \@ifx [1]{%
 \ifx #1\expandafter \@firstoftwo
 \else \expandafter \@secondoftwo
 \fi
}%
\providecommand \natexlab [1]{#1}%
\providecommand \enquote  [1]{``#1''}%
\providecommand \bibnamefont  [1]{#1}%
\providecommand \bibfnamefont [1]{#1}%
\providecommand \citenamefont [1]{#1}%
\providecommand \href@noop [0]{\@secondoftwo}%
\providecommand \href [0]{\begingroup \@sanitize@url \@href}%
\providecommand \@href[1]{\@@startlink{#1}\@@href}%
\providecommand \@@href[1]{\endgroup#1\@@endlink}%
\providecommand \@sanitize@url [0]{\catcode `\\12\catcode `\$12\catcode
  `\&12\catcode `\#12\catcode `\^12\catcode `\_12\catcode `\%12\relax}%
\providecommand \@@startlink[1]{}%
\providecommand \@@endlink[0]{}%
\providecommand \url  [0]{\begingroup\@sanitize@url \@url }%
\providecommand \@url [1]{\endgroup\@href {#1}{\urlprefix }}%
\providecommand \urlprefix  [0]{URL }%
\providecommand \Eprint [0]{\href }%
\providecommand \doibase [0]{http://dx.doi.org/}%
\providecommand \selectlanguage [0]{\@gobble}%
\providecommand \bibinfo  [0]{\@secondoftwo}%
\providecommand \bibfield  [0]{\@secondoftwo}%
\providecommand \translation [1]{[#1]}%
\providecommand \BibitemOpen [0]{}%
\providecommand \bibitemStop [0]{}%
\providecommand \bibitemNoStop [0]{.\EOS\space}%
\providecommand \EOS [0]{\spacefactor3000\relax}%
\providecommand \BibitemShut  [1]{\csname bibitem#1\endcsname}%
\let\auto@bib@innerbib\@empty
\bibitem [{\citenamefont {Philcox}(2025{\natexlab{a}})}]{Philcox4pt1}%
  \BibitemOpen
  \bibfield  {author} {\bibinfo {author} {\bibfnamefont {O.~H.~E.}\
  \bibnamefont {Philcox}},\ }\href@noop {} {\  (\bibinfo {year}
  {2025}{\natexlab{a}})},\ \Eprint {http://arxiv.org/abs/2502.04434}
  {arXiv:2502.04434 [astro-ph.CO]} \BibitemShut {NoStop}%
\bibitem [{\citenamefont {Philcox}(2025{\natexlab{b}})}]{Philcox4pt2}%
  \BibitemOpen
  \bibfield  {author} {\bibinfo {author} {\bibfnamefont {O.~H.~E.}\
  \bibnamefont {Philcox}},\ }\href@noop {} {\  (\bibinfo {year}
  {2025}{\natexlab{b}})},\ \Eprint {http://arxiv.org/abs/2502.05258}
  {arXiv:2502.05258 [astro-ph.CO]} \BibitemShut {NoStop}%
\bibitem [{\citenamefont {Guth}(1981)}]{Guth:1980zm}%
  \BibitemOpen
  \bibfield  {author} {\bibinfo {author} {\bibfnamefont {A.~H.}\ \bibnamefont
  {Guth}},\ }\href {\doibase 10.1103/PhysRevD.23.347} {\bibfield  {journal}
  {\bibinfo  {journal} {Phys. Rev. D}\ }\textbf {\bibinfo {volume} {23}},\
  \bibinfo {pages} {347} (\bibinfo {year} {1981})}\BibitemShut {NoStop}%
\bibitem [{\citenamefont {Linde}(1982)}]{Linde:1981mu}%
  \BibitemOpen
  \bibfield  {author} {\bibinfo {author} {\bibfnamefont {A.~D.}\ \bibnamefont
  {Linde}},\ }\href {\doibase 10.1016/0370-2693(82)91219-9} {\bibfield
  {journal} {\bibinfo  {journal} {Phys. Lett. B}\ }\textbf {\bibinfo {volume}
  {108}},\ \bibinfo {pages} {389} (\bibinfo {year} {1982})}\BibitemShut
  {NoStop}%
\bibitem [{\citenamefont {{Linde}}(1982)}]{1982PhLB..116..335L}%
  \BibitemOpen
  \bibfield  {author} {\bibinfo {author} {\bibfnamefont {A.~D.}\ \bibnamefont
  {{Linde}}},\ }\href {\doibase 10.1016/0370-2693(82)90293-3} {\bibfield
  {journal} {\bibinfo  {journal} {Physics Letters B}\ }\textbf {\bibinfo
  {volume} {116}},\ \bibinfo {pages} {335} (\bibinfo {year}
  {1982})}\BibitemShut {NoStop}%
\bibitem [{\citenamefont {Ratra}\ and\ \citenamefont
  {Peebles}(1988)}]{Ratra:1987rm}%
  \BibitemOpen
  \bibfield  {author} {\bibinfo {author} {\bibfnamefont {B.}~\bibnamefont
  {Ratra}}\ and\ \bibinfo {author} {\bibfnamefont {P.~J.~E.}\ \bibnamefont
  {Peebles}},\ }\href {\doibase 10.1103/PhysRevD.37.3406} {\bibfield  {journal}
  {\bibinfo  {journal} {Phys. Rev. D}\ }\textbf {\bibinfo {volume} {37}},\
  \bibinfo {pages} {3406} (\bibinfo {year} {1988})}\BibitemShut {NoStop}%
\bibitem [{\citenamefont {Starobinsky}(1982)}]{Starobinsky:1982ee}%
  \BibitemOpen
  \bibfield  {author} {\bibinfo {author} {\bibfnamefont {A.~A.}\ \bibnamefont
  {Starobinsky}},\ }\href {\doibase 10.1016/0370-2693(82)90541-X} {\bibfield
  {journal} {\bibinfo  {journal} {Phys. Lett. B}\ }\textbf {\bibinfo {volume}
  {117}},\ \bibinfo {pages} {175} (\bibinfo {year} {1982})}\BibitemShut
  {NoStop}%
\bibitem [{\citenamefont {Mukhanov}\ \emph {et~al.}(1992)\citenamefont
  {Mukhanov}, \citenamefont {Feldman},\ and\ \citenamefont
  {Brandenberger}}]{Mukhanov:1990me}%
  \BibitemOpen
  \bibfield  {author} {\bibinfo {author} {\bibfnamefont {V.~F.}\ \bibnamefont
  {Mukhanov}}, \bibinfo {author} {\bibfnamefont {H.~A.}\ \bibnamefont
  {Feldman}}, \ and\ \bibinfo {author} {\bibfnamefont {R.~H.}\ \bibnamefont
  {Brandenberger}},\ }\href {\doibase 10.1016/0370-1573(92)90044-Z} {\bibfield
  {journal} {\bibinfo  {journal} {Phys. Rept.}\ }\textbf {\bibinfo {volume}
  {215}},\ \bibinfo {pages} {203} (\bibinfo {year} {1992})}\BibitemShut
  {NoStop}%
\bibitem [{\citenamefont {Maldacena}(2003)}]{Maldacena:2002vr}%
  \BibitemOpen
  \bibfield  {author} {\bibinfo {author} {\bibfnamefont {J.~M.}\ \bibnamefont
  {Maldacena}},\ }\href {\doibase 10.1088/1126-6708/2003/05/013} {\bibfield
  {journal} {\bibinfo  {journal} {JHEP}\ }\textbf {\bibinfo {volume} {05}},\
  \bibinfo {pages} {013} (\bibinfo {year} {2003})},\ \Eprint
  {http://arxiv.org/abs/astro-ph/0210603} {arXiv:astro-ph/0210603} \BibitemShut
  {NoStop}%
\bibitem [{\citenamefont {Bartolo}\ \emph
  {et~al.}(2004{\natexlab{a}})\citenamefont {Bartolo}, \citenamefont {Komatsu},
  \citenamefont {Matarrese},\ and\ \citenamefont {Riotto}}]{Bartolo:2004if}%
  \BibitemOpen
  \bibfield  {author} {\bibinfo {author} {\bibfnamefont {N.}~\bibnamefont
  {Bartolo}}, \bibinfo {author} {\bibfnamefont {E.}~\bibnamefont {Komatsu}},
  \bibinfo {author} {\bibfnamefont {S.}~\bibnamefont {Matarrese}}, \ and\
  \bibinfo {author} {\bibfnamefont {A.}~\bibnamefont {Riotto}},\ }\href
  {\doibase 10.1016/j.physrep.2004.08.022} {\bibfield  {journal} {\bibinfo
  {journal} {Phys. Rept.}\ }\textbf {\bibinfo {volume} {402}},\ \bibinfo
  {pages} {103} (\bibinfo {year} {2004}{\natexlab{a}})},\ \Eprint
  {http://arxiv.org/abs/astro-ph/0406398} {arXiv:astro-ph/0406398} \BibitemShut
  {NoStop}%
\bibitem [{\citenamefont {Komatsu}(2010)}]{Komatsu:2010hc}%
  \BibitemOpen
  \bibfield  {author} {\bibinfo {author} {\bibfnamefont {E.}~\bibnamefont
  {Komatsu}},\ }\href {\doibase 10.1088/0264-9381/27/12/124010} {\bibfield
  {journal} {\bibinfo  {journal} {Class. Quant. Grav.}\ }\textbf {\bibinfo
  {volume} {27}},\ \bibinfo {pages} {124010} (\bibinfo {year} {2010})},\
  \Eprint {http://arxiv.org/abs/1003.6097} {arXiv:1003.6097 [astro-ph.CO]}
  \BibitemShut {NoStop}%
\bibitem [{\citenamefont {Komatsu}\ \emph {et~al.}(2005)\citenamefont
  {Komatsu}, \citenamefont {Spergel},\ and\ \citenamefont
  {Wandelt}}]{Komatsu:2003iq}%
  \BibitemOpen
  \bibfield  {author} {\bibinfo {author} {\bibfnamefont {E.}~\bibnamefont
  {Komatsu}}, \bibinfo {author} {\bibfnamefont {D.~N.}\ \bibnamefont
  {Spergel}}, \ and\ \bibinfo {author} {\bibfnamefont {B.~D.}\ \bibnamefont
  {Wandelt}},\ }\href {\doibase 10.1086/491724} {\bibfield  {journal} {\bibinfo
   {journal} {Astrophys. J.}\ }\textbf {\bibinfo {volume} {634}},\ \bibinfo
  {pages} {14} (\bibinfo {year} {2005})},\ \Eprint
  {http://arxiv.org/abs/astro-ph/0305189} {arXiv:astro-ph/0305189} \BibitemShut
  {NoStop}%
\bibitem [{\citenamefont {{Creminelli}}\ \emph {et~al.}(2006)\citenamefont
  {{Creminelli}}, \citenamefont {{Nicolis}}, \citenamefont {{Senatore}},
  \citenamefont {{Tegmark}},\ and\ \citenamefont
  {{Zaldarriaga}}}]{2006JCAP...05..004C}%
  \BibitemOpen
  \bibfield  {author} {\bibinfo {author} {\bibfnamefont {P.}~\bibnamefont
  {{Creminelli}}}, \bibinfo {author} {\bibfnamefont {A.}~\bibnamefont
  {{Nicolis}}}, \bibinfo {author} {\bibfnamefont {L.}~\bibnamefont
  {{Senatore}}}, \bibinfo {author} {\bibfnamefont {M.}~\bibnamefont
  {{Tegmark}}}, \ and\ \bibinfo {author} {\bibfnamefont {M.}~\bibnamefont
  {{Zaldarriaga}}},\ }\href {\doibase 10.1088/1475-7516/2006/05/004} {\bibfield
   {journal} {\bibinfo  {journal} {\jcap}\ }\textbf {\bibinfo {volume}
  {2006}},\ \bibinfo {eid} {004} (\bibinfo {year} {2006})},\ \Eprint
  {http://arxiv.org/abs/astro-ph/0509029} {arXiv:astro-ph/0509029 [astro-ph]}
  \BibitemShut {NoStop}%
\bibitem [{\citenamefont {Creminelli}\ \emph {et~al.}(2006)\citenamefont
  {Creminelli}, \citenamefont {Nicolis}, \citenamefont {Senatore},
  \citenamefont {Tegmark},\ and\ \citenamefont
  {Zaldarriaga}}]{Creminelli:2005hu}%
  \BibitemOpen
  \bibfield  {author} {\bibinfo {author} {\bibfnamefont {P.}~\bibnamefont
  {Creminelli}}, \bibinfo {author} {\bibfnamefont {A.}~\bibnamefont {Nicolis}},
  \bibinfo {author} {\bibfnamefont {L.}~\bibnamefont {Senatore}}, \bibinfo
  {author} {\bibfnamefont {M.}~\bibnamefont {Tegmark}}, \ and\ \bibinfo
  {author} {\bibfnamefont {M.}~\bibnamefont {Zaldarriaga}},\ }\href {\doibase
  10.1088/1475-7516/2006/05/004} {\bibfield  {journal} {\bibinfo  {journal}
  {JCAP}\ }\textbf {\bibinfo {volume} {05}},\ \bibinfo {pages} {004} (\bibinfo
  {year} {2006})},\ \Eprint {http://arxiv.org/abs/astro-ph/0509029}
  {arXiv:astro-ph/0509029} \BibitemShut {NoStop}%
\bibitem [{\citenamefont {Senatore}\ \emph {et~al.}(2010)\citenamefont
  {Senatore}, \citenamefont {Smith},\ and\ \citenamefont
  {Zaldarriaga}}]{Senatore:2009gt}%
  \BibitemOpen
  \bibfield  {author} {\bibinfo {author} {\bibfnamefont {L.}~\bibnamefont
  {Senatore}}, \bibinfo {author} {\bibfnamefont {K.~M.}\ \bibnamefont {Smith}},
  \ and\ \bibinfo {author} {\bibfnamefont {M.}~\bibnamefont {Zaldarriaga}},\
  }\href {\doibase 10.1088/1475-7516/2010/01/028} {\bibfield  {journal}
  {\bibinfo  {journal} {JCAP}\ }\textbf {\bibinfo {volume} {01}},\ \bibinfo
  {pages} {028} (\bibinfo {year} {2010})},\ \Eprint
  {http://arxiv.org/abs/0905.3746} {arXiv:0905.3746 [astro-ph.CO]} \BibitemShut
  {NoStop}%
\bibitem [{\citenamefont {Philcox}\ and\ \citenamefont
  {Shiraishi}(2024{\natexlab{a}})}]{Philcox:2023xxk}%
  \BibitemOpen
  \bibfield  {author} {\bibinfo {author} {\bibfnamefont {O.~H.~E.}\
  \bibnamefont {Philcox}}\ and\ \bibinfo {author} {\bibfnamefont
  {M.}~\bibnamefont {Shiraishi}},\ }\href {\doibase
  10.1103/PhysRevD.109.063522} {\bibfield  {journal} {\bibinfo  {journal}
  {Phys. Rev. D}\ }\textbf {\bibinfo {volume} {109}},\ \bibinfo {pages}
  {063522} (\bibinfo {year} {2024}{\natexlab{a}})},\ \Eprint
  {http://arxiv.org/abs/2312.12498} {arXiv:2312.12498 [astro-ph.CO]}
  \BibitemShut {NoStop}%
\bibitem [{\citenamefont {Komatsu}\ \emph {et~al.}(2002)\citenamefont
  {Komatsu}, \citenamefont {Wandelt}, \citenamefont {Spergel}, \citenamefont
  {Banday},\ and\ \citenamefont {Gorski}}]{Komatsu:2001wu}%
  \BibitemOpen
  \bibfield  {author} {\bibinfo {author} {\bibfnamefont {E.}~\bibnamefont
  {Komatsu}}, \bibinfo {author} {\bibfnamefont {B.~D.}\ \bibnamefont
  {Wandelt}}, \bibinfo {author} {\bibfnamefont {D.~N.}\ \bibnamefont
  {Spergel}}, \bibinfo {author} {\bibfnamefont {A.~J.}\ \bibnamefont {Banday}},
  \ and\ \bibinfo {author} {\bibfnamefont {K.~M.}\ \bibnamefont {Gorski}},\
  }\href {\doibase 10.1086/337963} {\bibfield  {journal} {\bibinfo  {journal}
  {Astrophys. J.}\ }\textbf {\bibinfo {volume} {566}},\ \bibinfo {pages} {19}
  (\bibinfo {year} {2002})},\ \Eprint {http://arxiv.org/abs/astro-ph/0107605}
  {arXiv:astro-ph/0107605} \BibitemShut {NoStop}%
\bibitem [{\citenamefont {Philcox}\ and\ \citenamefont
  {Shiraishi}(2024{\natexlab{b}})}]{Philcox:2024wqx}%
  \BibitemOpen
  \bibfield  {author} {\bibinfo {author} {\bibfnamefont {O.~H.~E.}\
  \bibnamefont {Philcox}}\ and\ \bibinfo {author} {\bibfnamefont
  {M.}~\bibnamefont {Shiraishi}},\ }\href@noop {} {\  (\bibinfo {year}
  {2024}{\natexlab{b}})},\ \Eprint {http://arxiv.org/abs/2409.10595}
  {arXiv:2409.10595 [astro-ph.CO]} \BibitemShut {NoStop}%
\bibitem [{\citenamefont {{Santos}}\ \emph {et~al.}(2003)\citenamefont
  {{Santos}}, \citenamefont {{Heavens}}, \citenamefont {{Balbi}}, \citenamefont
  {{Borrill}}, \citenamefont {{Ferreira}}, \citenamefont {{Hanany}},
  \citenamefont {{Jaffe}}, \citenamefont {{Lee}}, \citenamefont {{Rabii}},
  \citenamefont {{Richards}}, \citenamefont {{Smoot}}, \citenamefont
  {{Stompor}}, \citenamefont {{Winant}},\ and\ \citenamefont
  {{Wu}}}]{2003MNRAS.341..623S}%
  \BibitemOpen
  \bibfield  {author} {\bibinfo {author} {\bibfnamefont {M.~G.}\ \bibnamefont
  {{Santos}}}, \bibinfo {author} {\bibfnamefont {A.}~\bibnamefont {{Heavens}}},
  \bibinfo {author} {\bibfnamefont {A.}~\bibnamefont {{Balbi}}}, \bibinfo
  {author} {\bibfnamefont {J.}~\bibnamefont {{Borrill}}}, \bibinfo {author}
  {\bibfnamefont {P.~G.}\ \bibnamefont {{Ferreira}}}, \bibinfo {author}
  {\bibfnamefont {S.}~\bibnamefont {{Hanany}}}, \bibinfo {author}
  {\bibfnamefont {A.~H.}\ \bibnamefont {{Jaffe}}}, \bibinfo {author}
  {\bibfnamefont {A.~T.}\ \bibnamefont {{Lee}}}, \bibinfo {author}
  {\bibfnamefont {B.}~\bibnamefont {{Rabii}}}, \bibinfo {author} {\bibfnamefont
  {P.~L.}\ \bibnamefont {{Richards}}}, \bibinfo {author} {\bibfnamefont
  {G.~F.}\ \bibnamefont {{Smoot}}}, \bibinfo {author} {\bibfnamefont
  {R.}~\bibnamefont {{Stompor}}}, \bibinfo {author} {\bibfnamefont {C.~D.}\
  \bibnamefont {{Winant}}}, \ and\ \bibinfo {author} {\bibfnamefont {J.~H.~P.}\
  \bibnamefont {{Wu}}},\ }\href {\doibase 10.1046/j.1365-8711.2003.06438.x}
  {\bibfield  {journal} {\bibinfo  {journal} {\mnras}\ }\textbf {\bibinfo
  {volume} {341}},\ \bibinfo {pages} {623} (\bibinfo {year} {2003})},\ \Eprint
  {http://arxiv.org/abs/astro-ph/0211123} {arXiv:astro-ph/0211123 [astro-ph]}
  \BibitemShut {NoStop}%
\bibitem [{\citenamefont {{Planck Collaboration}}\ \emph
  {et~al.}(2014)\citenamefont {{Planck Collaboration}}, \citenamefont {{Ade}},
  \citenamefont {{Aghanim}}, \citenamefont {{Armitage-Caplan}}, \citenamefont
  {{Arnaud}}, \citenamefont {{Ashdown}}, \citenamefont {{Atrio-Barandela}},
  \citenamefont {{Aumont}}, \citenamefont {{Baccigalupi}}, \citenamefont
  {{Banday}} \emph {et~al.}}]{2014A&A...571A..24P}%
  \BibitemOpen
  \bibfield  {author} {\bibinfo {author} {\bibnamefont {{Planck
  Collaboration}}}, \bibinfo {author} {\bibfnamefont {P.~A.~R.}\ \bibnamefont
  {{Ade}}}, \bibinfo {author} {\bibfnamefont {N.}~\bibnamefont {{Aghanim}}},
  \bibinfo {author} {\bibfnamefont {C.}~\bibnamefont {{Armitage-Caplan}}},
  \bibinfo {author} {\bibfnamefont {M.}~\bibnamefont {{Arnaud}}}, \bibinfo
  {author} {\bibfnamefont {M.}~\bibnamefont {{Ashdown}}}, \bibinfo {author}
  {\bibfnamefont {F.}~\bibnamefont {{Atrio-Barandela}}}, \bibinfo {author}
  {\bibfnamefont {J.}~\bibnamefont {{Aumont}}}, \bibinfo {author}
  {\bibfnamefont {C.}~\bibnamefont {{Baccigalupi}}}, \bibinfo {author}
  {\bibfnamefont {A.~J.}\ \bibnamefont {{Banday}}},  \emph {et~al.},\ }\href
  {\doibase 10.1051/0004-6361/201321554} {\bibfield  {journal} {\bibinfo
  {journal} {\aap}\ }\textbf {\bibinfo {volume} {571}},\ \bibinfo {eid} {A24}
  (\bibinfo {year} {2014})},\ \Eprint {http://arxiv.org/abs/1303.5084}
  {arXiv:1303.5084 [astro-ph.CO]} \BibitemShut {NoStop}%
\bibitem [{\citenamefont {Ade}\ \emph {et~al.}(2016{\natexlab{a}})\citenamefont
  {Ade} \emph {et~al.}}]{Planck:2015zfm}%
  \BibitemOpen
  \bibfield  {author} {\bibinfo {author} {\bibfnamefont {P.~A.~R.}\
  \bibnamefont {Ade}} \emph {et~al.} (\bibinfo {collaboration} {Planck}),\
  }\href {\doibase 10.1051/0004-6361/201525836} {\bibfield  {journal} {\bibinfo
   {journal} {Astron. Astrophys.}\ }\textbf {\bibinfo {volume} {594}},\
  \bibinfo {pages} {A17} (\bibinfo {year} {2016}{\natexlab{a}})},\ \Eprint
  {http://arxiv.org/abs/1502.01592} {arXiv:1502.01592 [astro-ph.CO]}
  \BibitemShut {NoStop}%
\bibitem [{\citenamefont {Akrami}\ \emph
  {et~al.}(2020{\natexlab{a}})\citenamefont {Akrami} \emph
  {et~al.}}]{Planck:2019kim}%
  \BibitemOpen
  \bibfield  {author} {\bibinfo {author} {\bibfnamefont {Y.}~\bibnamefont
  {Akrami}} \emph {et~al.} (\bibinfo {collaboration} {Planck}),\ }\href
  {\doibase 10.1051/0004-6361/201935891} {\bibfield  {journal} {\bibinfo
  {journal} {Astron. Astrophys.}\ }\textbf {\bibinfo {volume} {641}},\ \bibinfo
  {pages} {A9} (\bibinfo {year} {2020}{\natexlab{a}})},\ \Eprint
  {http://arxiv.org/abs/1905.05697} {arXiv:1905.05697 [astro-ph.CO]}
  \BibitemShut {NoStop}%
\bibitem [{\citenamefont {Sohn}\ \emph {et~al.}(2024)\citenamefont {Sohn},
  \citenamefont {Wang}, \citenamefont {Fergusson},\ and\ \citenamefont
  {Shellard}}]{Sohn:2024xzd}%
  \BibitemOpen
  \bibfield  {author} {\bibinfo {author} {\bibfnamefont {W.}~\bibnamefont
  {Sohn}}, \bibinfo {author} {\bibfnamefont {D.-G.}\ \bibnamefont {Wang}},
  \bibinfo {author} {\bibfnamefont {J.~R.}\ \bibnamefont {Fergusson}}, \ and\
  \bibinfo {author} {\bibfnamefont {E.~P.~S.}\ \bibnamefont {Shellard}},\
  }\href {\doibase 10.1088/1475-7516/2024/09/016} {\bibfield  {journal}
  {\bibinfo  {journal} {JCAP}\ }\textbf {\bibinfo {volume} {09}},\ \bibinfo
  {pages} {016} (\bibinfo {year} {2024})},\ \Eprint
  {http://arxiv.org/abs/2404.07203} {arXiv:2404.07203 [astro-ph.CO]}
  \BibitemShut {NoStop}%
\bibitem [{\citenamefont {Smidt}\ \emph {et~al.}(2010)\citenamefont {Smidt},
  \citenamefont {Amblard}, \citenamefont {Byrnes}, \citenamefont {Cooray},
  \citenamefont {Heavens},\ and\ \citenamefont {Munshi}}]{Smidt:2010ra}%
  \BibitemOpen
  \bibfield  {author} {\bibinfo {author} {\bibfnamefont {J.}~\bibnamefont
  {Smidt}}, \bibinfo {author} {\bibfnamefont {A.}~\bibnamefont {Amblard}},
  \bibinfo {author} {\bibfnamefont {C.~T.}\ \bibnamefont {Byrnes}}, \bibinfo
  {author} {\bibfnamefont {A.}~\bibnamefont {Cooray}}, \bibinfo {author}
  {\bibfnamefont {A.}~\bibnamefont {Heavens}}, \ and\ \bibinfo {author}
  {\bibfnamefont {D.}~\bibnamefont {Munshi}},\ }\href {\doibase
  10.1103/PhysRevD.81.123007} {\bibfield  {journal} {\bibinfo  {journal} {Phys.
  Rev. D}\ }\textbf {\bibinfo {volume} {81}},\ \bibinfo {pages} {123007}
  (\bibinfo {year} {2010})},\ \Eprint {http://arxiv.org/abs/1004.1409}
  {arXiv:1004.1409 [astro-ph.CO]} \BibitemShut {NoStop}%
\bibitem [{\citenamefont {Salcedo}\ \emph {et~al.}(2024)\citenamefont
  {Salcedo}, \citenamefont {Colas},\ and\ \citenamefont
  {Pajer}}]{Salcedo:2024smn}%
  \BibitemOpen
  \bibfield  {author} {\bibinfo {author} {\bibfnamefont {S.~A.}\ \bibnamefont
  {Salcedo}}, \bibinfo {author} {\bibfnamefont {T.}~\bibnamefont {Colas}}, \
  and\ \bibinfo {author} {\bibfnamefont {E.}~\bibnamefont {Pajer}},\
  }\href@noop {} {\  (\bibinfo {year} {2024})},\ \Eprint
  {http://arxiv.org/abs/2404.15416} {arXiv:2404.15416 [hep-th]} \BibitemShut
  {NoStop}%
\bibitem [{\citenamefont {McCulloch}\ \emph {et~al.}(2024)\citenamefont
  {McCulloch}, \citenamefont {Pajer},\ and\ \citenamefont
  {Tong}}]{McCulloch:2024hiz}%
  \BibitemOpen
  \bibfield  {author} {\bibinfo {author} {\bibfnamefont {C.}~\bibnamefont
  {McCulloch}}, \bibinfo {author} {\bibfnamefont {E.}~\bibnamefont {Pajer}}, \
  and\ \bibinfo {author} {\bibfnamefont {X.}~\bibnamefont {Tong}},\ }\href
  {\doibase 10.1007/JHEP05(2024)262} {\bibfield  {journal} {\bibinfo  {journal}
  {JHEP}\ }\textbf {\bibinfo {volume} {05}},\ \bibinfo {pages} {262} (\bibinfo
  {year} {2024})},\ \Eprint {http://arxiv.org/abs/2401.11009} {arXiv:2401.11009
  [hep-th]} \BibitemShut {NoStop}%
\bibitem [{\citenamefont {Jazayeri}\ \emph
  {et~al.}(2023{\natexlab{a}})\citenamefont {Jazayeri}, \citenamefont
  {Renaux-Petel},\ and\ \citenamefont {Werth}}]{Jazayeri:2023xcj}%
  \BibitemOpen
  \bibfield  {author} {\bibinfo {author} {\bibfnamefont {S.}~\bibnamefont
  {Jazayeri}}, \bibinfo {author} {\bibfnamefont {S.}~\bibnamefont
  {Renaux-Petel}}, \ and\ \bibinfo {author} {\bibfnamefont {D.}~\bibnamefont
  {Werth}},\ }\href@noop {} {\  (\bibinfo {year} {2023}{\natexlab{a}})},\
  \Eprint {http://arxiv.org/abs/2307.01751} {arXiv:2307.01751 [hep-th]}
  \BibitemShut {NoStop}%
\bibitem [{\citenamefont {Flauger}\ \emph {et~al.}(2017)\citenamefont
  {Flauger}, \citenamefont {Mirbabayi}, \citenamefont {Senatore},\ and\
  \citenamefont {Silverstein}}]{Flauger:2016idt}%
  \BibitemOpen
  \bibfield  {author} {\bibinfo {author} {\bibfnamefont {R.}~\bibnamefont
  {Flauger}}, \bibinfo {author} {\bibfnamefont {M.}~\bibnamefont {Mirbabayi}},
  \bibinfo {author} {\bibfnamefont {L.}~\bibnamefont {Senatore}}, \ and\
  \bibinfo {author} {\bibfnamefont {E.}~\bibnamefont {Silverstein}},\ }\href
  {\doibase 10.1088/1475-7516/2017/10/058} {\bibfield  {journal} {\bibinfo
  {journal} {JCAP}\ }\textbf {\bibinfo {volume} {10}},\ \bibinfo {pages} {058}
  (\bibinfo {year} {2017})},\ \Eprint {http://arxiv.org/abs/1606.00513}
  {arXiv:1606.00513 [hep-th]} \BibitemShut {NoStop}%
\bibitem [{\citenamefont {Kim}\ \emph {et~al.}(2021)\citenamefont {Kim},
  \citenamefont {Kumar}, \citenamefont {Martin},\ and\ \citenamefont
  {Tsai}}]{Kim:2021ida}%
  \BibitemOpen
  \bibfield  {author} {\bibinfo {author} {\bibfnamefont {J.~H.}\ \bibnamefont
  {Kim}}, \bibinfo {author} {\bibfnamefont {S.}~\bibnamefont {Kumar}}, \bibinfo
  {author} {\bibfnamefont {A.}~\bibnamefont {Martin}}, \ and\ \bibinfo {author}
  {\bibfnamefont {Y.}~\bibnamefont {Tsai}},\ }\href {\doibase
  10.1007/JHEP11(2021)158} {\bibfield  {journal} {\bibinfo  {journal} {JHEP}\
  }\textbf {\bibinfo {volume} {11}},\ \bibinfo {pages} {158} (\bibinfo {year}
  {2021})},\ \Eprint {http://arxiv.org/abs/2107.09061} {arXiv:2107.09061
  [hep-ph]} \BibitemShut {NoStop}%
\bibitem [{\citenamefont {Martin}\ \emph {et~al.}(2013)\citenamefont {Martin},
  \citenamefont {Motohashi},\ and\ \citenamefont {Suyama}}]{Martin:2012pe}%
  \BibitemOpen
  \bibfield  {author} {\bibinfo {author} {\bibfnamefont {J.}~\bibnamefont
  {Martin}}, \bibinfo {author} {\bibfnamefont {H.}~\bibnamefont {Motohashi}}, \
  and\ \bibinfo {author} {\bibfnamefont {T.}~\bibnamefont {Suyama}},\ }\href
  {\doibase 10.1103/PhysRevD.87.023514} {\bibfield  {journal} {\bibinfo
  {journal} {Phys. Rev. D}\ }\textbf {\bibinfo {volume} {87}},\ \bibinfo
  {pages} {023514} (\bibinfo {year} {2013})},\ \Eprint
  {http://arxiv.org/abs/1211.0083} {arXiv:1211.0083 [astro-ph.CO]} \BibitemShut
  {NoStop}%
\bibitem [{\citenamefont {Tada}\ and\ \citenamefont
  {Vennin}(2022)}]{Tada:2021zzj}%
  \BibitemOpen
  \bibfield  {author} {\bibinfo {author} {\bibfnamefont {Y.}~\bibnamefont
  {Tada}}\ and\ \bibinfo {author} {\bibfnamefont {V.}~\bibnamefont {Vennin}},\
  }\href {\doibase 10.1088/1475-7516/2022/02/021} {\bibfield  {journal}
  {\bibinfo  {journal} {JCAP}\ }\textbf {\bibinfo {volume} {02}},\ \bibinfo
  {pages} {021} (\bibinfo {year} {2022})},\ \Eprint
  {http://arxiv.org/abs/2111.15280} {arXiv:2111.15280 [astro-ph.CO]}
  \BibitemShut {NoStop}%
\bibitem [{\citenamefont {M\"unchmeyer}\ and\ \citenamefont
  {Smith}(2019{\natexlab{a}})}]{Munchmeyer:2019wlh}%
  \BibitemOpen
  \bibfield  {author} {\bibinfo {author} {\bibfnamefont {M.}~\bibnamefont
  {M\"unchmeyer}}\ and\ \bibinfo {author} {\bibfnamefont {K.~M.}\ \bibnamefont
  {Smith}},\ }\href {\doibase 10.1103/PhysRevD.100.123511} {\bibfield
  {journal} {\bibinfo  {journal} {Phys. Rev. D}\ }\textbf {\bibinfo {volume}
  {100}},\ \bibinfo {pages} {123511} (\bibinfo {year} {2019}{\natexlab{a}})},\
  \Eprint {http://arxiv.org/abs/1910.00596} {arXiv:1910.00596 [astro-ph.CO]}
  \BibitemShut {NoStop}%
\bibitem [{\citenamefont {Philcox}\ \emph {et~al.}(2024)\citenamefont
  {Philcox}, \citenamefont {Kumar},\ and\ \citenamefont
  {Hill}}]{Philcox:2024jpd}%
  \BibitemOpen
  \bibfield  {author} {\bibinfo {author} {\bibfnamefont {O.~H.~E.}\
  \bibnamefont {Philcox}}, \bibinfo {author} {\bibfnamefont {S.}~\bibnamefont
  {Kumar}}, \ and\ \bibinfo {author} {\bibfnamefont {J.~C.}\ \bibnamefont
  {Hill}},\ }\href@noop {} {\  (\bibinfo {year} {2024})},\ \Eprint
  {http://arxiv.org/abs/2405.03738} {arXiv:2405.03738 [astro-ph.CO]}
  \BibitemShut {NoStop}%
\bibitem [{\citenamefont {Coulton}\ \emph {et~al.}(2024)\citenamefont
  {Coulton}, \citenamefont {Philcox},\ and\ \citenamefont
  {Villaescusa-Navarro}}]{Coulton:2024vot}%
  \BibitemOpen
  \bibfield  {author} {\bibinfo {author} {\bibfnamefont {W.~R.}\ \bibnamefont
  {Coulton}}, \bibinfo {author} {\bibfnamefont {O.~H.~E.}\ \bibnamefont
  {Philcox}}, \ and\ \bibinfo {author} {\bibfnamefont {F.}~\bibnamefont
  {Villaescusa-Navarro}},\ }\href@noop {} {\  (\bibinfo {year} {2024})},\
  \Eprint {http://arxiv.org/abs/2406.15546} {arXiv:2406.15546 [astro-ph.CO]}
  \BibitemShut {NoStop}%
\bibitem [{\citenamefont {Senatore}\ and\ \citenamefont
  {Zaldarriaga}(2012)}]{Senatore:2010wk}%
  \BibitemOpen
  \bibfield  {author} {\bibinfo {author} {\bibfnamefont {L.}~\bibnamefont
  {Senatore}}\ and\ \bibinfo {author} {\bibfnamefont {M.}~\bibnamefont
  {Zaldarriaga}},\ }\href {\doibase 10.1007/JHEP04(2012)024} {\bibfield
  {journal} {\bibinfo  {journal} {JHEP}\ }\textbf {\bibinfo {volume} {04}},\
  \bibinfo {pages} {024} (\bibinfo {year} {2012})},\ \Eprint
  {http://arxiv.org/abs/1009.2093} {arXiv:1009.2093 [hep-th]} \BibitemShut
  {NoStop}%
\bibitem [{\citenamefont {Arkani-Hamed}\ and\ \citenamefont
  {Maldacena}(2015)}]{Arkani-Hamed:2015bza}%
  \BibitemOpen
  \bibfield  {author} {\bibinfo {author} {\bibfnamefont {N.}~\bibnamefont
  {Arkani-Hamed}}\ and\ \bibinfo {author} {\bibfnamefont {J.}~\bibnamefont
  {Maldacena}},\ }\href@noop {} {\  (\bibinfo {year} {2015})},\ \Eprint
  {http://arxiv.org/abs/1503.08043} {arXiv:1503.08043 [hep-th]} \BibitemShut
  {NoStop}%
\bibitem [{\citenamefont {{Smith}}\ \emph {et~al.}(2015)\citenamefont
  {{Smith}}, \citenamefont {{Senatore}},\ and\ \citenamefont
  {{Zaldarriaga}}}]{2015arXiv150200635S}%
  \BibitemOpen
  \bibfield  {author} {\bibinfo {author} {\bibfnamefont {K.~M.}\ \bibnamefont
  {{Smith}}}, \bibinfo {author} {\bibfnamefont {L.}~\bibnamefont {{Senatore}}},
  \ and\ \bibinfo {author} {\bibfnamefont {M.}~\bibnamefont {{Zaldarriaga}}},\
  }\href@noop {} {\bibfield  {journal} {\bibinfo  {journal} {arXiv e-prints}\
  ,\ \bibinfo {eid} {arXiv:1502.00635}} (\bibinfo {year} {2015})},\ \Eprint
  {http://arxiv.org/abs/1502.00635} {arXiv:1502.00635 [astro-ph.CO]}
  \BibitemShut {NoStop}%
\bibitem [{\citenamefont {Bartolo}\ \emph {et~al.}(2010)\citenamefont
  {Bartolo}, \citenamefont {Fasiello}, \citenamefont {Matarrese},\ and\
  \citenamefont {Riotto}}]{Bartolo:2010di}%
  \BibitemOpen
  \bibfield  {author} {\bibinfo {author} {\bibfnamefont {N.}~\bibnamefont
  {Bartolo}}, \bibinfo {author} {\bibfnamefont {M.}~\bibnamefont {Fasiello}},
  \bibinfo {author} {\bibfnamefont {S.}~\bibnamefont {Matarrese}}, \ and\
  \bibinfo {author} {\bibfnamefont {A.}~\bibnamefont {Riotto}},\ }\href
  {\doibase 10.1088/1475-7516/2010/09/035} {\bibfield  {journal} {\bibinfo
  {journal} {JCAP}\ }\textbf {\bibinfo {volume} {09}},\ \bibinfo {pages} {035}
  (\bibinfo {year} {2010})},\ \Eprint {http://arxiv.org/abs/1006.5411}
  {arXiv:1006.5411 [astro-ph.CO]} \BibitemShut {NoStop}%
\bibitem [{\citenamefont {Bartolo}\ \emph
  {et~al.}(2013{\natexlab{a}})\citenamefont {Bartolo}, \citenamefont
  {Dimastrogiovanni},\ and\ \citenamefont {Fasiello}}]{Bartolo:2013eka}%
  \BibitemOpen
  \bibfield  {author} {\bibinfo {author} {\bibfnamefont {N.}~\bibnamefont
  {Bartolo}}, \bibinfo {author} {\bibfnamefont {E.}~\bibnamefont
  {Dimastrogiovanni}}, \ and\ \bibinfo {author} {\bibfnamefont
  {M.}~\bibnamefont {Fasiello}},\ }\href {\doibase
  10.1088/1475-7516/2013/09/037} {\bibfield  {journal} {\bibinfo  {journal}
  {JCAP}\ }\textbf {\bibinfo {volume} {09}},\ \bibinfo {pages} {037} (\bibinfo
  {year} {2013}{\natexlab{a}})},\ \Eprint {http://arxiv.org/abs/1305.0812}
  {arXiv:1305.0812 [astro-ph.CO]} \BibitemShut {NoStop}%
\bibitem [{\citenamefont {Arroja}\ and\ \citenamefont
  {Koyama}(2008)}]{Arroja:2008ga}%
  \BibitemOpen
  \bibfield  {author} {\bibinfo {author} {\bibfnamefont {F.}~\bibnamefont
  {Arroja}}\ and\ \bibinfo {author} {\bibfnamefont {K.}~\bibnamefont
  {Koyama}},\ }\href {\doibase 10.1103/PhysRevD.77.083517} {\bibfield
  {journal} {\bibinfo  {journal} {Phys. Rev. D}\ }\textbf {\bibinfo {volume}
  {77}},\ \bibinfo {pages} {083517} (\bibinfo {year} {2008})},\ \Eprint
  {http://arxiv.org/abs/0802.1167} {arXiv:0802.1167 [hep-th]} \BibitemShut
  {NoStop}%
\bibitem [{\citenamefont {Arroja}\ \emph {et~al.}(2009)\citenamefont {Arroja},
  \citenamefont {Mizuno}, \citenamefont {Koyama},\ and\ \citenamefont
  {Tanaka}}]{Arroja:2009pd}%
  \BibitemOpen
  \bibfield  {author} {\bibinfo {author} {\bibfnamefont {F.}~\bibnamefont
  {Arroja}}, \bibinfo {author} {\bibfnamefont {S.}~\bibnamefont {Mizuno}},
  \bibinfo {author} {\bibfnamefont {K.}~\bibnamefont {Koyama}}, \ and\ \bibinfo
  {author} {\bibfnamefont {T.}~\bibnamefont {Tanaka}},\ }\href {\doibase
  10.1103/PhysRevD.80.043527} {\bibfield  {journal} {\bibinfo  {journal} {Phys.
  Rev. D}\ }\textbf {\bibinfo {volume} {80}},\ \bibinfo {pages} {043527}
  (\bibinfo {year} {2009})},\ \Eprint {http://arxiv.org/abs/0905.3641}
  {arXiv:0905.3641 [hep-th]} \BibitemShut {NoStop}%
\bibitem [{\citenamefont {Chen}\ \emph {et~al.}(2022)\citenamefont {Chen},
  \citenamefont {Ebadi},\ and\ \citenamefont {Kumar}}]{Chen:2022vzh}%
  \BibitemOpen
  \bibfield  {author} {\bibinfo {author} {\bibfnamefont {X.}~\bibnamefont
  {Chen}}, \bibinfo {author} {\bibfnamefont {R.}~\bibnamefont {Ebadi}}, \ and\
  \bibinfo {author} {\bibfnamefont {S.}~\bibnamefont {Kumar}},\ }\href
  {\doibase 10.1088/1475-7516/2022/08/083} {\bibfield  {journal} {\bibinfo
  {journal} {JCAP}\ }\textbf {\bibinfo {volume} {08}},\ \bibinfo {pages} {083}
  (\bibinfo {year} {2022})},\ \Eprint {http://arxiv.org/abs/2205.01107}
  {arXiv:2205.01107 [hep-ph]} \BibitemShut {NoStop}%
\bibitem [{\citenamefont {Chen}\ \emph {et~al.}(2009)\citenamefont {Chen},
  \citenamefont {Hu}, \citenamefont {Huang}, \citenamefont {Shiu},\ and\
  \citenamefont {Wang}}]{Chen:2009bc}%
  \BibitemOpen
  \bibfield  {author} {\bibinfo {author} {\bibfnamefont {X.}~\bibnamefont
  {Chen}}, \bibinfo {author} {\bibfnamefont {B.}~\bibnamefont {Hu}}, \bibinfo
  {author} {\bibfnamefont {M.-x.}\ \bibnamefont {Huang}}, \bibinfo {author}
  {\bibfnamefont {G.}~\bibnamefont {Shiu}}, \ and\ \bibinfo {author}
  {\bibfnamefont {Y.}~\bibnamefont {Wang}},\ }\href {\doibase
  10.1088/1475-7516/2009/08/008} {\bibfield  {journal} {\bibinfo  {journal}
  {JCAP}\ }\textbf {\bibinfo {volume} {08}},\ \bibinfo {pages} {008} (\bibinfo
  {year} {2009})},\ \Eprint {http://arxiv.org/abs/0905.3494} {arXiv:0905.3494
  [astro-ph.CO]} \BibitemShut {NoStop}%
\bibitem [{\citenamefont {Meerburg}\ \emph {et~al.}(2017)\citenamefont
  {Meerburg}, \citenamefont {M\"unchmeyer}, \citenamefont {Mu\~noz},\ and\
  \citenamefont {Chen}}]{Meerburg:2016zdz}%
  \BibitemOpen
  \bibfield  {author} {\bibinfo {author} {\bibfnamefont {P.~D.}\ \bibnamefont
  {Meerburg}}, \bibinfo {author} {\bibfnamefont {M.}~\bibnamefont
  {M\"unchmeyer}}, \bibinfo {author} {\bibfnamefont {J.~B.}\ \bibnamefont
  {Mu\~noz}}, \ and\ \bibinfo {author} {\bibfnamefont {X.}~\bibnamefont
  {Chen}},\ }\href {\doibase 10.1088/1475-7516/2017/03/050} {\bibfield
  {journal} {\bibinfo  {journal} {JCAP}\ }\textbf {\bibinfo {volume} {03}},\
  \bibinfo {pages} {050} (\bibinfo {year} {2017})},\ \Eprint
  {http://arxiv.org/abs/1610.06559} {arXiv:1610.06559 [astro-ph.CO]}
  \BibitemShut {NoStop}%
\bibitem [{\citenamefont {Okamoto}\ and\ \citenamefont
  {Hu}(2002)}]{Okamoto:2002ik}%
  \BibitemOpen
  \bibfield  {author} {\bibinfo {author} {\bibfnamefont {T.}~\bibnamefont
  {Okamoto}}\ and\ \bibinfo {author} {\bibfnamefont {W.}~\bibnamefont {Hu}},\
  }\href {\doibase 10.1103/PhysRevD.66.063008} {\bibfield  {journal} {\bibinfo
  {journal} {Phys. Rev. D}\ }\textbf {\bibinfo {volume} {66}},\ \bibinfo
  {pages} {063008} (\bibinfo {year} {2002})},\ \Eprint
  {http://arxiv.org/abs/astro-ph/0206155} {arXiv:astro-ph/0206155} \BibitemShut
  {NoStop}%
\bibitem [{\citenamefont {Lee}\ and\ \citenamefont
  {Dvorkin}(2020)}]{Lee:2020ebj}%
  \BibitemOpen
  \bibfield  {author} {\bibinfo {author} {\bibfnamefont {H.}~\bibnamefont
  {Lee}}\ and\ \bibinfo {author} {\bibfnamefont {C.}~\bibnamefont {Dvorkin}},\
  }\href {\doibase 10.1088/1475-7516/2020/05/044} {\bibfield  {journal}
  {\bibinfo  {journal} {JCAP}\ }\textbf {\bibinfo {volume} {05}},\ \bibinfo
  {pages} {044} (\bibinfo {year} {2020})},\ \Eprint
  {http://arxiv.org/abs/2001.00584} {arXiv:2001.00584 [astro-ph.CO]}
  \BibitemShut {NoStop}%
\bibitem [{\citenamefont {{Munshi}}\ \emph {et~al.}(2011)\citenamefont
  {{Munshi}}, \citenamefont {{Heavens}}, \citenamefont {{Cooray}},
  \citenamefont {{Smidt}}, \citenamefont {{Coles}},\ and\ \citenamefont
  {{Serra}}}]{2011MNRAS.412.1993M}%
  \BibitemOpen
  \bibfield  {author} {\bibinfo {author} {\bibfnamefont {D.}~\bibnamefont
  {{Munshi}}}, \bibinfo {author} {\bibfnamefont {A.}~\bibnamefont {{Heavens}}},
  \bibinfo {author} {\bibfnamefont {A.}~\bibnamefont {{Cooray}}}, \bibinfo
  {author} {\bibfnamefont {J.}~\bibnamefont {{Smidt}}}, \bibinfo {author}
  {\bibfnamefont {P.}~\bibnamefont {{Coles}}}, \ and\ \bibinfo {author}
  {\bibfnamefont {P.}~\bibnamefont {{Serra}}},\ }\href {\doibase
  10.1111/j.1365-2966.2010.18035.x} {\bibfield  {journal} {\bibinfo  {journal}
  {\mnras}\ }\textbf {\bibinfo {volume} {412}},\ \bibinfo {pages} {1993}
  (\bibinfo {year} {2011})},\ \Eprint {http://arxiv.org/abs/0910.3693}
  {arXiv:0910.3693 [astro-ph.CO]} \BibitemShut {NoStop}%
\bibitem [{\citenamefont {Kogo}\ and\ \citenamefont
  {Komatsu}(2006)}]{Kogo:2006kh}%
  \BibitemOpen
  \bibfield  {author} {\bibinfo {author} {\bibfnamefont {N.}~\bibnamefont
  {Kogo}}\ and\ \bibinfo {author} {\bibfnamefont {E.}~\bibnamefont {Komatsu}},\
  }\href {\doibase 10.1103/PhysRevD.73.083007} {\bibfield  {journal} {\bibinfo
  {journal} {Phys. Rev. D}\ }\textbf {\bibinfo {volume} {73}},\ \bibinfo
  {pages} {083007} (\bibinfo {year} {2006})},\ \Eprint
  {http://arxiv.org/abs/astro-ph/0602099} {arXiv:astro-ph/0602099} \BibitemShut
  {NoStop}%
\bibitem [{\citenamefont {Hu}(2001)}]{Hu:2001fa}%
  \BibitemOpen
  \bibfield  {author} {\bibinfo {author} {\bibfnamefont {W.}~\bibnamefont
  {Hu}},\ }\href {\doibase 10.1103/PhysRevD.64.083005} {\bibfield  {journal}
  {\bibinfo  {journal} {Phys. Rev. D}\ }\textbf {\bibinfo {volume} {64}},\
  \bibinfo {pages} {083005} (\bibinfo {year} {2001})},\ \Eprint
  {http://arxiv.org/abs/astro-ph/0105117} {arXiv:astro-ph/0105117} \BibitemShut
  {NoStop}%
\bibitem [{\citenamefont {Chen}\ \emph {et~al.}(2006)\citenamefont {Chen},
  \citenamefont {Huang},\ and\ \citenamefont {Shiu}}]{Chen:2006dfn}%
  \BibitemOpen
  \bibfield  {author} {\bibinfo {author} {\bibfnamefont {X.}~\bibnamefont
  {Chen}}, \bibinfo {author} {\bibfnamefont {M.-x.}\ \bibnamefont {Huang}}, \
  and\ \bibinfo {author} {\bibfnamefont {G.}~\bibnamefont {Shiu}},\ }\href
  {\doibase 10.1103/PhysRevD.74.121301} {\bibfield  {journal} {\bibinfo
  {journal} {Phys. Rev. D}\ }\textbf {\bibinfo {volume} {74}},\ \bibinfo
  {pages} {121301} (\bibinfo {year} {2006})},\ \Eprint
  {http://arxiv.org/abs/hep-th/0610235} {arXiv:hep-th/0610235} \BibitemShut
  {NoStop}%
\bibitem [{\citenamefont {Regan}\ \emph {et~al.}(2010)\citenamefont {Regan},
  \citenamefont {Shellard},\ and\ \citenamefont {Fergusson}}]{Regan:2010cn}%
  \BibitemOpen
  \bibfield  {author} {\bibinfo {author} {\bibfnamefont {D.~M.}\ \bibnamefont
  {Regan}}, \bibinfo {author} {\bibfnamefont {E.~P.~S.}\ \bibnamefont
  {Shellard}}, \ and\ \bibinfo {author} {\bibfnamefont {J.~R.}\ \bibnamefont
  {Fergusson}},\ }\href {\doibase 10.1103/PhysRevD.82.023520} {\bibfield
  {journal} {\bibinfo  {journal} {Phys. Rev. D}\ }\textbf {\bibinfo {volume}
  {82}},\ \bibinfo {pages} {023520} (\bibinfo {year} {2010})},\ \Eprint
  {http://arxiv.org/abs/1004.2915} {arXiv:1004.2915 [astro-ph.CO]} \BibitemShut
  {NoStop}%
\bibitem [{\citenamefont {Kamionkowski}\ \emph {et~al.}(2011)\citenamefont
  {Kamionkowski}, \citenamefont {Smith},\ and\ \citenamefont
  {Heavens}}]{Kamionkowski:2010me}%
  \BibitemOpen
  \bibfield  {author} {\bibinfo {author} {\bibfnamefont {M.}~\bibnamefont
  {Kamionkowski}}, \bibinfo {author} {\bibfnamefont {T.~L.}\ \bibnamefont
  {Smith}}, \ and\ \bibinfo {author} {\bibfnamefont {A.}~\bibnamefont
  {Heavens}},\ }\href {\doibase 10.1103/PhysRevD.83.023007} {\bibfield
  {journal} {\bibinfo  {journal} {Phys. Rev. D}\ }\textbf {\bibinfo {volume}
  {83}},\ \bibinfo {pages} {023007} (\bibinfo {year} {2011})},\ \Eprint
  {http://arxiv.org/abs/1010.0251} {arXiv:1010.0251 [astro-ph.CO]} \BibitemShut
  {NoStop}%
\bibitem [{\citenamefont {Fergusson}\ \emph {et~al.}(2010)\citenamefont
  {Fergusson}, \citenamefont {Regan},\ and\ \citenamefont
  {Shellard}}]{Fergusson:2010gn}%
  \BibitemOpen
  \bibfield  {author} {\bibinfo {author} {\bibfnamefont {J.~R.}\ \bibnamefont
  {Fergusson}}, \bibinfo {author} {\bibfnamefont {D.~M.}\ \bibnamefont
  {Regan}}, \ and\ \bibinfo {author} {\bibfnamefont {E.~P.~S.}\ \bibnamefont
  {Shellard}},\ }\href@noop {} {\  (\bibinfo {year} {2010})},\ \Eprint
  {http://arxiv.org/abs/1012.6039} {arXiv:1012.6039 [astro-ph.CO]} \BibitemShut
  {NoStop}%
\bibitem [{\citenamefont {Feng}\ \emph {et~al.}(2015)\citenamefont {Feng},
  \citenamefont {Cooray}, \citenamefont {Smidt}, \citenamefont {O'Bryan},
  \citenamefont {Keating},\ and\ \citenamefont {Regan}}]{Feng:2015pva}%
  \BibitemOpen
  \bibfield  {author} {\bibinfo {author} {\bibfnamefont {C.}~\bibnamefont
  {Feng}}, \bibinfo {author} {\bibfnamefont {A.}~\bibnamefont {Cooray}},
  \bibinfo {author} {\bibfnamefont {J.}~\bibnamefont {Smidt}}, \bibinfo
  {author} {\bibfnamefont {J.}~\bibnamefont {O'Bryan}}, \bibinfo {author}
  {\bibfnamefont {B.}~\bibnamefont {Keating}}, \ and\ \bibinfo {author}
  {\bibfnamefont {D.}~\bibnamefont {Regan}},\ }\href {\doibase
  10.1103/PhysRevD.92.043509} {\bibfield  {journal} {\bibinfo  {journal} {Phys.
  Rev. D}\ }\textbf {\bibinfo {volume} {92}},\ \bibinfo {pages} {043509}
  (\bibinfo {year} {2015})},\ \Eprint {http://arxiv.org/abs/1502.00585}
  {arXiv:1502.00585 [astro-ph.CO]} \BibitemShut {NoStop}%
\bibitem [{\citenamefont {Munshi}\ \emph
  {et~al.}(2011{\natexlab{a}})\citenamefont {Munshi}, \citenamefont {Coles},
  \citenamefont {Cooray}, \citenamefont {Heavens},\ and\ \citenamefont
  {Smidt}}]{Munshi:2010bh}%
  \BibitemOpen
  \bibfield  {author} {\bibinfo {author} {\bibfnamefont {D.}~\bibnamefont
  {Munshi}}, \bibinfo {author} {\bibfnamefont {P.}~\bibnamefont {Coles}},
  \bibinfo {author} {\bibfnamefont {A.}~\bibnamefont {Cooray}}, \bibinfo
  {author} {\bibfnamefont {A.}~\bibnamefont {Heavens}}, \ and\ \bibinfo
  {author} {\bibfnamefont {J.}~\bibnamefont {Smidt}},\ }\href {\doibase
  10.1111/j.1365-2966.2010.17527.x} {\bibfield  {journal} {\bibinfo  {journal}
  {Mon. Not. Roy. Astron. Soc.}\ }\textbf {\bibinfo {volume} {410}},\ \bibinfo
  {pages} {1295} (\bibinfo {year} {2011}{\natexlab{a}})},\ \Eprint
  {http://arxiv.org/abs/1002.4998} {arXiv:1002.4998 [astro-ph.CO]} \BibitemShut
  {NoStop}%
\bibitem [{\citenamefont {Vielva}\ and\ \citenamefont
  {Sanz}(2010)}]{Vielva:2009jz}%
  \BibitemOpen
  \bibfield  {author} {\bibinfo {author} {\bibfnamefont {P.}~\bibnamefont
  {Vielva}}\ and\ \bibinfo {author} {\bibfnamefont {J.~L.}\ \bibnamefont
  {Sanz}},\ }\href {\doibase 10.1111/j.1365-2966.2010.16318.x} {\bibfield
  {journal} {\bibinfo  {journal} {Mon. Not. Roy. Astron. Soc.}\ }\textbf
  {\bibinfo {volume} {404}},\ \bibinfo {pages} {895} (\bibinfo {year}
  {2010})},\ \Eprint {http://arxiv.org/abs/0910.3196} {arXiv:0910.3196
  [astro-ph.CO]} \BibitemShut {NoStop}%
\bibitem [{\citenamefont {Hikage}\ and\ \citenamefont
  {Matsubara}(2012)}]{Hikage:2012bs}%
  \BibitemOpen
  \bibfield  {author} {\bibinfo {author} {\bibfnamefont {C.}~\bibnamefont
  {Hikage}}\ and\ \bibinfo {author} {\bibfnamefont {T.}~\bibnamefont
  {Matsubara}},\ }\href {\doibase 10.1111/j.1365-2966.2012.21572.x} {\bibfield
  {journal} {\bibinfo  {journal} {Mon. Not. Roy. Astron. Soc.}\ }\textbf
  {\bibinfo {volume} {425}},\ \bibinfo {pages} {2187} (\bibinfo {year}
  {2012})},\ \Eprint {http://arxiv.org/abs/1207.1183} {arXiv:1207.1183
  [astro-ph.CO]} \BibitemShut {NoStop}%
\bibitem [{\citenamefont {Sekiguchi}\ and\ \citenamefont
  {Sugiyama}(2013)}]{Sekiguchi:2013hza}%
  \BibitemOpen
  \bibfield  {author} {\bibinfo {author} {\bibfnamefont {T.}~\bibnamefont
  {Sekiguchi}}\ and\ \bibinfo {author} {\bibfnamefont {N.}~\bibnamefont
  {Sugiyama}},\ }\href {\doibase 10.1088/1475-7516/2013/09/002} {\bibfield
  {journal} {\bibinfo  {journal} {JCAP}\ }\textbf {\bibinfo {volume} {09}},\
  \bibinfo {pages} {002} (\bibinfo {year} {2013})},\ \Eprint
  {http://arxiv.org/abs/1303.4626} {arXiv:1303.4626 [astro-ph.CO]} \BibitemShut
  {NoStop}%
\bibitem [{\citenamefont {Spergel}\ \emph {et~al.}(2007)\citenamefont {Spergel}
  \emph {et~al.}}]{WMAP:2006bqn}%
  \BibitemOpen
  \bibfield  {author} {\bibinfo {author} {\bibfnamefont {D.~N.}\ \bibnamefont
  {Spergel}} \emph {et~al.} (\bibinfo {collaboration} {WMAP}),\ }\href
  {\doibase 10.1086/513700} {\bibfield  {journal} {\bibinfo  {journal}
  {Astrophys. J. Suppl.}\ }\textbf {\bibinfo {volume} {170}},\ \bibinfo {pages}
  {377} (\bibinfo {year} {2007})},\ \Eprint
  {http://arxiv.org/abs/astro-ph/0603449} {arXiv:astro-ph/0603449} \BibitemShut
  {NoStop}%
\bibitem [{\citenamefont {Marzouk}\ \emph {et~al.}(2022)\citenamefont
  {Marzouk}, \citenamefont {Lewis},\ and\ \citenamefont
  {Carron}}]{Marzouk:2022utf}%
  \BibitemOpen
  \bibfield  {author} {\bibinfo {author} {\bibfnamefont {K.}~\bibnamefont
  {Marzouk}}, \bibinfo {author} {\bibfnamefont {A.}~\bibnamefont {Lewis}}, \
  and\ \bibinfo {author} {\bibfnamefont {J.}~\bibnamefont {Carron}},\ }\href
  {\doibase 10.1088/1475-7516/2022/08/015} {\bibfield  {journal} {\bibinfo
  {journal} {JCAP}\ }\textbf {\bibinfo {volume} {08}},\ \bibinfo {pages} {015}
  (\bibinfo {year} {2022})},\ \Eprint {http://arxiv.org/abs/2205.14408}
  {arXiv:2205.14408 [astro-ph.CO]} \BibitemShut {NoStop}%
\bibitem [{\citenamefont {Kunz}\ \emph {et~al.}(2001)\citenamefont {Kunz},
  \citenamefont {Banday}, \citenamefont {Castro}, \citenamefont {Ferreira},\
  and\ \citenamefont {Gorski}}]{Kunz:2001ym}%
  \BibitemOpen
  \bibfield  {author} {\bibinfo {author} {\bibfnamefont {M.}~\bibnamefont
  {Kunz}}, \bibinfo {author} {\bibfnamefont {A.~J.}\ \bibnamefont {Banday}},
  \bibinfo {author} {\bibfnamefont {P.~G.}\ \bibnamefont {Castro}}, \bibinfo
  {author} {\bibfnamefont {P.~G.}\ \bibnamefont {Ferreira}}, \ and\ \bibinfo
  {author} {\bibfnamefont {K.~M.}\ \bibnamefont {Gorski}},\ }\href {\doibase
  10.1086/338602} {\bibfield  {journal} {\bibinfo  {journal} {Astrophys. J.
  Lett.}\ }\textbf {\bibinfo {volume} {563}},\ \bibinfo {pages} {L99} (\bibinfo
  {year} {2001})},\ \Eprint {http://arxiv.org/abs/astro-ph/0111250}
  {arXiv:astro-ph/0111250} \BibitemShut {NoStop}%
\bibitem [{\citenamefont {Akrami}\ \emph
  {et~al.}(2020{\natexlab{b}})\citenamefont {Akrami} \emph
  {et~al.}}]{Planck:2020olo}%
  \BibitemOpen
  \bibfield  {author} {\bibinfo {author} {\bibfnamefont {Y.}~\bibnamefont
  {Akrami}} \emph {et~al.} (\bibinfo {collaboration} {Planck}),\ }\href
  {\doibase 10.1051/0004-6361/202038073} {\bibfield  {journal} {\bibinfo
  {journal} {Astron. Astrophys.}\ }\textbf {\bibinfo {volume} {643}},\ \bibinfo
  {pages} {A42} (\bibinfo {year} {2020}{\natexlab{b}})},\ \Eprint
  {http://arxiv.org/abs/2007.04997} {arXiv:2007.04997 [astro-ph.CO]}
  \BibitemShut {NoStop}%
\bibitem [{\citenamefont {Chen}\ and\ \citenamefont
  {Wang}(2010)}]{Chen:2009zp}%
  \BibitemOpen
  \bibfield  {author} {\bibinfo {author} {\bibfnamefont {X.}~\bibnamefont
  {Chen}}\ and\ \bibinfo {author} {\bibfnamefont {Y.}~\bibnamefont {Wang}},\
  }\href {\doibase 10.1088/1475-7516/2010/04/027} {\bibfield  {journal}
  {\bibinfo  {journal} {JCAP}\ }\textbf {\bibinfo {volume} {04}},\ \bibinfo
  {pages} {027} (\bibinfo {year} {2010})},\ \Eprint
  {http://arxiv.org/abs/0911.3380} {arXiv:0911.3380 [hep-th]} \BibitemShut
  {NoStop}%
\bibitem [{\citenamefont {Suyama}\ and\ \citenamefont
  {Yamaguchi}(2008)}]{Suyama:2007bg}%
  \BibitemOpen
  \bibfield  {author} {\bibinfo {author} {\bibfnamefont {T.}~\bibnamefont
  {Suyama}}\ and\ \bibinfo {author} {\bibfnamefont {M.}~\bibnamefont
  {Yamaguchi}},\ }\href {\doibase 10.1103/PhysRevD.77.023505} {\bibfield
  {journal} {\bibinfo  {journal} {Phys. Rev. D}\ }\textbf {\bibinfo {volume}
  {77}},\ \bibinfo {pages} {023505} (\bibinfo {year} {2008})},\ \Eprint
  {http://arxiv.org/abs/0709.2545} {arXiv:0709.2545 [astro-ph]} \BibitemShut
  {NoStop}%
\bibitem [{\citenamefont {Chen}\ \emph {et~al.}(2007)\citenamefont {Chen},
  \citenamefont {Huang}, \citenamefont {Kachru},\ and\ \citenamefont
  {Shiu}}]{Chen:2006nt}%
  \BibitemOpen
  \bibfield  {author} {\bibinfo {author} {\bibfnamefont {X.}~\bibnamefont
  {Chen}}, \bibinfo {author} {\bibfnamefont {M.-x.}\ \bibnamefont {Huang}},
  \bibinfo {author} {\bibfnamefont {S.}~\bibnamefont {Kachru}}, \ and\ \bibinfo
  {author} {\bibfnamefont {G.}~\bibnamefont {Shiu}},\ }\href {\doibase
  10.1088/1475-7516/2007/01/002} {\bibfield  {journal} {\bibinfo  {journal}
  {JCAP}\ }\textbf {\bibinfo {volume} {01}},\ \bibinfo {pages} {002} (\bibinfo
  {year} {2007})},\ \Eprint {http://arxiv.org/abs/hep-th/0605045}
  {arXiv:hep-th/0605045} \BibitemShut {NoStop}%
\bibitem [{\citenamefont {Senatore}\ and\ \citenamefont
  {Zaldarriaga}(2011)}]{Senatore:2010jy}%
  \BibitemOpen
  \bibfield  {author} {\bibinfo {author} {\bibfnamefont {L.}~\bibnamefont
  {Senatore}}\ and\ \bibinfo {author} {\bibfnamefont {M.}~\bibnamefont
  {Zaldarriaga}},\ }\href {\doibase 10.1088/1475-7516/2011/01/003} {\bibfield
  {journal} {\bibinfo  {journal} {JCAP}\ }\textbf {\bibinfo {volume} {01}},\
  \bibinfo {pages} {003} (\bibinfo {year} {2011})},\ \Eprint
  {http://arxiv.org/abs/1004.1201} {arXiv:1004.1201 [hep-th]} \BibitemShut
  {NoStop}%
\bibitem [{\citenamefont {Cheung}\ \emph {et~al.}(2008)\citenamefont {Cheung},
  \citenamefont {Creminelli}, \citenamefont {Fitzpatrick}, \citenamefont
  {Kaplan},\ and\ \citenamefont {Senatore}}]{Cheung:2007st}%
  \BibitemOpen
  \bibfield  {author} {\bibinfo {author} {\bibfnamefont {C.}~\bibnamefont
  {Cheung}}, \bibinfo {author} {\bibfnamefont {P.}~\bibnamefont {Creminelli}},
  \bibinfo {author} {\bibfnamefont {A.~L.}\ \bibnamefont {Fitzpatrick}},
  \bibinfo {author} {\bibfnamefont {J.}~\bibnamefont {Kaplan}}, \ and\ \bibinfo
  {author} {\bibfnamefont {L.}~\bibnamefont {Senatore}},\ }\href {\doibase
  10.1088/1126-6708/2008/03/014} {\bibfield  {journal} {\bibinfo  {journal}
  {JHEP}\ }\textbf {\bibinfo {volume} {03}},\ \bibinfo {pages} {014} (\bibinfo
  {year} {2008})},\ \Eprint {http://arxiv.org/abs/0709.0293} {arXiv:0709.0293
  [hep-th]} \BibitemShut {NoStop}%
\bibitem [{\citenamefont {Mizuno}\ \emph {et~al.}(2009)\citenamefont {Mizuno},
  \citenamefont {Arroja},\ and\ \citenamefont {Koyama}}]{Mizuno:2009mv}%
  \BibitemOpen
  \bibfield  {author} {\bibinfo {author} {\bibfnamefont {S.}~\bibnamefont
  {Mizuno}}, \bibinfo {author} {\bibfnamefont {F.}~\bibnamefont {Arroja}}, \
  and\ \bibinfo {author} {\bibfnamefont {K.}~\bibnamefont {Koyama}},\ }\href
  {\doibase 10.1103/PhysRevD.80.083517} {\bibfield  {journal} {\bibinfo
  {journal} {Phys. Rev. D}\ }\textbf {\bibinfo {volume} {80}},\ \bibinfo
  {pages} {083517} (\bibinfo {year} {2009})},\ \Eprint
  {http://arxiv.org/abs/0907.2439} {arXiv:0907.2439 [hep-th]} \BibitemShut
  {NoStop}%
\bibitem [{\citenamefont {Langlois}\ \emph
  {et~al.}(2008{\natexlab{a}})\citenamefont {Langlois}, \citenamefont
  {Renaux-Petel}, \citenamefont {Steer},\ and\ \citenamefont
  {Tanaka}}]{Langlois:2008qf}%
  \BibitemOpen
  \bibfield  {author} {\bibinfo {author} {\bibfnamefont {D.}~\bibnamefont
  {Langlois}}, \bibinfo {author} {\bibfnamefont {S.}~\bibnamefont
  {Renaux-Petel}}, \bibinfo {author} {\bibfnamefont {D.~A.}\ \bibnamefont
  {Steer}}, \ and\ \bibinfo {author} {\bibfnamefont {T.}~\bibnamefont
  {Tanaka}},\ }\href {\doibase 10.1103/PhysRevD.78.063523} {\bibfield
  {journal} {\bibinfo  {journal} {Phys. Rev. D}\ }\textbf {\bibinfo {volume}
  {78}},\ \bibinfo {pages} {063523} (\bibinfo {year} {2008}{\natexlab{a}})},\
  \Eprint {http://arxiv.org/abs/0806.0336} {arXiv:0806.0336 [hep-th]}
  \BibitemShut {NoStop}%
\bibitem [{\citenamefont {Langlois}\ \emph
  {et~al.}(2008{\natexlab{b}})\citenamefont {Langlois}, \citenamefont
  {Renaux-Petel}, \citenamefont {Steer},\ and\ \citenamefont
  {Tanaka}}]{Langlois:2008wt}%
  \BibitemOpen
  \bibfield  {author} {\bibinfo {author} {\bibfnamefont {D.}~\bibnamefont
  {Langlois}}, \bibinfo {author} {\bibfnamefont {S.}~\bibnamefont
  {Renaux-Petel}}, \bibinfo {author} {\bibfnamefont {D.~A.}\ \bibnamefont
  {Steer}}, \ and\ \bibinfo {author} {\bibfnamefont {T.}~\bibnamefont
  {Tanaka}},\ }\href {\doibase 10.1103/PhysRevLett.101.061301} {\bibfield
  {journal} {\bibinfo  {journal} {Phys. Rev. Lett.}\ }\textbf {\bibinfo
  {volume} {101}},\ \bibinfo {pages} {061301} (\bibinfo {year}
  {2008}{\natexlab{b}})},\ \Eprint {http://arxiv.org/abs/0804.3139}
  {arXiv:0804.3139 [hep-th]} \BibitemShut {NoStop}%
\bibitem [{\citenamefont {Shiraishi}\ \emph {et~al.}(2014)\citenamefont
  {Shiraishi}, \citenamefont {Komatsu},\ and\ \citenamefont
  {Peloso}}]{Shiraishi:2013oqa}%
  \BibitemOpen
  \bibfield  {author} {\bibinfo {author} {\bibfnamefont {M.}~\bibnamefont
  {Shiraishi}}, \bibinfo {author} {\bibfnamefont {E.}~\bibnamefont {Komatsu}},
  \ and\ \bibinfo {author} {\bibfnamefont {M.}~\bibnamefont {Peloso}},\ }\href
  {\doibase 10.1088/1475-7516/2014/04/027} {\bibfield  {journal} {\bibinfo
  {journal} {JCAP}\ }\textbf {\bibinfo {volume} {04}},\ \bibinfo {pages} {027}
  (\bibinfo {year} {2014})},\ \Eprint {http://arxiv.org/abs/1312.5221}
  {arXiv:1312.5221 [astro-ph.CO]} \BibitemShut {NoStop}%
\bibitem [{\citenamefont {Shiraishi}(2016)}]{Shiraishi:2016mok}%
  \BibitemOpen
  \bibfield  {author} {\bibinfo {author} {\bibfnamefont {M.}~\bibnamefont
  {Shiraishi}},\ }\href {\doibase 10.1103/PhysRevD.94.083503} {\bibfield
  {journal} {\bibinfo  {journal} {Phys. Rev. D}\ }\textbf {\bibinfo {volume}
  {94}},\ \bibinfo {pages} {083503} (\bibinfo {year} {2016})},\ \Eprint
  {http://arxiv.org/abs/1608.00368} {arXiv:1608.00368 [astro-ph.CO]}
  \BibitemShut {NoStop}%
\bibitem [{\citenamefont {Lee}\ \emph {et~al.}(2016)\citenamefont {Lee},
  \citenamefont {Baumann},\ and\ \citenamefont {Pimentel}}]{Lee:2016vti}%
  \BibitemOpen
  \bibfield  {author} {\bibinfo {author} {\bibfnamefont {H.}~\bibnamefont
  {Lee}}, \bibinfo {author} {\bibfnamefont {D.}~\bibnamefont {Baumann}}, \ and\
  \bibinfo {author} {\bibfnamefont {G.~L.}\ \bibnamefont {Pimentel}},\ }\href
  {\doibase 10.1007/JHEP12(2016)040} {\bibfield  {journal} {\bibinfo  {journal}
  {JHEP}\ }\textbf {\bibinfo {volume} {12}},\ \bibinfo {pages} {040} (\bibinfo
  {year} {2016})},\ \Eprint {http://arxiv.org/abs/1607.03735} {arXiv:1607.03735
  [hep-th]} \BibitemShut {NoStop}%
\bibitem [{\citenamefont {Lewis}\ and\ \citenamefont
  {Challinor}(2006)}]{Lewis:2006fu}%
  \BibitemOpen
  \bibfield  {author} {\bibinfo {author} {\bibfnamefont {A.}~\bibnamefont
  {Lewis}}\ and\ \bibinfo {author} {\bibfnamefont {A.}~\bibnamefont
  {Challinor}},\ }\href {\doibase 10.1016/j.physrep.2006.03.002} {\bibfield
  {journal} {\bibinfo  {journal} {Phys. Rept.}\ }\textbf {\bibinfo {volume}
  {429}},\ \bibinfo {pages} {1} (\bibinfo {year} {2006})},\ \Eprint
  {http://arxiv.org/abs/astro-ph/0601594} {arXiv:astro-ph/0601594} \BibitemShut
  {NoStop}%
\bibitem [{\citenamefont {Babich}\ \emph {et~al.}(2004)\citenamefont {Babich},
  \citenamefont {Creminelli},\ and\ \citenamefont
  {Zaldarriaga}}]{Babich:2004gb}%
  \BibitemOpen
  \bibfield  {author} {\bibinfo {author} {\bibfnamefont {D.}~\bibnamefont
  {Babich}}, \bibinfo {author} {\bibfnamefont {P.}~\bibnamefont {Creminelli}},
  \ and\ \bibinfo {author} {\bibfnamefont {M.}~\bibnamefont {Zaldarriaga}},\
  }\href {\doibase 10.1088/1475-7516/2004/08/009} {\bibfield  {journal}
  {\bibinfo  {journal} {JCAP}\ }\textbf {\bibinfo {volume} {08}},\ \bibinfo
  {pages} {009} (\bibinfo {year} {2004})},\ \Eprint
  {http://arxiv.org/abs/astro-ph/0405356} {arXiv:astro-ph/0405356} \BibitemShut
  {NoStop}%
\bibitem [{\citenamefont {Qu}\ \emph {et~al.}(2024)\citenamefont {Qu} \emph
  {et~al.}}]{ACT:2023dou}%
  \BibitemOpen
  \bibfield  {author} {\bibinfo {author} {\bibfnamefont {F.~J.}\ \bibnamefont
  {Qu}} \emph {et~al.} (\bibinfo {collaboration} {ACT}),\ }\href {\doibase
  10.3847/1538-4357/acfe06} {\bibfield  {journal} {\bibinfo  {journal}
  {Astrophys. J.}\ }\textbf {\bibinfo {volume} {962}},\ \bibinfo {pages} {112}
  (\bibinfo {year} {2024})},\ \Eprint {http://arxiv.org/abs/2304.05202}
  {arXiv:2304.05202 [astro-ph.CO]} \BibitemShut {NoStop}%
\bibitem [{\citenamefont {Ade}\ \emph {et~al.}(2014)\citenamefont {Ade} \emph
  {et~al.}}]{Planck:2013mth}%
  \BibitemOpen
  \bibfield  {author} {\bibinfo {author} {\bibfnamefont {P.~A.~R.}\
  \bibnamefont {Ade}} \emph {et~al.} (\bibinfo {collaboration} {Planck}),\
  }\href {\doibase 10.1051/0004-6361/201321543} {\bibfield  {journal} {\bibinfo
   {journal} {Astron. Astrophys.}\ }\textbf {\bibinfo {volume} {571}},\
  \bibinfo {pages} {A17} (\bibinfo {year} {2014})},\ \Eprint
  {http://arxiv.org/abs/1303.5077} {arXiv:1303.5077 [astro-ph.CO]} \BibitemShut
  {NoStop}%
\bibitem [{\citenamefont {Hobson}\ \emph {et~al.}(1999)\citenamefont {Hobson},
  \citenamefont {Barreiro}, \citenamefont {Toffolatti}, \citenamefont
  {Lasenby}, \citenamefont {Sanz}, \citenamefont {Jones},\ and\ \citenamefont
  {Bouchet}}]{Hobson:1998sc}%
  \BibitemOpen
  \bibfield  {author} {\bibinfo {author} {\bibfnamefont {M.~P.}\ \bibnamefont
  {Hobson}}, \bibinfo {author} {\bibfnamefont {R.~B.}\ \bibnamefont
  {Barreiro}}, \bibinfo {author} {\bibfnamefont {L.}~\bibnamefont
  {Toffolatti}}, \bibinfo {author} {\bibfnamefont {A.~N.}\ \bibnamefont
  {Lasenby}}, \bibinfo {author} {\bibfnamefont {J.~L.}\ \bibnamefont {Sanz}},
  \bibinfo {author} {\bibfnamefont {A.~W.}\ \bibnamefont {Jones}}, \ and\
  \bibinfo {author} {\bibfnamefont {F.~R.}\ \bibnamefont {Bouchet}},\ }\href
  {\doibase 10.1046/j.1365-8711.1999.02546.x} {\bibfield  {journal} {\bibinfo
  {journal} {Mon. Not. Roy. Astron. Soc.}\ }\textbf {\bibinfo {volume} {306}},\
  \bibinfo {pages} {232} (\bibinfo {year} {1999})},\ \Eprint
  {http://arxiv.org/abs/astro-ph/9810241} {arXiv:astro-ph/9810241} \BibitemShut
  {NoStop}%
\bibitem [{\citenamefont {Coulton}\ \emph {et~al.}(2023)\citenamefont
  {Coulton}, \citenamefont {Miranthis},\ and\ \citenamefont
  {Challinor}}]{Coulton:2022wln}%
  \BibitemOpen
  \bibfield  {author} {\bibinfo {author} {\bibfnamefont {W.}~\bibnamefont
  {Coulton}}, \bibinfo {author} {\bibfnamefont {A.}~\bibnamefont {Miranthis}},
  \ and\ \bibinfo {author} {\bibfnamefont {A.}~\bibnamefont {Challinor}},\
  }\href {\doibase 10.1093/mnras/stad1305} {\bibfield  {journal} {\bibinfo
  {journal} {Mon. Not. Roy. Astron. Soc.}\ }\textbf {\bibinfo {volume} {523}},\
  \bibinfo {pages} {825} (\bibinfo {year} {2023})},\ \Eprint
  {http://arxiv.org/abs/2208.12270} {arXiv:2208.12270 [astro-ph.CO]}
  \BibitemShut {NoStop}%
\bibitem [{\citenamefont {Wang}\ and\ \citenamefont
  {Kamionkowski}(2000)}]{Wang:1999vf}%
  \BibitemOpen
  \bibfield  {author} {\bibinfo {author} {\bibfnamefont {L.-M.}\ \bibnamefont
  {Wang}}\ and\ \bibinfo {author} {\bibfnamefont {M.}~\bibnamefont
  {Kamionkowski}},\ }\href {\doibase 10.1103/PhysRevD.61.063504} {\bibfield
  {journal} {\bibinfo  {journal} {Phys. Rev. D}\ }\textbf {\bibinfo {volume}
  {61}},\ \bibinfo {pages} {063504} (\bibinfo {year} {2000})},\ \Eprint
  {http://arxiv.org/abs/astro-ph/9907431} {arXiv:astro-ph/9907431} \BibitemShut
  {NoStop}%
\bibitem [{\citenamefont {Chen}\ \emph {et~al.}(2008)\citenamefont {Chen},
  \citenamefont {Easther},\ and\ \citenamefont {Lim}}]{Chen:2008wn}%
  \BibitemOpen
  \bibfield  {author} {\bibinfo {author} {\bibfnamefont {X.}~\bibnamefont
  {Chen}}, \bibinfo {author} {\bibfnamefont {R.}~\bibnamefont {Easther}}, \
  and\ \bibinfo {author} {\bibfnamefont {E.~A.}\ \bibnamefont {Lim}},\ }\href
  {\doibase 10.1088/1475-7516/2008/04/010} {\bibfield  {journal} {\bibinfo
  {journal} {JCAP}\ }\textbf {\bibinfo {volume} {04}},\ \bibinfo {pages} {010}
  (\bibinfo {year} {2008})},\ \Eprint {http://arxiv.org/abs/0801.3295}
  {arXiv:0801.3295 [astro-ph]} \BibitemShut {NoStop}%
\bibitem [{\citenamefont {Flauger}\ \emph {et~al.}(2010)\citenamefont
  {Flauger}, \citenamefont {McAllister}, \citenamefont {Pajer}, \citenamefont
  {Westphal},\ and\ \citenamefont {Xu}}]{Flauger:2009ab}%
  \BibitemOpen
  \bibfield  {author} {\bibinfo {author} {\bibfnamefont {R.}~\bibnamefont
  {Flauger}}, \bibinfo {author} {\bibfnamefont {L.}~\bibnamefont {McAllister}},
  \bibinfo {author} {\bibfnamefont {E.}~\bibnamefont {Pajer}}, \bibinfo
  {author} {\bibfnamefont {A.}~\bibnamefont {Westphal}}, \ and\ \bibinfo
  {author} {\bibfnamefont {G.}~\bibnamefont {Xu}},\ }\href {\doibase
  10.1088/1475-7516/2010/06/009} {\bibfield  {journal} {\bibinfo  {journal}
  {JCAP}\ }\textbf {\bibinfo {volume} {06}},\ \bibinfo {pages} {009} (\bibinfo
  {year} {2010})},\ \Eprint {http://arxiv.org/abs/0907.2916} {arXiv:0907.2916
  [hep-th]} \BibitemShut {NoStop}%
\bibitem [{\citenamefont {Flauger}\ and\ \citenamefont
  {Pajer}(2011)}]{Flauger:2010ja}%
  \BibitemOpen
  \bibfield  {author} {\bibinfo {author} {\bibfnamefont {R.}~\bibnamefont
  {Flauger}}\ and\ \bibinfo {author} {\bibfnamefont {E.}~\bibnamefont
  {Pajer}},\ }\href {\doibase 10.1088/1475-7516/2011/01/017} {\bibfield
  {journal} {\bibinfo  {journal} {JCAP}\ }\textbf {\bibinfo {volume} {01}},\
  \bibinfo {pages} {017} (\bibinfo {year} {2011})},\ \Eprint
  {http://arxiv.org/abs/1002.0833} {arXiv:1002.0833 [hep-th]} \BibitemShut
  {NoStop}%
\bibitem [{\citenamefont {Baumann}\ \emph {et~al.}(2018)\citenamefont
  {Baumann}, \citenamefont {Goon}, \citenamefont {Lee},\ and\ \citenamefont
  {Pimentel}}]{Baumann:2017jvh}%
  \BibitemOpen
  \bibfield  {author} {\bibinfo {author} {\bibfnamefont {D.}~\bibnamefont
  {Baumann}}, \bibinfo {author} {\bibfnamefont {G.}~\bibnamefont {Goon}},
  \bibinfo {author} {\bibfnamefont {H.}~\bibnamefont {Lee}}, \ and\ \bibinfo
  {author} {\bibfnamefont {G.~L.}\ \bibnamefont {Pimentel}},\ }\href {\doibase
  10.1007/JHEP04(2018)140} {\bibfield  {journal} {\bibinfo  {journal} {JHEP}\
  }\textbf {\bibinfo {volume} {04}},\ \bibinfo {pages} {140} (\bibinfo {year}
  {2018})},\ \Eprint {http://arxiv.org/abs/1712.06624} {arXiv:1712.06624
  [hep-th]} \BibitemShut {NoStop}%
\bibitem [{\citenamefont {Bartolo}\ \emph {et~al.}(2002)\citenamefont
  {Bartolo}, \citenamefont {Matarrese},\ and\ \citenamefont
  {Riotto}}]{Bartolo:2001cw}%
  \BibitemOpen
  \bibfield  {author} {\bibinfo {author} {\bibfnamefont {N.}~\bibnamefont
  {Bartolo}}, \bibinfo {author} {\bibfnamefont {S.}~\bibnamefont {Matarrese}},
  \ and\ \bibinfo {author} {\bibfnamefont {A.}~\bibnamefont {Riotto}},\ }\href
  {\doibase 10.1103/PhysRevD.65.103505} {\bibfield  {journal} {\bibinfo
  {journal} {Phys. Rev. D}\ }\textbf {\bibinfo {volume} {65}},\ \bibinfo
  {pages} {103505} (\bibinfo {year} {2002})},\ \Eprint
  {http://arxiv.org/abs/hep-ph/0112261} {arXiv:hep-ph/0112261} \BibitemShut
  {NoStop}%
\bibitem [{\citenamefont {Cabass}\ \emph
  {et~al.}(2022{\natexlab{a}})\citenamefont {Cabass}, \citenamefont {Jazayeri},
  \citenamefont {Pajer},\ and\ \citenamefont {Stefanyszyn}}]{Cabass:2022rhr}%
  \BibitemOpen
  \bibfield  {author} {\bibinfo {author} {\bibfnamefont {G.}~\bibnamefont
  {Cabass}}, \bibinfo {author} {\bibfnamefont {S.}~\bibnamefont {Jazayeri}},
  \bibinfo {author} {\bibfnamefont {E.}~\bibnamefont {Pajer}}, \ and\ \bibinfo
  {author} {\bibfnamefont {D.}~\bibnamefont {Stefanyszyn}},\ }\href@noop {} {\
  (\bibinfo {year} {2022}{\natexlab{a}})},\ \Eprint
  {http://arxiv.org/abs/2210.02907} {arXiv:2210.02907 [hep-th]} \BibitemShut
  {NoStop}%
\bibitem [{\citenamefont {Creque-Sarbinowski}\ \emph
  {et~al.}(2023)\citenamefont {Creque-Sarbinowski}, \citenamefont {Alexander},
  \citenamefont {Kamionkowski},\ and\ \citenamefont {Philcox}}]{CyrilCS}%
  \BibitemOpen
  \bibfield  {author} {\bibinfo {author} {\bibfnamefont {C.}~\bibnamefont
  {Creque-Sarbinowski}}, \bibinfo {author} {\bibfnamefont {S.}~\bibnamefont
  {Alexander}}, \bibinfo {author} {\bibfnamefont {M.}~\bibnamefont
  {Kamionkowski}}, \ and\ \bibinfo {author} {\bibfnamefont {O.}~\bibnamefont
  {Philcox}},\ }\href@noop {} {\  (\bibinfo {year} {2023})},\ \Eprint
  {http://arxiv.org/abs/2303.04815} {arXiv:2303.04815 [astro-ph.CO]}
  \BibitemShut {NoStop}%
\bibitem [{\citenamefont {Bartolo}\ \emph {et~al.}(2021)\citenamefont
  {Bartolo}, \citenamefont {Caloni}, \citenamefont {Orlando},\ and\
  \citenamefont {Ricciardone}}]{Bartolo:2020gsh}%
  \BibitemOpen
  \bibfield  {author} {\bibinfo {author} {\bibfnamefont {N.}~\bibnamefont
  {Bartolo}}, \bibinfo {author} {\bibfnamefont {L.}~\bibnamefont {Caloni}},
  \bibinfo {author} {\bibfnamefont {G.}~\bibnamefont {Orlando}}, \ and\
  \bibinfo {author} {\bibfnamefont {A.}~\bibnamefont {Ricciardone}},\ }\href
  {\doibase 10.1088/1475-7516/2021/03/073} {\bibfield  {journal} {\bibinfo
  {journal} {JCAP}\ }\textbf {\bibinfo {volume} {03}},\ \bibinfo {pages} {073}
  (\bibinfo {year} {2021})},\ \Eprint {http://arxiv.org/abs/2008.01715}
  {arXiv:2008.01715 [astro-ph.CO]} \BibitemShut {NoStop}%
\bibitem [{\citenamefont {Moretti}\ \emph {et~al.}(2024)\citenamefont
  {Moretti}, \citenamefont {Bartolo},\ and\ \citenamefont
  {Greco}}]{Moretti:2024fzb}%
  \BibitemOpen
  \bibfield  {author} {\bibinfo {author} {\bibfnamefont {T.}~\bibnamefont
  {Moretti}}, \bibinfo {author} {\bibfnamefont {N.}~\bibnamefont {Bartolo}}, \
  and\ \bibinfo {author} {\bibfnamefont {A.}~\bibnamefont {Greco}},\
  }\href@noop {} {\  (\bibinfo {year} {2024})},\ \Eprint
  {http://arxiv.org/abs/2410.11801} {arXiv:2410.11801 [astro-ph.CO]}
  \BibitemShut {NoStop}%
\bibitem [{\citenamefont {Alexander}\ \emph {et~al.}(2019)\citenamefont
  {Alexander}, \citenamefont {Gates}, \citenamefont {Jenks}, \citenamefont
  {Koutrolikos},\ and\ \citenamefont {McDonough}}]{Alexander:2019vtb}%
  \BibitemOpen
  \bibfield  {author} {\bibinfo {author} {\bibfnamefont {S.}~\bibnamefont
  {Alexander}}, \bibinfo {author} {\bibfnamefont {S.~J.}\ \bibnamefont
  {Gates}}, \bibinfo {author} {\bibfnamefont {L.}~\bibnamefont {Jenks}},
  \bibinfo {author} {\bibfnamefont {K.}~\bibnamefont {Koutrolikos}}, \ and\
  \bibinfo {author} {\bibfnamefont {E.}~\bibnamefont {McDonough}},\ }\href
  {\doibase 10.1007/JHEP10(2019)156} {\bibfield  {journal} {\bibinfo  {journal}
  {JHEP}\ }\textbf {\bibinfo {volume} {10}},\ \bibinfo {pages} {156} (\bibinfo
  {year} {2019})},\ \Eprint {http://arxiv.org/abs/1907.05829} {arXiv:1907.05829
  [hep-th]} \BibitemShut {NoStop}%
\bibitem [{\citenamefont {Byrnes}\ \emph {et~al.}(2010)\citenamefont {Byrnes},
  \citenamefont {Gerstenlauer}, \citenamefont {Nurmi}, \citenamefont
  {Tasinato},\ and\ \citenamefont {Wands}}]{Byrnes:2010ft}%
  \BibitemOpen
  \bibfield  {author} {\bibinfo {author} {\bibfnamefont {C.~T.}\ \bibnamefont
  {Byrnes}}, \bibinfo {author} {\bibfnamefont {M.}~\bibnamefont
  {Gerstenlauer}}, \bibinfo {author} {\bibfnamefont {S.}~\bibnamefont {Nurmi}},
  \bibinfo {author} {\bibfnamefont {G.}~\bibnamefont {Tasinato}}, \ and\
  \bibinfo {author} {\bibfnamefont {D.}~\bibnamefont {Wands}},\ }\href
  {\doibase 10.1088/1475-7516/2010/10/004} {\bibfield  {journal} {\bibinfo
  {journal} {JCAP}\ }\textbf {\bibinfo {volume} {10}},\ \bibinfo {pages} {004}
  (\bibinfo {year} {2010})},\ \Eprint {http://arxiv.org/abs/1007.4277}
  {arXiv:1007.4277 [astro-ph.CO]} \BibitemShut {NoStop}%
\bibitem [{\citenamefont {Wang}\ \emph {et~al.}(2023)\citenamefont {Wang},
  \citenamefont {Pimentel},\ and\ \citenamefont {Ach\'ucarro}}]{Wang:2022eop}%
  \BibitemOpen
  \bibfield  {author} {\bibinfo {author} {\bibfnamefont {D.-G.}\ \bibnamefont
  {Wang}}, \bibinfo {author} {\bibfnamefont {G.~L.}\ \bibnamefont {Pimentel}},
  \ and\ \bibinfo {author} {\bibfnamefont {A.}~\bibnamefont {Ach\'ucarro}},\
  }\href {\doibase 10.1088/1475-7516/2023/05/043} {\bibfield  {journal}
  {\bibinfo  {journal} {JCAP}\ }\textbf {\bibinfo {volume} {05}},\ \bibinfo
  {pages} {043} (\bibinfo {year} {2023})},\ \Eprint
  {http://arxiv.org/abs/2212.14035} {arXiv:2212.14035 [astro-ph.CO]}
  \BibitemShut {NoStop}%
\bibitem [{\citenamefont {Jazayeri}\ and\ \citenamefont
  {Renaux-Petel}(2022)}]{Jazayeri:2022kjy}%
  \BibitemOpen
  \bibfield  {author} {\bibinfo {author} {\bibfnamefont {S.}~\bibnamefont
  {Jazayeri}}\ and\ \bibinfo {author} {\bibfnamefont {S.}~\bibnamefont
  {Renaux-Petel}},\ }\href {\doibase 10.1007/JHEP12(2022)137} {\bibfield
  {journal} {\bibinfo  {journal} {JHEP}\ }\textbf {\bibinfo {volume} {12}},\
  \bibinfo {pages} {137} (\bibinfo {year} {2022})},\ \Eprint
  {http://arxiv.org/abs/2205.10340} {arXiv:2205.10340 [hep-th]} \BibitemShut
  {NoStop}%
\bibitem [{\citenamefont {Meerburg}\ \emph {et~al.}(2009)\citenamefont
  {Meerburg}, \citenamefont {van~der Schaar},\ and\ \citenamefont
  {Corasaniti}}]{Meerburg:2009ys}%
  \BibitemOpen
  \bibfield  {author} {\bibinfo {author} {\bibfnamefont {P.~D.}\ \bibnamefont
  {Meerburg}}, \bibinfo {author} {\bibfnamefont {J.~P.}\ \bibnamefont {van~der
  Schaar}}, \ and\ \bibinfo {author} {\bibfnamefont {P.~S.}\ \bibnamefont
  {Corasaniti}},\ }\href {\doibase 10.1088/1475-7516/2009/05/018} {\bibfield
  {journal} {\bibinfo  {journal} {JCAP}\ }\textbf {\bibinfo {volume} {05}},\
  \bibinfo {pages} {018} (\bibinfo {year} {2009})},\ \Eprint
  {http://arxiv.org/abs/0901.4044} {arXiv:0901.4044 [hep-th]} \BibitemShut
  {NoStop}%
\bibitem [{\citenamefont {Mylova}\ \emph {et~al.}(2022)\citenamefont {Mylova},
  \citenamefont {Moschou}, \citenamefont {Afshordi},\ and\ \citenamefont
  {Magueijo}}]{Mylova:2021eld}%
  \BibitemOpen
  \bibfield  {author} {\bibinfo {author} {\bibfnamefont {M.}~\bibnamefont
  {Mylova}}, \bibinfo {author} {\bibfnamefont {M.}~\bibnamefont {Moschou}},
  \bibinfo {author} {\bibfnamefont {N.}~\bibnamefont {Afshordi}}, \ and\
  \bibinfo {author} {\bibfnamefont {J.~a.}\ \bibnamefont {Magueijo}},\ }\href
  {\doibase 10.1088/1475-7516/2022/07/005} {\bibfield  {journal} {\bibinfo
  {journal} {JCAP}\ }\textbf {\bibinfo {volume} {07}},\ \bibinfo {pages} {005}
  (\bibinfo {year} {2022})},\ \Eprint {http://arxiv.org/abs/2112.08179}
  {arXiv:2112.08179 [hep-th]} \BibitemShut {NoStop}%
\bibitem [{\citenamefont {Grin}\ \emph {et~al.}(2014)\citenamefont {Grin},
  \citenamefont {Hanson}, \citenamefont {Holder}, \citenamefont {Dor\'e},\ and\
  \citenamefont {Kamionkowski}}]{Grin:2013uya}%
  \BibitemOpen
  \bibfield  {author} {\bibinfo {author} {\bibfnamefont {D.}~\bibnamefont
  {Grin}}, \bibinfo {author} {\bibfnamefont {D.}~\bibnamefont {Hanson}},
  \bibinfo {author} {\bibfnamefont {G.~P.}\ \bibnamefont {Holder}}, \bibinfo
  {author} {\bibfnamefont {O.}~\bibnamefont {Dor\'e}}, \ and\ \bibinfo {author}
  {\bibfnamefont {M.}~\bibnamefont {Kamionkowski}},\ }\href {\doibase
  10.1103/PhysRevD.89.023006} {\bibfield  {journal} {\bibinfo  {journal} {Phys.
  Rev. D}\ }\textbf {\bibinfo {volume} {89}},\ \bibinfo {pages} {023006}
  (\bibinfo {year} {2014})},\ \Eprint {http://arxiv.org/abs/1306.4319}
  {arXiv:1306.4319 [astro-ph.CO]} \BibitemShut {NoStop}%
\bibitem [{\citenamefont {Hill}(2018)}]{Hill:2018ypf}%
  \BibitemOpen
  \bibfield  {author} {\bibinfo {author} {\bibfnamefont {J.~C.}\ \bibnamefont
  {Hill}},\ }\href {\doibase 10.1103/PhysRevD.98.083542} {\bibfield  {journal}
  {\bibinfo  {journal} {Phys. Rev. D}\ }\textbf {\bibinfo {volume} {98}},\
  \bibinfo {pages} {083542} (\bibinfo {year} {2018})},\ \Eprint
  {http://arxiv.org/abs/1807.07324} {arXiv:1807.07324 [astro-ph.CO]}
  \BibitemShut {NoStop}%
\bibitem [{\citenamefont {Philcox}\ and\ \citenamefont
  {Hill}(2025)}]{Philcox:2025lxt}%
  \BibitemOpen
  \bibfield  {author} {\bibinfo {author} {\bibfnamefont {O.~H.~E.}\
  \bibnamefont {Philcox}}\ and\ \bibinfo {author} {\bibfnamefont {J.~C.}\
  \bibnamefont {Hill}},\ }\href@noop {} {\  (\bibinfo {year} {2025})},\ \Eprint
  {http://arxiv.org/abs/2504.03826} {arXiv:2504.03826 [astro-ph.CO]}
  \BibitemShut {NoStop}%
\bibitem [{\citenamefont {Philcox}(2023{\natexlab{a}})}]{PhilcoxCMB}%
  \BibitemOpen
  \bibfield  {author} {\bibinfo {author} {\bibfnamefont {O.~H.~E.}\
  \bibnamefont {Philcox}},\ }\href@noop {} {\  (\bibinfo {year}
  {2023}{\natexlab{a}})},\ \Eprint {http://arxiv.org/abs/2303.12106}
  {arXiv:2303.12106 [astro-ph.CO]} \BibitemShut {NoStop}%
\bibitem [{\citenamefont {Philcox}\ and\ \citenamefont
  {Shiraishi}(2023)}]{Philcox:2023ypl}%
  \BibitemOpen
  \bibfield  {author} {\bibinfo {author} {\bibfnamefont {O.~H.~E.}\
  \bibnamefont {Philcox}}\ and\ \bibinfo {author} {\bibfnamefont
  {M.}~\bibnamefont {Shiraishi}},\ }\href@noop {} {\  (\bibinfo {year}
  {2023})},\ \Eprint {http://arxiv.org/abs/2308.03831} {arXiv:2308.03831
  [astro-ph.CO]} \BibitemShut {NoStop}%
\bibitem [{\citenamefont {Adam}\ \emph {et~al.}(2016)\citenamefont {Adam} \emph
  {et~al.}}]{Planck:2015mis}%
  \BibitemOpen
  \bibfield  {author} {\bibinfo {author} {\bibfnamefont {R.}~\bibnamefont
  {Adam}} \emph {et~al.} (\bibinfo {collaboration} {Planck}),\ }\href {\doibase
  10.1051/0004-6361/201525936} {\bibfield  {journal} {\bibinfo  {journal}
  {Astron. Astrophys.}\ }\textbf {\bibinfo {volume} {594}},\ \bibinfo {pages}
  {A9} (\bibinfo {year} {2016})},\ \Eprint {http://arxiv.org/abs/1502.05956}
  {arXiv:1502.05956 [astro-ph.CO]} \BibitemShut {NoStop}%
\bibitem [{\citenamefont {Akrami}\ \emph
  {et~al.}(2020{\natexlab{c}})\citenamefont {Akrami} \emph
  {et~al.}}]{Planck:2018yye}%
  \BibitemOpen
  \bibfield  {author} {\bibinfo {author} {\bibfnamefont {Y.}~\bibnamefont
  {Akrami}} \emph {et~al.} (\bibinfo {collaboration} {Planck}),\ }\href
  {\doibase 10.1051/0004-6361/201833881} {\bibfield  {journal} {\bibinfo
  {journal} {Astron. Astrophys.}\ }\textbf {\bibinfo {volume} {641}},\ \bibinfo
  {pages} {A4} (\bibinfo {year} {2020}{\natexlab{c}})},\ \Eprint
  {http://arxiv.org/abs/1807.06208} {arXiv:1807.06208 [astro-ph.CO]}
  \BibitemShut {NoStop}%
\bibitem [{\citenamefont {Carron}\ \emph {et~al.}(2022)\citenamefont {Carron},
  \citenamefont {Mirmelstein},\ and\ \citenamefont {Lewis}}]{Carron:2022eyg}%
  \BibitemOpen
  \bibfield  {author} {\bibinfo {author} {\bibfnamefont {J.}~\bibnamefont
  {Carron}}, \bibinfo {author} {\bibfnamefont {M.}~\bibnamefont {Mirmelstein}},
  \ and\ \bibinfo {author} {\bibfnamefont {A.}~\bibnamefont {Lewis}},\ }\href
  {\doibase 10.1088/1475-7516/2022/09/039} {\bibfield  {journal} {\bibinfo
  {journal} {JCAP}\ }\textbf {\bibinfo {volume} {09}},\ \bibinfo {pages} {039}
  (\bibinfo {year} {2022})},\ \Eprint {http://arxiv.org/abs/2206.07773}
  {arXiv:2206.07773 [astro-ph.CO]} \BibitemShut {NoStop}%
\bibitem [{\citenamefont {G\'orski}\ \emph {et~al.}(2005)\citenamefont
  {G\'orski}, \citenamefont {Hivon}, \citenamefont {Banday}, \citenamefont
  {Wandelt}, \citenamefont {Hansen}, \citenamefont {Reinecke},\ and\
  \citenamefont {Bartelman}}]{Gorski:2004by}%
  \BibitemOpen
  \bibfield  {author} {\bibinfo {author} {\bibfnamefont {K.~M.}\ \bibnamefont
  {G\'orski}}, \bibinfo {author} {\bibfnamefont {E.}~\bibnamefont {Hivon}},
  \bibinfo {author} {\bibfnamefont {A.~J.}\ \bibnamefont {Banday}}, \bibinfo
  {author} {\bibfnamefont {B.~D.}\ \bibnamefont {Wandelt}}, \bibinfo {author}
  {\bibfnamefont {F.~K.}\ \bibnamefont {Hansen}}, \bibinfo {author}
  {\bibfnamefont {M.}~\bibnamefont {Reinecke}}, \ and\ \bibinfo {author}
  {\bibfnamefont {M.}~\bibnamefont {Bartelman}},\ }\href {\doibase
  10.1086/427976} {\bibfield  {journal} {\bibinfo  {journal} {Astrophys. J.}\
  }\textbf {\bibinfo {volume} {622}},\ \bibinfo {pages} {759} (\bibinfo {year}
  {2005})},\ \Eprint {http://arxiv.org/abs/astro-ph/0409513}
  {arXiv:astro-ph/0409513} \BibitemShut {NoStop}%
\bibitem [{\citenamefont {Ade}\ \emph {et~al.}(2016{\natexlab{b}})\citenamefont
  {Ade} \emph {et~al.}}]{Planck:2015txa}%
  \BibitemOpen
  \bibfield  {author} {\bibinfo {author} {\bibfnamefont {P.~A.~R.}\
  \bibnamefont {Ade}} \emph {et~al.} (\bibinfo {collaboration} {Planck}),\
  }\href {\doibase 10.1051/0004-6361/201527103} {\bibfield  {journal} {\bibinfo
   {journal} {Astron. Astrophys.}\ }\textbf {\bibinfo {volume} {594}},\
  \bibinfo {pages} {A12} (\bibinfo {year} {2016}{\natexlab{b}})},\ \Eprint
  {http://arxiv.org/abs/1509.06348} {arXiv:1509.06348 [astro-ph.CO]}
  \BibitemShut {NoStop}%
\bibitem [{\citenamefont {{Oh}}\ \emph {et~al.}(1999)\citenamefont {{Oh}},
  \citenamefont {{Spergel}},\ and\ \citenamefont
  {{Hinshaw}}}]{1999ApJ...510..551O}%
  \BibitemOpen
  \bibfield  {author} {\bibinfo {author} {\bibfnamefont {S.~P.}\ \bibnamefont
  {{Oh}}}, \bibinfo {author} {\bibfnamefont {D.~N.}\ \bibnamefont {{Spergel}}},
  \ and\ \bibinfo {author} {\bibfnamefont {G.}~\bibnamefont {{Hinshaw}}},\
  }\href {\doibase 10.1086/306629} {\bibfield  {journal} {\bibinfo  {journal}
  {\apj}\ }\textbf {\bibinfo {volume} {510}},\ \bibinfo {pages} {551} (\bibinfo
  {year} {1999})},\ \Eprint {http://arxiv.org/abs/astro-ph/9805339}
  {arXiv:astro-ph/9805339 [astro-ph]} \BibitemShut {NoStop}%
\bibitem [{\citenamefont {Smith}\ \emph {et~al.}(2007)\citenamefont {Smith},
  \citenamefont {Zahn},\ and\ \citenamefont {Dore}}]{Smith:2007rg}%
  \BibitemOpen
  \bibfield  {author} {\bibinfo {author} {\bibfnamefont {K.~M.}\ \bibnamefont
  {Smith}}, \bibinfo {author} {\bibfnamefont {O.}~\bibnamefont {Zahn}}, \ and\
  \bibinfo {author} {\bibfnamefont {O.}~\bibnamefont {Dore}},\ }\href {\doibase
  10.1103/PhysRevD.76.043510} {\bibfield  {journal} {\bibinfo  {journal} {Phys.
  Rev. D}\ }\textbf {\bibinfo {volume} {76}},\ \bibinfo {pages} {043510}
  (\bibinfo {year} {2007})},\ \Eprint {http://arxiv.org/abs/0705.3980}
  {arXiv:0705.3980 [astro-ph]} \BibitemShut {NoStop}%
\bibitem [{\citenamefont {Oh}\ \emph {et~al.}(1999)\citenamefont {Oh},
  \citenamefont {Spergel},\ and\ \citenamefont {Hinshaw}}]{Oh:1998sr}%
  \BibitemOpen
  \bibfield  {author} {\bibinfo {author} {\bibfnamefont {S.~P.}\ \bibnamefont
  {Oh}}, \bibinfo {author} {\bibfnamefont {D.~N.}\ \bibnamefont {Spergel}}, \
  and\ \bibinfo {author} {\bibfnamefont {G.}~\bibnamefont {Hinshaw}},\ }\href
  {\doibase 10.1086/306629} {\bibfield  {journal} {\bibinfo  {journal}
  {Astrophys. J.}\ }\textbf {\bibinfo {volume} {510}},\ \bibinfo {pages} {551}
  (\bibinfo {year} {1999})},\ \Eprint {http://arxiv.org/abs/astro-ph/9805339}
  {arXiv:astro-ph/9805339} \BibitemShut {NoStop}%
\bibitem [{\citenamefont {M\"unchmeyer}\ and\ \citenamefont
  {Smith}(2019{\natexlab{b}})}]{Munchmeyer:2019kng}%
  \BibitemOpen
  \bibfield  {author} {\bibinfo {author} {\bibfnamefont {M.}~\bibnamefont
  {M\"unchmeyer}}\ and\ \bibinfo {author} {\bibfnamefont {K.~M.}\ \bibnamefont
  {Smith}},\ }\href@noop {} {\  (\bibinfo {year} {2019}{\natexlab{b}})},\
  \Eprint {http://arxiv.org/abs/1905.05846} {arXiv:1905.05846 [astro-ph.CO]}
  \BibitemShut {NoStop}%
\bibitem [{\citenamefont {Costanza}\ \emph
  {et~al.}(2024{\natexlab{a}})\citenamefont {Costanza}, \citenamefont
  {Sc\'occola},\ and\ \citenamefont {Zaldarriaga}}]{Costanza:2023cgt}%
  \BibitemOpen
  \bibfield  {author} {\bibinfo {author} {\bibfnamefont {B.}~\bibnamefont
  {Costanza}}, \bibinfo {author} {\bibfnamefont {C.~G.}\ \bibnamefont
  {Sc\'occola}}, \ and\ \bibinfo {author} {\bibfnamefont {M.}~\bibnamefont
  {Zaldarriaga}},\ }\href {\doibase 10.1088/1475-7516/2024/04/041} {\bibfield
  {journal} {\bibinfo  {journal} {JCAP}\ }\textbf {\bibinfo {volume} {04}},\
  \bibinfo {pages} {041} (\bibinfo {year} {2024}{\natexlab{a}})},\ \Eprint
  {http://arxiv.org/abs/2312.09943} {arXiv:2312.09943 [astro-ph.CO]}
  \BibitemShut {NoStop}%
\bibitem [{\citenamefont {Costanza}\ \emph
  {et~al.}(2024{\natexlab{b}})\citenamefont {Costanza}, \citenamefont
  {Sc\'occola},\ and\ \citenamefont {Zaldarriaga}}]{Costanza:2024rut}%
  \BibitemOpen
  \bibfield  {author} {\bibinfo {author} {\bibfnamefont {B.}~\bibnamefont
  {Costanza}}, \bibinfo {author} {\bibfnamefont {C.~G.}\ \bibnamefont
  {Sc\'occola}}, \ and\ \bibinfo {author} {\bibfnamefont {M.}~\bibnamefont
  {Zaldarriaga}},\ }\href@noop {} {\  (\bibinfo {year} {2024}{\natexlab{b}})},\
  \Eprint {http://arxiv.org/abs/2412.10580} {arXiv:2412.10580 [astro-ph.CO]}
  \BibitemShut {NoStop}%
\bibitem [{\citenamefont {Philcox}\ and\ \citenamefont
  {Fl\"oss}(2024)}]{Philcox:2024rqr}%
  \BibitemOpen
  \bibfield  {author} {\bibinfo {author} {\bibfnamefont {O.~H.~E.}\
  \bibnamefont {Philcox}}\ and\ \bibinfo {author} {\bibfnamefont
  {T.}~\bibnamefont {Fl\"oss}},\ }\href@noop {} {\  (\bibinfo {year} {2024})},\
  \Eprint {http://arxiv.org/abs/2404.07249} {arXiv:2404.07249 [astro-ph.CO]}
  \BibitemShut {NoStop}%
\bibitem [{\citenamefont {Philcox}(2023{\natexlab{b}})}]{Philcox:2023uwe}%
  \BibitemOpen
  \bibfield  {author} {\bibinfo {author} {\bibfnamefont {O.~H.~E.}\
  \bibnamefont {Philcox}},\ }\href {\doibase 10.1103/PhysRevD.107.123516}
  {\bibfield  {journal} {\bibinfo  {journal} {Phys. Rev. D}\ }\textbf {\bibinfo
  {volume} {107}},\ \bibinfo {pages} {123516} (\bibinfo {year}
  {2023}{\natexlab{b}})},\ \Eprint {http://arxiv.org/abs/2303.08828}
  {arXiv:2303.08828 [astro-ph.CO]} \BibitemShut {NoStop}%
\bibitem [{\citenamefont {Philcox}(2023{\natexlab{c}})}]{Philcox:2023psd}%
  \BibitemOpen
  \bibfield  {author} {\bibinfo {author} {\bibfnamefont {O.~H.~E.}\
  \bibnamefont {Philcox}},\ }\href@noop {} {\  (\bibinfo {year}
  {2023}{\natexlab{c}})},\ \Eprint {http://arxiv.org/abs/2306.03915}
  {arXiv:2306.03915 [astro-ph.CO]} \BibitemShut {NoStop}%
\bibitem [{\citenamefont {{Philcox}}(2023)}]{PolyBin}%
  \BibitemOpen
  \bibfield  {author} {\bibinfo {author} {\bibfnamefont {O.~H.~E.}\
  \bibnamefont {{Philcox}}},\ }\href@noop {} {\enquote {\bibinfo {title}
  {{PolyBin: Binned polyspectrum estimation on the full sky}},}\ }\bibinfo
  {howpublished} {Astrophysics Source Code Library, record ascl:2307.020}
  (\bibinfo {year} {2023}),\ \Eprint {http://arxiv.org/abs/2307.020}
  {ascl:2307.020} \BibitemShut {NoStop}%
\bibitem [{\citenamefont {Reinecke}\ and\ \citenamefont
  {Seljebotn}(2013)}]{Reinecke_2013}%
  \BibitemOpen
  \bibfield  {author} {\bibinfo {author} {\bibfnamefont {M.}~\bibnamefont
  {Reinecke}}\ and\ \bibinfo {author} {\bibfnamefont {D.~S.}\ \bibnamefont
  {Seljebotn}},\ }\href {\doibase 10.1051/0004-6361/201321494} {\bibfield
  {journal} {\bibinfo  {journal} {Astronomy \& Astrophysics}\ }\textbf
  {\bibinfo {volume} {554}},\ \bibinfo {pages} {A112} (\bibinfo {year}
  {2013})}\BibitemShut {NoStop}%
\bibitem [{\citenamefont {Zonca}\ \emph {et~al.}(2019)\citenamefont {Zonca},
  \citenamefont {Singer}, \citenamefont {Lenz}, \citenamefont {Reinecke},
  \citenamefont {Rosset}, \citenamefont {Hivon},\ and\ \citenamefont
  {Gorski}}]{Zonca:2019vzt}%
  \BibitemOpen
  \bibfield  {author} {\bibinfo {author} {\bibfnamefont {A.}~\bibnamefont
  {Zonca}}, \bibinfo {author} {\bibfnamefont {L.}~\bibnamefont {Singer}},
  \bibinfo {author} {\bibfnamefont {D.}~\bibnamefont {Lenz}}, \bibinfo {author}
  {\bibfnamefont {M.}~\bibnamefont {Reinecke}}, \bibinfo {author}
  {\bibfnamefont {C.}~\bibnamefont {Rosset}}, \bibinfo {author} {\bibfnamefont
  {E.}~\bibnamefont {Hivon}}, \ and\ \bibinfo {author} {\bibfnamefont
  {K.}~\bibnamefont {Gorski}},\ }\href {\doibase 10.21105/joss.01298}
  {\bibfield  {journal} {\bibinfo  {journal} {Journal of Open Source Software}\
  }\textbf {\bibinfo {volume} {4}},\ \bibinfo {pages} {1298} (\bibinfo {year}
  {2019})}\BibitemShut {NoStop}%
\bibitem [{\citenamefont {{Smith}}\ and\ \citenamefont
  {{Zaldarriaga}}(2011)}]{2011MNRAS.417....2S}%
  \BibitemOpen
  \bibfield  {author} {\bibinfo {author} {\bibfnamefont {K.~M.}\ \bibnamefont
  {{Smith}}}\ and\ \bibinfo {author} {\bibfnamefont {M.}~\bibnamefont
  {{Zaldarriaga}}},\ }\href {\doibase 10.1111/j.1365-2966.2010.18175.x}
  {\bibfield  {journal} {\bibinfo  {journal} {\mnras}\ }\textbf {\bibinfo
  {volume} {417}},\ \bibinfo {pages} {2} (\bibinfo {year} {2011})},\ \Eprint
  {http://arxiv.org/abs/astro-ph/0612571} {arXiv:astro-ph/0612571 [astro-ph]}
  \BibitemShut {NoStop}%
\bibitem [{\citenamefont {Namikawa}\ \emph {et~al.}(2013)\citenamefont
  {Namikawa}, \citenamefont {Hanson},\ and\ \citenamefont
  {Takahashi}}]{Namikawa:2012pe}%
  \BibitemOpen
  \bibfield  {author} {\bibinfo {author} {\bibfnamefont {T.}~\bibnamefont
  {Namikawa}}, \bibinfo {author} {\bibfnamefont {D.}~\bibnamefont {Hanson}}, \
  and\ \bibinfo {author} {\bibfnamefont {R.}~\bibnamefont {Takahashi}},\ }\href
  {\doibase 10.1093/mnras/stt195} {\bibfield  {journal} {\bibinfo  {journal}
  {Mon. Not. Roy. Astron. Soc.}\ }\textbf {\bibinfo {volume} {431}},\ \bibinfo
  {pages} {609} (\bibinfo {year} {2013})},\ \Eprint
  {http://arxiv.org/abs/1209.0091} {arXiv:1209.0091 [astro-ph.CO]} \BibitemShut
  {NoStop}%
\bibitem [{\citenamefont {Carron}(2023)}]{Carron:2022edh}%
  \BibitemOpen
  \bibfield  {author} {\bibinfo {author} {\bibfnamefont {J.}~\bibnamefont
  {Carron}},\ }\href {\doibase 10.1088/1475-7516/2023/02/057} {\bibfield
  {journal} {\bibinfo  {journal} {JCAP}\ }\textbf {\bibinfo {volume} {02}},\
  \bibinfo {pages} {057} (\bibinfo {year} {2023})},\ \Eprint
  {http://arxiv.org/abs/2210.05449} {arXiv:2210.05449 [astro-ph.CO]}
  \BibitemShut {NoStop}%
\bibitem [{\citenamefont {Maniyar}\ \emph {et~al.}(2021)\citenamefont
  {Maniyar}, \citenamefont {Ali-Ha\"\i{}moud}, \citenamefont {Carron},
  \citenamefont {Lewis},\ and\ \citenamefont
  {Madhavacheril}}]{Maniyar:2021msb}%
  \BibitemOpen
  \bibfield  {author} {\bibinfo {author} {\bibfnamefont {A.~S.}\ \bibnamefont
  {Maniyar}}, \bibinfo {author} {\bibfnamefont {Y.}~\bibnamefont
  {Ali-Ha\"\i{}moud}}, \bibinfo {author} {\bibfnamefont {J.}~\bibnamefont
  {Carron}}, \bibinfo {author} {\bibfnamefont {A.}~\bibnamefont {Lewis}}, \
  and\ \bibinfo {author} {\bibfnamefont {M.~S.}\ \bibnamefont
  {Madhavacheril}},\ }\href {\doibase 10.1103/PhysRevD.103.083524} {\bibfield
  {journal} {\bibinfo  {journal} {Phys. Rev. D}\ }\textbf {\bibinfo {volume}
  {103}},\ \bibinfo {pages} {083524} (\bibinfo {year} {2021})},\ \Eprint
  {http://arxiv.org/abs/2101.12193} {arXiv:2101.12193 [astro-ph.CO]}
  \BibitemShut {NoStop}%
\bibitem [{\citenamefont {Kalaja}\ \emph {et~al.}(2021)\citenamefont {Kalaja},
  \citenamefont {Meerburg}, \citenamefont {Pimentel},\ and\ \citenamefont
  {Coulton}}]{Kalaja:2020mkq}%
  \BibitemOpen
  \bibfield  {author} {\bibinfo {author} {\bibfnamefont {A.}~\bibnamefont
  {Kalaja}}, \bibinfo {author} {\bibfnamefont {P.~D.}\ \bibnamefont
  {Meerburg}}, \bibinfo {author} {\bibfnamefont {G.~L.}\ \bibnamefont
  {Pimentel}}, \ and\ \bibinfo {author} {\bibfnamefont {W.~R.}\ \bibnamefont
  {Coulton}},\ }\href {\doibase 10.1088/1475-7516/2021/04/050} {\bibfield
  {journal} {\bibinfo  {journal} {JCAP}\ }\textbf {\bibinfo {volume} {04}},\
  \bibinfo {pages} {050} (\bibinfo {year} {2021})},\ \Eprint
  {http://arxiv.org/abs/2011.09461} {arXiv:2011.09461 [astro-ph.CO]}
  \BibitemShut {NoStop}%
\bibitem [{\citenamefont {Smith}\ and\ \citenamefont
  {Kamionkowski}(2012)}]{Smith:2012ty}%
  \BibitemOpen
  \bibfield  {author} {\bibinfo {author} {\bibfnamefont {T.~L.}\ \bibnamefont
  {Smith}}\ and\ \bibinfo {author} {\bibfnamefont {M.}~\bibnamefont
  {Kamionkowski}},\ }\href {\doibase 10.1103/PhysRevD.86.063009} {\bibfield
  {journal} {\bibinfo  {journal} {Phys. Rev. D}\ }\textbf {\bibinfo {volume}
  {86}},\ \bibinfo {pages} {063009} (\bibinfo {year} {2012})},\ \Eprint
  {http://arxiv.org/abs/1203.6654} {arXiv:1203.6654 [astro-ph.CO]} \BibitemShut
  {NoStop}%
\bibitem [{\citenamefont {Hamimeche}\ and\ \citenamefont
  {Lewis}(2008)}]{Hamimeche:2008ai}%
  \BibitemOpen
  \bibfield  {author} {\bibinfo {author} {\bibfnamefont {S.}~\bibnamefont
  {Hamimeche}}\ and\ \bibinfo {author} {\bibfnamefont {A.}~\bibnamefont
  {Lewis}},\ }\href {\doibase 10.1103/PhysRevD.77.103013} {\bibfield  {journal}
  {\bibinfo  {journal} {Phys. Rev. D}\ }\textbf {\bibinfo {volume} {77}},\
  \bibinfo {pages} {103013} (\bibinfo {year} {2008})},\ \Eprint
  {http://arxiv.org/abs/0801.0554} {arXiv:0801.0554 [astro-ph]} \BibitemShut
  {NoStop}%
\bibitem [{\citenamefont {Coulton}\ and\ \citenamefont
  {Spergel}(2019)}]{Coulton:2019bnz}%
  \BibitemOpen
  \bibfield  {author} {\bibinfo {author} {\bibfnamefont {W.~R.}\ \bibnamefont
  {Coulton}}\ and\ \bibinfo {author} {\bibfnamefont {D.~N.}\ \bibnamefont
  {Spergel}},\ }\href {\doibase 10.1088/1475-7516/2019/10/056} {\bibfield
  {journal} {\bibinfo  {journal} {JCAP}\ }\textbf {\bibinfo {volume} {10}},\
  \bibinfo {pages} {056} (\bibinfo {year} {2019})},\ \Eprint
  {http://arxiv.org/abs/1901.04515} {arXiv:1901.04515 [astro-ph.CO]}
  \BibitemShut {NoStop}%
\bibitem [{\citenamefont {Coulton}\ \emph {et~al.}(2020)\citenamefont
  {Coulton}, \citenamefont {Meerburg}, \citenamefont {Baker}, \citenamefont
  {Hotinli}, \citenamefont {Duivenvoorden},\ and\ \citenamefont {van
  Engelen}}]{Coulton:2019odk}%
  \BibitemOpen
  \bibfield  {author} {\bibinfo {author} {\bibfnamefont {W.~R.}\ \bibnamefont
  {Coulton}}, \bibinfo {author} {\bibfnamefont {P.~D.}\ \bibnamefont
  {Meerburg}}, \bibinfo {author} {\bibfnamefont {D.~G.}\ \bibnamefont {Baker}},
  \bibinfo {author} {\bibfnamefont {S.}~\bibnamefont {Hotinli}}, \bibinfo
  {author} {\bibfnamefont {A.~J.}\ \bibnamefont {Duivenvoorden}}, \ and\
  \bibinfo {author} {\bibfnamefont {A.}~\bibnamefont {van Engelen}},\ }\href
  {\doibase 10.1103/PhysRevD.101.123504} {\bibfield  {journal} {\bibinfo
  {journal} {Phys. Rev. D}\ }\textbf {\bibinfo {volume} {101}},\ \bibinfo
  {pages} {123504} (\bibinfo {year} {2020})},\ \Eprint
  {http://arxiv.org/abs/1912.07619} {arXiv:1912.07619 [astro-ph.CO]}
  \BibitemShut {NoStop}%
\bibitem [{\citenamefont {Shiraishi}(2019)}]{Shiraishi:2019yux}%
  \BibitemOpen
  \bibfield  {author} {\bibinfo {author} {\bibfnamefont {M.}~\bibnamefont
  {Shiraishi}},\ }\href {\doibase 10.3389/fspas.2019.00049} {\bibfield
  {journal} {\bibinfo  {journal} {Front. Astron. Space Sci.}\ }\textbf
  {\bibinfo {volume} {6}},\ \bibinfo {pages} {49} (\bibinfo {year} {2019})},\
  \Eprint {http://arxiv.org/abs/1905.12485} {arXiv:1905.12485 [astro-ph.CO]}
  \BibitemShut {NoStop}%
\bibitem [{\citenamefont {Green}\ \emph {et~al.}(2017)\citenamefont {Green},
  \citenamefont {Meyers},\ and\ \citenamefont {van Engelen}}]{Green:2016cjr}%
  \BibitemOpen
  \bibfield  {author} {\bibinfo {author} {\bibfnamefont {D.}~\bibnamefont
  {Green}}, \bibinfo {author} {\bibfnamefont {J.}~\bibnamefont {Meyers}}, \
  and\ \bibinfo {author} {\bibfnamefont {A.}~\bibnamefont {van Engelen}},\
  }\href {\doibase 10.1088/1475-7516/2017/12/005} {\bibfield  {journal}
  {\bibinfo  {journal} {JCAP}\ }\textbf {\bibinfo {volume} {12}},\ \bibinfo
  {pages} {005} (\bibinfo {year} {2017})},\ \Eprint
  {http://arxiv.org/abs/1609.08143} {arXiv:1609.08143 [astro-ph.CO]}
  \BibitemShut {NoStop}%
\bibitem [{\citenamefont {Trendafilova}\ \emph {et~al.}(2024)\citenamefont
  {Trendafilova}, \citenamefont {Hotinli},\ and\ \citenamefont
  {Meyers}}]{Trendafilova:2023xtq}%
  \BibitemOpen
  \bibfield  {author} {\bibinfo {author} {\bibfnamefont {C.}~\bibnamefont
  {Trendafilova}}, \bibinfo {author} {\bibfnamefont {S.~C.}\ \bibnamefont
  {Hotinli}}, \ and\ \bibinfo {author} {\bibfnamefont {J.}~\bibnamefont
  {Meyers}},\ }\href {\doibase 10.1088/1475-7516/2024/06/017} {\bibfield
  {journal} {\bibinfo  {journal} {JCAP}\ }\textbf {\bibinfo {volume} {06}},\
  \bibinfo {pages} {017} (\bibinfo {year} {2024})},\ \Eprint
  {http://arxiv.org/abs/2312.02954} {arXiv:2312.02954 [astro-ph.CO]}
  \BibitemShut {NoStop}%
\bibitem [{\citenamefont {Higuchi}(1987)}]{Higuchi:1986py}%
  \BibitemOpen
  \bibfield  {author} {\bibinfo {author} {\bibfnamefont {A.}~\bibnamefont
  {Higuchi}},\ }\href {\doibase 10.1016/0550-3213(87)90691-2} {\bibfield
  {journal} {\bibinfo  {journal} {Nucl. Phys. B}\ }\textbf {\bibinfo {volume}
  {282}},\ \bibinfo {pages} {397} (\bibinfo {year} {1987})}\BibitemShut
  {NoStop}%
\bibitem [{\citenamefont {Bordin}\ and\ \citenamefont
  {Cabass}(2019)}]{Bordin:2019tyb}%
  \BibitemOpen
  \bibfield  {author} {\bibinfo {author} {\bibfnamefont {L.}~\bibnamefont
  {Bordin}}\ and\ \bibinfo {author} {\bibfnamefont {G.}~\bibnamefont
  {Cabass}},\ }\href {\doibase 10.1088/1475-7516/2019/06/050} {\bibfield
  {journal} {\bibinfo  {journal} {JCAP}\ }\textbf {\bibinfo {volume} {06}},\
  \bibinfo {pages} {050} (\bibinfo {year} {2019})},\ \Eprint
  {http://arxiv.org/abs/1902.09519} {arXiv:1902.09519 [astro-ph.CO]}
  \BibitemShut {NoStop}%
\bibitem [{\citenamefont {Moradinezhad~Dizgah}\ \emph
  {et~al.}(2018)\citenamefont {Moradinezhad~Dizgah}, \citenamefont {Lee},
  \citenamefont {Mu\~noz},\ and\ \citenamefont
  {Dvorkin}}]{MoradinezhadDizgah:2018ssw}%
  \BibitemOpen
  \bibfield  {author} {\bibinfo {author} {\bibfnamefont {A.}~\bibnamefont
  {Moradinezhad~Dizgah}}, \bibinfo {author} {\bibfnamefont {H.}~\bibnamefont
  {Lee}}, \bibinfo {author} {\bibfnamefont {J.~B.}\ \bibnamefont {Mu\~noz}}, \
  and\ \bibinfo {author} {\bibfnamefont {C.}~\bibnamefont {Dvorkin}},\ }\href
  {\doibase 10.1088/1475-7516/2018/05/013} {\bibfield  {journal} {\bibinfo
  {journal} {JCAP}\ }\textbf {\bibinfo {volume} {05}},\ \bibinfo {pages} {013}
  (\bibinfo {year} {2018})},\ \Eprint {http://arxiv.org/abs/1801.07265}
  {arXiv:1801.07265 [astro-ph.CO]} \BibitemShut {NoStop}%
\bibitem [{\citenamefont {Munshi}\ \emph
  {et~al.}(2011{\natexlab{b}})\citenamefont {Munshi}, \citenamefont {Heavens},
  \citenamefont {Cooray}, \citenamefont {Smidt}, \citenamefont {Coles},\ and\
  \citenamefont {Serra}}]{Munshi:2009wy}%
  \BibitemOpen
  \bibfield  {author} {\bibinfo {author} {\bibfnamefont {D.}~\bibnamefont
  {Munshi}}, \bibinfo {author} {\bibfnamefont {A.}~\bibnamefont {Heavens}},
  \bibinfo {author} {\bibfnamefont {A.}~\bibnamefont {Cooray}}, \bibinfo
  {author} {\bibfnamefont {J.}~\bibnamefont {Smidt}}, \bibinfo {author}
  {\bibfnamefont {P.}~\bibnamefont {Coles}}, \ and\ \bibinfo {author}
  {\bibfnamefont {P.}~\bibnamefont {Serra}},\ }\href {\doibase
  10.1111/j.1365-2966.2010.18035.x} {\bibfield  {journal} {\bibinfo  {journal}
  {Mon. Not. Roy. Astron. Soc.}\ }\textbf {\bibinfo {volume} {412}},\ \bibinfo
  {pages} {1993} (\bibinfo {year} {2011}{\natexlab{b}})},\ \Eprint
  {http://arxiv.org/abs/0910.3693} {arXiv:0910.3693 [astro-ph.CO]} \BibitemShut
  {NoStop}%
\bibitem [{\citenamefont {Hanson}\ and\ \citenamefont
  {Lewis}(2009)}]{Hanson:2009gu}%
  \BibitemOpen
  \bibfield  {author} {\bibinfo {author} {\bibfnamefont {D.}~\bibnamefont
  {Hanson}}\ and\ \bibinfo {author} {\bibfnamefont {A.}~\bibnamefont {Lewis}},\
  }\href {\doibase 10.1103/PhysRevD.80.063004} {\bibfield  {journal} {\bibinfo
  {journal} {Phys. Rev. D}\ }\textbf {\bibinfo {volume} {80}},\ \bibinfo
  {pages} {063004} (\bibinfo {year} {2009})},\ \Eprint
  {http://arxiv.org/abs/0908.0963} {arXiv:0908.0963 [astro-ph.CO]} \BibitemShut
  {NoStop}%
\bibitem [{\citenamefont {Desjacques}\ and\ \citenamefont
  {Seljak}(2010)}]{Desjacques:2009jb}%
  \BibitemOpen
  \bibfield  {author} {\bibinfo {author} {\bibfnamefont {V.}~\bibnamefont
  {Desjacques}}\ and\ \bibinfo {author} {\bibfnamefont {U.}~\bibnamefont
  {Seljak}},\ }\href {\doibase 10.1103/PhysRevD.81.023006} {\bibfield
  {journal} {\bibinfo  {journal} {Phys. Rev. D}\ }\textbf {\bibinfo {volume}
  {81}},\ \bibinfo {pages} {023006} (\bibinfo {year} {2010})},\ \Eprint
  {http://arxiv.org/abs/0907.2257} {arXiv:0907.2257 [astro-ph.CO]} \BibitemShut
  {NoStop}%
\bibitem [{\citenamefont {Smith}\ \emph {et~al.}(2012)\citenamefont {Smith},
  \citenamefont {Ferraro},\ and\ \citenamefont {LoVerde}}]{Smith:2011ub}%
  \BibitemOpen
  \bibfield  {author} {\bibinfo {author} {\bibfnamefont {K.~M.}\ \bibnamefont
  {Smith}}, \bibinfo {author} {\bibfnamefont {S.}~\bibnamefont {Ferraro}}, \
  and\ \bibinfo {author} {\bibfnamefont {M.}~\bibnamefont {LoVerde}},\ }\href
  {\doibase 10.1088/1475-7516/2012/03/032} {\bibfield  {journal} {\bibinfo
  {journal} {JCAP}\ }\textbf {\bibinfo {volume} {03}},\ \bibinfo {pages} {032}
  (\bibinfo {year} {2012})},\ \Eprint {http://arxiv.org/abs/1106.0503}
  {arXiv:1106.0503 [astro-ph.CO]} \BibitemShut {NoStop}%
\bibitem [{\citenamefont {LoVerde}\ and\ \citenamefont
  {Smith}(2011)}]{LoVerde:2011iz}%
  \BibitemOpen
  \bibfield  {author} {\bibinfo {author} {\bibfnamefont {M.}~\bibnamefont
  {LoVerde}}\ and\ \bibinfo {author} {\bibfnamefont {K.~M.}\ \bibnamefont
  {Smith}},\ }\href {\doibase 10.1088/1475-7516/2011/08/003} {\bibfield
  {journal} {\bibinfo  {journal} {JCAP}\ }\textbf {\bibinfo {volume} {08}},\
  \bibinfo {pages} {003} (\bibinfo {year} {2011})},\ \Eprint
  {http://arxiv.org/abs/1102.1439} {arXiv:1102.1439 [astro-ph.CO]} \BibitemShut
  {NoStop}%
\bibitem [{\citenamefont {Jeong}\ and\ \citenamefont
  {Komatsu}(2009)}]{Jeong:2009vd}%
  \BibitemOpen
  \bibfield  {author} {\bibinfo {author} {\bibfnamefont {D.}~\bibnamefont
  {Jeong}}\ and\ \bibinfo {author} {\bibfnamefont {E.}~\bibnamefont
  {Komatsu}},\ }\href {\doibase 10.1088/0004-637X/703/2/1230} {\bibfield
  {journal} {\bibinfo  {journal} {Astrophys. J.}\ }\textbf {\bibinfo {volume}
  {703}},\ \bibinfo {pages} {1230} (\bibinfo {year} {2009})},\ \Eprint
  {http://arxiv.org/abs/0904.0497} {arXiv:0904.0497 [astro-ph.CO]} \BibitemShut
  {NoStop}%
\bibitem [{\citenamefont {Chongchitnan}\ and\ \citenamefont
  {Silk}(2010)}]{Chongchitnan:2010xz}%
  \BibitemOpen
  \bibfield  {author} {\bibinfo {author} {\bibfnamefont {S.}~\bibnamefont
  {Chongchitnan}}\ and\ \bibinfo {author} {\bibfnamefont {J.}~\bibnamefont
  {Silk}},\ }\href {\doibase 10.1088/0004-637X/724/1/285} {\bibfield  {journal}
  {\bibinfo  {journal} {Astrophys. J.}\ }\textbf {\bibinfo {volume} {724}},\
  \bibinfo {pages} {285} (\bibinfo {year} {2010})},\ \Eprint
  {http://arxiv.org/abs/1007.1230} {arXiv:1007.1230 [astro-ph.CO]} \BibitemShut
  {NoStop}%
\bibitem [{\citenamefont {Slosar}\ \emph {et~al.}(2008)\citenamefont {Slosar},
  \citenamefont {Hirata}, \citenamefont {Seljak}, \citenamefont {Ho},\ and\
  \citenamefont {Padmanabhan}}]{Slosar:2008hx}%
  \BibitemOpen
  \bibfield  {author} {\bibinfo {author} {\bibfnamefont {A.}~\bibnamefont
  {Slosar}}, \bibinfo {author} {\bibfnamefont {C.}~\bibnamefont {Hirata}},
  \bibinfo {author} {\bibfnamefont {U.}~\bibnamefont {Seljak}}, \bibinfo
  {author} {\bibfnamefont {S.}~\bibnamefont {Ho}}, \ and\ \bibinfo {author}
  {\bibfnamefont {N.}~\bibnamefont {Padmanabhan}},\ }\href {\doibase
  10.1088/1475-7516/2008/08/031} {\bibfield  {journal} {\bibinfo  {journal}
  {JCAP}\ }\textbf {\bibinfo {volume} {08}},\ \bibinfo {pages} {031} (\bibinfo
  {year} {2008})},\ \Eprint {http://arxiv.org/abs/0805.3580} {arXiv:0805.3580
  [astro-ph]} \BibitemShut {NoStop}%
\bibitem [{\citenamefont {Leistedt}\ \emph {et~al.}(2014)\citenamefont
  {Leistedt}, \citenamefont {Peiris},\ and\ \citenamefont
  {Roth}}]{Leistedt:2014zqa}%
  \BibitemOpen
  \bibfield  {author} {\bibinfo {author} {\bibfnamefont {B.}~\bibnamefont
  {Leistedt}}, \bibinfo {author} {\bibfnamefont {H.~V.}\ \bibnamefont
  {Peiris}}, \ and\ \bibinfo {author} {\bibfnamefont {N.}~\bibnamefont
  {Roth}},\ }\href {\doibase 10.1103/PhysRevLett.113.221301} {\bibfield
  {journal} {\bibinfo  {journal} {Phys. Rev. Lett.}\ }\textbf {\bibinfo
  {volume} {113}},\ \bibinfo {pages} {221301} (\bibinfo {year} {2014})},\
  \Eprint {http://arxiv.org/abs/1405.4315} {arXiv:1405.4315 [astro-ph.CO]}
  \BibitemShut {NoStop}%
\bibitem [{\citenamefont {Giannantonio}\ \emph {et~al.}(2014)\citenamefont
  {Giannantonio}, \citenamefont {Ross}, \citenamefont {Percival}, \citenamefont
  {Crittenden}, \citenamefont {Bacher}, \citenamefont {Kilbinger},
  \citenamefont {Nichol},\ and\ \citenamefont {Weller}}]{Giannantonio:2013uqa}%
  \BibitemOpen
  \bibfield  {author} {\bibinfo {author} {\bibfnamefont {T.}~\bibnamefont
  {Giannantonio}}, \bibinfo {author} {\bibfnamefont {A.~J.}\ \bibnamefont
  {Ross}}, \bibinfo {author} {\bibfnamefont {W.~J.}\ \bibnamefont {Percival}},
  \bibinfo {author} {\bibfnamefont {R.}~\bibnamefont {Crittenden}}, \bibinfo
  {author} {\bibfnamefont {D.}~\bibnamefont {Bacher}}, \bibinfo {author}
  {\bibfnamefont {M.}~\bibnamefont {Kilbinger}}, \bibinfo {author}
  {\bibfnamefont {R.}~\bibnamefont {Nichol}}, \ and\ \bibinfo {author}
  {\bibfnamefont {J.}~\bibnamefont {Weller}},\ }\href {\doibase
  10.1103/PhysRevD.89.023511} {\bibfield  {journal} {\bibinfo  {journal} {Phys.
  Rev. D}\ }\textbf {\bibinfo {volume} {89}},\ \bibinfo {pages} {023511}
  (\bibinfo {year} {2014})},\ \Eprint {http://arxiv.org/abs/1303.1349}
  {arXiv:1303.1349 [astro-ph.CO]} \BibitemShut {NoStop}%
\bibitem [{\citenamefont {Ferraro}\ and\ \citenamefont
  {Smith}(2015)}]{Ferraro:2014jba}%
  \BibitemOpen
  \bibfield  {author} {\bibinfo {author} {\bibfnamefont {S.}~\bibnamefont
  {Ferraro}}\ and\ \bibinfo {author} {\bibfnamefont {K.~M.}\ \bibnamefont
  {Smith}},\ }\href {\doibase 10.1103/PhysRevD.91.043506} {\bibfield  {journal}
  {\bibinfo  {journal} {Phys. Rev. D}\ }\textbf {\bibinfo {volume} {91}},\
  \bibinfo {pages} {043506} (\bibinfo {year} {2015})},\ \Eprint
  {http://arxiv.org/abs/1408.3126} {arXiv:1408.3126 [astro-ph.CO]} \BibitemShut
  {NoStop}%
\bibitem [{\citenamefont {Biagetti}\ \emph {et~al.}(2013)\citenamefont
  {Biagetti}, \citenamefont {Desjacques},\ and\ \citenamefont
  {Riotto}}]{Biagetti:2012xy}%
  \BibitemOpen
  \bibfield  {author} {\bibinfo {author} {\bibfnamefont {M.}~\bibnamefont
  {Biagetti}}, \bibinfo {author} {\bibfnamefont {V.}~\bibnamefont
  {Desjacques}}, \ and\ \bibinfo {author} {\bibfnamefont {A.}~\bibnamefont
  {Riotto}},\ }\href {\doibase 10.1093/mnras/sts467} {\bibfield  {journal}
  {\bibinfo  {journal} {Mon. Not. Roy. Astron. Soc.}\ }\textbf {\bibinfo
  {volume} {429}},\ \bibinfo {pages} {1774} (\bibinfo {year} {2013})},\ \Eprint
  {http://arxiv.org/abs/1208.1616} {arXiv:1208.1616 [astro-ph.CO]} \BibitemShut
  {NoStop}%
\bibitem [{\citenamefont {Assassi}\ \emph
  {et~al.}(2015{\natexlab{a}})\citenamefont {Assassi}, \citenamefont
  {Baumann},\ and\ \citenamefont {Schmidt}}]{Assassi:2015fma}%
  \BibitemOpen
  \bibfield  {author} {\bibinfo {author} {\bibfnamefont {V.}~\bibnamefont
  {Assassi}}, \bibinfo {author} {\bibfnamefont {D.}~\bibnamefont {Baumann}}, \
  and\ \bibinfo {author} {\bibfnamefont {F.}~\bibnamefont {Schmidt}},\ }\href
  {\doibase 10.1088/1475-7516/2015/12/043} {\bibfield  {journal} {\bibinfo
  {journal} {JCAP}\ }\textbf {\bibinfo {volume} {12}},\ \bibinfo {pages} {043}
  (\bibinfo {year} {2015}{\natexlab{a}})},\ \Eprint
  {http://arxiv.org/abs/1510.03723} {arXiv:1510.03723 [astro-ph.CO]}
  \BibitemShut {NoStop}%
\bibitem [{\citenamefont {Assassi}\ \emph
  {et~al.}(2015{\natexlab{b}})\citenamefont {Assassi}, \citenamefont {Baumann},
  \citenamefont {Pajer}, \citenamefont {Welling},\ and\ \citenamefont {van~der
  Woude}}]{Assassi:2015jqa}%
  \BibitemOpen
  \bibfield  {author} {\bibinfo {author} {\bibfnamefont {V.}~\bibnamefont
  {Assassi}}, \bibinfo {author} {\bibfnamefont {D.}~\bibnamefont {Baumann}},
  \bibinfo {author} {\bibfnamefont {E.}~\bibnamefont {Pajer}}, \bibinfo
  {author} {\bibfnamefont {Y.}~\bibnamefont {Welling}}, \ and\ \bibinfo
  {author} {\bibfnamefont {D.}~\bibnamefont {van~der Woude}},\ }\href {\doibase
  10.1088/1475-7516/2015/11/024} {\bibfield  {journal} {\bibinfo  {journal}
  {JCAP}\ }\textbf {\bibinfo {volume} {11}},\ \bibinfo {pages} {024} (\bibinfo
  {year} {2015}{\natexlab{b}})},\ \Eprint {http://arxiv.org/abs/1505.06668}
  {arXiv:1505.06668 [astro-ph.CO]} \BibitemShut {NoStop}%
\bibitem [{\citenamefont {Cabass}\ \emph
  {et~al.}(2022{\natexlab{b}})\citenamefont {Cabass}, \citenamefont {Ivanov},
  \citenamefont {Philcox}, \citenamefont {Simonovi\'c},\ and\ \citenamefont
  {Zaldarriaga}}]{Cabass:2022ymb}%
  \BibitemOpen
  \bibfield  {author} {\bibinfo {author} {\bibfnamefont {G.}~\bibnamefont
  {Cabass}}, \bibinfo {author} {\bibfnamefont {M.~M.}\ \bibnamefont {Ivanov}},
  \bibinfo {author} {\bibfnamefont {O.~H.~E.}\ \bibnamefont {Philcox}},
  \bibinfo {author} {\bibfnamefont {M.}~\bibnamefont {Simonovi\'c}}, \ and\
  \bibinfo {author} {\bibfnamefont {M.}~\bibnamefont {Zaldarriaga}},\ }\href
  {\doibase 10.1103/PhysRevD.106.043506} {\bibfield  {journal} {\bibinfo
  {journal} {Phys. Rev. D}\ }\textbf {\bibinfo {volume} {106}},\ \bibinfo
  {pages} {043506} (\bibinfo {year} {2022}{\natexlab{b}})},\ \Eprint
  {http://arxiv.org/abs/2204.01781} {arXiv:2204.01781 [astro-ph.CO]}
  \BibitemShut {NoStop}%
\bibitem [{\citenamefont {D'Amico}\ \emph {et~al.}(2022)\citenamefont
  {D'Amico}, \citenamefont {Lewandowski}, \citenamefont {Senatore},\ and\
  \citenamefont {Zhang}}]{DAmico:2022gki}%
  \BibitemOpen
  \bibfield  {author} {\bibinfo {author} {\bibfnamefont {G.}~\bibnamefont
  {D'Amico}}, \bibinfo {author} {\bibfnamefont {M.}~\bibnamefont
  {Lewandowski}}, \bibinfo {author} {\bibfnamefont {L.}~\bibnamefont
  {Senatore}}, \ and\ \bibinfo {author} {\bibfnamefont {P.}~\bibnamefont
  {Zhang}},\ }\href@noop {} {\  (\bibinfo {year} {2022})},\ \Eprint
  {http://arxiv.org/abs/2201.11518} {arXiv:2201.11518 [astro-ph.CO]}
  \BibitemShut {NoStop}%
\bibitem [{\citenamefont {Goldstein}\ \emph
  {et~al.}(2024{\natexlab{a}})\citenamefont {Goldstein}, \citenamefont
  {Philcox}, \citenamefont {Hill},\ and\ \citenamefont
  {Hui}}]{Goldstein:2024bky}%
  \BibitemOpen
  \bibfield  {author} {\bibinfo {author} {\bibfnamefont {S.}~\bibnamefont
  {Goldstein}}, \bibinfo {author} {\bibfnamefont {O.~H.~E.}\ \bibnamefont
  {Philcox}}, \bibinfo {author} {\bibfnamefont {J.~C.}\ \bibnamefont {Hill}}, \
  and\ \bibinfo {author} {\bibfnamefont {L.}~\bibnamefont {Hui}},\ }\href@noop
  {} {\  (\bibinfo {year} {2024}{\natexlab{a}})},\ \Eprint
  {http://arxiv.org/abs/2407.08731} {arXiv:2407.08731 [astro-ph.CO]}
  \BibitemShut {NoStop}%
\bibitem [{\citenamefont {Anil~Kumar}\ \emph {et~al.}(2022)\citenamefont
  {Anil~Kumar}, \citenamefont {Sato-Polito}, \citenamefont {Kamionkowski},\
  and\ \citenamefont {Hotinli}}]{AnilKumar:2022flx}%
  \BibitemOpen
  \bibfield  {author} {\bibinfo {author} {\bibfnamefont {N.}~\bibnamefont
  {Anil~Kumar}}, \bibinfo {author} {\bibfnamefont {G.}~\bibnamefont
  {Sato-Polito}}, \bibinfo {author} {\bibfnamefont {M.}~\bibnamefont
  {Kamionkowski}}, \ and\ \bibinfo {author} {\bibfnamefont {S.~C.}\
  \bibnamefont {Hotinli}},\ }\href {\doibase 10.1103/PhysRevD.106.063533}
  {\bibfield  {journal} {\bibinfo  {journal} {Phys. Rev. D}\ }\textbf {\bibinfo
  {volume} {106}},\ \bibinfo {pages} {063533} (\bibinfo {year} {2022})},\
  \Eprint {http://arxiv.org/abs/2205.03423} {arXiv:2205.03423 [astro-ph.CO]}
  \BibitemShut {NoStop}%
\bibitem [{\citenamefont {Bartolo}\ \emph {et~al.}(2016)\citenamefont
  {Bartolo}, \citenamefont {Liguori},\ and\ \citenamefont
  {Shiraishi}}]{Bartolo:2015fqz}%
  \BibitemOpen
  \bibfield  {author} {\bibinfo {author} {\bibfnamefont {N.}~\bibnamefont
  {Bartolo}}, \bibinfo {author} {\bibfnamefont {M.}~\bibnamefont {Liguori}}, \
  and\ \bibinfo {author} {\bibfnamefont {M.}~\bibnamefont {Shiraishi}},\ }\href
  {\doibase 10.1088/1475-7516/2016/03/029} {\bibfield  {journal} {\bibinfo
  {journal} {JCAP}\ }\textbf {\bibinfo {volume} {03}},\ \bibinfo {pages} {029}
  (\bibinfo {year} {2016})},\ \Eprint {http://arxiv.org/abs/1511.01474}
  {arXiv:1511.01474 [astro-ph.CO]} \BibitemShut {NoStop}%
\bibitem [{\citenamefont {Khatri}\ and\ \citenamefont
  {Sunyaev}(2015)}]{Khatri:2015tla}%
  \BibitemOpen
  \bibfield  {author} {\bibinfo {author} {\bibfnamefont {R.}~\bibnamefont
  {Khatri}}\ and\ \bibinfo {author} {\bibfnamefont {R.}~\bibnamefont
  {Sunyaev}},\ }\href {\doibase 10.1088/1475-7516/2015/9/026} {\bibfield
  {journal} {\bibinfo  {journal} {JCAP}\ }\textbf {\bibinfo {volume} {09}},\
  \bibinfo {pages} {026} (\bibinfo {year} {2015})},\ \Eprint
  {http://arxiv.org/abs/1507.05615} {arXiv:1507.05615 [astro-ph.CO]}
  \BibitemShut {NoStop}%
\bibitem [{\citenamefont {Fl\"oss}\ \emph {et~al.}(2022)\citenamefont
  {Fl\"oss}, \citenamefont {de~Wild}, \citenamefont {Meerburg},\ and\
  \citenamefont {Koopmans}}]{Floss:2022grj}%
  \BibitemOpen
  \bibfield  {author} {\bibinfo {author} {\bibfnamefont {T.}~\bibnamefont
  {Fl\"oss}}, \bibinfo {author} {\bibfnamefont {T.}~\bibnamefont {de~Wild}},
  \bibinfo {author} {\bibfnamefont {P.~D.}\ \bibnamefont {Meerburg}}, \ and\
  \bibinfo {author} {\bibfnamefont {L.~V.~E.}\ \bibnamefont {Koopmans}},\
  }\href {\doibase 10.1088/1475-7516/2022/06/020} {\bibfield  {journal}
  {\bibinfo  {journal} {JCAP}\ }\textbf {\bibinfo {volume} {06}},\ \bibinfo
  {pages} {020} (\bibinfo {year} {2022})},\ \Eprint
  {http://arxiv.org/abs/2201.08843} {arXiv:2201.08843 [astro-ph.CO]}
  \BibitemShut {NoStop}%
\bibitem [{\citenamefont {Sailer}\ \emph {et~al.}(2021)\citenamefont {Sailer},
  \citenamefont {Castorina}, \citenamefont {Ferraro},\ and\ \citenamefont
  {White}}]{Sailer:2021yzm}%
  \BibitemOpen
  \bibfield  {author} {\bibinfo {author} {\bibfnamefont {N.}~\bibnamefont
  {Sailer}}, \bibinfo {author} {\bibfnamefont {E.}~\bibnamefont {Castorina}},
  \bibinfo {author} {\bibfnamefont {S.}~\bibnamefont {Ferraro}}, \ and\
  \bibinfo {author} {\bibfnamefont {M.}~\bibnamefont {White}},\ }\href
  {\doibase 10.1088/1475-7516/2021/12/049} {\bibfield  {journal} {\bibinfo
  {journal} {JCAP}\ }\textbf {\bibinfo {volume} {12}},\ \bibinfo {pages} {049}
  (\bibinfo {year} {2021})},\ \Eprint {http://arxiv.org/abs/2106.09713}
  {arXiv:2106.09713 [astro-ph.CO]} \BibitemShut {NoStop}%
\bibitem [{\citenamefont {Cabass}\ \emph
  {et~al.}(2022{\natexlab{c}})\citenamefont {Cabass}, \citenamefont {Ivanov},
  \citenamefont {Philcox}, \citenamefont {Simonovic},\ and\ \citenamefont
  {Zaldarriaga}}]{Cabass:2022epm}%
  \BibitemOpen
  \bibfield  {author} {\bibinfo {author} {\bibfnamefont {G.}~\bibnamefont
  {Cabass}}, \bibinfo {author} {\bibfnamefont {M.~M.}\ \bibnamefont {Ivanov}},
  \bibinfo {author} {\bibfnamefont {O.~H.~E.}\ \bibnamefont {Philcox}},
  \bibinfo {author} {\bibfnamefont {M.}~\bibnamefont {Simonovic}}, \ and\
  \bibinfo {author} {\bibfnamefont {M.}~\bibnamefont {Zaldarriaga}},\
  }\href@noop {} {\  (\bibinfo {year} {2022}{\natexlab{c}})},\ \Eprint
  {http://arxiv.org/abs/2211.14899} {arXiv:2211.14899 [astro-ph.CO]}
  \BibitemShut {NoStop}%
\bibitem [{\citenamefont {Bartolo}\ \emph
  {et~al.}(2004{\natexlab{b}})\citenamefont {Bartolo}, \citenamefont
  {Matarrese},\ and\ \citenamefont {Riotto}}]{Bartolo:2003jx}%
  \BibitemOpen
  \bibfield  {author} {\bibinfo {author} {\bibfnamefont {N.}~\bibnamefont
  {Bartolo}}, \bibinfo {author} {\bibfnamefont {S.}~\bibnamefont {Matarrese}},
  \ and\ \bibinfo {author} {\bibfnamefont {A.}~\bibnamefont {Riotto}},\ }\href
  {\doibase 10.1103/PhysRevD.69.043503} {\bibfield  {journal} {\bibinfo
  {journal} {Phys. Rev. D}\ }\textbf {\bibinfo {volume} {69}},\ \bibinfo
  {pages} {043503} (\bibinfo {year} {2004}{\natexlab{b}})},\ \Eprint
  {http://arxiv.org/abs/hep-ph/0309033} {arXiv:hep-ph/0309033} \BibitemShut
  {NoStop}%
\bibitem [{\citenamefont {Sasaki}\ \emph {et~al.}(2006)\citenamefont {Sasaki},
  \citenamefont {Valiviita},\ and\ \citenamefont {Wands}}]{Sasaki:2006kq}%
  \BibitemOpen
  \bibfield  {author} {\bibinfo {author} {\bibfnamefont {M.}~\bibnamefont
  {Sasaki}}, \bibinfo {author} {\bibfnamefont {J.}~\bibnamefont {Valiviita}}, \
  and\ \bibinfo {author} {\bibfnamefont {D.}~\bibnamefont {Wands}},\ }\href
  {\doibase 10.1103/PhysRevD.74.103003} {\bibfield  {journal} {\bibinfo
  {journal} {Phys. Rev. D}\ }\textbf {\bibinfo {volume} {74}},\ \bibinfo
  {pages} {103003} (\bibinfo {year} {2006})},\ \Eprint
  {http://arxiv.org/abs/astro-ph/0607627} {arXiv:astro-ph/0607627} \BibitemShut
  {NoStop}%
\bibitem [{\citenamefont {Byrnes}\ and\ \citenamefont
  {Choi}(2010)}]{Byrnes:2010em}%
  \BibitemOpen
  \bibfield  {author} {\bibinfo {author} {\bibfnamefont {C.~T.}\ \bibnamefont
  {Byrnes}}\ and\ \bibinfo {author} {\bibfnamefont {K.-Y.}\ \bibnamefont
  {Choi}},\ }\href {\doibase 10.1155/2010/724525} {\bibfield  {journal}
  {\bibinfo  {journal} {Adv. Astron.}\ }\textbf {\bibinfo {volume} {2010}},\
  \bibinfo {pages} {724525} (\bibinfo {year} {2010})},\ \Eprint
  {http://arxiv.org/abs/1002.3110} {arXiv:1002.3110 [astro-ph.CO]} \BibitemShut
  {NoStop}%
\bibitem [{\citenamefont {Huang}(2013)}]{Huang:2013yla}%
  \BibitemOpen
  \bibfield  {author} {\bibinfo {author} {\bibfnamefont {Q.-G.}\ \bibnamefont
  {Huang}},\ }\href {\doibase 10.1088/1475-7516/2013/05/030} {\bibfield
  {journal} {\bibinfo  {journal} {JCAP}\ }\textbf {\bibinfo {volume} {05}},\
  \bibinfo {pages} {030} (\bibinfo {year} {2013})},\ \Eprint
  {http://arxiv.org/abs/1303.6084} {arXiv:1303.6084 [astro-ph.CO]} \BibitemShut
  {NoStop}%
\bibitem [{\citenamefont {Lehners}\ and\ \citenamefont
  {Steinhardt}(2013)}]{Lehners:2013cka}%
  \BibitemOpen
  \bibfield  {author} {\bibinfo {author} {\bibfnamefont {J.-L.}\ \bibnamefont
  {Lehners}}\ and\ \bibinfo {author} {\bibfnamefont {P.~J.}\ \bibnamefont
  {Steinhardt}},\ }\href {\doibase 10.1103/PhysRevD.87.123533} {\bibfield
  {journal} {\bibinfo  {journal} {Phys. Rev. D}\ }\textbf {\bibinfo {volume}
  {87}},\ \bibinfo {pages} {123533} (\bibinfo {year} {2013})},\ \Eprint
  {http://arxiv.org/abs/1304.3122} {arXiv:1304.3122 [astro-ph.CO]} \BibitemShut
  {NoStop}%
\bibitem [{\citenamefont {Fertig}\ and\ \citenamefont
  {Lehners}(2016)}]{Fertig:2015ola}%
  \BibitemOpen
  \bibfield  {author} {\bibinfo {author} {\bibfnamefont {A.}~\bibnamefont
  {Fertig}}\ and\ \bibinfo {author} {\bibfnamefont {J.-L.}\ \bibnamefont
  {Lehners}},\ }\href {\doibase 10.1088/1475-7516/2016/01/026} {\bibfield
  {journal} {\bibinfo  {journal} {JCAP}\ }\textbf {\bibinfo {volume} {01}},\
  \bibinfo {pages} {026} (\bibinfo {year} {2016})},\ \Eprint
  {http://arxiv.org/abs/1510.03439} {arXiv:1510.03439 [hep-th]} \BibitemShut
  {NoStop}%
\bibitem [{\citenamefont {Suyama}\ \emph {et~al.}(2013)\citenamefont {Suyama},
  \citenamefont {Takahashi}, \citenamefont {Yamaguchi},\ and\ \citenamefont
  {Yokoyama}}]{Suyama:2013nva}%
  \BibitemOpen
  \bibfield  {author} {\bibinfo {author} {\bibfnamefont {T.}~\bibnamefont
  {Suyama}}, \bibinfo {author} {\bibfnamefont {T.}~\bibnamefont {Takahashi}},
  \bibinfo {author} {\bibfnamefont {M.}~\bibnamefont {Yamaguchi}}, \ and\
  \bibinfo {author} {\bibfnamefont {S.}~\bibnamefont {Yokoyama}},\ }\href
  {\doibase 10.1088/1475-7516/2013/06/012} {\bibfield  {journal} {\bibinfo
  {journal} {JCAP}\ }\textbf {\bibinfo {volume} {06}},\ \bibinfo {pages} {012}
  (\bibinfo {year} {2013})},\ \Eprint {http://arxiv.org/abs/1303.5374}
  {arXiv:1303.5374 [astro-ph.CO]} \BibitemShut {NoStop}%
\bibitem [{\citenamefont {Dimopoulos}(2006)}]{Dimopoulos:2006ms}%
  \BibitemOpen
  \bibfield  {author} {\bibinfo {author} {\bibfnamefont {K.}~\bibnamefont
  {Dimopoulos}},\ }\href {\doibase 10.1103/PhysRevD.74.083502} {\bibfield
  {journal} {\bibinfo  {journal} {Phys. Rev. D}\ }\textbf {\bibinfo {volume}
  {74}},\ \bibinfo {pages} {083502} (\bibinfo {year} {2006})},\ \Eprint
  {http://arxiv.org/abs/hep-ph/0607229} {arXiv:hep-ph/0607229} \BibitemShut
  {NoStop}%
\bibitem [{\citenamefont {Valenzuela-Toledo}\ and\ \citenamefont
  {Rodriguez}(2010)}]{Valenzuela-Toledo:2009bzd}%
  \BibitemOpen
  \bibfield  {author} {\bibinfo {author} {\bibfnamefont {C.~A.}\ \bibnamefont
  {Valenzuela-Toledo}}\ and\ \bibinfo {author} {\bibfnamefont {Y.}~\bibnamefont
  {Rodriguez}},\ }\href {\doibase 10.1016/j.physletb.2010.01.060} {\bibfield
  {journal} {\bibinfo  {journal} {Phys. Lett. B}\ }\textbf {\bibinfo {volume}
  {685}},\ \bibinfo {pages} {120} (\bibinfo {year} {2010})},\ \Eprint
  {http://arxiv.org/abs/0910.4208} {arXiv:0910.4208 [astro-ph.CO]} \BibitemShut
  {NoStop}%
\bibitem [{\citenamefont {Akrami}\ \emph
  {et~al.}(2020{\natexlab{d}})\citenamefont {Akrami} \emph
  {et~al.}}]{Planck:2019evm}%
  \BibitemOpen
  \bibfield  {author} {\bibinfo {author} {\bibfnamefont {Y.}~\bibnamefont
  {Akrami}} \emph {et~al.} (\bibinfo {collaboration} {Planck}),\ }\href
  {\doibase 10.1051/0004-6361/201935201} {\bibfield  {journal} {\bibinfo
  {journal} {Astron. Astrophys.}\ }\textbf {\bibinfo {volume} {641}},\ \bibinfo
  {pages} {A7} (\bibinfo {year} {2020}{\natexlab{d}})},\ \Eprint
  {http://arxiv.org/abs/1906.02552} {arXiv:1906.02552 [astro-ph.CO]}
  \BibitemShut {NoStop}%
\bibitem [{\citenamefont {Smith}\ \emph {et~al.}(2011)\citenamefont {Smith},
  \citenamefont {LoVerde},\ and\ \citenamefont {Zaldarriaga}}]{Smith:2011if}%
  \BibitemOpen
  \bibfield  {author} {\bibinfo {author} {\bibfnamefont {K.~M.}\ \bibnamefont
  {Smith}}, \bibinfo {author} {\bibfnamefont {M.}~\bibnamefont {LoVerde}}, \
  and\ \bibinfo {author} {\bibfnamefont {M.}~\bibnamefont {Zaldarriaga}},\
  }\href {\doibase 10.1103/PhysRevLett.107.191301} {\bibfield  {journal}
  {\bibinfo  {journal} {Phys. Rev. Lett.}\ }\textbf {\bibinfo {volume} {107}},\
  \bibinfo {pages} {191301} (\bibinfo {year} {2011})},\ \Eprint
  {http://arxiv.org/abs/1108.1805} {arXiv:1108.1805 [astro-ph.CO]} \BibitemShut
  {NoStop}%
\bibitem [{\citenamefont {Izumi}\ \emph {et~al.}(2012)\citenamefont {Izumi},
  \citenamefont {Mizuno},\ and\ \citenamefont {Koyama}}]{Izumi:2011di}%
  \BibitemOpen
  \bibfield  {author} {\bibinfo {author} {\bibfnamefont {K.}~\bibnamefont
  {Izumi}}, \bibinfo {author} {\bibfnamefont {S.}~\bibnamefont {Mizuno}}, \
  and\ \bibinfo {author} {\bibfnamefont {K.}~\bibnamefont {Koyama}},\ }\href
  {\doibase 10.1103/PhysRevD.85.023521} {\bibfield  {journal} {\bibinfo
  {journal} {Phys. Rev. D}\ }\textbf {\bibinfo {volume} {85}},\ \bibinfo
  {pages} {023521} (\bibinfo {year} {2012})},\ \Eprint
  {http://arxiv.org/abs/1109.3746} {arXiv:1109.3746 [astro-ph.CO]} \BibitemShut
  {NoStop}%
\bibitem [{\citenamefont {Mizuno}\ and\ \citenamefont
  {Yokoyama}(2015)}]{Mizuno:2015qma}%
  \BibitemOpen
  \bibfield  {author} {\bibinfo {author} {\bibfnamefont {S.}~\bibnamefont
  {Mizuno}}\ and\ \bibinfo {author} {\bibfnamefont {S.}~\bibnamefont
  {Yokoyama}},\ }\href {\doibase 10.1103/PhysRevD.91.123521} {\bibfield
  {journal} {\bibinfo  {journal} {Phys. Rev. D}\ }\textbf {\bibinfo {volume}
  {91}},\ \bibinfo {pages} {123521} (\bibinfo {year} {2015})},\ \Eprint
  {http://arxiv.org/abs/1504.05505} {arXiv:1504.05505 [astro-ph.CO]}
  \BibitemShut {NoStop}%
\bibitem [{\citenamefont {Cabass}\ \emph
  {et~al.}(2022{\natexlab{d}})\citenamefont {Cabass}, \citenamefont {Ivanov},
  \citenamefont {Philcox}, \citenamefont {Simonovi\'c},\ and\ \citenamefont
  {Zaldarriaga}}]{Cabass:2022wjy}%
  \BibitemOpen
  \bibfield  {author} {\bibinfo {author} {\bibfnamefont {G.}~\bibnamefont
  {Cabass}}, \bibinfo {author} {\bibfnamefont {M.~M.}\ \bibnamefont {Ivanov}},
  \bibinfo {author} {\bibfnamefont {O.~H.~E.}\ \bibnamefont {Philcox}},
  \bibinfo {author} {\bibfnamefont {M.}~\bibnamefont {Simonovi\'c}}, \ and\
  \bibinfo {author} {\bibfnamefont {M.}~\bibnamefont {Zaldarriaga}},\ }\href
  {\doibase 10.1103/PhysRevLett.129.021301} {\bibfield  {journal} {\bibinfo
  {journal} {Phys. Rev. Lett.}\ }\textbf {\bibinfo {volume} {129}},\ \bibinfo
  {pages} {021301} (\bibinfo {year} {2022}{\natexlab{d}})},\ \Eprint
  {http://arxiv.org/abs/2201.07238} {arXiv:2201.07238 [astro-ph.CO]}
  \BibitemShut {NoStop}%
\bibitem [{\citenamefont {Chen}\ \emph {et~al.}(2013)\citenamefont {Chen},
  \citenamefont {Firouzjahi}, \citenamefont {Namjoo},\ and\ \citenamefont
  {Sasaki}}]{Chen:2013aj}%
  \BibitemOpen
  \bibfield  {author} {\bibinfo {author} {\bibfnamefont {X.}~\bibnamefont
  {Chen}}, \bibinfo {author} {\bibfnamefont {H.}~\bibnamefont {Firouzjahi}},
  \bibinfo {author} {\bibfnamefont {M.~H.}\ \bibnamefont {Namjoo}}, \ and\
  \bibinfo {author} {\bibfnamefont {M.}~\bibnamefont {Sasaki}},\ }\href
  {\doibase 10.1209/0295-5075/102/59001} {\bibfield  {journal} {\bibinfo
  {journal} {EPL}\ }\textbf {\bibinfo {volume} {102}},\ \bibinfo {pages}
  {59001} (\bibinfo {year} {2013})},\ \Eprint {http://arxiv.org/abs/1301.5699}
  {arXiv:1301.5699 [hep-th]} \BibitemShut {NoStop}%
\bibitem [{\citenamefont {Alishahiha}\ \emph {et~al.}(2004)\citenamefont
  {Alishahiha}, \citenamefont {Silverstein},\ and\ \citenamefont
  {Tong}}]{Alishahiha:2004eh}%
  \BibitemOpen
  \bibfield  {author} {\bibinfo {author} {\bibfnamefont {M.}~\bibnamefont
  {Alishahiha}}, \bibinfo {author} {\bibfnamefont {E.}~\bibnamefont
  {Silverstein}}, \ and\ \bibinfo {author} {\bibfnamefont {D.}~\bibnamefont
  {Tong}},\ }\href {\doibase 10.1103/PhysRevD.70.123505} {\bibfield  {journal}
  {\bibinfo  {journal} {Phys. Rev. D}\ }\textbf {\bibinfo {volume} {70}},\
  \bibinfo {pages} {123505} (\bibinfo {year} {2004})},\ \Eprint
  {http://arxiv.org/abs/hep-th/0404084} {arXiv:hep-th/0404084} \BibitemShut
  {NoStop}%
\bibitem [{\citenamefont {Silverstein}\ and\ \citenamefont
  {Tong}(2004)}]{Silverstein:2003hf}%
  \BibitemOpen
  \bibfield  {author} {\bibinfo {author} {\bibfnamefont {E.}~\bibnamefont
  {Silverstein}}\ and\ \bibinfo {author} {\bibfnamefont {D.}~\bibnamefont
  {Tong}},\ }\href {\doibase 10.1103/PhysRevD.70.103505} {\bibfield  {journal}
  {\bibinfo  {journal} {Phys. Rev. D}\ }\textbf {\bibinfo {volume} {70}},\
  \bibinfo {pages} {103505} (\bibinfo {year} {2004})},\ \Eprint
  {http://arxiv.org/abs/hep-th/0310221} {arXiv:hep-th/0310221} \BibitemShut
  {NoStop}%
\bibitem [{\citenamefont {Arkani-Hamed}\ \emph {et~al.}(2004)\citenamefont
  {Arkani-Hamed}, \citenamefont {Creminelli}, \citenamefont {Mukohyama},\ and\
  \citenamefont {Zaldarriaga}}]{Arkani-Hamed:2003juy}%
  \BibitemOpen
  \bibfield  {author} {\bibinfo {author} {\bibfnamefont {N.}~\bibnamefont
  {Arkani-Hamed}}, \bibinfo {author} {\bibfnamefont {P.}~\bibnamefont
  {Creminelli}}, \bibinfo {author} {\bibfnamefont {S.}~\bibnamefont
  {Mukohyama}}, \ and\ \bibinfo {author} {\bibfnamefont {M.}~\bibnamefont
  {Zaldarriaga}},\ }\href {\doibase 10.1088/1475-7516/2004/04/001} {\bibfield
  {journal} {\bibinfo  {journal} {JCAP}\ }\textbf {\bibinfo {volume} {04}},\
  \bibinfo {pages} {001} (\bibinfo {year} {2004})},\ \Eprint
  {http://arxiv.org/abs/hep-th/0312100} {arXiv:hep-th/0312100} \BibitemShut
  {NoStop}%
\bibitem [{\citenamefont {Cabass}\ \emph {et~al.}(2023)\citenamefont {Cabass},
  \citenamefont {Ivanov},\ and\ \citenamefont {Philcox}}]{Cabass:2022oap}%
  \BibitemOpen
  \bibfield  {author} {\bibinfo {author} {\bibfnamefont {G.}~\bibnamefont
  {Cabass}}, \bibinfo {author} {\bibfnamefont {M.~M.}\ \bibnamefont {Ivanov}},
  \ and\ \bibinfo {author} {\bibfnamefont {O.~H.~E.}\ \bibnamefont {Philcox}},\
  }\href {\doibase 10.1103/PhysRevD.107.023523} {\bibfield  {journal} {\bibinfo
   {journal} {Phys. Rev. D}\ }\textbf {\bibinfo {volume} {107}},\ \bibinfo
  {pages} {023523} (\bibinfo {year} {2023})},\ \Eprint
  {http://arxiv.org/abs/2210.16320} {arXiv:2210.16320 [astro-ph.CO]}
  \BibitemShut {NoStop}%
\bibitem [{\citenamefont {Philcox}(2022)}]{Philcox:2022hkh}%
  \BibitemOpen
  \bibfield  {author} {\bibinfo {author} {\bibfnamefont {O.~H.~E.}\
  \bibnamefont {Philcox}},\ }\href {\doibase 10.1103/PhysRevD.106.063501}
  {\bibfield  {journal} {\bibinfo  {journal} {Phys. Rev. D}\ }\textbf {\bibinfo
  {volume} {106}},\ \bibinfo {pages} {063501} (\bibinfo {year} {2022})},\
  \Eprint {http://arxiv.org/abs/2206.04227} {arXiv:2206.04227 [astro-ph.CO]}
  \BibitemShut {NoStop}%
\bibitem [{\citenamefont {Shiraishi}\ \emph {et~al.}(2016)\citenamefont
  {Shiraishi}, \citenamefont {Bartolo},\ and\ \citenamefont
  {Liguori}}]{Shiraishi:2016hjd}%
  \BibitemOpen
  \bibfield  {author} {\bibinfo {author} {\bibfnamefont {M.}~\bibnamefont
  {Shiraishi}}, \bibinfo {author} {\bibfnamefont {N.}~\bibnamefont {Bartolo}},
  \ and\ \bibinfo {author} {\bibfnamefont {M.}~\bibnamefont {Liguori}},\ }\href
  {\doibase 10.1088/1475-7516/2016/10/015} {\bibfield  {journal} {\bibinfo
  {journal} {JCAP}\ }\textbf {\bibinfo {volume} {10}},\ \bibinfo {pages} {015}
  (\bibinfo {year} {2016})},\ \Eprint {http://arxiv.org/abs/1607.01363}
  {arXiv:1607.01363 [astro-ph.CO]} \BibitemShut {NoStop}%
\bibitem [{\citenamefont {Bartolo}\ \emph
  {et~al.}(2015{\natexlab{a}})\citenamefont {Bartolo}, \citenamefont
  {Matarrese}, \citenamefont {Peloso},\ and\ \citenamefont
  {Shiraishi}}]{Bartolo:2015dga}%
  \BibitemOpen
  \bibfield  {author} {\bibinfo {author} {\bibfnamefont {N.}~\bibnamefont
  {Bartolo}}, \bibinfo {author} {\bibfnamefont {S.}~\bibnamefont {Matarrese}},
  \bibinfo {author} {\bibfnamefont {M.}~\bibnamefont {Peloso}}, \ and\ \bibinfo
  {author} {\bibfnamefont {M.}~\bibnamefont {Shiraishi}},\ }\href {\doibase
  10.1088/1475-7516/2015/07/039} {\bibfield  {journal} {\bibinfo  {journal}
  {JCAP}\ }\textbf {\bibinfo {volume} {07}},\ \bibinfo {pages} {039} (\bibinfo
  {year} {2015}{\natexlab{a}})},\ \Eprint {http://arxiv.org/abs/1505.02193}
  {arXiv:1505.02193 [astro-ph.CO]} \BibitemShut {NoStop}%
\bibitem [{\citenamefont {Bartolo}\ \emph
  {et~al.}(2015{\natexlab{b}})\citenamefont {Bartolo}, \citenamefont
  {Matarrese}, \citenamefont {Peloso},\ and\ \citenamefont
  {Shiraishi}}]{Bartolo:2014hwa}%
  \BibitemOpen
  \bibfield  {author} {\bibinfo {author} {\bibfnamefont {N.}~\bibnamefont
  {Bartolo}}, \bibinfo {author} {\bibfnamefont {S.}~\bibnamefont {Matarrese}},
  \bibinfo {author} {\bibfnamefont {M.}~\bibnamefont {Peloso}}, \ and\ \bibinfo
  {author} {\bibfnamefont {M.}~\bibnamefont {Shiraishi}},\ }\href {\doibase
  10.1088/1475-7516/2015/01/027} {\bibfield  {journal} {\bibinfo  {journal}
  {JCAP}\ }\textbf {\bibinfo {volume} {01}},\ \bibinfo {pages} {027} (\bibinfo
  {year} {2015}{\natexlab{b}})},\ \Eprint {http://arxiv.org/abs/1411.2521}
  {arXiv:1411.2521 [astro-ph.CO]} \BibitemShut {NoStop}%
\bibitem [{\citenamefont {Shiraishi}\ \emph {et~al.}(2013)\citenamefont
  {Shiraishi}, \citenamefont {Komatsu}, \citenamefont {Peloso},\ and\
  \citenamefont {Barnaby}}]{Shiraishi:2013vja}%
  \BibitemOpen
  \bibfield  {author} {\bibinfo {author} {\bibfnamefont {M.}~\bibnamefont
  {Shiraishi}}, \bibinfo {author} {\bibfnamefont {E.}~\bibnamefont {Komatsu}},
  \bibinfo {author} {\bibfnamefont {M.}~\bibnamefont {Peloso}}, \ and\ \bibinfo
  {author} {\bibfnamefont {N.}~\bibnamefont {Barnaby}},\ }\href {\doibase
  10.1088/1475-7516/2013/05/002} {\bibfield  {journal} {\bibinfo  {journal}
  {JCAP}\ }\textbf {\bibinfo {volume} {05}},\ \bibinfo {pages} {002} (\bibinfo
  {year} {2013})},\ \Eprint {http://arxiv.org/abs/1302.3056} {arXiv:1302.3056
  [astro-ph.CO]} \BibitemShut {NoStop}%
\bibitem [{\citenamefont {Naruko}\ \emph {et~al.}(2015)\citenamefont {Naruko},
  \citenamefont {Komatsu},\ and\ \citenamefont {Yamaguchi}}]{Naruko:2014bxa}%
  \BibitemOpen
  \bibfield  {author} {\bibinfo {author} {\bibfnamefont {A.}~\bibnamefont
  {Naruko}}, \bibinfo {author} {\bibfnamefont {E.}~\bibnamefont {Komatsu}}, \
  and\ \bibinfo {author} {\bibfnamefont {M.}~\bibnamefont {Yamaguchi}},\ }\href
  {\doibase 10.1088/1475-7516/2015/04/045} {\bibfield  {journal} {\bibinfo
  {journal} {JCAP}\ }\textbf {\bibinfo {volume} {04}},\ \bibinfo {pages} {045}
  (\bibinfo {year} {2015})},\ \Eprint {http://arxiv.org/abs/1411.5489}
  {arXiv:1411.5489 [astro-ph.CO]} \BibitemShut {NoStop}%
\bibitem [{\citenamefont {Bartolo}\ \emph
  {et~al.}(2013{\natexlab{b}})\citenamefont {Bartolo}, \citenamefont
  {Matarrese}, \citenamefont {Peloso},\ and\ \citenamefont
  {Ricciardone}}]{Bartolo:2012sd}%
  \BibitemOpen
  \bibfield  {author} {\bibinfo {author} {\bibfnamefont {N.}~\bibnamefont
  {Bartolo}}, \bibinfo {author} {\bibfnamefont {S.}~\bibnamefont {Matarrese}},
  \bibinfo {author} {\bibfnamefont {M.}~\bibnamefont {Peloso}}, \ and\ \bibinfo
  {author} {\bibfnamefont {A.}~\bibnamefont {Ricciardone}},\ }\href {\doibase
  10.1103/PhysRevD.87.023504} {\bibfield  {journal} {\bibinfo  {journal} {Phys.
  Rev. D}\ }\textbf {\bibinfo {volume} {87}},\ \bibinfo {pages} {023504}
  (\bibinfo {year} {2013}{\natexlab{b}})},\ \Eprint
  {http://arxiv.org/abs/1210.3257} {arXiv:1210.3257 [astro-ph.CO]} \BibitemShut
  {NoStop}%
\bibitem [{\citenamefont {Dimastrogiovanni}\ \emph {et~al.}(2010)\citenamefont
  {Dimastrogiovanni}, \citenamefont {Bartolo}, \citenamefont {Matarrese},\ and\
  \citenamefont {Riotto}}]{Dimastrogiovanni:2010sm}%
  \BibitemOpen
  \bibfield  {author} {\bibinfo {author} {\bibfnamefont {E.}~\bibnamefont
  {Dimastrogiovanni}}, \bibinfo {author} {\bibfnamefont {N.}~\bibnamefont
  {Bartolo}}, \bibinfo {author} {\bibfnamefont {S.}~\bibnamefont {Matarrese}},
  \ and\ \bibinfo {author} {\bibfnamefont {A.}~\bibnamefont {Riotto}},\ }\href
  {\doibase 10.1155/2010/752670} {\bibfield  {journal} {\bibinfo  {journal}
  {Adv. Astron.}\ }\textbf {\bibinfo {volume} {2010}},\ \bibinfo {pages}
  {752670} (\bibinfo {year} {2010})},\ \Eprint {http://arxiv.org/abs/1001.4049}
  {arXiv:1001.4049 [astro-ph.CO]} \BibitemShut {NoStop}%
\bibitem [{\citenamefont {Bartolo}\ and\ \citenamefont
  {Orlando}(2017)}]{Bartolo:2017szm}%
  \BibitemOpen
  \bibfield  {author} {\bibinfo {author} {\bibfnamefont {N.}~\bibnamefont
  {Bartolo}}\ and\ \bibinfo {author} {\bibfnamefont {G.}~\bibnamefont
  {Orlando}},\ }\href {\doibase 10.1088/1475-7516/2017/07/034} {\bibfield
  {journal} {\bibinfo  {journal} {JCAP}\ }\textbf {\bibinfo {volume} {07}},\
  \bibinfo {pages} {034} (\bibinfo {year} {2017})},\ \Eprint
  {http://arxiv.org/abs/1706.04627} {arXiv:1706.04627 [astro-ph.CO]}
  \BibitemShut {NoStop}%
\bibitem [{\citenamefont {Bartolo}\ \emph {et~al.}(2019)\citenamefont
  {Bartolo}, \citenamefont {Orlando},\ and\ \citenamefont
  {Shiraishi}}]{Bartolo:2018elp}%
  \BibitemOpen
  \bibfield  {author} {\bibinfo {author} {\bibfnamefont {N.}~\bibnamefont
  {Bartolo}}, \bibinfo {author} {\bibfnamefont {G.}~\bibnamefont {Orlando}}, \
  and\ \bibinfo {author} {\bibfnamefont {M.}~\bibnamefont {Shiraishi}},\ }\href
  {\doibase 10.1088/1475-7516/2019/01/050} {\bibfield  {journal} {\bibinfo
  {journal} {JCAP}\ }\textbf {\bibinfo {volume} {01}},\ \bibinfo {pages} {050}
  (\bibinfo {year} {2019})},\ \Eprint {http://arxiv.org/abs/1809.11170}
  {arXiv:1809.11170 [astro-ph.CO]} \BibitemShut {NoStop}%
\bibitem [{\citenamefont {Salvarese}(2022)}]{Salvarese}%
  \BibitemOpen
  \bibfield  {author} {\bibinfo {author} {\bibfnamefont {A.}~\bibnamefont
  {Salvarese}},\ }\emph {\bibinfo {title} {Probing parity violation in the
  Early Universe}},\ \href@noop {} {Master's thesis},\ \bibinfo  {school}
  {University of Padova} (\bibinfo {year} {2022})\BibitemShut {NoStop}%
\bibitem [{\citenamefont {Akrami}\ \emph
  {et~al.}(2020{\natexlab{e}})\citenamefont {Akrami} \emph
  {et~al.}}]{Planck:2018jri}%
  \BibitemOpen
  \bibfield  {author} {\bibinfo {author} {\bibfnamefont {Y.}~\bibnamefont
  {Akrami}} \emph {et~al.} (\bibinfo {collaboration} {Planck}),\ }\href
  {\doibase 10.1051/0004-6361/201833887} {\bibfield  {journal} {\bibinfo
  {journal} {Astron. Astrophys.}\ }\textbf {\bibinfo {volume} {641}},\ \bibinfo
  {pages} {A10} (\bibinfo {year} {2020}{\natexlab{e}})},\ \Eprint
  {http://arxiv.org/abs/1807.06211} {arXiv:1807.06211 [astro-ph.CO]}
  \BibitemShut {NoStop}%
\bibitem [{\citenamefont {Endlich}\ \emph {et~al.}(2013)\citenamefont
  {Endlich}, \citenamefont {Nicolis},\ and\ \citenamefont
  {Wang}}]{Endlich:2012pz}%
  \BibitemOpen
  \bibfield  {author} {\bibinfo {author} {\bibfnamefont {S.}~\bibnamefont
  {Endlich}}, \bibinfo {author} {\bibfnamefont {A.}~\bibnamefont {Nicolis}}, \
  and\ \bibinfo {author} {\bibfnamefont {J.}~\bibnamefont {Wang}},\ }\href
  {\doibase 10.1088/1475-7516/2013/10/011} {\bibfield  {journal} {\bibinfo
  {journal} {JCAP}\ }\textbf {\bibinfo {volume} {10}},\ \bibinfo {pages} {011}
  (\bibinfo {year} {2013})},\ \Eprint {http://arxiv.org/abs/1210.0569}
  {arXiv:1210.0569 [hep-th]} \BibitemShut {NoStop}%
\bibitem [{\citenamefont {Endlich}\ \emph {et~al.}(2014)\citenamefont
  {Endlich}, \citenamefont {Horn}, \citenamefont {Nicolis},\ and\ \citenamefont
  {Wang}}]{Endlich:2013jia}%
  \BibitemOpen
  \bibfield  {author} {\bibinfo {author} {\bibfnamefont {S.}~\bibnamefont
  {Endlich}}, \bibinfo {author} {\bibfnamefont {B.}~\bibnamefont {Horn}},
  \bibinfo {author} {\bibfnamefont {A.}~\bibnamefont {Nicolis}}, \ and\
  \bibinfo {author} {\bibfnamefont {J.}~\bibnamefont {Wang}},\ }\href {\doibase
  10.1103/PhysRevD.90.063506} {\bibfield  {journal} {\bibinfo  {journal} {Phys.
  Rev. D}\ }\textbf {\bibinfo {volume} {90}},\ \bibinfo {pages} {063506}
  (\bibinfo {year} {2014})},\ \Eprint {http://arxiv.org/abs/1307.8114}
  {arXiv:1307.8114 [hep-th]} \BibitemShut {NoStop}%
\bibitem [{\citenamefont {Gruzinov}(2004)}]{Gruzinov:2004ty}%
  \BibitemOpen
  \bibfield  {author} {\bibinfo {author} {\bibfnamefont {A.}~\bibnamefont
  {Gruzinov}},\ }\href {\doibase 10.1103/PhysRevD.70.063518} {\bibfield
  {journal} {\bibinfo  {journal} {Phys. Rev. D}\ }\textbf {\bibinfo {volume}
  {70}},\ \bibinfo {pages} {063518} (\bibinfo {year} {2004})},\ \Eprint
  {http://arxiv.org/abs/astro-ph/0404548} {arXiv:astro-ph/0404548} \BibitemShut
  {NoStop}%
\bibitem [{\citenamefont {Bartolo}\ \emph {et~al.}(2014)\citenamefont
  {Bartolo}, \citenamefont {Peloso}, \citenamefont {Ricciardone},\ and\
  \citenamefont {Unal}}]{Bartolo:2014xfa}%
  \BibitemOpen
  \bibfield  {author} {\bibinfo {author} {\bibfnamefont {N.}~\bibnamefont
  {Bartolo}}, \bibinfo {author} {\bibfnamefont {M.}~\bibnamefont {Peloso}},
  \bibinfo {author} {\bibfnamefont {A.}~\bibnamefont {Ricciardone}}, \ and\
  \bibinfo {author} {\bibfnamefont {C.}~\bibnamefont {Unal}},\ }\href {\doibase
  10.1088/1475-7516/2014/11/009} {\bibfield  {journal} {\bibinfo  {journal}
  {JCAP}\ }\textbf {\bibinfo {volume} {11}},\ \bibinfo {pages} {009} (\bibinfo
  {year} {2014})},\ \Eprint {http://arxiv.org/abs/1407.8053} {arXiv:1407.8053
  [astro-ph.CO]} \BibitemShut {NoStop}%
\bibitem [{\citenamefont {Shaw}\ and\ \citenamefont
  {Lewis}(2010)}]{Shaw:2009nf}%
  \BibitemOpen
  \bibfield  {author} {\bibinfo {author} {\bibfnamefont {J.~R.}\ \bibnamefont
  {Shaw}}\ and\ \bibinfo {author} {\bibfnamefont {A.}~\bibnamefont {Lewis}},\
  }\href {\doibase 10.1103/PhysRevD.81.043517} {\bibfield  {journal} {\bibinfo
  {journal} {Phys. Rev. D}\ }\textbf {\bibinfo {volume} {81}},\ \bibinfo
  {pages} {043517} (\bibinfo {year} {2010})},\ \Eprint
  {http://arxiv.org/abs/0911.2714} {arXiv:0911.2714 [astro-ph.CO]} \BibitemShut
  {NoStop}%
\bibitem [{\citenamefont {Shiraishi}(2012)}]{Shiraishi:2012sn}%
  \BibitemOpen
  \bibfield  {author} {\bibinfo {author} {\bibfnamefont {M.}~\bibnamefont
  {Shiraishi}},\ }\href {\doibase 10.1088/1475-7516/2012/06/015} {\bibfield
  {journal} {\bibinfo  {journal} {JCAP}\ }\textbf {\bibinfo {volume} {06}},\
  \bibinfo {pages} {015} (\bibinfo {year} {2012})},\ \Eprint
  {http://arxiv.org/abs/1202.2847} {arXiv:1202.2847 [astro-ph.CO]} \BibitemShut
  {NoStop}%
\bibitem [{\citenamefont {Shiraishi}(2013)}]{Shiraishi:2013vha}%
  \BibitemOpen
  \bibfield  {author} {\bibinfo {author} {\bibfnamefont {M.}~\bibnamefont
  {Shiraishi}},\ }\href {\doibase 10.1088/1475-7516/2013/11/006} {\bibfield
  {journal} {\bibinfo  {journal} {JCAP}\ }\textbf {\bibinfo {volume} {11}},\
  \bibinfo {pages} {006} (\bibinfo {year} {2013})},\ \Eprint
  {http://arxiv.org/abs/1308.2531} {arXiv:1308.2531 [astro-ph.CO]} \BibitemShut
  {NoStop}%
\bibitem [{\citenamefont {Ade}\ \emph {et~al.}(2016{\natexlab{c}})\citenamefont
  {Ade} \emph {et~al.}}]{Planck:2015zrl}%
  \BibitemOpen
  \bibfield  {author} {\bibinfo {author} {\bibfnamefont {P.~A.~R.}\
  \bibnamefont {Ade}} \emph {et~al.} (\bibinfo {collaboration} {Planck}),\
  }\href {\doibase 10.1051/0004-6361/201525821} {\bibfield  {journal} {\bibinfo
   {journal} {Astron. Astrophys.}\ }\textbf {\bibinfo {volume} {594}},\
  \bibinfo {pages} {A19} (\bibinfo {year} {2016}{\natexlab{c}})},\ \Eprint
  {http://arxiv.org/abs/1502.01594} {arXiv:1502.01594 [astro-ph.CO]}
  \BibitemShut {NoStop}%
\bibitem [{\citenamefont {Shiraishi}\ \emph {et~al.}(2012)\citenamefont
  {Shiraishi}, \citenamefont {Nitta}, \citenamefont {Yokoyama},\ and\
  \citenamefont {Ichiki}}]{Shiraishi:2012rm}%
  \BibitemOpen
  \bibfield  {author} {\bibinfo {author} {\bibfnamefont {M.}~\bibnamefont
  {Shiraishi}}, \bibinfo {author} {\bibfnamefont {D.}~\bibnamefont {Nitta}},
  \bibinfo {author} {\bibfnamefont {S.}~\bibnamefont {Yokoyama}}, \ and\
  \bibinfo {author} {\bibfnamefont {K.}~\bibnamefont {Ichiki}},\ }\href
  {\doibase 10.1088/1475-7516/2012/03/041} {\bibfield  {journal} {\bibinfo
  {journal} {JCAP}\ }\textbf {\bibinfo {volume} {03}},\ \bibinfo {pages} {041}
  (\bibinfo {year} {2012})},\ \Eprint {http://arxiv.org/abs/1201.0376}
  {arXiv:1201.0376 [astro-ph.CO]} \BibitemShut {NoStop}%
\bibitem [{\citenamefont {Trivedi}\ \emph {et~al.}(2012)\citenamefont
  {Trivedi}, \citenamefont {Seshadri},\ and\ \citenamefont
  {Subramanian}}]{Trivedi:2011vt}%
  \BibitemOpen
  \bibfield  {author} {\bibinfo {author} {\bibfnamefont {P.}~\bibnamefont
  {Trivedi}}, \bibinfo {author} {\bibfnamefont {T.~R.}\ \bibnamefont
  {Seshadri}}, \ and\ \bibinfo {author} {\bibfnamefont {K.}~\bibnamefont
  {Subramanian}},\ }\href {\doibase 10.1103/PhysRevLett.108.231301} {\bibfield
  {journal} {\bibinfo  {journal} {Phys. Rev. Lett.}\ }\textbf {\bibinfo
  {volume} {108}},\ \bibinfo {pages} {231301} (\bibinfo {year} {2012})},\
  \Eprint {http://arxiv.org/abs/1111.0744} {arXiv:1111.0744 [astro-ph.CO]}
  \BibitemShut {NoStop}%
\bibitem [{\citenamefont {Trivedi}\ \emph {et~al.}(2014)\citenamefont
  {Trivedi}, \citenamefont {Subramanian},\ and\ \citenamefont
  {Seshadri}}]{Trivedi:2013wqa}%
  \BibitemOpen
  \bibfield  {author} {\bibinfo {author} {\bibfnamefont {P.}~\bibnamefont
  {Trivedi}}, \bibinfo {author} {\bibfnamefont {K.}~\bibnamefont
  {Subramanian}}, \ and\ \bibinfo {author} {\bibfnamefont {T.~R.}\ \bibnamefont
  {Seshadri}},\ }\href {\doibase 10.1103/PhysRevD.89.043523} {\bibfield
  {journal} {\bibinfo  {journal} {Phys. Rev. D}\ }\textbf {\bibinfo {volume}
  {89}},\ \bibinfo {pages} {043523} (\bibinfo {year} {2014})},\ \Eprint
  {http://arxiv.org/abs/1312.5308} {arXiv:1312.5308 [astro-ph.CO]} \BibitemShut
  {NoStop}%
\bibitem [{\citenamefont {Pimentel}\ and\ \citenamefont
  {Wang}(2022)}]{Pimentel:2022fsc}%
  \BibitemOpen
  \bibfield  {author} {\bibinfo {author} {\bibfnamefont {G.~L.}\ \bibnamefont
  {Pimentel}}\ and\ \bibinfo {author} {\bibfnamefont {D.-G.}\ \bibnamefont
  {Wang}},\ }\href {\doibase 10.1007/JHEP10(2022)177} {\bibfield  {journal}
  {\bibinfo  {journal} {JHEP}\ }\textbf {\bibinfo {volume} {10}},\ \bibinfo
  {pages} {177} (\bibinfo {year} {2022})},\ \Eprint
  {http://arxiv.org/abs/2205.00013} {arXiv:2205.00013 [hep-th]} \BibitemShut
  {NoStop}%
\bibitem [{\citenamefont {Kumar}\ and\ \citenamefont
  {Sundrum}(2020)}]{Kumar:2019ebj}%
  \BibitemOpen
  \bibfield  {author} {\bibinfo {author} {\bibfnamefont {S.}~\bibnamefont
  {Kumar}}\ and\ \bibinfo {author} {\bibfnamefont {R.}~\bibnamefont
  {Sundrum}},\ }\href {\doibase 10.1007/JHEP04(2020)077} {\bibfield  {journal}
  {\bibinfo  {journal} {JHEP}\ }\textbf {\bibinfo {volume} {04}},\ \bibinfo
  {pages} {077} (\bibinfo {year} {2020})},\ \Eprint
  {http://arxiv.org/abs/1908.11378} {arXiv:1908.11378 [hep-ph]} \BibitemShut
  {NoStop}%
\bibitem [{\citenamefont {Reece}\ \emph {et~al.}(2023)\citenamefont {Reece},
  \citenamefont {Wang},\ and\ \citenamefont {Xianyu}}]{Reece:2022soh}%
  \BibitemOpen
  \bibfield  {author} {\bibinfo {author} {\bibfnamefont {M.}~\bibnamefont
  {Reece}}, \bibinfo {author} {\bibfnamefont {L.-T.}\ \bibnamefont {Wang}}, \
  and\ \bibinfo {author} {\bibfnamefont {Z.-Z.}\ \bibnamefont {Xianyu}},\
  }\href {\doibase 10.1103/PhysRevD.107.L101304} {\bibfield  {journal}
  {\bibinfo  {journal} {Phys. Rev. D}\ }\textbf {\bibinfo {volume} {107}},\
  \bibinfo {pages} {L101304} (\bibinfo {year} {2023})},\ \Eprint
  {http://arxiv.org/abs/2204.11869} {arXiv:2204.11869 [hep-ph]} \BibitemShut
  {NoStop}%
\bibitem [{\citenamefont {Liu}\ \emph {et~al.}(2020)\citenamefont {Liu},
  \citenamefont {Tong}, \citenamefont {Wang},\ and\ \citenamefont
  {Xianyu}}]{Liu:2019fag}%
  \BibitemOpen
  \bibfield  {author} {\bibinfo {author} {\bibfnamefont {T.}~\bibnamefont
  {Liu}}, \bibinfo {author} {\bibfnamefont {X.}~\bibnamefont {Tong}}, \bibinfo
  {author} {\bibfnamefont {Y.}~\bibnamefont {Wang}}, \ and\ \bibinfo {author}
  {\bibfnamefont {Z.-Z.}\ \bibnamefont {Xianyu}},\ }\href {\doibase
  10.1007/JHEP04(2020)189} {\bibfield  {journal} {\bibinfo  {journal} {JHEP}\
  }\textbf {\bibinfo {volume} {04}},\ \bibinfo {pages} {189} (\bibinfo {year}
  {2020})},\ \Eprint {http://arxiv.org/abs/1909.01819} {arXiv:1909.01819
  [hep-ph]} \BibitemShut {NoStop}%
\bibitem [{\citenamefont {Wang}\ and\ \citenamefont
  {Xianyu}(2020{\natexlab{a}})}]{Wang:2019gbi}%
  \BibitemOpen
  \bibfield  {author} {\bibinfo {author} {\bibfnamefont {L.-T.}\ \bibnamefont
  {Wang}}\ and\ \bibinfo {author} {\bibfnamefont {Z.-Z.}\ \bibnamefont
  {Xianyu}},\ }\href {\doibase 10.1007/JHEP02(2020)044} {\bibfield  {journal}
  {\bibinfo  {journal} {JHEP}\ }\textbf {\bibinfo {volume} {02}},\ \bibinfo
  {pages} {044} (\bibinfo {year} {2020}{\natexlab{a}})},\ \Eprint
  {http://arxiv.org/abs/1910.12876} {arXiv:1910.12876 [hep-ph]} \BibitemShut
  {NoStop}%
\bibitem [{\citenamefont {Tong}\ and\ \citenamefont
  {Xianyu}(2022)}]{Tong:2022cdz}%
  \BibitemOpen
  \bibfield  {author} {\bibinfo {author} {\bibfnamefont {X.}~\bibnamefont
  {Tong}}\ and\ \bibinfo {author} {\bibfnamefont {Z.-Z.}\ \bibnamefont
  {Xianyu}},\ }\href {\doibase 10.1007/JHEP10(2022)194} {\bibfield  {journal}
  {\bibinfo  {journal} {JHEP}\ }\textbf {\bibinfo {volume} {10}},\ \bibinfo
  {pages} {194} (\bibinfo {year} {2022})},\ \Eprint
  {http://arxiv.org/abs/2203.06349} {arXiv:2203.06349 [hep-ph]} \BibitemShut
  {NoStop}%
\bibitem [{\citenamefont {Cabass}\ \emph {et~al.}(2024)\citenamefont {Cabass},
  \citenamefont {Philcox}, \citenamefont {Ivanov}, \citenamefont {Akitsu},
  \citenamefont {Chen}, \citenamefont {Simonovi\'c},\ and\ \citenamefont
  {Zaldarriaga}}]{Cabass:2024wob}%
  \BibitemOpen
  \bibfield  {author} {\bibinfo {author} {\bibfnamefont {G.}~\bibnamefont
  {Cabass}}, \bibinfo {author} {\bibfnamefont {O.~H.~E.}\ \bibnamefont
  {Philcox}}, \bibinfo {author} {\bibfnamefont {M.~M.}\ \bibnamefont {Ivanov}},
  \bibinfo {author} {\bibfnamefont {K.}~\bibnamefont {Akitsu}}, \bibinfo
  {author} {\bibfnamefont {S.-F.}\ \bibnamefont {Chen}}, \bibinfo {author}
  {\bibfnamefont {M.}~\bibnamefont {Simonovi\'c}}, \ and\ \bibinfo {author}
  {\bibfnamefont {M.}~\bibnamefont {Zaldarriaga}},\ }\href@noop {} {\
  (\bibinfo {year} {2024})},\ \Eprint {http://arxiv.org/abs/2404.01894}
  {arXiv:2404.01894 [astro-ph.CO]} \BibitemShut {NoStop}%
\bibitem [{\citenamefont {Chen}\ \emph {et~al.}(2017)\citenamefont {Chen},
  \citenamefont {Wang},\ and\ \citenamefont {Xianyu}}]{Chen:2016uwp}%
  \BibitemOpen
  \bibfield  {author} {\bibinfo {author} {\bibfnamefont {X.}~\bibnamefont
  {Chen}}, \bibinfo {author} {\bibfnamefont {Y.}~\bibnamefont {Wang}}, \ and\
  \bibinfo {author} {\bibfnamefont {Z.-Z.}\ \bibnamefont {Xianyu}},\ }\href
  {\doibase 10.1103/PhysRevLett.118.261302} {\bibfield  {journal} {\bibinfo
  {journal} {Phys. Rev. Lett.}\ }\textbf {\bibinfo {volume} {118}},\ \bibinfo
  {pages} {261302} (\bibinfo {year} {2017})},\ \Eprint
  {http://arxiv.org/abs/1610.06597} {arXiv:1610.06597 [hep-th]} \BibitemShut
  {NoStop}%
\bibitem [{\citenamefont {Chen}\ \emph
  {et~al.}(2018{\natexlab{a}})\citenamefont {Chen}, \citenamefont {Wang},\ and\
  \citenamefont {Xianyu}}]{Chen:2018xck}%
  \BibitemOpen
  \bibfield  {author} {\bibinfo {author} {\bibfnamefont {X.}~\bibnamefont
  {Chen}}, \bibinfo {author} {\bibfnamefont {Y.}~\bibnamefont {Wang}}, \ and\
  \bibinfo {author} {\bibfnamefont {Z.-Z.}\ \bibnamefont {Xianyu}},\ }\href
  {\doibase 10.1007/JHEP09(2018)022} {\bibfield  {journal} {\bibinfo  {journal}
  {JHEP}\ }\textbf {\bibinfo {volume} {09}},\ \bibinfo {pages} {022} (\bibinfo
  {year} {2018}{\natexlab{a}})},\ \Eprint {http://arxiv.org/abs/1805.02656}
  {arXiv:1805.02656 [hep-ph]} \BibitemShut {NoStop}%
\bibitem [{\citenamefont {Lu}\ \emph {et~al.}(2020)\citenamefont {Lu},
  \citenamefont {Wang},\ and\ \citenamefont {Xianyu}}]{Lu:2019tjj}%
  \BibitemOpen
  \bibfield  {author} {\bibinfo {author} {\bibfnamefont {S.}~\bibnamefont
  {Lu}}, \bibinfo {author} {\bibfnamefont {Y.}~\bibnamefont {Wang}}, \ and\
  \bibinfo {author} {\bibfnamefont {Z.-Z.}\ \bibnamefont {Xianyu}},\ }\href
  {\doibase 10.1007/JHEP02(2020)011} {\bibfield  {journal} {\bibinfo  {journal}
  {JHEP}\ }\textbf {\bibinfo {volume} {02}},\ \bibinfo {pages} {011} (\bibinfo
  {year} {2020})},\ \Eprint {http://arxiv.org/abs/1907.07390} {arXiv:1907.07390
  [hep-th]} \BibitemShut {NoStop}%
\bibitem [{\citenamefont {Wang}\ and\ \citenamefont
  {Xianyu}(2020{\natexlab{b}})}]{Wang:2020ioa}%
  \BibitemOpen
  \bibfield  {author} {\bibinfo {author} {\bibfnamefont {L.-T.}\ \bibnamefont
  {Wang}}\ and\ \bibinfo {author} {\bibfnamefont {Z.-Z.}\ \bibnamefont
  {Xianyu}},\ }\href {\doibase 10.1007/JHEP11(2020)082} {\bibfield  {journal}
  {\bibinfo  {journal} {JHEP}\ }\textbf {\bibinfo {volume} {11}},\ \bibinfo
  {pages} {082} (\bibinfo {year} {2020}{\natexlab{b}})},\ \Eprint
  {http://arxiv.org/abs/2004.02887} {arXiv:2004.02887 [hep-ph]} \BibitemShut
  {NoStop}%
\bibitem [{\citenamefont {Bodas}\ \emph {et~al.}(2021)\citenamefont {Bodas},
  \citenamefont {Kumar},\ and\ \citenamefont {Sundrum}}]{Bodas:2020yho}%
  \BibitemOpen
  \bibfield  {author} {\bibinfo {author} {\bibfnamefont {A.}~\bibnamefont
  {Bodas}}, \bibinfo {author} {\bibfnamefont {S.}~\bibnamefont {Kumar}}, \ and\
  \bibinfo {author} {\bibfnamefont {R.}~\bibnamefont {Sundrum}},\ }\href
  {\doibase 10.1007/JHEP02(2021)079} {\bibfield  {journal} {\bibinfo  {journal}
  {JHEP}\ }\textbf {\bibinfo {volume} {02}},\ \bibinfo {pages} {079} (\bibinfo
  {year} {2021})},\ \Eprint {http://arxiv.org/abs/2010.04727} {arXiv:2010.04727
  [hep-ph]} \BibitemShut {NoStop}%
\bibitem [{\citenamefont {Kim}\ \emph {et~al.}(2019)\citenamefont {Kim},
  \citenamefont {Noumi}, \citenamefont {Takeuchi},\ and\ \citenamefont
  {Zhou}}]{Kim:2019wjo}%
  \BibitemOpen
  \bibfield  {author} {\bibinfo {author} {\bibfnamefont {S.}~\bibnamefont
  {Kim}}, \bibinfo {author} {\bibfnamefont {T.}~\bibnamefont {Noumi}}, \bibinfo
  {author} {\bibfnamefont {K.}~\bibnamefont {Takeuchi}}, \ and\ \bibinfo
  {author} {\bibfnamefont {S.}~\bibnamefont {Zhou}},\ }\href {\doibase
  10.1007/JHEP12(2019)107} {\bibfield  {journal} {\bibinfo  {journal} {JHEP}\
  }\textbf {\bibinfo {volume} {12}},\ \bibinfo {pages} {107} (\bibinfo {year}
  {2019})},\ \Eprint {http://arxiv.org/abs/1906.11840} {arXiv:1906.11840
  [hep-th]} \BibitemShut {NoStop}%
\bibitem [{\citenamefont {Lu}\ \emph {et~al.}(2021)\citenamefont {Lu},
  \citenamefont {Reece},\ and\ \citenamefont {Xianyu}}]{Lu:2021wxu}%
  \BibitemOpen
  \bibfield  {author} {\bibinfo {author} {\bibfnamefont {Q.}~\bibnamefont
  {Lu}}, \bibinfo {author} {\bibfnamefont {M.}~\bibnamefont {Reece}}, \ and\
  \bibinfo {author} {\bibfnamefont {Z.-Z.}\ \bibnamefont {Xianyu}},\ }\href
  {\doibase 10.1007/JHEP12(2021)098} {\bibfield  {journal} {\bibinfo  {journal}
  {JHEP}\ }\textbf {\bibinfo {volume} {12}},\ \bibinfo {pages} {098} (\bibinfo
  {year} {2021})},\ \Eprint {http://arxiv.org/abs/2108.11385} {arXiv:2108.11385
  [hep-ph]} \BibitemShut {NoStop}%
\bibitem [{\citenamefont {Cui}\ and\ \citenamefont
  {Xianyu}(2022)}]{Cui:2021iie}%
  \BibitemOpen
  \bibfield  {author} {\bibinfo {author} {\bibfnamefont {Y.}~\bibnamefont
  {Cui}}\ and\ \bibinfo {author} {\bibfnamefont {Z.-Z.}\ \bibnamefont
  {Xianyu}},\ }\href {\doibase 10.1103/PhysRevLett.129.111301} {\bibfield
  {journal} {\bibinfo  {journal} {Phys. Rev. Lett.}\ }\textbf {\bibinfo
  {volume} {129}},\ \bibinfo {pages} {111301} (\bibinfo {year} {2022})},\
  \Eprint {http://arxiv.org/abs/2112.10793} {arXiv:2112.10793 [hep-ph]}
  \BibitemShut {NoStop}%
\bibitem [{\citenamefont {Qin}\ and\ \citenamefont
  {Xianyu}(2022)}]{Qin:2022lva}%
  \BibitemOpen
  \bibfield  {author} {\bibinfo {author} {\bibfnamefont {Z.}~\bibnamefont
  {Qin}}\ and\ \bibinfo {author} {\bibfnamefont {Z.-Z.}\ \bibnamefont
  {Xianyu}},\ }\href {\doibase 10.1007/JHEP10(2022)192} {\bibfield  {journal}
  {\bibinfo  {journal} {JHEP}\ }\textbf {\bibinfo {volume} {10}},\ \bibinfo
  {pages} {192} (\bibinfo {year} {2022})},\ \Eprint
  {http://arxiv.org/abs/2205.01692} {arXiv:2205.01692 [hep-th]} \BibitemShut
  {NoStop}%
\bibitem [{\citenamefont {Werth}\ \emph {et~al.}(2024)\citenamefont {Werth},
  \citenamefont {Pinol},\ and\ \citenamefont {Renaux-Petel}}]{Werth:2023pfl}%
  \BibitemOpen
  \bibfield  {author} {\bibinfo {author} {\bibfnamefont {D.}~\bibnamefont
  {Werth}}, \bibinfo {author} {\bibfnamefont {L.}~\bibnamefont {Pinol}}, \ and\
  \bibinfo {author} {\bibfnamefont {S.}~\bibnamefont {Renaux-Petel}},\ }\href
  {\doibase 10.1103/PhysRevLett.133.141002} {\bibfield  {journal} {\bibinfo
  {journal} {Phys. Rev. Lett.}\ }\textbf {\bibinfo {volume} {133}},\ \bibinfo
  {pages} {141002} (\bibinfo {year} {2024})},\ \Eprint
  {http://arxiv.org/abs/2302.00655} {arXiv:2302.00655 [hep-th]} \BibitemShut
  {NoStop}%
\bibitem [{\citenamefont {Xianyu}\ and\ \citenamefont
  {Zang}(2024)}]{Xianyu:2023ytd}%
  \BibitemOpen
  \bibfield  {author} {\bibinfo {author} {\bibfnamefont {Z.-Z.}\ \bibnamefont
  {Xianyu}}\ and\ \bibinfo {author} {\bibfnamefont {J.}~\bibnamefont {Zang}},\
  }\href {\doibase 10.1007/JHEP03(2024)070} {\bibfield  {journal} {\bibinfo
  {journal} {JHEP}\ }\textbf {\bibinfo {volume} {03}},\ \bibinfo {pages} {070}
  (\bibinfo {year} {2024})},\ \Eprint {http://arxiv.org/abs/2309.10849}
  {arXiv:2309.10849 [hep-th]} \BibitemShut {NoStop}%
\bibitem [{\citenamefont {Pinol}\ \emph {et~al.}(2023)\citenamefont {Pinol},
  \citenamefont {Renaux-Petel},\ and\ \citenamefont {Werth}}]{Pinol:2023oux}%
  \BibitemOpen
  \bibfield  {author} {\bibinfo {author} {\bibfnamefont {L.}~\bibnamefont
  {Pinol}}, \bibinfo {author} {\bibfnamefont {S.}~\bibnamefont {Renaux-Petel}},
  \ and\ \bibinfo {author} {\bibfnamefont {D.}~\bibnamefont {Werth}},\
  }\href@noop {} {\  (\bibinfo {year} {2023})},\ \Eprint
  {http://arxiv.org/abs/2312.06559} {arXiv:2312.06559 [astro-ph.CO]}
  \BibitemShut {NoStop}%
\bibitem [{\citenamefont {Chakraborty}\ and\ \citenamefont
  {Stout}(2024)}]{Chakraborty:2023qbp}%
  \BibitemOpen
  \bibfield  {author} {\bibinfo {author} {\bibfnamefont {P.}~\bibnamefont
  {Chakraborty}}\ and\ \bibinfo {author} {\bibfnamefont {J.}~\bibnamefont
  {Stout}},\ }\href {\doibase 10.1007/JHEP02(2024)021} {\bibfield  {journal}
  {\bibinfo  {journal} {JHEP}\ }\textbf {\bibinfo {volume} {02}},\ \bibinfo
  {pages} {021} (\bibinfo {year} {2024})},\ \Eprint
  {http://arxiv.org/abs/2310.01494} {arXiv:2310.01494 [hep-th]} \BibitemShut
  {NoStop}%
\bibitem [{\citenamefont {Craig}\ \emph {et~al.}(2024)\citenamefont {Craig},
  \citenamefont {Kumar},\ and\ \citenamefont {McCune}}]{Craig:2024qgy}%
  \BibitemOpen
  \bibfield  {author} {\bibinfo {author} {\bibfnamefont {N.}~\bibnamefont
  {Craig}}, \bibinfo {author} {\bibfnamefont {S.}~\bibnamefont {Kumar}}, \ and\
  \bibinfo {author} {\bibfnamefont {A.}~\bibnamefont {McCune}},\ }\href
  {\doibase 10.1007/JHEP07(2024)108} {\bibfield  {journal} {\bibinfo  {journal}
  {JHEP}\ }\textbf {\bibinfo {volume} {07}},\ \bibinfo {pages} {108} (\bibinfo
  {year} {2024})},\ \Eprint {http://arxiv.org/abs/2401.10976} {arXiv:2401.10976
  [hep-ph]} \BibitemShut {NoStop}%
\bibitem [{\citenamefont {Yin}(2024)}]{Yin:2023jlv}%
  \BibitemOpen
  \bibfield  {author} {\bibinfo {author} {\bibfnamefont {Y.}~\bibnamefont
  {Yin}},\ }\href {\doibase 10.1103/PhysRevD.109.043535} {\bibfield  {journal}
  {\bibinfo  {journal} {Phys. Rev. D}\ }\textbf {\bibinfo {volume} {109}},\
  \bibinfo {pages} {043535} (\bibinfo {year} {2024})},\ \Eprint
  {http://arxiv.org/abs/2309.05244} {arXiv:2309.05244 [hep-ph]} \BibitemShut
  {NoStop}%
\bibitem [{\citenamefont {Baumann}\ and\ \citenamefont
  {Green}(2012)}]{Baumann:2011nk}%
  \BibitemOpen
  \bibfield  {author} {\bibinfo {author} {\bibfnamefont {D.}~\bibnamefont
  {Baumann}}\ and\ \bibinfo {author} {\bibfnamefont {D.}~\bibnamefont
  {Green}},\ }\href {\doibase 10.1103/PhysRevD.85.103520} {\bibfield  {journal}
  {\bibinfo  {journal} {Phys. Rev. D}\ }\textbf {\bibinfo {volume} {85}},\
  \bibinfo {pages} {103520} (\bibinfo {year} {2012})},\ \Eprint
  {http://arxiv.org/abs/1109.0292} {arXiv:1109.0292 [hep-th]} \BibitemShut
  {NoStop}%
\bibitem [{\citenamefont {Assassi}\ \emph {et~al.}(2012)\citenamefont
  {Assassi}, \citenamefont {Baumann},\ and\ \citenamefont
  {Green}}]{Assassi:2012zq}%
  \BibitemOpen
  \bibfield  {author} {\bibinfo {author} {\bibfnamefont {V.}~\bibnamefont
  {Assassi}}, \bibinfo {author} {\bibfnamefont {D.}~\bibnamefont {Baumann}}, \
  and\ \bibinfo {author} {\bibfnamefont {D.}~\bibnamefont {Green}},\ }\href
  {\doibase 10.1088/1475-7516/2012/11/047} {\bibfield  {journal} {\bibinfo
  {journal} {JCAP}\ }\textbf {\bibinfo {volume} {11}},\ \bibinfo {pages} {047}
  (\bibinfo {year} {2012})},\ \Eprint {http://arxiv.org/abs/1204.4207}
  {arXiv:1204.4207 [hep-th]} \BibitemShut {NoStop}%
\bibitem [{\citenamefont {Arkani-Hamed}\ \emph {et~al.}(2020)\citenamefont
  {Arkani-Hamed}, \citenamefont {Baumann}, \citenamefont {Lee},\ and\
  \citenamefont {Pimentel}}]{Arkani-Hamed:2018kmz}%
  \BibitemOpen
  \bibfield  {author} {\bibinfo {author} {\bibfnamefont {N.}~\bibnamefont
  {Arkani-Hamed}}, \bibinfo {author} {\bibfnamefont {D.}~\bibnamefont
  {Baumann}}, \bibinfo {author} {\bibfnamefont {H.}~\bibnamefont {Lee}}, \ and\
  \bibinfo {author} {\bibfnamefont {G.~L.}\ \bibnamefont {Pimentel}},\ }\href
  {\doibase 10.1007/JHEP04(2020)105} {\bibfield  {journal} {\bibinfo  {journal}
  {JHEP}\ }\textbf {\bibinfo {volume} {04}},\ \bibinfo {pages} {105} (\bibinfo
  {year} {2020})},\ \Eprint {http://arxiv.org/abs/1811.00024} {arXiv:1811.00024
  [hep-th]} \BibitemShut {NoStop}%
\bibitem [{\citenamefont {Bordin}\ \emph {et~al.}(2018)\citenamefont {Bordin},
  \citenamefont {Creminelli}, \citenamefont {Khmelnitsky},\ and\ \citenamefont
  {Senatore}}]{Bordin:2018pca}%
  \BibitemOpen
  \bibfield  {author} {\bibinfo {author} {\bibfnamefont {L.}~\bibnamefont
  {Bordin}}, \bibinfo {author} {\bibfnamefont {P.}~\bibnamefont {Creminelli}},
  \bibinfo {author} {\bibfnamefont {A.}~\bibnamefont {Khmelnitsky}}, \ and\
  \bibinfo {author} {\bibfnamefont {L.}~\bibnamefont {Senatore}},\ }\href
  {\doibase 10.1088/1475-7516/2018/10/013} {\bibfield  {journal} {\bibinfo
  {journal} {JCAP}\ }\textbf {\bibinfo {volume} {10}},\ \bibinfo {pages} {013}
  (\bibinfo {year} {2018})},\ \Eprint {http://arxiv.org/abs/1806.10587}
  {arXiv:1806.10587 [hep-th]} \BibitemShut {NoStop}%
\bibitem [{\citenamefont {Jazayeri}\ \emph
  {et~al.}(2023{\natexlab{b}})\citenamefont {Jazayeri}, \citenamefont
  {Renaux-Petel}, \citenamefont {Tong}, \citenamefont {Werth},\ and\
  \citenamefont {Zhu}}]{Jazayeri:2023kji}%
  \BibitemOpen
  \bibfield  {author} {\bibinfo {author} {\bibfnamefont {S.}~\bibnamefont
  {Jazayeri}}, \bibinfo {author} {\bibfnamefont {S.}~\bibnamefont
  {Renaux-Petel}}, \bibinfo {author} {\bibfnamefont {X.}~\bibnamefont {Tong}},
  \bibinfo {author} {\bibfnamefont {D.}~\bibnamefont {Werth}}, \ and\ \bibinfo
  {author} {\bibfnamefont {Y.}~\bibnamefont {Zhu}},\ }\href@noop {} {\
  (\bibinfo {year} {2023}{\natexlab{b}})},\ \Eprint
  {http://arxiv.org/abs/2308.11315} {arXiv:2308.11315 [hep-th]} \BibitemShut
  {NoStop}%
\bibitem [{\citenamefont {Chen}\ \emph
  {et~al.}(2018{\natexlab{b}})\citenamefont {Chen}, \citenamefont {Chua},
  \citenamefont {Guo}, \citenamefont {Wang}, \citenamefont {Xianyu},\ and\
  \citenamefont {Xie}}]{Chen:2018sce}%
  \BibitemOpen
  \bibfield  {author} {\bibinfo {author} {\bibfnamefont {X.}~\bibnamefont
  {Chen}}, \bibinfo {author} {\bibfnamefont {W.~Z.}\ \bibnamefont {Chua}},
  \bibinfo {author} {\bibfnamefont {Y.}~\bibnamefont {Guo}}, \bibinfo {author}
  {\bibfnamefont {Y.}~\bibnamefont {Wang}}, \bibinfo {author} {\bibfnamefont
  {Z.-Z.}\ \bibnamefont {Xianyu}}, \ and\ \bibinfo {author} {\bibfnamefont
  {T.}~\bibnamefont {Xie}},\ }\href {\doibase 10.1088/1475-7516/2018/05/049}
  {\bibfield  {journal} {\bibinfo  {journal} {JCAP}\ }\textbf {\bibinfo
  {volume} {05}},\ \bibinfo {pages} {049} (\bibinfo {year}
  {2018}{\natexlab{b}})},\ \Eprint {http://arxiv.org/abs/1803.04412}
  {arXiv:1803.04412 [hep-th]} \BibitemShut {NoStop}%
\bibitem [{\citenamefont {Green}\ \emph
  {et~al.}(2024{\natexlab{a}})\citenamefont {Green}, \citenamefont {Huang},
  \citenamefont {Shen},\ and\ \citenamefont {Baumann}}]{Green:2023ids}%
  \BibitemOpen
  \bibfield  {author} {\bibinfo {author} {\bibfnamefont {D.}~\bibnamefont
  {Green}}, \bibinfo {author} {\bibfnamefont {Y.}~\bibnamefont {Huang}},
  \bibinfo {author} {\bibfnamefont {C.-H.}\ \bibnamefont {Shen}}, \ and\
  \bibinfo {author} {\bibfnamefont {D.}~\bibnamefont {Baumann}},\ }\href
  {\doibase 10.1007/JHEP04(2024)034} {\bibfield  {journal} {\bibinfo  {journal}
  {JHEP}\ }\textbf {\bibinfo {volume} {04}},\ \bibinfo {pages} {034} (\bibinfo
  {year} {2024}{\natexlab{a}})},\ \Eprint {http://arxiv.org/abs/2310.02490}
  {arXiv:2310.02490 [hep-th]} \BibitemShut {NoStop}%
\bibitem [{\citenamefont {Noumi}\ \emph {et~al.}(2013)\citenamefont {Noumi},
  \citenamefont {Yamaguchi},\ and\ \citenamefont {Yokoyama}}]{Noumi:2012vr}%
  \BibitemOpen
  \bibfield  {author} {\bibinfo {author} {\bibfnamefont {T.}~\bibnamefont
  {Noumi}}, \bibinfo {author} {\bibfnamefont {M.}~\bibnamefont {Yamaguchi}}, \
  and\ \bibinfo {author} {\bibfnamefont {D.}~\bibnamefont {Yokoyama}},\ }\href
  {\doibase 10.1007/JHEP06(2013)051} {\bibfield  {journal} {\bibinfo  {journal}
  {JHEP}\ }\textbf {\bibinfo {volume} {06}},\ \bibinfo {pages} {051} (\bibinfo
  {year} {2013})},\ \Eprint {http://arxiv.org/abs/1211.1624} {arXiv:1211.1624
  [hep-th]} \BibitemShut {NoStop}%
\bibitem [{\citenamefont {Moradinezhad~Dizgah}\ and\ \citenamefont
  {Dvorkin}(2018)}]{MoradinezhadDizgah:2017szk}%
  \BibitemOpen
  \bibfield  {author} {\bibinfo {author} {\bibfnamefont {A.}~\bibnamefont
  {Moradinezhad~Dizgah}}\ and\ \bibinfo {author} {\bibfnamefont
  {C.}~\bibnamefont {Dvorkin}},\ }\href {\doibase
  10.1088/1475-7516/2018/01/010} {\bibfield  {journal} {\bibinfo  {journal}
  {JCAP}\ }\textbf {\bibinfo {volume} {01}},\ \bibinfo {pages} {010} (\bibinfo
  {year} {2018})},\ \Eprint {http://arxiv.org/abs/1708.06473} {arXiv:1708.06473
  [astro-ph.CO]} \BibitemShut {NoStop}%
\bibitem [{\citenamefont {Cabass}\ \emph {et~al.}(2018)\citenamefont {Cabass},
  \citenamefont {Pajer},\ and\ \citenamefont {Schmidt}}]{Cabass:2018roz}%
  \BibitemOpen
  \bibfield  {author} {\bibinfo {author} {\bibfnamefont {G.}~\bibnamefont
  {Cabass}}, \bibinfo {author} {\bibfnamefont {E.}~\bibnamefont {Pajer}}, \
  and\ \bibinfo {author} {\bibfnamefont {F.}~\bibnamefont {Schmidt}},\ }\href
  {\doibase 10.1088/1475-7516/2018/09/003} {\bibfield  {journal} {\bibinfo
  {journal} {JCAP}\ }\textbf {\bibinfo {volume} {09}},\ \bibinfo {pages} {003}
  (\bibinfo {year} {2018})},\ \Eprint {http://arxiv.org/abs/1804.07295}
  {arXiv:1804.07295 [astro-ph.CO]} \BibitemShut {NoStop}%
\bibitem [{\citenamefont {Green}\ \emph
  {et~al.}(2024{\natexlab{b}})\citenamefont {Green}, \citenamefont {Guo},
  \citenamefont {Han},\ and\ \citenamefont {Wallisch}}]{Green:2023uyz}%
  \BibitemOpen
  \bibfield  {author} {\bibinfo {author} {\bibfnamefont {D.}~\bibnamefont
  {Green}}, \bibinfo {author} {\bibfnamefont {Y.}~\bibnamefont {Guo}}, \bibinfo
  {author} {\bibfnamefont {J.}~\bibnamefont {Han}}, \ and\ \bibinfo {author}
  {\bibfnamefont {B.}~\bibnamefont {Wallisch}},\ }\href {\doibase
  10.1088/1475-7516/2024/05/090} {\bibfield  {journal} {\bibinfo  {journal}
  {JCAP}\ }\textbf {\bibinfo {volume} {05}},\ \bibinfo {pages} {090} (\bibinfo
  {year} {2024}{\natexlab{b}})},\ \Eprint {http://arxiv.org/abs/2311.04882}
  {arXiv:2311.04882 [astro-ph.CO]} \BibitemShut {NoStop}%
\bibitem [{\citenamefont {Schmidt}\ \emph {et~al.}(2015)\citenamefont
  {Schmidt}, \citenamefont {Chisari},\ and\ \citenamefont
  {Dvorkin}}]{Schmidt:2015xka}%
  \BibitemOpen
  \bibfield  {author} {\bibinfo {author} {\bibfnamefont {F.}~\bibnamefont
  {Schmidt}}, \bibinfo {author} {\bibfnamefont {N.~E.}\ \bibnamefont
  {Chisari}}, \ and\ \bibinfo {author} {\bibfnamefont {C.}~\bibnamefont
  {Dvorkin}},\ }\href {\doibase 10.1088/1475-7516/2015/10/032} {\bibfield
  {journal} {\bibinfo  {journal} {JCAP}\ }\textbf {\bibinfo {volume} {10}},\
  \bibinfo {pages} {032} (\bibinfo {year} {2015})},\ \Eprint
  {http://arxiv.org/abs/1506.02671} {arXiv:1506.02671 [astro-ph.CO]}
  \BibitemShut {NoStop}%
\bibitem [{\citenamefont {Kogai}\ \emph {et~al.}(2021)\citenamefont {Kogai},
  \citenamefont {Akitsu}, \citenamefont {Schmidt},\ and\ \citenamefont
  {Urakawa}}]{Kogai:2020vzz}%
  \BibitemOpen
  \bibfield  {author} {\bibinfo {author} {\bibfnamefont {K.}~\bibnamefont
  {Kogai}}, \bibinfo {author} {\bibfnamefont {K.}~\bibnamefont {Akitsu}},
  \bibinfo {author} {\bibfnamefont {F.}~\bibnamefont {Schmidt}}, \ and\
  \bibinfo {author} {\bibfnamefont {Y.}~\bibnamefont {Urakawa}},\ }\href
  {\doibase 10.1088/1475-7516/2021/03/060} {\bibfield  {journal} {\bibinfo
  {journal} {JCAP}\ }\textbf {\bibinfo {volume} {03}},\ \bibinfo {pages} {060}
  (\bibinfo {year} {2021})},\ \Eprint {http://arxiv.org/abs/2009.05517}
  {arXiv:2009.05517 [astro-ph.CO]} \BibitemShut {NoStop}%
\bibitem [{\citenamefont {Goldstein}\ \emph
  {et~al.}(2024{\natexlab{b}})\citenamefont {Goldstein}, \citenamefont
  {Philcox}, \citenamefont {Hill}, \citenamefont {Esposito},\ and\
  \citenamefont {Hui}}]{Goldstein:2023brb}%
  \BibitemOpen
  \bibfield  {author} {\bibinfo {author} {\bibfnamefont {S.}~\bibnamefont
  {Goldstein}}, \bibinfo {author} {\bibfnamefont {O.~H.~E.}\ \bibnamefont
  {Philcox}}, \bibinfo {author} {\bibfnamefont {J.~C.}\ \bibnamefont {Hill}},
  \bibinfo {author} {\bibfnamefont {A.}~\bibnamefont {Esposito}}, \ and\
  \bibinfo {author} {\bibfnamefont {L.}~\bibnamefont {Hui}},\ }\href {\doibase
  10.1103/PhysRevD.109.043515} {\bibfield  {journal} {\bibinfo  {journal}
  {Phys. Rev. D}\ }\textbf {\bibinfo {volume} {109}},\ \bibinfo {pages}
  {043515} (\bibinfo {year} {2024}{\natexlab{b}})},\ \Eprint
  {http://arxiv.org/abs/2310.12959} {arXiv:2310.12959 [astro-ph.CO]}
  \BibitemShut {NoStop}%
\bibitem [{\citenamefont {Dimastrogiovanni}\ \emph {et~al.}(2016)\citenamefont
  {Dimastrogiovanni}, \citenamefont {Fasiello},\ and\ \citenamefont
  {Kamionkowski}}]{Dimastrogiovanni:2015pla}%
  \BibitemOpen
  \bibfield  {author} {\bibinfo {author} {\bibfnamefont {E.}~\bibnamefont
  {Dimastrogiovanni}}, \bibinfo {author} {\bibfnamefont {M.}~\bibnamefont
  {Fasiello}}, \ and\ \bibinfo {author} {\bibfnamefont {M.}~\bibnamefont
  {Kamionkowski}},\ }\href {\doibase 10.1088/1475-7516/2016/02/017} {\bibfield
  {journal} {\bibinfo  {journal} {JCAP}\ }\textbf {\bibinfo {volume} {02}},\
  \bibinfo {pages} {017} (\bibinfo {year} {2016})},\ \Eprint
  {http://arxiv.org/abs/1504.05993} {arXiv:1504.05993 [astro-ph.CO]}
  \BibitemShut {NoStop}%
\bibitem [{\citenamefont {Ade}\ \emph {et~al.}(2016{\natexlab{d}})\citenamefont
  {Ade} \emph {et~al.}}]{Planck:2015mym}%
  \BibitemOpen
  \bibfield  {author} {\bibinfo {author} {\bibfnamefont {P.~A.~R.}\
  \bibnamefont {Ade}} \emph {et~al.} (\bibinfo {collaboration} {Planck}),\
  }\href {\doibase 10.1051/0004-6361/201525941} {\bibfield  {journal} {\bibinfo
   {journal} {Astron. Astrophys.}\ }\textbf {\bibinfo {volume} {594}},\
  \bibinfo {pages} {A15} (\bibinfo {year} {2016}{\natexlab{d}})},\ \Eprint
  {http://arxiv.org/abs/1502.01591} {arXiv:1502.01591 [astro-ph.CO]}
  \BibitemShut {NoStop}%
\bibitem [{\citenamefont {Aghanim}\ \emph {et~al.}(2020)\citenamefont {Aghanim}
  \emph {et~al.}}]{Planck:2018lbu}%
  \BibitemOpen
  \bibfield  {author} {\bibinfo {author} {\bibfnamefont {N.}~\bibnamefont
  {Aghanim}} \emph {et~al.} (\bibinfo {collaboration} {Planck}),\ }\href
  {\doibase 10.1051/0004-6361/201833886} {\bibfield  {journal} {\bibinfo
  {journal} {Astron. Astrophys.}\ }\textbf {\bibinfo {volume} {641}},\ \bibinfo
  {pages} {A8} (\bibinfo {year} {2020})},\ \Eprint
  {http://arxiv.org/abs/1807.06210} {arXiv:1807.06210 [astro-ph.CO]}
  \BibitemShut {NoStop}%
\bibitem [{\citenamefont {Lyons}(2008)}]{look-elsewhere}%
  \BibitemOpen
  \bibfield  {author} {\bibinfo {author} {\bibfnamefont {L.}~\bibnamefont
  {Lyons}},\ }\href {\doibase 10.1214/08-AOAS163} {\bibfield  {journal}
  {\bibinfo  {journal} {The Annals of Applied Statistics}\ }\textbf {\bibinfo
  {volume} {2}},\ \bibinfo {pages} {887 } (\bibinfo {year} {2008})}\BibitemShut
  {NoStop}%
\bibitem [{\citenamefont {Bayer}\ and\ \citenamefont
  {Seljak}(2020)}]{Bayer:2020pva}%
  \BibitemOpen
  \bibfield  {author} {\bibinfo {author} {\bibfnamefont {A.~E.}\ \bibnamefont
  {Bayer}}\ and\ \bibinfo {author} {\bibfnamefont {U.}~\bibnamefont {Seljak}},\
  }\href {\doibase 10.1088/1475-7516/2020/10/009} {\bibfield  {journal}
  {\bibinfo  {journal} {JCAP}\ }\textbf {\bibinfo {volume} {10}},\ \bibinfo
  {pages} {009} (\bibinfo {year} {2020})},\ \Eprint
  {http://arxiv.org/abs/2007.13821} {arXiv:2007.13821 [physics.data-an]}
  \BibitemShut {NoStop}%
\bibitem [{\citenamefont {Bayer}\ \emph {et~al.}(2021)\citenamefont {Bayer},
  \citenamefont {Seljak},\ and\ \citenamefont {Robnik}}]{Bayer:2021lhk}%
  \BibitemOpen
  \bibfield  {author} {\bibinfo {author} {\bibfnamefont {A.~E.}\ \bibnamefont
  {Bayer}}, \bibinfo {author} {\bibfnamefont {U.}~\bibnamefont {Seljak}}, \
  and\ \bibinfo {author} {\bibfnamefont {J.}~\bibnamefont {Robnik}},\ }\href
  {\doibase 10.1093/mnras/stab2331} {\bibfield  {journal} {\bibinfo  {journal}
  {Mon. Not. Roy. Astron. Soc.}\ }\textbf {\bibinfo {volume} {508}},\ \bibinfo
  {pages} {1346} (\bibinfo {year} {2021})},\ \Eprint
  {http://arxiv.org/abs/2108.06333} {arXiv:2108.06333 [astro-ph.IM]}
  \BibitemShut {NoStop}%
\bibitem [{\citenamefont {Izumi}\ and\ \citenamefont
  {Mukohyama}(2010)}]{Izumi:2010wm}%
  \BibitemOpen
  \bibfield  {author} {\bibinfo {author} {\bibfnamefont {K.}~\bibnamefont
  {Izumi}}\ and\ \bibinfo {author} {\bibfnamefont {S.}~\bibnamefont
  {Mukohyama}},\ }\href {\doibase 10.1088/1475-7516/2010/06/016} {\bibfield
  {journal} {\bibinfo  {journal} {JCAP}\ }\textbf {\bibinfo {volume} {06}},\
  \bibinfo {pages} {016} (\bibinfo {year} {2010})},\ \Eprint
  {http://arxiv.org/abs/1004.1776} {arXiv:1004.1776 [hep-th]} \BibitemShut
  {NoStop}%
\bibitem [{\citenamefont {Huang}(2010)}]{Huang:2010ab}%
  \BibitemOpen
  \bibfield  {author} {\bibinfo {author} {\bibfnamefont {Q.-G.}\ \bibnamefont
  {Huang}},\ }\href {\doibase 10.1088/1475-7516/2010/07/025} {\bibfield
  {journal} {\bibinfo  {journal} {JCAP}\ }\textbf {\bibinfo {volume} {07}},\
  \bibinfo {pages} {025} (\bibinfo {year} {2010})},\ \Eprint
  {http://arxiv.org/abs/1004.0808} {arXiv:1004.0808 [astro-ph.CO]} \BibitemShut
  {NoStop}%
\bibitem [{\citenamefont {Hazumi}\ \emph {et~al.}(2019)\citenamefont {Hazumi}
  \emph {et~al.}}]{Hazumi:2019lys}%
  \BibitemOpen
  \bibfield  {author} {\bibinfo {author} {\bibfnamefont {M.}~\bibnamefont
  {Hazumi}} \emph {et~al.},\ }\href {\doibase 10.1007/s10909-019-02150-5}
  {\bibfield  {journal} {\bibinfo  {journal} {J. Low Temp. Phys.}\ }\textbf
  {\bibinfo {volume} {194}},\ \bibinfo {pages} {443} (\bibinfo {year}
  {2019})}\BibitemShut {NoStop}%
\bibitem [{\citenamefont {Ade}\ \emph {et~al.}(2019)\citenamefont {Ade} \emph
  {et~al.}}]{SimonsObservatory:2018koc}%
  \BibitemOpen
  \bibfield  {author} {\bibinfo {author} {\bibfnamefont {P.}~\bibnamefont
  {Ade}} \emph {et~al.} (\bibinfo {collaboration} {Simons Observatory}),\
  }\href {\doibase 10.1088/1475-7516/2019/02/056} {\bibfield  {journal}
  {\bibinfo  {journal} {JCAP}\ }\textbf {\bibinfo {volume} {02}},\ \bibinfo
  {pages} {056} (\bibinfo {year} {2019})},\ \Eprint
  {http://arxiv.org/abs/1808.07445} {arXiv:1808.07445 [astro-ph.CO]}
  \BibitemShut {NoStop}%
\bibitem [{\citenamefont {Abazajian}\ \emph {et~al.}(2016)\citenamefont
  {Abazajian} \emph {et~al.}}]{CMB-S4:2016ple}%
  \BibitemOpen
  \bibfield  {author} {\bibinfo {author} {\bibfnamefont {K.~N.}\ \bibnamefont
  {Abazajian}} \emph {et~al.} (\bibinfo {collaboration} {CMB-S4}),\ }\href@noop
  {} {\  (\bibinfo {year} {2016})},\ \Eprint {http://arxiv.org/abs/1610.02743}
  {arXiv:1610.02743 [astro-ph.CO]} \BibitemShut {NoStop}%
\end{thebibliography}%

\end{document}